\documentclass[aps,twocolumn,floats,nofootinbib]{revtex4}
\usepackage{graphicx}
\usepackage{latexsym}

\def\ea{{\it et al.}\ }
\def\g0{\,$G_0$}

\begin{document}

\title{Quantum properties of atomic-sized conductors}
\author{Nicol\'as Agra{\"\i}t}
\affiliation{Laboratorio de Bajas Temperaturas, Departamento de
F\'{\i}sica de la Materia Condensada C-III, and Instituto
Universitario de Ciencia de Materiales ``Nicol\'{a}s Cabrera'',
Universidad Aut\'{o}noma de Madrid, E-28049 Madrid, Spain}
\author{Alfredo Levy Yeyati}
\affiliation{Departamento de F\'{\i}sica Te\'{o}rica de la Materia
Condensada C-V, and Instituto Universitario de Ciencia de
Materiales ``Nicol\'{a}s Cabrera'', Universidad Aut\'{o}noma de
Madrid, E-28049 Madrid, Spain}
\author{Jan M. van Ruitenbeek}
\affiliation{Kamerlingh Onnes Laboratorium, Universiteit Leiden,
Postbus 9504, 2300 RA Leiden, The Netherlands}
\date{\today}

\begin{abstract}
Using remarkably simple experimental techniques it is possible to
gently break a metallic  contact and thus form conducting
nanowires. During the last stages of the pulling a neck-shaped
wire connects the two electrodes, the diameter of which is reduced
to single atom upon further stretching. For some metals it is even
possible to form a chain of individual atoms in this fashion.
Although the atomic structure of contacts can be quite
complicated, as soon as the weakest point is reduced to just a
single atom the complexity is removed. The properties of the
contact are then dominantly determined by the nature of this atom.
This has allowed for quantitative comparison of theory and
experiment for many properties, and atomic contacts have proven to
form a rich test-bed for concepts from mesoscopic physics.
Properties investigated include multiple Andreev reflection, shot
noise, conductance quantization, conductance fluctuations, and
dynamical Coulomb blockade. In addition, pronounced quantum
effects show up in the mechanical properties of the contacts, as
seen in the force and cohesion energy of the nanowires. We review
this research, which has been performed mainly during the past
decade, and we discuss the results in the context of related
developments. {\hfil\bf Preprint, to be published in Physics
Reports (2003).}
\end{abstract}

\maketitle

\tableofcontents

\section{Introduction}\label{s.introduction}

The electrical and mechanical properties of a piece of any metal
are not different, whether its size is millimeters or kilometers.
However, as soon as its size approaches the atomic scale all
common knowledge about material properties becomes invalid. The
familiar Ohm's Law, from which we learn that the resistance of a
conductor scales proportional to its length, breaks down. The
reason is that the distance an electron travels between two
scattering events is typically much larger than the atomic size.
The electrons traverse an atomic-sized conductor ballistically,
and the resistance becomes independent of its length. In fact, the
character of the resistance changes conceptually and it will be
necessary to invoke the wave nature of the electrons in the
conductor for a proper description. The energy scales involved are
so large than quantum effects are visible at room temperature. The
chemical nature of the metallic element starts to play an
essential role. As a consequence, while in the macroscopic world
gold is a better conductor than lead by an order of magnitude, for
conduction through a single atom, lead beats gold by a factor of
three. The mechanical properties are quite unusual: plastic
deformation in a macroscopic metal occurs via dislocation motion.
On the other hand, atomic-sized metal wires flow in response to
applied stresses via structural rearrangements and their yield
strength is one or two orders of magnitude larger than for bulk
materials. Not just the electronic properties are to be described
in terms of electron waves, but also understanding metallic
cohesion of nanometer-size wires requires taking electron waves
into account that extend over the entire conductor.

The experimental investigation of these phenomena requires tools
for manipulation and characterization of structures at the atomic
and molecular scale. In laboratories worldwide there is rapid
progress in this area. The field is known as nanophysics, or
nanoscience, where the prefix `nano' refers to the size scale of
nanometers. By its very nature, the boundaries of the field of
physics of very small objects with the field of chemistry are
fading. Indeed, in parallel, chemists are striving to make
ever-larger molecules and metal cluster compounds that start to
have bulk material properties. From a third direction, biology has
developed to the point where we are able to scrutinize the
function and properties of the individual molecular building
blocks of living organisms.

An important tool that has stimulated these developments is the
Scanning Tunneling Microscope (STM), developed by Gerd Binnig and
Heinrich Rohrer, for which they were awarded the Nobel prize in
1986. Over the past two decades the STM has inspired many related
scanning probe microscopy tools, which measure a great variety of
properties with atomic resolution \cite{wiesendanger94}. By far
the most important probe is the Atomic Force Microscope (AFM),
which allows the study of poorly conducting surfaces and has been
used for the study of such problems as the forces required for
unfolding an individual protein molecule \cite{rief97}. The latter
example also illustrates an important aspect of these tools: apart
from imaging atoms at the surface of a solid, it is possible to
manipulate individual atoms and molecules. Very appealing examples
of the possibility to position atoms at pre-designed positions on
a surface have been given by Don Eigler and his coworkers
\cite{crommie93}.

A second ingredient, which has greatly contributed to the rapid
developments in nanophysics, is the wide body of knowledge
obtained in the field of mesoscopic physics \cite{imry97}.
Mesoscopic physics studies effects of quantum coherence in the
properties of conductors that are large on the scale of atoms but
small compared to everyday (macroscopic) dimensions. One of the
concepts developed in mesoscopic physics which is directly
applicable at the atomic scale is the notion that electrical
conductance is equivalent to the transmission probability for
incoming waves. This idea, which goes back to Rolf Landauer
\cite{landauer57}, forms one of the central themes of this review,
where we discuss conductance in the quantum regime. This applies
to atomic-sized metallic contacts and wires, as well as to
molecules. A much studied example of the latter is conductance
through carbon nanotubes \cite{dekker99}, long cylindrical
molecules of exclusively carbon atoms with a diameter of order of
1\,nm. Even applications to biological problems have appeared,
where the techniques of mesoscopic physics and nanophysics have
been exploited to study the conductance of individual DNA
molecules \cite{fink99,porath00,kasumov01}. There is, however, a
characteristic distinction between mesoscopic physics and
nanophysics. While the former field concentrates on `universal'
features relating to the wave character of the electrons, to the
quantization of charge in units of the electron charge, and the
like, at the nanometer scale the composition and properties of the
materials play an important role. In nanophysics the phenomena
observed are often non-generic and the rich variety of chemistry
enters.

The attention to mesoscopic physics, and more recently to
nanophysics, is strongly encouraged by the ongoing miniaturization
in the microelectronics industry. At the time of this writing the
smallest size of components on a mass-fabricated integrated
circuit amounts to only 110\,nm. It is expected that this trend
towards further miniaturization continues for at least another
decade and will then reach the level of 30\,nm, which corresponds
to only $\sim$100 atoms in a row. Nanophysics takes these
developments to the ultimate size limit: the size of atoms and
molecules. One should not be over-optimistic about the chances of
this research leading to large-scale fabrication in the
foreseeable future of atomically engineered circuits replacing
present day silicon technology. Many barriers would have to be
taken, including problems of long-term stability at room
temperature and the time required for fabrication and design of
giga-component circuits.  Smaller scale applications may be
expected from intrinsically stable structures such as carbon
nanotubes. However, the research is most important for
understanding what modified properties may be met upon further
size reduction, and searching for new principles to be exploited.

Finally, a third field of research with intimate connections to
the work described in this review is related to materials science,
where the fundamentals of adhesion, friction and wear are being
rebuilt upon the mechanical properties of materials at the atomic
scale \cite{persson96,bhushan98}. This involves, among many other
aspects, large-scale computer simulations of atomistic models
under applied stress, which allows the macroscopic material
properties to be traced to microscopic processes.

\subsection{The scope of this review}\label{ss.scope}

Having sketched the outlines of the field of nanophysics, which
forms the natural habitat for our work, we will now limit the
scope of what will be discussed in this review. We will discuss
the electrical and mechanical properties of atomic sized metallic
conductors. The central theme is the question as to what
determines the electrical conductance of a single atom. The answer
involves concepts from mesoscopic physics and chemistry, and
suggests a new way of thinking about conductance in general. In
metals the Fermi wavelength is comparable to the size of the atom,
which immediately implies that a full quantum mechanical
description is required. The consequences of this picture for
other transport properties will be explored, including those with
the leads in the superconducting state, thermal transport,
non-linear conductance and noise. It will be shown that the
quantum mechanics giving rise to the conductance cannot be
separated from the question of the mechanical cohesion of the
contact, which naturally leads us to discuss problems of forces
and mechanical stability.

Before we present a logical discussion of the concepts and results
it is useful to give a brief account of the history of the
developments. This is a most delicate task, since we have all been
heavily involved in this work, which will make the account
unavoidably personally colored. The following may be the least
scientific part of this paper, but may be of interest to some as
our personal perspective of the events.

\subsubsection{A brief history of the field}\label{sss.history}
The developments of three fields come together around 1990. To
start with, briefly after the invention of the STM in 1986
Gimzewski and M{\"o}ller \cite{gimzewski87} were the first to
employ an STM to study the conductance in atomic-sized contacts
and the forces were measured using an AFM by D{\"u}rig \ea
\cite{durig90}. They observed a transition between contact and
vacuum tunneling at a resistance of about 20\,k$\Omega$ and the
adhesion forces when approaching contact from the tunneling
regime. Second, shortly afterwards, in 1988, the quantization of
conductance was discovered in two-dimensional electron gas devices
\cite{wees88,wharam88}. The theory describing this new quantum
phenomenon has provided the conceptual framework for discussing
transport for contacts that have a width comparable to the Fermi
wavelength. Although the connection between these two developments
was made in a few theoretical papers
\cite{lang87,ferrer88,ciraci89,garcia89,zagoskin90,tekman91} it
took a few years before new experiments on the conductance of
atomic-sized contacts appeared. As a third ingredient, the
mechanical properties of atomic-sized metallic contacts were
discussed in two seminal papers which appeared in 1990
\cite{landman90,sutton90}. Here it was shown, using molecular
dynamics computer simulations of the contact between an atomically
sharp metallic STM tip and a flat surface, that upon stretching
the contact is expected to go through successive stages of elastic
deformation and sudden rearrangements of the atomic structure.

These three developments led up to a surge of activity in the
beginning of the nineties. In 1992 in Leiden \cite{muller92a} a
new techniques was introduced by Muller {\it et al.} dedicated to
the study of atomic sized junctions, baptized the Mechanically
Controllable Break Junction (MCBJ) technique, based on an earlier
design by Moreland and Ekin \cite{moreland85}. First results were
shown for Nb and Pt contacts \cite{muller92}, with steps in the
conductance and supercurrent. The former have a magnitude of order
of the conductance quantum, $G_0=2e^2/h$, and the connection with
quantization of the conductance was discussed. However, the
authors argued that the steps should be explained by the atomic
structural rearrangement mechanisms of Landman {\it et al.}
\cite{landman90} and Sutton and Pethica \cite{sutton90}. This was
clearly illustrated in a calculation by Todorov and Sutton
\cite{todorov93a}, which combines molecular dynamics for
calculation of the atomic structure and a tight binding
calculation of the electronic structure at each step in the
evolution of the structure to evaluate the conductance. This
subtle interplay between atomic structure and quantization of the
conductance would fire a lively debate for a few years to come.
This debate started with the appearance, at about the same time,
of experimental results for atomic sized contacts obtained using
various methods by four different groups
\cite{agrait93,pascual93,krans93,olesen94}.

The experiments involve a recording of the conductance of
atomic-sized contacts while the contacts are stretched to the
point of breaking. The conductance is seen to decrease in a
stepwise fashion, with steps of order of the quantum unit of
conductance. Each curve has a different appearance due to the many
possible atomic configurations that the contact may assume. By far
not all plateaus in the conductance curves could be unambiguously
identified with an integer multiple of the conductance quantum,
$nG_0$. In an attempt at an objective analysis of the data,
histograms of conductance values were introduced, constructed from
a large number of individual conductance curves
\cite{olesen95,krans95,gai96}. These demonstrated, for gold and
sodium, that the conductance has a certain preference for
multiples of $G_0$, after correction for a phenomenological
`series resistance'. When it was shown that similar results can be
obtained under ambient conditions by simply touching two bulk gold
wires \cite{costa95} many more results were published on a wide
variety of metals under various conditions. However, a
straightforward interpretation of the conductance behavior in
terms of free-electron waves inside smooth contact walls giving
rise to quantization of the conductance continued to be
challenged. The dynamical behavior of the conductance steps
suggested strongly that the allowed diameters of the contact are
restricted by atomic size constraints \cite{krans96}. Also, the
free electron model could not account for the differences in
results for different materials. Convincing proof for the atomic
rearrangements at the conductance steps was finally presented in a
paper by Rubio {\it et al.} in 1996 \cite{rubio96}, where they
combine conductance and force measurements to show that jumps in
the conductance are associated with distinct jumps in the force.

As a parallel development there was an increasing interest in the
structure of the current-voltage (I-V) characteristics for quantum
point contacts between superconducting leads. I-V curves for
contacts between superconductors show rich structure in the region
of voltages below $2\Delta/e$, where $\Delta$ is the
superconducting gap energy. The basis for the interpretation had
been given by Klapwijk, Blonder and Tinkham in 1982
\cite{klapwijk82} in terms of multiple Andreev reflection.
However, this description did not take the full phase coherence
between the scattering events into account. The first full quantum
description of current-voltage curves was given by Arnold
\cite{arnold87}. Independently, three groups applied these
concepts to contacts with a single conductance channel
\cite{averin95,cuevas96,bratus95,bratus97}. A quantitative
experimental confirmation of this description was obtained using
niobium atomic-sized vacuum tunnel junctions \cite{post94}.

This led to a breakthrough in the understanding of conductance at
the atomic scale. In 1997 Scheer {\it et al.} \cite{scheer97}
published a study of the current-voltage relation in
superconducting single-atom contacts. They discovered that the I-V
curves did not fit the predicted shape for a single conductance
channel \cite{averin95,cuevas96,bratus95,bratus97}, although the
conductance was close to one conductance unit, $G_0$. Instead, a
good fit was obtained when allowing for several independent
conductance channels, with transmission probabilities $\tau_n <
1$. For a single-atom contact of aluminum three channels turned
out to be sufficient to describe the data. The interpretation of
the results by Scheer {\it et al.} was provided by an analysis of
a tight binding model of atomic size geometries by Cuevas {\it et
al.} \cite{cuevas98a}. This picture agrees with earlier first
principles calculations \cite{lang87,lang95}, where the
conductance is discussed in terms of `resonances' in the local
density of states. The picture by Cuevas {\it et al.} has the
advantage that it can be understood on the basis of a very simple
concept: the number of conductance channels is determined by the
number of valence orbitals of the atom. This view was confirmed
experimentally by a subsequent systematic study for various
superconductors \cite{scheer98}.

\begin{figure}[!t]
\begin{center}
\includegraphics[width=80mm]{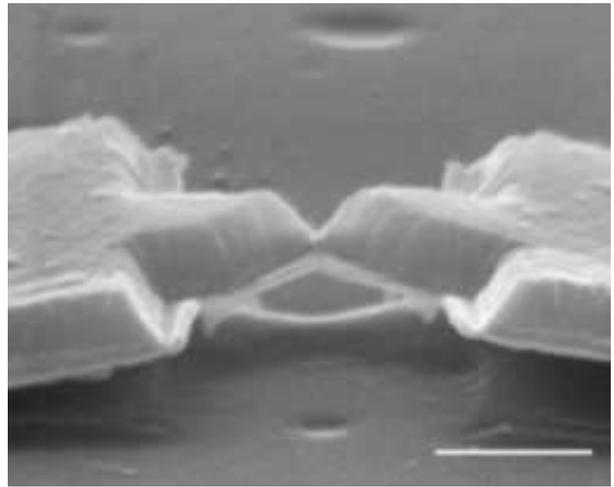}
\end{center}
\caption{A lithographically fabricated MCBJ device for gold. The
image has been taken with a scanning electron microscope. The
contact at the narrowest part is formed in a thin 20~nm gold
layer. The gold layer is in intimate contact with a thick 400~nm
aluminum layer. The bridge is freely suspended above the
substrate, and only anchored to the substrate at the wider regions
left and right. When bending the substrate the wire breaks at the
narrowest part, and a single gold atom contact can be adjusted by
relaxing the bending force. The close proximity of the thick
aluminum layer to the contact induces superconducting properties
into the atomic sized contact. The horizontal scale bar is
$\sim$1~$\mu$m. Courtesy E. Scheer \protect\cite{scheer98}.}
\label{f.turtles}
\end{figure}

Within this picture it is still possible to apply free-electron
like models of conductance, provided we restrict ourselves
primarily to monovalent metals. When one evaluates the total
energy of the occupied states within a constriction\footnote{The
words `contact' and `constriction' are used throughout this paper
as equivalent in describing a fine wire-shaped connection between
two bulk electrodes, with usually a smooth change in
cross-section.}, using an independent electron model one finds
that the energy has distinct minima for certain cross-sections of
the constriction \cite{stafford97,ruitenbeek97,yannouleas97}. The
energy minima are associated with the position of the bottom of
the subbands for each of the quantum modes. This suggests that the
cohesion force of the constriction is at least partly determined
by the delocalized electronic quantum modes. Experimental evidence
for this quantum-mode-based picture of the cohesive force was
obtained for sodium point contacts, which show enhance mechanical
stability at `magic radii'  as a result of the quantum mode
structure in the density of states \cite{yanson99}.

Another discovery of unusual mechanical behavior was found for
gold contacts, which were shown to allow stretching into
conducting chains of individual atoms. This was inferred from the
response of the conductance upon stretching of the contacts
\cite{yanson98}, and was directly observed in a room temperature
experiment of an STM constructed at the focal point of a
high-resolution transmission electron microscope (HRTEM)
\cite{ohnishi98}. It was suggested that the exceptional stability
of these chains may derive from the quantum-mode based mechanism
mentioned above \cite{sanchez99}. The atomic chains form
one-dimensional conductors, with a conductance very close to
1\,$G_0$. This connects the research to the active research on
one-dimensional conductors, with notably the carbon nanotubes as
the prime system of interest \cite{dekker99}. A chain of atoms,
consisting of two non-metallic Xe atoms, has been constructed by
STM manipulation by Yazdani {\it et al.} \cite{yazdani96}, and
this may point the way to future studies using ultimate
atomic-scale control over the construction of the conductors.

\subsection{Outline of this review}\label{ss.outline}
Although the field is still rapidly evolving, a number of new
discoveries, concepts and insights have been established and
deserve to be clearly presented in a comprehensive review. This
should provide an introduction of the concepts for those
interested in entering the field, and a reference source and guide
into the literature for those already active. A few reviews on the
subject with a more limited scope have been published recently
\cite{houten96,ruitenbeek97a,ruitenbeek00} and one conference
proceedings was dedicated to conductance in nanowires
\cite{serena97}. In the following we will attempt to give a
systematic presentation of the theoretical concepts and
experimental results, and try to be as nearly complete in
discussion of the relevant literature on this subject as
practically possible.

We start in Sect.~\ref{s.fabrication} by introducing the
experimental techniques for studying atomic-sized metallic
conductors. Some examples of results obtained by the techniques
will be shown, and these will be used to point out the interesting
aspects that require explanation. The theoretical basis for
conductance at the atomic scale will be explained in detail in
Sect.~\ref{s.transport_theory}. As pointed out above,
superconductivity has played an essential role in the discussion
on quantum point contacts. Therefore, before we introduce the
experimental results, in Sect.~\ref{s.sctransporttheory} the
various theoretical approaches are reviewed to calculate the
current-voltage characteristics for quantum point contacts between
superconductors. Then we turn to experiment and begin the
discussion with the linear conductance. The behavior of the
conductance of the contacts is described as a function of the
stretching of the contact. The conductance steps and plateaus, and
the conductance histograms are presented. Results for the various
experimental techniques, for a range of metallic elements, and the
interpretation of the data are critically evaluated. The last
conductance plateau before breaking of the contact is usually
interpreted as the last-atom contact, and the evidence for this
interpretation is presented. Although it will become clear that
electrical transport and mechanical properties of the contacts are
intimately related, we choose to present the experimental results
for the mechanical properties separately in
Sect.~\ref{s.mechanical}. The relation between the two aspects is
discussed in next section. For the interpretation of the
experimental results computer simulations have been indispensable,
and this forms the subject of Sect.~\ref{s.models}. Molecular
dynamics simulations are introduced and the results for the
evolution of the structure of atomic-scale contacts are presented.
Various approaches to calculate the conductance are discussed,
with an emphasis on free-electron gas calculations and t he
effects of the conductance modes on the cohesive force in these
models. The valence-orbitals basis of the conductance modes
follows from a discussion of tight-binding models and the results
of these are compared to {\it ab initio}, density functional
calculations.

Sect.~\ref{s.exp_modes} presents the experimental evidence for the
valence-orbitals interpretation of the conductance modes.
Analyzing the superconducting subgap structure forms the central
technique, but additional evidence is obtained from shot noise
experiments, and from de strain dependence of the conductance. The
next two sections discuss special electrical properties of
metallic quantum point contacts, including conductance
fluctuations, inelastic scattering of the conduction electrons,
and the Josephson current for contacts between superconducting
leads. Sect.~\ref{s.chains} presents the evidence for the
spontaneous formation of chains of single atoms, notably for gold
contacts, and the relevant model calculations for this problem. A
second unusual mechanical effect is discussed in
Sect.~\ref{s.shells}, which presents  the evidence for shell
structure in alkali nanowires. We end our review with a few
summary remarks and an outlook on further research and unsolved
problems.

There are two features that make the subject discussed here
particularly attractive. The first is the fact that by reducing
the cross section of the conductors to a single atom one
eliminates a lot of the complexity of solid state physics, which
makes the problem amenable to direct and quantitative comparison
with theory. This is a field of solid state physics where theory
and experiment meet: all can be very well characterized and theory
fits extremely well. A second attractive aspect lies in the fact
that many experiments can be performed with simple means. Although
many advanced and complex measurements have been performed, some
aspects are simple enough that they can be performed in class-room
experiments by undergraduate students. A description of a
class-room experiment can be found in Ref. \cite{foley99}.

\section{Fabrication of metallic point contacts}\label{s.fabrication}

A  wide variety of tools have been employed during the last decade
to study the mechanical and transport properties of atomic-sized
contacts, and many of these are extremely simple. Before we start
a description of the main techniques it is important to stress the
great difference between room temperature and helium temperature
experiments. At low temperatures atomic-sized contacts can be held
stable for any desired length of time, allowing detailed
investigation of the conductance properties. The low-temperature
environment at the same time prevents adsorption of contaminating
gasses on the metal surface. At room temperature, on the other
hand, the thermal diffusion of the atoms prevents long-term
stability of a contact of single atom, and ultra-high vacuum (UHV)
conditions are required for a clean metal junction. However, using
fast scan techniques for the study of the noble metals, in
particular gold, a lot of information has been obtained by very
simple means.

\subsection{Early developments: spear-anvil and related techniques} \label{ss.spear-anvil}

\begin{figure}
\centerline{\includegraphics[width= 8cm]{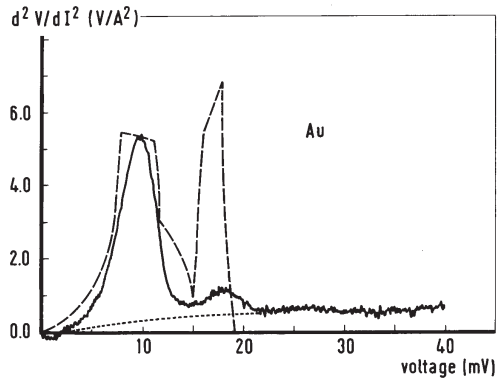}} \caption{An
example of an electron-phonon spectrum measured for a gold point
contact by taking the second derivative of the voltage with
respect to the current. The long-dashed curve represents the
phonon density of states obtained from inelastic neutron
scattering. Courtesy A.G.M. Jansen, \protect\cite{jansen80}.}
\label{f.PCS}
\end{figure}

Many  years before the rise of nanofabrication, ballistic metallic
point contacts were widely studied, and many beautiful experiments
have been performed \cite{khotkevich95,ruitenbeek96a}. The
principle was discovered by Yanson \cite{yanson74} and later
developed by his group and by Jansen {\it et al.}\
\cite{jansen80}. The technique has been worked out with various
refinements for a range of applications, but essentially it
consists of bringing a needle of a metal gently into contact with
a metal surface. This is known as the spear-anvil technique.
Usually, some type of differential-screw mechanism is used to
manually adjust the contact. With this technique stable contacts
are typically formed having resistances in the range from
$\sim$0.1 to $\sim$10~$\Omega$, which corresponds (see
Sect.~\ref{ss.Sharvin}) to contact diameters between $d\simeq10$
and 100~nm. The elastic and inelastic mean free path of the charge
carriers can be much longer than this length $d$, when working
with clean metals at low temperatures, and the ballistic nature of
the transport in such contacts has been convincingly demonstrated
in many experiments. The main application of the technique has
been to study the electron-phonon interaction in metals. Here, one
makes use of the fact that the (small but finite) probability for
back-scattering through the contact is enhanced as soon as the
electrons acquire sufficient energy from the electric potential
difference over the contact that they are able to excite the main
phonon modes of the material. The differential resistance,
$dV/dI$, of the contact is seen to increase at the characteristic
phonon energies of the material. A spectrum of the
energy-dependent electron-phonon scattering can be directly
obtained by measuring the second derivative of the voltage with
current, $d^2V/dI^2$, as a function of the applied bias voltage.
An example is given in Fig.\,\ref{f.PCS}.
Peaks  in the spectra are typically observed between 10 and 30~mV,
and are generally in excellent agreement with spectral information
from other experiments, and with calculated spectra. The
application of electron-phonon spectroscopy in atomic-sized
contacts will be discussed in Sect.~\ref{sss.phonons}.

The ballistic  character of the transport has been exploited in
even more ingenious experiments such as the focusing of the
electron trajectories onto a second point contact by the
application of a perpendicular magnetic field \cite{tsoi74} and
the injection of ballistic electrons onto a normal
metal-superconductor interface for a direct observation of Andreev
reflection \cite{benistant83}.

Contacts of the  spear-anvil type are not suitable for the study
of the quantum regime, which requires contact diameters comparable
to the Fermi wavelength, i.e. contacts of the size of atoms. For
smaller contacts (higher resistances) the above-described
technique is not sufficiently stable for measurement. What is more
important, most of the experiments in the quantum regime need some
means fine control over the contact size. These requirements can
be met using the scanning tunneling microscope (STM) or the
mechanically controllable break junction technique (MCBJ).

\subsection{The use of scanning tunneling microscopes}\label{ss.STM}

The scanning tunneling microscope  (STM) \footnote{For an overview
of STM see e.g. Ref.~\cite{wiesendanger94}.} is a versatile tool
that allows studying the topography and electronic properties of a
metal or semiconductor surface with atomic resolution, and it is
also ideal for studying atomic-sized contacts. In its normal
topographic mode a sharp needle (the tip) is scanned over the
sample to be studied without making contact. The tip-sample
separation is maintained constant by controlling the current that
flows between them due to the tunneling effect when applying a
constant bias voltage. The control signal gives a topographic
image of the sample surface. It is possible to achieve atomic
resolution because of the exponential dependence of the tunneling
current on the tip-sample separation: only the foremost atom of
the tip will {\em see} the sample. Typical operating currents are
of the order of nanoamperes and the tip-sample separation is just
a few angstroms. Evidently, the sample must be conducting.
Essential for the operation of the STM is the control of the
relative position of the tip and sample with subnanometer
accuracy, which is possible using piezoelectric ceramics.
Conventionally the lateral scan directions are termed $x$- and
$y$-directions and the vertical direction is the $z$-direction.

The distance between tip and sample is  so small that accidental
contact between them is quite possible in normal STM work, and
should usually be avoided. However, it soon became evident that
the STM tip could be used to modify the sample on a nanometer
scale. The first report of the formation and study of a metallic
contact of atomic dimensions with STM is that of Gimzewski and
M\"{o}ller \cite{gimzewski87}. In contrast to previous works that
were more aimed at
 surface modifications \cite{abraham86,vankempen86},
the surface was gently touched and the  transition from the
tunneling regime to metallic contact was observed as an abrupt
jump in the conductance. From the magnitude of the resistance at
the jump ($\sim 10$~k$\Omega$) using the semi-classical Sharvin
formula (see Sect.~\ref{ss.Sharvin}), the contact diameter was
estimated to be 0.15\,nm, which suggested that the contact should
consist of one or two atoms.

Different groups have performed STM experiments on the conductance
of atomic-sized contacts in  different experimental conditions:
at cryogenic temperatures
\cite{agrait93,agrait93a,sirvent96,sirvent96a}; at room
temperature
 under ambient conditions
\cite{pascual93,pascual95,dremov95,landman96a}, and UHV
\cite{olesen94,gai96,brandbyge95}.

\begin{figure}
\begin{center}
\includegraphics[width=80mm]{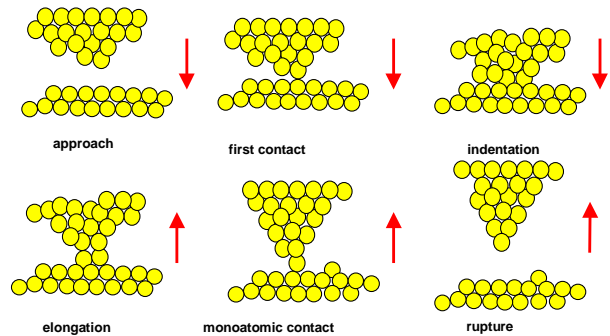}
\end{center}
\caption{\label{f.contact-formation}Cartoon representation of
contact fabrication using an STM.}
\end{figure}

The presence of adsorbates, contamination, and oxides on the
contacting surfaces can prevent the formation of small metallic
contacts, and also produce spurious experimental results. This
problem can be avoided, in principle, by performing the
experiments in UHV with {\em in situ} cleaning procedures for both
tip and sample \cite{gimzewski87}. However, it is also possible to
fabricate clean metallic contacts in non-UHV conditions. After
conventional cleaning of tip and sample prior to mounting in the
STM, the contacting surfaces of tip and sample are cleaned {\em in
situ} by repeatedly crashing the tip on the spot of the sample
where the contact is to be formed (see
Fig.\,\ref{f.contact-formation}). This procedure pushes the
adsorbates aside making metal-metal contact possible. The tip and
sample are bonded (that is, cold welded) and as the tip is
retracted and contact is broken fresh surfaces are exposed.
Evidence of this welding in clean contacts is the observation of a
protrusion at the spot where the contact was formed
\cite{gimzewski87}. This cleaning procedure works particularly
well at low temperatures where the surfaces can stay clean for
long periods of time since all reactive gasses are frozen. This is
adequate if the contact to be studied is homogeneous, since
otherwise there will be transfer of material from one electrode to
the other. On the other hand, this wetting behavior of the sample
metal (e.g. Ni or Au) onto a hard metal (W or PtIr) tip has also
been exploited for the study of homogeneous contacts, assuming
full coverage of the tip by the sample material \cite{olesen94}.

\begin{figure}[!t]
\begin{center}
\includegraphics[width=80mm]{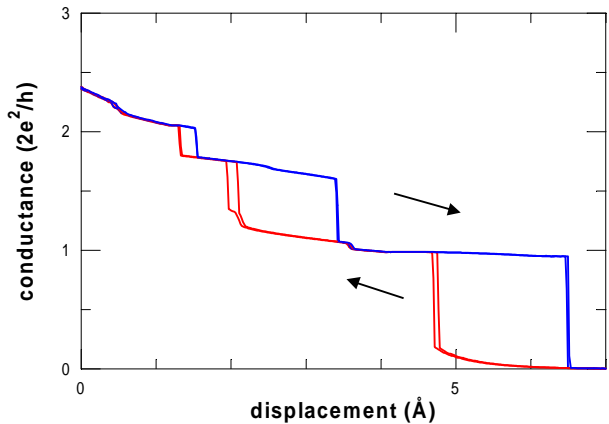}
\end{center}
\caption{\label{f.STM_curve_example} Conductance curves of a gold
nanocontact at low temperature (4.2 K) using a stable STM. Two
complete consecutive cycles of approach-retraction are shown. The
lower and higher curves correspond to approach and retraction,
respectively.}
\end{figure}

In  an STM experiment on metallic contacts, the bias voltage is
kept fixed (at a low value, say 10 mV) and the current is recorded
as the tip-sample distance is varied by ramping the
$z$-piezovoltage. The results are typically presented as a plot of
the conductance (or current) versus $z$-piezovoltage (or time).
Fig.\,\ref{f.STM_curve_example} shows a typical STM conductance
curve for a clean Au contact at low temperatures. Before contact
the current depends exponentially on the distance with an apparent
tunneling barrier of the order of the work function of the
material. Such a high value of the apparent tunneling barrier is a
signature of a clean contact, since adsorbates lower the tunneling
barrier dramatically \cite{gimzewski87} (with the exception of the
inert helium gas, see Sect.\,\ref{sss.calibration} below).
Metallic contact takes place as an abrupt jump in the conductance,
or the current. After this jump the conductance increases in a
stepwise manner as the size of the contact increases. Reversing
the motion of the tip shows that these steps are hysteretic. In
the case of Au, as we will see, the conductance of the first
plateau is quite well defined with a value of approximately
$2e^2/h$ and corresponds to a one-atom contact. For other metals
the conductance curves will look somewhat different, depending on
the electronic structure of the metal.  It is important to note
that the $z$-piezovoltage in STM experiments is not directly
related to the size of the contact: as the contact is submitted to
strain its atomic configuration changes in a stepwise manner, as
will be discussed in detail in Sect.\,\ref{s.conductance}.

For studying nanocontacts a standard STM can be used, but it must
be taken into account that currents to be  measured are about 2--3
orders of magnitude larger than in usual STM operation. Mechanical
stability  of the STM setup is an important factor. Careful design
makes possible to achieve noise vibration amplitudes of the order
of a few picometers at low temperatures.

\begin{figure}[!b]
\centerline{ \includegraphics[width= 8cm]{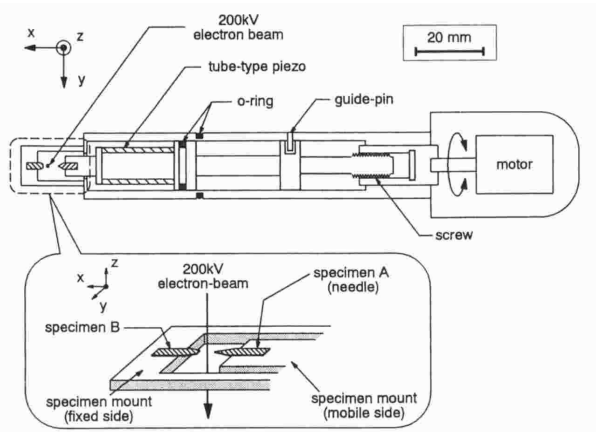}}
\caption{Schematic representation of the sample holder allowing
piezo control of the tip-sample distance in the very limited space
available in a HRTEM. Reprinted with permission from
\protect\cite{kizuka97}. \copyright 1997 American Physical
Society.} \label{f.Kizuka}
\end{figure}

Direct  observation of metallic nanocontacts is possible using
high-resolution transmission electron microscopy (HRTEM). Several
groups have constructed an STM with the tip apex at the focal
point of a HRTEM. Kizuka and co-workers have observed the atomic
contact formation processes in gold using a piezo-driven specimen
holder \cite{kizuka97,kizuka98,kizuka98a}, Fig.\,\ref{f.Kizuka},
and Takayanagi and co-workers have studied the structure of gold
nanowires \cite{kondo97}, and atomic wires \cite{ohnishi98}. The
conditions for the experiments are (ultra-)high vacuum and ambient
temperature and a time resolution of 1/60~s for the video frame
images.

\subsection{The mechanically controllable break junction technique}\label{ss.MCBJ}

In  1985 Moreland and Ekin \cite{moreland85} introduced ``break''
junctions for the study of the tunneling characteristics for
superconductors. They used a thin wire of the superconductor
mounted on top of a glass bending beam. An electromagnetic
actuator controlled the force on the bending beam. Several
extensions and modifications to this concept have been introduced
later, initially by Muller {\it et al.}\
\cite{muller92a,muller92}, who introduced the name Mechanically
Controllable Break Junction (MCBJ). The technique has proven to be
very fruitful for the study of atomic-sized metallic contacts.

\subsubsection{Description of the MCBJ technique}\label{sss.MCBJ technique}

\begin{figure}[!b]
\centerline{ \includegraphics[width= 8cm]{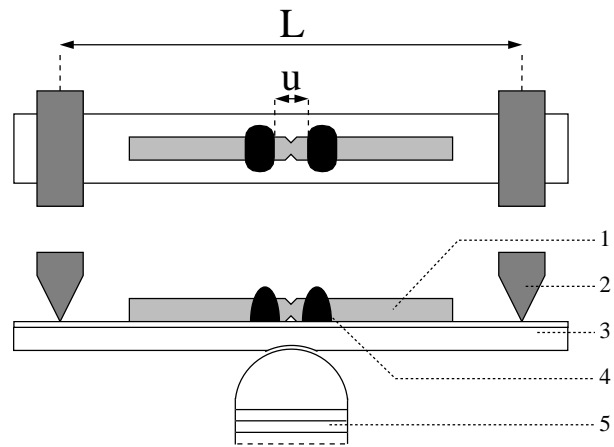}} \caption{
Schematic top and side view of the mounting of a MCBJ, with the
notched wire (1), two fixed counter supports (2), bending beam
(3), drops of epoxy adhesive (4) and the stacked piezo element
(5). }
\label{f.MCBJ}
\end{figure}

\begin{figure}[!t]
\centerline{\includegraphics[width=8cm ]{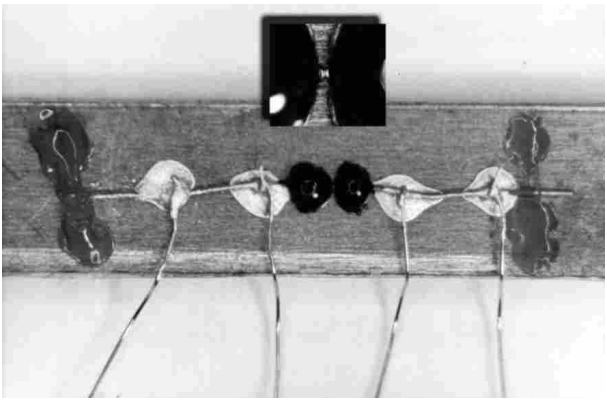}}
\caption{Top view of a MCBJ seen under an optical microscope. The
substrate is 4.5\,mm wide and the sample wire is a 0.1\,mm
diameter gold wire. The inset shows an enlargement of the wire
with the notch between the two drops of epoxy.  On each side of
the notch two wires make contact to the sample wire using silver
paint. } \label{f.photoMCBJ}
\end{figure}

The principle of the technique is illustrated in
Fig.\,\ref{f.MCBJ}. The figure shows a schematic top and side view
of the mounting of a MCBJ, where the metal to be studied has the
form of a notched wire, typically 0.1mm in diameter, which is
fixed onto an insulated elastic substrate with two drops of epoxy
adhesive (Stycast 2850FT and curing agent 24LV) very close to
either side of the notch. The notch is manually cut into the
center of a piece of wire of the metal to be studied. For most
metals, except the hardest, it is possible to roll the wire under
the tip of a surgical knife in order to obtain a diameter at the
notch of about one third of the original wire diameter. A
photograph of a mounted wire is shown in Fig.\,\ref{f.photoMCBJ}.
The distance between the drops of epoxy adhesive can be reduced to
only about 0.1mm by having the epoxy cure at ambient conditions
for about 3 hours before applying it. This prevents that the small
drops deposited at some distance from the notch flow together. The
epoxy is still malleable, and under a microscope the drops can be
gradually pushed towards the center.

The substrate is mounted in a three-point bending configuration
between the top of a stacked piezo-element and two fixed counter
supports. This set-up is mounted inside a vacuum can and cooled
down to liquid helium temperatures. Then the substrate is bent by
moving the piezo-element forward using a mechanical gear
arrangement. The bending causes the top surface of the substrate
to expand and the wire to break at the notch. By breaking the
metal, two clean fracture surfaces are exposed, which remain clean
due to the cryo-pumping action of the low-temperature vacuum can.
The fracture surfaces can be brought back into contact by relaxing
the force on the elastic substrate, where the piezoelectric
element is used for fine control. The roughness of the fracture
surfaces usually results in a first contact at one point.

In addition to a clean surface, a second advantage of the method
is the stability of the two electrodes with respect to each other.
From the noise in the current in the tunneling regime one obtains
an estimate of the vibration amplitude of the vacuum distance,
which is typically less than $10^{-4}$\,nm. The stability results
from the reduction of the mechanical loop which connects one
contact side to the other, from centimeters, in the case of an STM
scanner, to $\sim 0.1$~mm in the MCBJ.

The  most common choice for the bending beam is a plate of
phosphorous bronze, about 0.5--1mm thick, 20mm long and 3-5mm
wide. The top surface is usually insulated by covering it with a
thin polymer foil (Kapton) using regular epoxy. The advantage over
brittle materials, such as glass as was used in the experiments my
Moreland and Ekin \cite{moreland85}, is that one avoids the risk
of fracture of the bending beam. For brittle materials the maximum
strain before breaking is usually about 1\% . The principle of the
MCBJ lies in the concentration of the strain in the entire length
of the unglued section $u$ (Fig.\,\ref{f.MCBJ}) of the top surface
of the bending beam onto the notch of the wire. Since metals tend
to deform plastically this strain concentration is often still not
sufficient to break the wire, unless the notch is cut very deep.
Since the cutting of a deep notch without separating the wire ends
is not always very practical, one mostly chooses a metallic
bending beam such as phosphor bronze. The rate of success for this
arrangement is very good (of order 90\%), but one often needs to
bend the substrate beyond the elastic limit in order to obtain a
break in the wire. This poses no serious problems, except that the
displacement ratio $r_{\rm d}$, i.e. the ratio between the
distance over which the two wire ends are displaced with respect
to each other and the extension of the piezo-element, is reduced
and not very predictable.

For the ideal  case of homogenous strain in the bending beam the
displacement ratio can be expressed as
$$
r_{\rm d} = {3 u h \over L^2} ,
$$
where $u$  and $L$ are the unglued section and the distance
between the two counter supports, respectively, as indicated in
Fig.\,\ref{f.MCBJ}, and $h$ is the thickness of the bending beam.
For the dimensions indicated above for a typical MCBJ device we
obtain $r_{\rm d} \simeq 10^{-3}$. In practice, the plastic
deformation of the bending beam may result in a reduction of the
displacement ratio by about an order of magnitude. For experiments
where it is necessary to have a calibrated displacement scale, a
calibration is required for each new sample and the procedure is
described in Sect.~\ref{sss.calibration}. For optimal stability of
the atomic-sized junctions it is favorable to have a small
displacement ratio, since the external vibrations that couple in
through the sample mounting mechanism, are also reduced by this
ratio.

Although   it cannot be excluded that contacts are formed at
multiple locations on the fracture surfaces, experiments usually
give no evidence of multiple contacts. As will be explained in
more detail in Sect.~\ref{s.mechanical}, from the mechanical
response of the contacts one can deduce that upon stretching the
shape of the contact evolves plastically to form a connecting neck
between the two wire ends. The neck gradually thins down and
usually breaks at the level of a single atom.

\begin{figure}
\centerline{\includegraphics[width= 8cm]{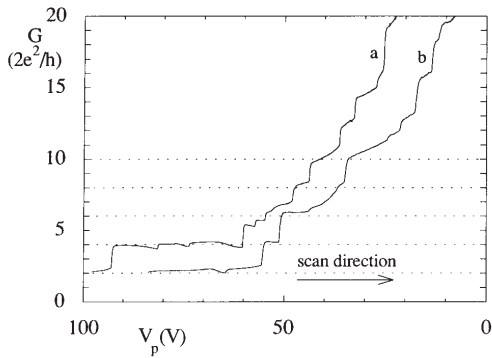}}
\caption{Two examples of traces of the conductance, $G$ , measured
on Pt contacts using the MCBJ technique at a temperature of
1.2\,K. The electrodes are pushed together by decreasing $V_p$. An
estimate for the corresponding displacement is 10V$\sim$0.1\,nm.
Reprinted with permission from \protect\cite{muller92}. Copyright
1992 American Physical Society. }
\label{f.MullerPt} 
\end{figure}

In the  first experiments using the MCBJ for a study of
conductance in atomic-sized metallic contacts
\cite{muller92a,muller92}, distinct steps were observed in the
conductance of Pt and Nb contacts. Figure~\ref{f.MullerPt} shows
two examples of recordings of the conductance as a function of the
voltage $V_p$ on the piezo-element, which is a measure of the
displacement of the two wire ends with respect to each other. In
the experiment, when coming from the tunneling regime, a contact
is formed by moving the electrodes together. The contact is then
slowly increased while recording the conductance. The scans as
given in Fig.\,\ref{f.MullerPt} are recorded in about 20 minutes
each. The conductance is observed to increase in a step-wise
fashion, which is different each time a new contact is made.
Although the steps  are of order of the conductance quantum, the
authors caution against a direct interpretation in terms of
conductance quantization. This point will be discussed in more
detail in Sect.~\ref{s.conductance}.

\subsubsection{Microfabrication of MCBJ devices}\label{sss.micro-MCBJ}

\begin{figure}[!t]
\centerline{ \includegraphics[width=7cm ]{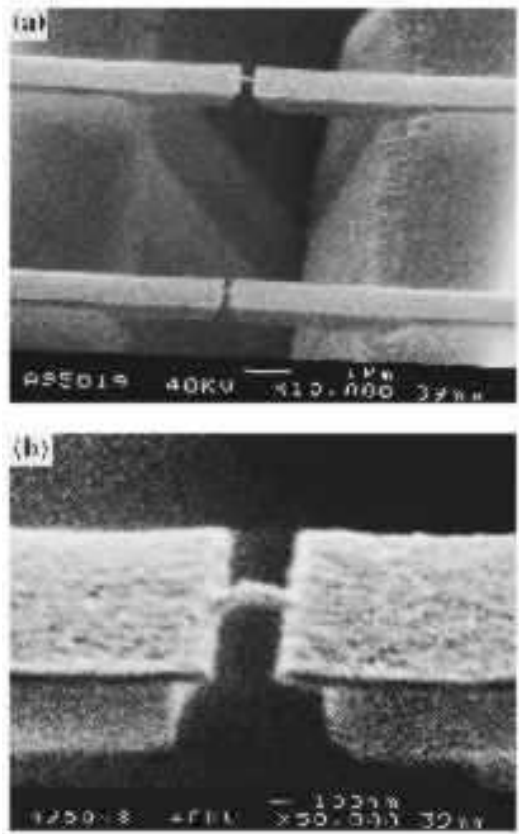} }
\caption{(a) Electron microscopy image of two microfabricated
bridges suspended above a triangular pit in the silicon substrate.
The close-up in (b) shows the two SiO$_2$ cantilevers, which are
about 700 nm apart. The cantilevers are covered by a gold layer
from which the final conducting bridge of about 100~nm wide is
formed, and which is broken by bending of the silicon substrate.
Reprinted with permission from \protect\cite{zhou95}. \copyright
1995 American Institute of Physics.} \label{f.lithoMCBJ}
\end{figure}

The principle  of the break junction technique can be refined by
employing microfabrication techniques to define the metal bridge.
The advantage is a further improved immunity to external
vibrations and the possibility to design the electronic and
electromagnetic environment of the junction.
Fig.\,\ref{f.lithoMCBJ} shows a lithographically fabricated MCBJ
device on a silicon substrate, as developed by Zhou {\it et al.}\
\cite{zhou95}. These authors used a $\langle 100
\rangle$-oriented, 250~$\mu$m thick silicon substrate, covered by
a 400~nm thick SiO$_2$ insulating oxide layer. On top of this they
deposit a 8~nm gold film, which is defined into the shape of a
100~nm wide bridge by standard electron beam lithography. Using
the metal film as a mask, they etch a triangular groove into the
silicon substrate below the bridge. The bridge can be broken at
the narrowest part, as for the regular MCBJ devices, by bending
the substrate. The parameters for the bridge need to be chosen
such that the metal bridge breaks before breaking the silicon
substrate itself.

Alternatively, one  can microfabricate the MCBJ device on a
phosphorous bronze substrate \cite{ruitenbeek96}. After polishing
the substrate it is covered with a polyimide layer by spin
coating, which serves to smoothen the substrate and insulate the
junction electrically. The metal bridge is defined into a metal
film deposited onto the polyimide layer by techniques similar to
those used by Zhou {\it et al.}, after which the polyimide layer
is carved in an reactive ion etcher, producing a freely suspended
bridge over a length of approximately 2~$\mu$m.

The displacement ratio  for the microfabricated MCBJ devices is
about two orders of magnitude smaller than that for a regular
device, $r_{\rm d} \sim 10^{-4}$. As a consequence the immunity to
vibrations and drift is such that the electrode distance changes
by less than 0.2~pm per hour and it is possible to manually adjust
the bending to form a single atom contact. On the other hand, a
drawback is the fact that the displacement of the electrodes that
can be controlled by the piezovoltage is limited to only a few
angstroms due to the small displacement ratio and the limited
range of expansion of the piezo-element. For larger displacements
a mechanical gearbox arrangement in combination with an
electromotor can be used, but such systems have a rather large
backlash, which hampers a smooth forward and backward sweep over
the contact size.

The major advantage of the microfabricated MCBJ devices  is the
possibility to define the environment of the atomic-sized contact.
An example was given in Fig.\,\ref{f.turtles} showing a device
that allows to form an atomic-sized contact for gold with
superconductivity induced through the close proximity of a thick
aluminum film. This experiment will be discussed in
Sect.~\ref{s.exp_modes} and in Sects.~\ref{ss.ECB} and
\ref{ss.cpr} some examples will be given of designing the
electromagnetic environment of the junction.

\subsubsection{Calibration of the displacement ratio}\label{sss.calibration}

Usually the displacement ratio  cannot be determined very
accurately from the design of the MCBJ and it is necessary to make
a calibration for each new device. The simplest approach is to
exploit the exponential dependence of the resistance $R$ with
distance $\delta$ between the electrodes in the tunneling regime,
\begin{equation}
R \propto \exp\bigl( {{2\sqrt{2m\Phi}}\over{\hbar} }\delta \bigr)
. \label{eq.exponential_tunneling}
\end{equation}
Here, $\Phi$ is the workfunction of the metal, and $m$ the
electron mass. Kolesnychenko {\it et al.} introduced a more
accurate method, using the Gundlach oscillations in the tunnel
current \cite{kolesnychenko99}. These oscillations arise from
resonances in the tunnel probability under conditions of field
emission. For bias voltages larger than the workfunction of the
metal the tunnel current increases exponentially, and on top of
this rapid increase a modulation can be observed resulting from
partial reflection of the electron wave in the vacuum between the
two electrodes. Several tens of oscillations can be observed,
allowing not only an accurate calibration of the displacement, but
also an independent measurement of the workfunction. Surprisingly,
from these studies it was found that the workfunction obtained is
strongly influence by the presence of helium at the surface of the
metal. Full helium coverage was found to {\it increase} the
workfunction by about a factor of two \cite{kolesnychenko99a}.
Since helium is often used as a thermal exchange gas for cooling
down to low temperatures, this result explains the rather large
variation obtained in previous work for the distance calibration
for any single device.

\subsubsection{Special sample preparations}\label{sss.special MCBJ techniques}

The principle of the MCBJ  technique, consisting of exposing clean
fracture surfaces by concentration of stress on a constriction in
a sample, can be exploited also for materials that cannot be
handled as described above. Delicate single crystals
\cite{kempen93,krans94} and hard metals can be studied with the
single modification of cutting the notch by spark erosion, rather
than with a knife.

\begin{figure}
\centerline{\includegraphics[width= 6.5cm]{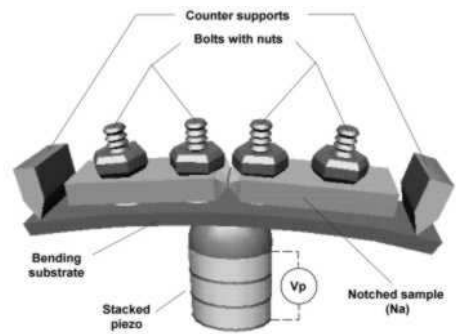}}
\caption{Principle of the MCBJ technique adapted for the reactive
alkali metals.}
\label{f.Alkali-MCBJ} 
\end{figure}

The alkali metals Li, Na, K, etc.,  form an important subject for
study, since they are nearly-free electron metals and most closely
approach the predictions of simple free-electron gas models. The
experiments will be discussed in Sects.~\ref{s.conductance} and
\ref{s.shells}. A schematic view of the MCBJ technique for alkali
metals is given in Fig.\,\ref{f.Alkali-MCBJ}
\cite{krans95,yanson99}. While immersed in paraffin oil for
protection against rapid oxidation, the sample is cut into the
shape of a long thin bar and pressed onto four, 1~mm diameter,
brass bolts, which are glued onto the isolated substrate, and
tightened by corresponding nuts. Current and voltage leads are
fixed to each of the bolts. A notch is cut into the sample at the
center. This assembly is taken out of the paraffin and quickly
mounted inside a vacuum can, which is then immersed in liquid
helium. By bending the substrate at 4.2\,K in vacuum, the sample
is broken at the notch. The oxidation of the surface and the
paraffin layer covering it simply break at his temperature and
contact can be established between two fresh metal surfaces. This
allows the study of clean metal contacts for the alkali metals for
up to three days, before signs of contamination are found.

One of the draw-backs of the MCBJ technique  compared to STM-based
techniques is that one has no information on the contact geometry.
Attempts have been made to resolve this problem by using a hybrid
technique consisting of an MCBJ device with additional thin
piezo-elements inserted under each of the two wire ends. One of
these thin piezo plates is used in regular extension-mode, and for
the other a shear-mode piezo is used. This allows scanning the two
fracture surfaces with respect to each other, as in the STM
\cite{post96,keijsers96a}. Although some successful experiments
have been performed, the fact that there is no well-define tip
geometry makes the images difficult to interpret, and atomic-like
features are only occasionally visible.

\subsection{Force measurements}\label{ss.force techniques}

The simultaneous  measurement of  conductance and forces in
metallic contacts of atomic size is not an easy task as evidenced
by the scarcity of experimental results. Conventional contact-mode
atomic force microscopy (AFM) sensors are not adequate for this
task because they are too compliant. In order not to perturb the
dynamics of the deformation process, the force sensor must be
stiffer than the contact itself and yet have sub-nanonewton
sensitivity.  The forces are of the order of 1\,nN, changing over
distances of less than 0.1\,nm. It is required to have the force
constant of the cantilever to be at least an order of magnitude
larger than that of the atomic structure, which imposes that it
should be at least several
 tens N/m. An additional complication is that the presence of
contamination or adsorbates on the contacting surfaces can cause
forces much larger than those due to metallic interaction.
Capillary forces due to water rule out experiments in ambient
conditions.

In all experimental measurements of forces in atomic-sized
contacts the force sensor is a cantilever beam onto which either
the sample or  tip is mounted, but different detection methods are
used.  Most of the experiments measure either the deflection of
the beam or the variation of its resonant frequency. The beam
deflection is directly proportional to the force exerted on the
contact, while the resonant frequency of the beam is influenced by
the gradient of this force. D\"{u}rig \ea \cite{durig90} used an
Ir foil beam with dimensions $7.5 \times 0.9 \times 0.05$ mm$^3$
and spring constant 36 N/m to measure the interaction forces
between tip and sample as a function of tip-sample
 separation up to the jump-to-contact in UHV. The changes in
the oscillation frequency were determined from the tunneling
current. Recently, quartz tuning fork sensors have been
implemented in MCBJ and STM. One electrode is attached to one of
the legs of the  tuning fork  while the other is fixed. These
sensors are very rigid with spring constants larger than several
thousands N/m,  can be excited mechanically or electrically, and
their motion is detected by measuring the piezoelectric current
\cite{tuningfork}.

\begin{figure}[!b]
\begin{center}
\includegraphics[width=70mm]{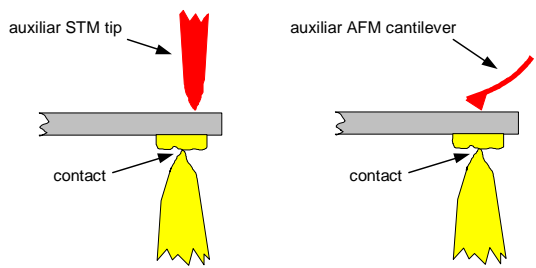}
\end{center}
\caption{\label{f.STM-AFM} Measuring the deflection of the
cantilever beam on which the sample is mounted using an auxiliar
STM tip or AFM cantilever.
 }
\end{figure}

The deflection of the cantilever can be measured directly using
various methods (see Fig.\,\ref{f.STM-AFM}). A second STM  acting
as deflection detector has been used to measure forces in
relatively large contacts using a phosphorous  bronze cantilever
beam of millimeter dimensions at low temperature  (spring constant
$\sim 700$ N/m) \cite{agrait94,agrait95}  and at room temperature
(spring constant $380$ N/m) \cite{agrait96}. Rubio-Bollinger \ea
\cite{rubio01} used the sample, a 0.125 mm diameter, 2 mm length,
cylindrical gold wire as a cantilever beam. They measured the
forces during the formation of an atomic chain at 4.2 K. In all
these experiments the auxiliary STM works on the constant current
mode, which implies that the tip-cantilever distance and
interaction are constant, minimizing the effect on the
measurement. A conventional AFM can also be used to measure the
deflection   of the cantilever beam on which the sample is
mounted. Rubio \ea \cite{rubio96}  measured the picometer
deflection of the 5 mm $\times$ 2 mm cantilever beam by
maintaining the 100 $\mu$m conventional AFM cantilever at constant
deflection. Metal-coated non-contact mode AFM cantilevers with
spring constants of 20-100 N/m have also been used in experiments
at room temperature in air \cite{marszalek00} and  UHV
\cite{jarvis99}in conventional AFM setups.

A different approach was followed by Stalder \ea \cite{stalder96a}
who detected the changes of resonant frequency of a tensioned
carbon-fiber coupled to the cantilever beam.

\begin{figure}[!t]
\begin{center}
\includegraphics[width=80mm]{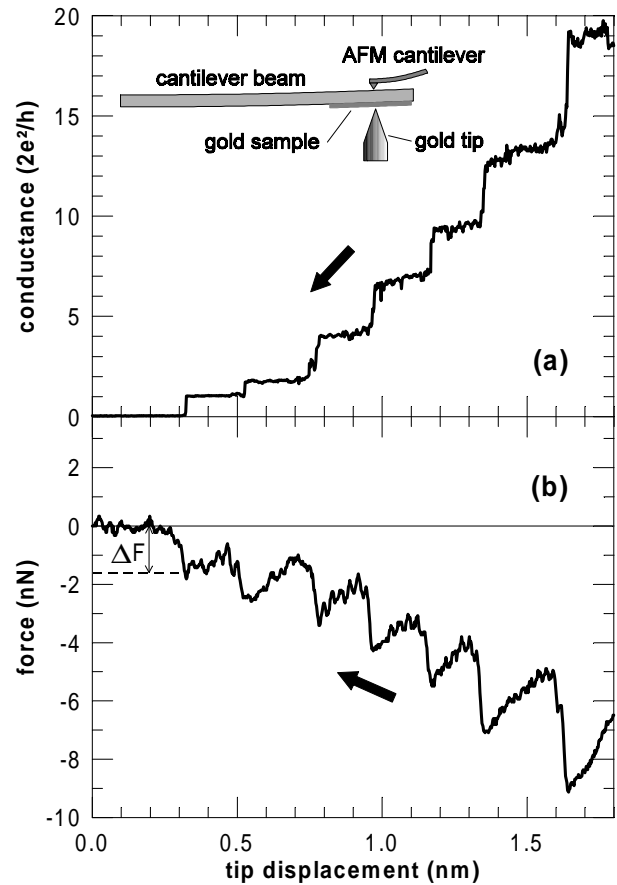}
\end{center}
\caption{\label{f.rubio1996_fig1} Simultaneous force and
conductance measurement in an atomic-sized contact at 300 K. The
inset shows the experimental setup. Reprinted with permission from
\protect\cite{rubio96}. \copyright 1996 American Physical Society.
}
\end{figure}

In Fig.\,\ref{f.rubio1996_fig1}, an example of simultaneous
measurement of forces and conductance is shown. The saw-tooth
shape of the force curve indicates that a mechanical deformation
process takes place in the form of elastic stages (the linear
portions of the curve), in which the nanostructure is deformed
elastically, followed by sudden relaxations (the vertical portions
of the curve) in which the system becomes unstable and its  atomic
configuration changes. It must be remarked that the slope of the
elastic stages measures the {\em local} effective spring constant
of the contact in series with the {\em macroscopic} effective
spring constant of the constriction, which depends on its shape
and could be comparable to the that of the contact itself in the
case of sharp, long tips, and the spring constant of the force
sensor. The system becomes unstable when the local gradient of the
force at the contact exceeds the effective spring constant of the
combined constriction-sensor system, with its strength given by
the maximum force before relaxation. Note that even in the absence
of a force sensor the  dynamical evolution of a long constriction
(lower effective elastic constant) will be different to that of a
short constriction (higher effective elastic constant). The
conductance curve in Fig.\,\ref{f.rubio1996_fig1}, shows that
during the elastic stages the conductance is almost constant,
changing suddenly as the system changes configuration, implying
that the sudden jumps in the conductance are due to the changes in
geometry.

\subsection{Nanofabricated contacts}\label{ss.nanofab contacts}

Various approaches have been explored to produce fixed contacts by
nanofabrication techniques. The first of these approaches was
introduced by Ralls and Buhrman \cite{ralls88}. They used
electron-beam lithography to fabricate a nanometer size hole in a
free standing thin film of Si$_3$N$_4$, and a metal film was
evaporated onto both sides of the silicon nitride film, filling up
the hole and forming a point contact between the metal films on
opposite sides. These structures are very stable, and contacts
only several nanometers wide can be produced. The great advantage,
here, is that the point contact can be cycled to room temperature
and be measured as a function of field or temperature without
influence on the contact size.

Such contacts are still fairly large compared to  atomic
dimensions and in order to reduce the size down to a single atom
one has used methods employing feedback during fabrication by
monitoring the contact resistance. There exist roughly two
approaches: anodic oxidation of a metal film and deposition from
an electrolytic solution. The first approach was introduced by
Snow {\it et al.} \cite{snow96,snow96a}. A metal film (Al or Ti,
$\sim$10\,nm thick) can be locally oxidized from the surface down
to the substrate induced by the current from an AFM tip operating
under slightly humid ambient conditions. By scanning the tip
current over the surface they produced a constriction in the metal
film, which they were able to gradually thin down to a single
atom. When the contact resistance comes in the range of kilo-ohms
the resistance is seen to change step-wise and the last steps are
of order of the conductance quantum, $2e^2/h$, which is an
indication that the contact is reduced to atomic size. At room
temperature such contacts usually reduce spontaneously into a
tunnel junction on a time scale of a few minutes. However, some
stabilize at a conductance value close to $2e^2/h$ for periods of
a day or more. The controlled thinning of the contact can also be
achieved by the current through the contact in the film itself
\cite{schmidt98}.

A second approach consists of controlled deposition  or
dissolution by feedback of the voltage polarity on the electrodes
immersed in the electrolyte. Li and Tao \cite{li98,li99} thinned
down a copper wire by electrochemical etching in a a copper
sulphate solution. Atomic-sized contacts are found to be stable
for many hours, before the conductance drops to zero. The
deposition or dissolution rate can be controlled by the
electrochemical potential of the wire and by feedback the contact
resistance can be held at a desired value. A further refinement of
the technique starts from a lithographically defined wire, but is
otherwise similar in procedure \cite{li99}, and gold junctions can
be fabricated from a potassium cyanaurate solution
\cite{morpurgo99}.

A hybrid technique was used by Junno {\it et al.}\ \cite{junno98}
who  first patterned two gold electrodes onto a SiO$_2$ layer on
top of a silicon wafer, with a gap of only 20 to 50~nm separating
the two electrodes. Subsequently a grid pattern of 30--100~nm gold
particles was formed on the same substrate in a second e-beam
lithography fabrication step. The particles were then imaged by an
AFM, and a proper particle was selected and manipulated into the
gap between the electrodes by the AFM tip. This process allows
{\it in situ} control over the contacts down to atomic size, and
would also be suitable for contacting other types of metal
particles or even molecules.

Recently Davidovic and coworkers produced atomic-sized contacts by
evaporating gold on a Si$_3$N$_4$ substrate that contains a slit
of 70\,nm, while monitoring the conductance across the slit
\cite{anaya02}. As soon as a tunneling current is detected the
evaporation is interrupted and a contact is allowed to form by
electric field induced surface migration with a bias of up to
10\,V applied over the contact. The contacts that were produced
appear to have nanometer-sized grains between the electrodes
giving rise to Coulomb-blockade features in the $IV$ curves.

Ohnishi \ea exploited the heating by the electron  beam in a HRTEM
to produce atomic-sized wires while imaging the structures
\cite{ohnishi98,rodrigues01}. The method starts from a thin metal
film and by focusing the electron beam on two nearby points one is
able to melt holes into the film. The thermal mobility of the
atoms results in a gradual thinning of the wire that separates the
two holes, down to a single atom or chain of atoms. The advantage
is that the structure is very stable and the process can be
followed with video-frame time resolution. The conductance cannot
be measured in this configuration.

\subsection{Relays}\label{ss.relays}

Under the name relay we refer to all techniques  bringing two
macroscopic metallic conductors into contact by some means of
mechanical control. In its simplest form it can be achieved by
lightly touching two wires which are allowed to vibrate in and out
of contact while measuring the conductance with a fast digital
recorder \cite{costa95}. For more reliable operation and in order
to obtain sufficient statistics over many contact breaking cycles
commercial or home made relays have been used
\cite{hansen97,costa97b,yasuda97}, based on electromagnetic or
piezo controlled operation. Gregory \cite{gregory90} used the
Lorentz force on a wire in a magnetic field to push it into
contact with a perpendicularly oriented wire. Although this
produces very stable tunnel junctions, atomic sized contacts have
not been demonstrated.

The relay techniques are suitable for averaging  the conductance
properties of atomic-sized contacts over large numbers of breaking
cycles at room temperature.

\section{Theory for the transport properties of normal metal point contacts}
\label{s.transport_theory}

\subsection{Introduction}
\label{ss.intro_theory}

Macroscopic conductors are characterized by Ohm's law, which
establishes that the conductance $G$ of a given sample is directly
proportional to its transverse area $S$ and inversely proportional
to its length $L$, i.e.
\begin{equation}
G = \sigma S/L  \, ,
\end{equation}
where $\sigma$ is the conductivity of the sample.

Although the conductance is also a key quantity for analyzing
atomic-sized conductors, simple concepts like Ohm's law are no
longer applicable at the atomic scale. Atomic-sized conductors are
a limiting case of mesoscopic systems in which quantum coherence
plays a central role in the transport properties.

In mesoscopic systems one can identify different transport regimes
according to the relative size of various length scales. These
scales are, in turn, determined by different scattering
mechanisms. A fundamental length scale is the {\it phase-coherence
length}, $L_{\varphi}$ which measures the distance over which
quantum coherence is preserved. Phase coherence can be destroyed
by electron-electron and electron-phonon collisions. Scattering of
electrons by magnetic impurities, with internal degrees of
freedom,  also degrades the phase but elastic scattering by
(static) non-magnetic impurities does not affect the coherence
length. Deriving the coherence length from microscopic
calculations is a very difficult task. One can, however, obtain
information on $L_{\varphi}$ indirectly from weak localization
experiments \cite{datta97a}. A typical value for Au at $T=1$\,K is
around 1\,$\mu$m \cite{washburn91}. The mesoscopic regime is
determined by the condition $L < L_{\varphi}$, where $L$ is a
typical length scale of our sample.

\begin{figure}[!t]
\centerline{\includegraphics[width=8cm] {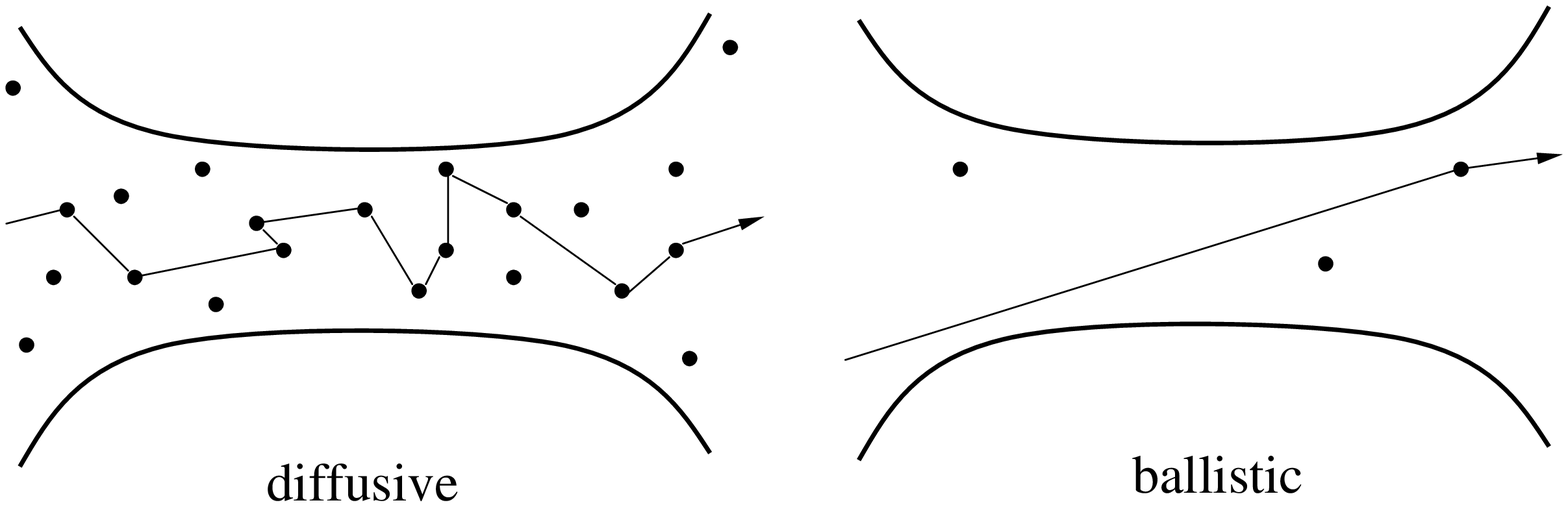}}
\caption{Schematic illustration of a diffusive (left) and
ballistic (right) conductor} \label{f.scheme of regimes}
\end{figure}

Another important length scale is the elastic mean free path
$\ell$, which roughly measures the distance between elastic
collisions with static impurities. The regime $\ell \ll L$ is
called {\it diffusive}. In a semi-classical picture the electron
motion in this regime can be viewed as a random walk of step size
$\ell$ among the impurities. On the other hand, when $ \ell > L $
we reach the {\it ballistic} regime in which the electron momentum
can be assumed to be constant and only limited by scattering with
the boundaries of the sample. These two regimes are illustrated in
Fig.~\ref{f.scheme of regimes}.

In the previous discussion we have implicitly assumed that the
typical dimensions of the sample are much larger than the Fermi
wavelength $\lambda_{\rm F}$. However, when dealing with
atomic-sized contacts the contact width $W$ is of the order of a
few nanometers or even less and thus we have $W \sim \lambda_{\rm
F}$. We thus enter into the {\it full quantum} limit which cannot
be described by semi-classical arguments. A main challenge for the
theory is to derive the conductance of an atomic-sized conductor
from microscopic principles.

The objective of this section will be to review the basic theory
for transport properties of small conductors. We find it
instructive to start first by discussing the classical and
semi-classical theories usually employed to analyze point contacts
which are large with respect to the atomic scale. We shall then
discuss the scattering approach pioneered by Landauer
\cite{landauer57} to describe electron transport in quantum
coherent structures and show its connection to other formalisms,
such as Kubo's linear response theory. The more specific
microscopic models for the calculation of conductance in
atomic-sized contacts will be presented in Sect.~\ref{s.models}.

\subsection{Classical Limit (Maxwell)} \label{ss.Maxwell}

Classically the current $I$ passing through a sample that is
submitted to a voltage drop $V$, depends on the conductivity of
the material $\sigma$ and on its geometrical shape. At each point
of the material the current density ${\bf j}$  is assumed to be
proportional to the local electric field ${\bf E}$, that is, ${\bf
j}({\bf r})=\sigma {\bf E}({\bf r})$, which is the microscopic
form of Ohm's law. The electric field satisfies Poisson's equation
and the boundary conditions specify that the current density
component normal to the surface of the conductor must be zero.

\begin{figure}[!t]
\centerline{\includegraphics[width=8cm] {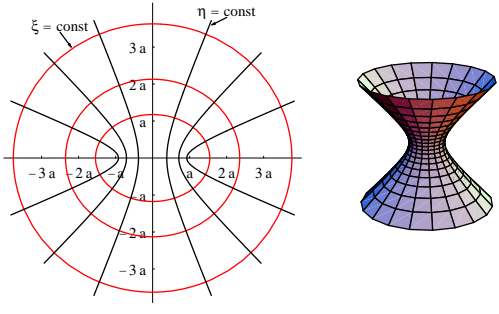}}
\caption{Oblate spheroidal coordinates, and $\eta=\mbox{const.}$
surface. } \label{coordenadas}
\end{figure}

To calculate the conductance of a point-contact, we can model the
contact as a constriction in the material. This problem was
already studied by Maxwell \cite{maxwell}, who considered a
constriction of hyperbolic geometry. Then it is possible to obtain
an analytic solution using oblate spheroidal coordinates
$(\xi,\eta,\varphi)$ defined as
\begin{eqnarray}
x & = & a \cosh \xi \cos \eta \cos \varphi , \nonumber \\ y & = &
a\cosh\xi
 \cos \eta \sin \varphi,\nonumber \\  z & = & a \sinh\xi\sin\eta,\nonumber
\end{eqnarray}
where $2a$ is the distance between the foci, and
$(0\leq\xi<\infty)$, ($-\pi/2\leq \eta \leq \pi/2)$,
$(-\pi<\varphi\leq\pi)$, see Fig.\ \ref{coordenadas}. The
constriction is then defined by the surface $\eta=\eta_{0}=
\mbox{const.}$  and the radius of the narrowest section is given
by $r_{0}=a\cos\eta_{0}$. Since a metal can be considered
effectively charge neutral, Poisson's equation reduces to
Laplace's equation
\begin{equation}
\nabla^{2}V({\bf r})=0,
\end{equation}
where $V({\bf r})$ is the electrostatic potential.  Anticipating a
solution that depends only on $\xi$, the equipotential surfaces
are ellipsoids and the boundary conditions are automatically
satisfied. The solution is then given by
\begin{equation}
V(\xi)=-\frac{V_0}{2}+\frac{2V_0}{\pi}\arctan(e^{\xi}),
\end{equation}
where $V_{0}$ is the voltage drop at the constriction.

The total current is then obtained using Ohm's law and integrating
over the constriction, and by dividing out the applied potential
$V_{0}$ we can express the conductance of the constriction as
\begin{equation}
G_{M}=2a\sigma\
(1-\sin\eta_{0})=2r_{0}\sigma\frac{1-\sin\eta_{0}}{\cos\eta_{0}}.
\end{equation}
This gives the so-called Maxwell conductance of the constriction.
In the limiting case $\eta_{0}=0$ the contact is simply an orifice
of radius $a$ in an otherwise non-conducting plate separating two
metallic half-spaces, and its conductance is
\begin{equation}
G_{M}=2a\sigma=2a/\rho, \label{eq.maxwell}
\end{equation}
where $\rho$ is the resistivity.

\subsection{Semiclassical approximation for ballistic contacts (Sharvin)} \label{ss.Sharvin}

When the dimensions of a contact are much smaller than their mean
free path $\ell$, the electrons will pass through ballistically.
In such contacts there will be a large  potential gradient near
the contact, causing the electrons to accelerate within a short
distance. The conduction through this type of contacts was first
considered by Sharvin \cite{sharvin65}, who pointed out the
resemblance to the problem of the flow of a dilute gas through a
small hole \cite{knudsen34}.

Semiclassically  the current density is written as
\begin{equation}
{\bf j}({\bf r})=\frac{2e}{L^3}\sum_{\bf k} {\bf v_{k}}f_{\bf
k}({\bf r}),
\end{equation}
where $ f_{\bf k}({\bf r})$ is the semiclassical  distribution
function and gives the occupation of state ${\bf k}$ at position
${\bf r}$ and ${\bf v_{k}}$ is the group velocity of the
electrons.  In the absence of collisions, the distribution
function at the contact is quite simple: for the right-moving
states the occupation is fixed by the electrochemical potential
within the left-hand-side electrode, and conversely for the
left-moving states. Thus for an applied voltage $V$, the
right-moving will be occupied to an energy $eV$ higher than the
left-moving states, which results in a net current density, $ j=e
\langle v_{z} \rangle \rho(\epsilon_{\rm F})eV/2 $ where
$\rho(\epsilon_{\rm F})=mk_{\rm F}/\pi^2\hbar^2$ is the density of
states at the Fermi level, and $\langle v_{z} \rangle={\hbar
k_{\rm F}/2m}$ is the average velocity in the positive
$z$-direction. The total current is obtained by integration over
the contact, and hence the conductance (the so-called Sharvin
conductance) is given by
\begin{equation}
G_{S}=\frac{2e^2}{h}\left( \frac{k_{\rm F}a}{2} \right)^{2},
\label{eq.sharvin}
\end{equation}
where $h$ is Planck's constant, $k_{\rm F}$ is the Fermi wave
vector, and $a$ is the contact radius. Note that the Sharvin
conductance depends only on the electron density (through $k_{\rm
F}$), and is totally independent of the conductivity $\sigma$ and
mean free path $\ell$. Quantum mechanics enters only through Fermi
statistics.

\begin{figure}[!t]
\centerline{\includegraphics[width=8cm] {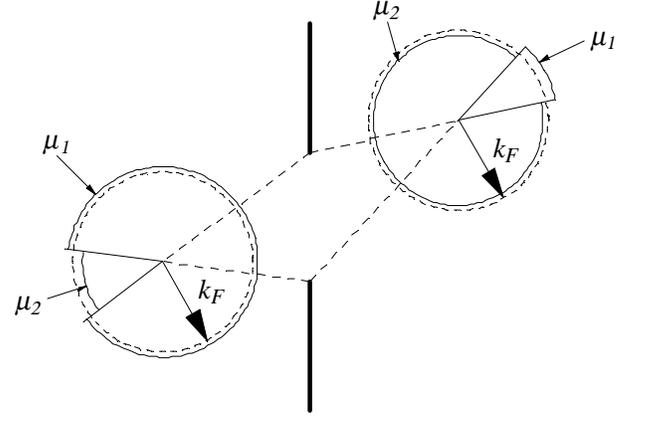}}
\caption{Electron distribution function in the vicinity of the
orifice. $k_{\rm F}$ is the equilibrium Fermi wavevector;
$\mu_{1}$ and $\mu_{2}$ are the chemical potentials for each side,
which far from the orifice, in the presence of an applied
potential $V$, are equal to $E_{\rm F}-eV/2$ and $E_{\rm F}+eV/2$,
respectively.} \label{distribution}
\end{figure}

In a more detailed calculation, the distribution  function $f_{\bf
k}({\bf r})$ is obtained from Boltzmann equation in the absence of
collisions \cite{jansen80},
\begin{equation}
{\bf v_{k}}\cdot  \nabla_{\bf r} f_{\bf k}({\bf r})- (e/\hbar){\bf
E}\cdot \nabla_{\bf k} f_{\bf k}({\bf r})=0. \label{Boltzmann}
\end{equation}
Far from the contact $f_{\bf k}$ is the Fermi distribution
function, $f_{0}(\epsilon_{\bf k})$, with the adequate
electrochemical potential in each electrode. Fig.\
\ref{distribution} shows how the distribution function in the
vicinity of the contact is modified with respect to the Fermi
sphere. The states at a given point on the left side of the
orifice are occupied to $ E_{\rm F}-eV/2$, unless they arrive from
the other electrode, which defines a `wedge-of-cake' in the
electron distribution. Now, the electrostatic potential at this
point is fixed by the requirement of charge neutrality, $\sum_{\bf
k} [f_{\bf k}({\bf r})-f_{0}(\epsilon_{\bf k})]=0$, i.e. the total
volume of the two parts must be equal to the equilibrium charge
density. Far away from the orifice the `wedge' vanishes and the
Fermi spheres in both electrodes are equal in magnitude, but  the
bottom of the conduction band differs by $eV$. At the point in the
middle of the orifice the electrons on the left hemisphere arrive
from the right, and vise versa, which implies that to first
approximation the number of excess electrons exactly balances the
number of deficit electrons on the other side, and this defines
the point $V=0$. The voltage away from this point changes
proportional to the solid angle of view of the orifice and
approaches the limiting values $\pm eV/2$. The important
conclusion is that the voltage drop is concentrated on a length
scale of order $a$ near the contact.

It is  instructive to point out that the power $P=IV$ is entirely
converted into kinetic energy of the electrons that are shot
ballistically into the other electrode. As for the full-quantum
point contacts that will be discussed next, energy relaxation of
the electrons is not taken into account. This is a good
approximation as long as the mean free path for inelastic
scattering is much longer than the dimensions of the contact.
Energy dissipation then takes place far away into the banks by
scattering with phonons. In reality a small but finite amount of
inelastic scattering takes place near the contact, which will be
discussed later.

\subsection{The scattering approach} \label{ss.scattering_approach}

In a typical transport experiment on a mesoscopic device, the
sample (which in our case is an atomic-sized constriction) is
connected to macroscopic electrodes by a set of {\it leads} which
allow us to inject currents and fix voltages. The electrodes act
as ideal electron reservoirs in thermal equilibrium with a
well-defined temperature and chemical potential. The basic idea of
the scattering approach is to relate the transport properties
(conductances) with the transmission and reflection probabilities
for carriers incident on the sample. In this one-electron approach
phase-coherence is assumed to be preserved on the entire sample
and inelastic scattering is restricted to the electron reservoirs
only. Instead of dealing with complex processes taking place
inside the reservoirs they enter into the description as a set of
boundary conditions \cite{buttiker90a}. In spite of its
simplicity, this approach has been very successful in explaining
many experiments on mesoscopic devices.

\begin{figure}[!t]
\centerline{\includegraphics[width=8cm] {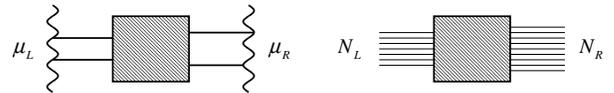}}
\caption{Schematic representation for a ballistic two-terminal
conductance problem. The gray box represents the sample, or the
scattering area. The reservoirs (or electrodes) the left and right
of the wiggly lines emit electrons onto the sample with an energy
distribution corresponding to the electrochemical potentials
$\mu_L$ and $\mu_R$, respectively. Electrons reflected from the
sample are perfectly absorbed by the reservoirs. The straight
sections connecting the reservoirs to the sample represent perfect
leads, where the number of modes in the left and right lead is
$N_L$ and $N_R$, respectively, and these numbers are not required
to be equal.} \label{f.two terminal}
\end{figure}

Let us consider a two terminal configuration as depicted in
Fig.~\ref{f.two terminal}. One can model the conductor as a
scattering region connected to the electron reservoirs by perfect
leads. On these leads the electrons propagate as plane waves along
the longitudinal direction, while its transverse momentum is
quantized due to the lateral confinement. As in the usual
wave-guide problem, the quantization of transverse momentum
defines a set of incoming and outgoing modes on each lead (let us
call $N_{\alpha}$ the number of modes on lead $\alpha$). Notice
that the perfect leads do not really exist in actual systems: they
are an auxiliary construction of the scattering approach which
greatly simplifies the formalism. The more general case can be
analyzed using Green function techniques (see
Sect.~\ref{sss.Green}). Nevertheless, as we discuss below, the use
of perfect leads does not affect the results as long as the number
of modes considered is sufficiently large.

Another hypothesis that considerably simplifies the scattering
approach is that there is a ``perfect" coupling between the leads
and the electron reservoirs. This perfect coupling fixes the
distribution of the incoming modes, which is determined by the
Fermi distribution on the corresponding electrode. On the other
hand, the outgoing modes on the leads are transmitted into the
electrodes with probability one. The boundary conditions on the
incoming and outgoing modes become thus very simple.

Before going further in the discussion of the general formalism,
it is instructive to consider the simple case where the sample is
just a perfect one-dimensional conductor, having a single mode
occupied. Let us assume that there is a voltage difference $V$
applied between the electrodes. A net current will arise from the
imbalance between the population of the mode moving from left to
right (fixed by the Fermi distribution on the left electrode,
$f_L$) and the population of the mode moving in the opposite sense
(fixed by $f_R$). The current is then simply given by
\begin{eqnarray}
I &=& \frac{e}{L} \sum_{k\sigma} v_k \left(f_L(\epsilon_k) -
f_R(\epsilon_k) \right) \nonumber
\\
 &=& \frac{e}{\pi} \int dk v_k
\left(f_L(\epsilon_k) - f_R(\epsilon_k) \right) ,
\label{e.Land1ch}
\end{eqnarray}
where $L$ is the length of the conductor and $\sigma$ is the
electron spin. For a long conductor one can replace the sum over
allowed $k$ values by an integral over $k$. As we are dealing with
a one-dimensional system the density of states is $\rho(\epsilon)
= 1/v_k \hbar$ and the current can be written as
\begin{equation}
I = \frac{2e}{h} \int d\epsilon \left(f_L(\epsilon) -
f_R(\epsilon) \right) .
\end{equation}
The factor 2 in this expression is due to spin-degeneracy. At zero
temperature $f_L(\epsilon)$ and $f_R(\epsilon)$ are step
functions, equal to 1 below $\epsilon_{\rm F} + eV/2$ and
$\epsilon_{\rm F} - eV/2$, respectively, and 0 above this energy.
Thus the expression leads to $I = G V$, where the conductance is
$G = 2e^2/h$.

This simple calculation demonstrates that a perfect single mode
conductor between two electrodes has a finite resistance, given by
the universal quantity $h/2e^2 \approx 12.9 {\rm k}\Omega$. This
is an important difference with respect to macroscopic leads,
where one expects to have zero resistance for the perfectly
conducting case. The proper interpretation of this result was
first pointed out by Imry \cite{imry86}, who associated the finite
resistance with the resistance arising at the interfaces between
the leads and the electrodes.

\subsubsection{The Landauer formula} \label{sss.Landauer}

Let us now discuss the general scattering formalism for the
two-terminal configuration. The amplitudes of incoming and
outgoing waves are related by a (energy dependent) scattering
matrix
\begin{equation}
 \hat{S} = \left( \begin{array}{cc} \hat{s}_{11} & \hat{s}_{12} \\
\hat{s}_{21} & \hat{s}_{22} \end{array} \right)  \equiv
\left( \begin{array}{cc} \hat{r} & \hat{t}^{\prime} \\
\hat{t} & \hat{r}^{\prime} \end{array} \right) ,
\end{equation}
where $\hat{s}_{\alpha \beta}$ is a $N_{\alpha} \times N_{\beta}$
matrix whose components $(\hat{s}_{\alpha \beta})_{mn}$ are the
ratio between the outgoing amplitude on mode $n$ in lead $\alpha$
and the incoming amplitude on mode $m$ in lead $\beta$.

Following Ref. \cite{buttiker92}, it is convenient to introduce
creation and annihilation operators $a_{m
\alpha}^{\dagger}(\epsilon)$ and $a_{m \alpha}(\epsilon)$ which
create and destroy an incoming electron on mode $m$ in lead
$\alpha$ with energy $\epsilon$. Similarly we introduce creation
and annihilation operators for the outgoing states $b_{m
\alpha}^{\dagger}(\epsilon)$ and $b_{m \alpha}(\epsilon)$. These
are naturally related to the $a_{m\alpha}$ operators by
\begin{equation}
b_{m \alpha} = \sum_{n\beta} (\hat{s}_{\alpha \beta})_{mn}
a_{n\beta} .
\end{equation}
According to the hypothesis of perfect coupling between leads and
electrodes, the population of the incoming modes is fixed by
$\langle a_{m \alpha}^{\dagger}(\epsilon) a_{n
\beta}(\epsilon)\rangle = \delta_{m n} \delta_{\alpha \beta}
f_{\alpha}(\epsilon)$, where $f_{\alpha}(\epsilon)$ is the Fermi
distribution function on the electron reservoir connected to lead
$\alpha$.

The current on mode $m$ in lead $\alpha$ is due to the imbalance
between the population of incoming and outgoing states and is
given by

\begin{equation}
I_{m\alpha} = \frac{2e}{h} \int_{-\infty}^{\infty} d\epsilon
\left[\langle a_{m \alpha}^{\dagger}(\epsilon) a_{m
\alpha}(\epsilon)\rangle  - \langle b_{m
\alpha}^{\dagger}(\epsilon) b_{m \alpha}(\epsilon)\rangle
\right].
\end{equation}
As in the single mode case, this expression arises due to the
exact cancellation between the density of states and the group
velocity on each mode. Using the scattering matrix this expression
can be reduced to

\begin{eqnarray}
I_{m\alpha} & = & \frac{2e}{h} \int_{-\infty}^{\infty} d\epsilon
\left[ \left(1 - \sum_n |\hat{r}_{mn}|^2 \right)  f_{\alpha}
\right. \nonumber \\
& - & \left. \sum_{\beta\ne\alpha} \sum_{n} |\hat{t}_{mn}|^2
f_{\beta} \right].
\end{eqnarray}
Adding the contribution of all modes, the net current on lead 1
will be given by
\begin{equation}
I_1 = \frac{2e}{h} \int_{-\infty}^{\infty} d\epsilon \left[(N_1 -
R_{11}) f_1 -  T_{12} f_2 \right] ,
\end{equation}
where $R_{11} = \mbox{Tr} (\hat{r}^{\dagger} \hat{r})$ and $T_{12}
= \mbox{Tr} (\hat{t}^{\dagger} \hat{t})$.
Unitarity of the scattering matrix (which is required by current
conservation) ensures that $\hat{r}^{\dagger}\hat{r} +
\hat{t}^{\dagger}\hat{t} = \hat{I}$ and, therefore, taking the
trace over all modes one has  $T_{12} + R_{11} = N_1$, and thus
\begin{equation}
I_1 = \frac{2e}{h} \int_{-\infty}^{\infty} d\epsilon T_{12} (f_1 -
f_2) . \label{e.twoterminal current}
\end{equation}
The linear response conductance is thus given by

\begin{equation} G = \frac{2e^2}{h} \int_{-\infty}^{\infty} d\epsilon
\left(-\frac{\partial f}{\partial \epsilon} \right) T_{12},
\label{e.Landauer}
\end{equation}
which at zero temperature reduces to the well known Landauer
formula $G = (2e^2/h) T_{12}$ \cite{landauer70}.

\subsubsection{The concept of eigenchannels} \label{sss.eigenchannels}

The Landauer formula teaches us that the linear conductance can be
evaluated from the coefficients $t_{nm}$ which give the outgoing
amplitude on mode $m$ in lead 2 for unity amplitude of the
incoming  mode $n$ in lead 1. Notice that although $\hat{t}$ is
not in general a square matrix (the number of modes on each lead
need not to be the same) the matrix $\hat{t}^{\dagger} \hat{t}$ is
always a $N_1 \times N_1$ square matrix. Current conservation
certainly requires that  $T_{12} = T_{21} = \mbox{Tr}
\left((\hat{t}^{\prime})^{\dagger} \hat{t}^{\prime}\right)$. This
property is a simple consequence of time reversal symmetry of the
Schr\"odinger equation  which ensures that $t_{nm} =
(t^{\prime}_{mn})^*$.

Being the trace of an Hermitian matrix, $T_{12}$ has certain
invariance properties. For instance, there exists a unitary
transformation $\hat{U}$ such that $\hat{U}^{-1} \hat{t}^{\dagger}
\hat{t} \hat{U}$ adopts a diagonal form. Due to hermiticity of
$\hat{t}^{\dagger} \hat{t}$ its eigenvalues ${\tau_i}, i=1,...,
N_1$ should be real. Moreover, due to the unitarity of the
scattering matrix one has $\hat{t}^{\dagger}\hat{t} +
\hat{r}^{\dagger} \hat{r} = \hat{I}$ and then both
$\hat{t}^{\dagger}\hat{t}$ and $\hat{r}^{\dagger}\hat{r}$ should
become diagonal under the same transformation $\hat{U}$. As also
both $\hat{t}^{\dagger}\hat{t}$ and $\hat{r}^{\dagger}\hat{r}$ are
positive definite it is then easy to show that $0 \le \tau_i \le
1$ for all $i$.

The eigenvectors of $\hat{t}^{\dagger} \hat{t}$ and
$\hat{r}^{\dagger} \hat{r}$ are called {\it eigenchannels}. They
correspond to a particular linear combination of the incoming
modes which remains invariant upon reflection on the sample. In
the basis of eigenchannels the transport problem becomes a simple
superposition of independent single mode problems without any
coupling, and the conductance can be written as
\begin{equation}
G = \frac{2e^2}{h} \sum_i \tau_i .
\end{equation}
\

At this point the definition of eigenchannels may seem somewhat
arbitrary and dependent on the number of channels of the perfect
leads attached to the sample. For instance, the dimension of the
transmission matrix $\hat{t}^{\dagger} \hat{t}$ can be arbitrarily
large depending on the number of modes introduced to represent the
leads, which suggests that the number of eigenchannels is not a
well defined quantity for a given sample.

\begin{figure}[!t]
\centerline{\includegraphics[width= 5cm] {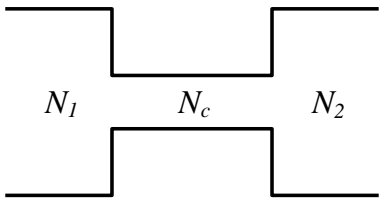}}
\caption{Two perfect cylindrical leads connecting to a sample in
the form of a narrow cylindrical conductor.} \label{f.narrow
constriction}
\end{figure}

In order to convince ourselves that this is not the case, let us
consider a situation where the sample is a narrow cylindrical
constriction between two wide cylindrical leads as shown in
Fig.\,\ref{f.narrow constriction}. Let us call $N_{\rm c}$ the
number of propagating modes at the Fermi energy on the
constriction. Clearly one has $N_{\rm c} \ll N_1 , N_2$. This
geometry can be analyzed as two `wide-narrow' interfaces connected
in series. In such an interface the number of conduction channels
with non-vanishing transmission is controlled by the number of
propagating modes in the narrowest cross-section. This property is
a simple consequence of current conservation along each conduction
channel. Mathematically, one can show that the non-vanishing
eigenvalues of $\hat{t}^{\dagger}\hat{t}$ (a $N_1 \times N_1$
matrix) should be the same as those of $\hat{t} \hat{t}^{\dagger}$
(a $N_{\rm c} \times N_{\rm c}$ matrix). Therefore, there should
be $N_1 - N_{\rm c}$ channels with zero transmission. By applying
the same reasoning to the second ``narrow-wide" interface we
conclude that only a small fraction of the incoming channels could
have a non-zero transmission. The number of relevant eigenchannels
is thus determined by the narrowest cross-section of the
constriction.

For a constriction of only one atom in cross section one can
estimate the number of conductance channels as $N_{\rm c} \simeq
(k_{\rm F}a/2)^2$, which is between 1 and 3 for most metals. We
shall see that the actual number of channels is determined by the
valence orbital structure of the atoms.

\subsubsection{Shot Noise} \label{sss.theory_shot_noise}

Shot noise is another important quantity for characterizing the
transport properties of mesoscopic systems
\cite{deJong97,blanter00}. It refers to the time-dependent current
fluctuations due to the discreteness of the electron charge. In a
mesoscopic conductor these fluctuations have a quantum origin,
arising from the quantum mechanical probability of electrons being
transmitted or reflected through the sample. In contrast to
thermal noise, shot noise only appears in the presence of
transport, i.e. in a non-equilibrium situation.

Shot noise measurements provide information on temporal
correlations between the electrons. In a tunnel junctions, where
the electrons are transmitted randomly and correlation effects can
be neglected, the transfer of carriers of charge $q$ is described
by Poisson statistics and the amplitude of the current
fluctuations is $2 q I$. In mesoscopic conductors correlations may
suppress the shot noise below this value. Even when
electron-electron interactions can be neglected the Pauli
principle provides a source for electron correlations.

The relation between shot noise and the transmitted charge unit
$q$ has allowed the detection of the fractional $q=e/3$ charge
carriers in the fractional quantum Hall regime
\cite{saminadayar97,depicciotto97}. It has also been proposed that
measurements of shot noise in superconducting atomic contacts
could give evidence of transmission of multiple $ne$ charges
associated with multiple Andreev reflection processes
\cite{cuevas99,naveh99}. This issue will be discussed in
Sect.~\ref{s.superconductors}.

Qualitatively, the shot noise in ballistic samples can be
understood from the diagram in Fig.~\ref{Shot noise from
Fermi-Dirac}. For the right moving states, which have been
transmitted through the contact with an excess energy between 0
and $eV$, the average occupation number, $\overline{n}$, is given
by the transmission probability $\tau_{n}$. For the fluctuations
in this number we find
\begin{equation}
\overline{\Delta n^{2}}=\overline{n^{2}}-\overline{n}^{2}=\tau_{n}(1-\tau_{n})%
\, , \label{e.t1-t}
\end{equation}
where in the last step we used the fact that
$\overline{n^{2}}=\overline{n}$, since for fermions $n$ is either
zero or one. Hence, the fluctuations in the current are suppressed
both for $\tau_n = 1$ and for $\tau_n=0$. According to
Eq.\,(\ref{e.t1-t}) the fluctuations will be maximal when the
electrons have a probability of one half to be transmitted. The
shot noise is thus a non-linear function of the transmission
coefficients, which can provide additional information on the
contact properties to that contained in the conductance, as will
be discussed in Sect.~\ref{s.exp_modes}.

\begin{figure}[!t]
\centerline{\includegraphics[height= 4cm] {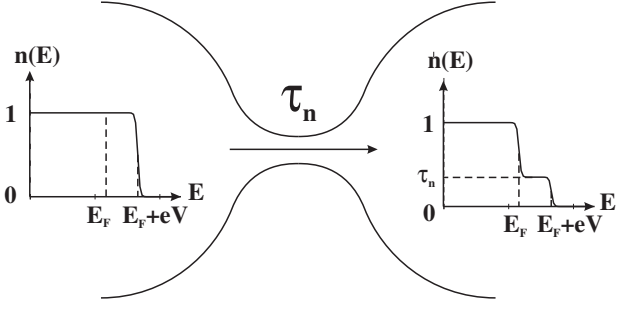}}
\caption{In a ballistic quantum point contact with bias voltage,
$V$, the transmission probability, $\tau_{n}$, determines the
distribution function, $n(E)$, of a transmitted state as a
function of its energy, $E$. In the right reservoir, states with
energy lower than the Fermi energy are all occupied, while
right-moving states with higher energy can only be coming from the
left reservoir, and therefore their average occupation is equal to
the transmission probability, $\tau_{n}$. This argument applies to
every individual eigenchannel} \label{Shot noise from Fermi-Dirac}
\end{figure}


Although there has recently appeared a specialized review on this
subject \cite{blanter00} we shall re-derive here the main results
concerning shot noise in quantum point-contacts for the sake of
completeness.

In noise measurements the quantity one is interested in is the
{\it noise power} spectrum given by the following current-current
correlation function

\begin{equation}
S_{\alpha \beta}(\omega) = \frac{1}{2} \int e^{i \omega t} \langle
\Delta \hat{I}_{\alpha}(t) \Delta \hat{I}_{\beta}(0) + \Delta
\hat{I}_{\beta}(0) \Delta \hat{I}_{\alpha}(t) \rangle  dt \,,
\end{equation}
where $\Delta \hat{I}_{\alpha} = \hat{I}_{\alpha}  - I_{\alpha}$
is the operator measuring the current fluctuations on lead
$\alpha$. This operator can be written in terms of creation and
annihilation operators on each channel as in
Sect.~\ref{sss.Landauer}, i.e.
\begin{eqnarray}
\hat{I}_{\alpha}(t) &=& \frac{2e}{h} \sum_m \int d\epsilon
d\epsilon^{\prime} e^{i(\epsilon-\epsilon^{\prime})t/\hbar}
\nonumber
\\
&\times& \left[a_{m \alpha}^{\dagger}(\epsilon) a_{m
\alpha}(\epsilon^{\prime}) - b_{m \alpha}^{\dagger}(\epsilon) b_{m
\alpha}(\epsilon^{\prime}) \right] .
\end{eqnarray}
Thus, to obtain the noise spectrum one has to evaluate the
expectation value of products of four operators. These products
can be decoupled into all possible contractions of creation and
annihilation operators taken by pairs, as dictated by Wick's
theorem. This decoupling is valid as far as we are dealing with
non-interacting electrons. In this way, a typical expectation
value contributing to the power spectrum can be reduced as follows

\begin{eqnarray}
& &\langle a_{m\alpha}^{\dagger}(\epsilon_1)
a_{n\beta}(\epsilon_2) a_{l\gamma}^{\dagger}(\epsilon_3)
a_{k\delta}(\epsilon_4)\rangle
\nonumber \\
& &- \langle a_{m\alpha}^{\dagger}(\epsilon_1)
a_{n\beta}(\epsilon_2)\rangle \langle
a_{l\gamma}^{\dagger}(\epsilon_3) a_{k\delta}(\epsilon_4)\rangle =
\nonumber
\\
& &\delta_{\alpha \delta} \delta_{\beta \gamma} \delta_{m k}
\delta_{n l} \delta(\epsilon_1 - \epsilon_4) \delta(\epsilon_2 -
\epsilon_3) f_{\alpha}(\epsilon_1) \left[1 - f_{\beta}(\epsilon_2)
\right] .\nonumber
\\
\end{eqnarray}

After some algebra one obtains the following general expression
for the noise spectrum \cite{buttiker92}

\begin{eqnarray}
& & S_{\alpha\beta}(\omega) = \frac{2e^2}{h} \sum_{\gamma \delta}
\sum_{m n} \int d\epsilon A^{mn}_{\gamma
\delta}(\alpha;\epsilon,\epsilon+\hbar \omega) \nonumber \\
& & \times A^{nm}_{\delta
\gamma}(\beta;\epsilon+\hbar \omega,\epsilon)\nonumber \\
& & \times \left\{f_{\gamma}(\epsilon) \left[1 -
f_{\delta}(\epsilon+\hbar \omega) \right] +
f_{\delta}(\epsilon+\hbar \omega) \left[1 - f_{\gamma}(\epsilon)
\right] \right\},
\nonumber \\
\end{eqnarray}
where

\[ A^{mn}_{\delta \gamma}(\beta;\epsilon,\epsilon^{\prime}) =
\delta_{mn} \delta_{\delta \beta} \delta_{\gamma \beta} - \sum_k
\left(\hat{s}^{\dagger}_{\beta \delta} \right)_{m k}(\epsilon)
\left(\hat{s}_{\beta \gamma}\right)_{k n}(\epsilon^{\prime}). \]

This general expression is more than we need to analyze a noise
experiment in atomic contacts. On the one hand, as we are dealing
with a two terminal geometry the expression can be considerably
simplified. In practice what is measured is the time dependent
current on one of the two leads which will be related to $S_{11}$
or $S_{22}$. On the other hand, typical frequencies in noise
experiments are of the order of 20~GHz or even much lower. For
these frequencies $\hbar \omega$ is much smaller than the typical
energy scale for variation of the transmission coefficients and
thus one can safely adopt the zero frequency limit. Under these
conditions the quantity that can be related to the experimental
results is

\begin{eqnarray}
S_{11}(0) & = & \frac{2e^2}{h} \int d\epsilon \left\{
\mbox{Tr}\left[\hat{t}^{\dagger}(\epsilon) \hat{t}(\epsilon)
\hat{t}^{\dagger}(\epsilon) \hat{t}(\epsilon) \right] \right.
\nonumber
\\
&\times& \left. \left[f_1 \left( 1 - f_1 \right) + f_2 \left( 1 -
f_2 \right) \right]
\right. \nonumber \\
& & \left. + \mbox{Tr}\left[\hat{t}^{\dagger}(\epsilon)
\hat{t}(\epsilon) \left( \hat{I} - \hat{t}^{\dagger}(\epsilon)
\hat{t}(\epsilon) \right) \right]
\right. \nonumber \\
&\times & \left.  \left[f_1 \left( 1 - f_2 \right) + f_2 \left( 1
- f_1 \right) \right] \right\}. \end{eqnarray}

As one can observe, this expression has two parts: the terms in
$f_1(1-f_1)$ and $f_2(1-f_2)$ vanish at zero temperature and
correspond to thermal fluctuations, and the terms in $f_1(1-f_2)$
and $f_2(1-f_1)$ which remain finite at zero temperature when
there is an applied bias voltage correspond to the shot noise. One
can further simplify this expression using the basis of
eigenchannels as

\begin{eqnarray}
& & S_{11}(0) = \frac{2e^2}{h} \sum_n \int d\epsilon \left\{
\tau_n(\epsilon)^2 \left[f_1 \left( 1 - f_1 \right) + f_2 \left( 1
- f_2 \right) \right] \right.
\nonumber \\
& & \left. + \tau_n(\epsilon) \left[1 - \tau_n(\epsilon) \right]
\left[f_1 \left( 1 - f_2 \right) + f_2 \left( 1 - f_1 \right)
\right] \right\}.
\end{eqnarray}

Moreover, as in general the energy scale for the variation of the
transmission coefficients is larger than both temperature and
applied voltage we can evaluate these coefficients at the Fermi
energy and perform integration over the energy taking into account
the Fermi factors, which yields \cite{khlus89}

\begin{eqnarray}
S_{11}(0) & = & \frac{2e^2}{h} \left[ 2 k_{\rm B} T \sum_n
\tau_n^2 \right.
\nonumber \\
& & \left. + eV \mbox{coth} \left(\frac{eV}{2k_{\rm B} T}\right)
\sum_n \tau_n\left(1-\tau_n \right) \right]. \label{eq.shotnoise}
\end{eqnarray}

\subsubsection{Thermal transport} \label{sss.theory_thermal}

The scattering approach can be extended to study thermoelectric
phenomena in mesoscopic systems \cite{sivan86,streda89}. In the
previous discussion it was implicitly assumed that the leads are
connected to electron reservoirs which have the same temperature.
If there is a temperature difference between the electrodes there
will be an energy flux in addition to the electric current. Let us
consider a two terminal geometry and call $T_1$ and $T_2$ the
temperature on the left and right electrode respectively. The
total entropy current moving to the right on the left lead will be
given by

\begin{equation}
J^{\rightarrow}_{1S} = -\frac{k_{\rm B}}{h} \int \left[ f_1 \ln
f_1 + (1 - f_1) \ln (1 - f_1) \right] dE , \label{e.Jsright}
\end{equation}
where $f_1 = f(E,\mu_1,T_1)$ denotes the Fermi function on the
left electrode. On the other hand the entropy current going to the
left on the same lead is given by

\begin{eqnarray}
J^{\leftarrow}_{1S} &=& -\frac{k_{\rm B}}{h} \int \left[ (R_{11}
f_1 + T_{12} f_2)
\ln (R_{11} f_1 + T_{12} f_2) + \right. \nonumber  \\
& & \left. (1 - R_{11} f_1 - T_{12} f_2) \ln (1 - R_{11} f_1 -
T_{12} f_2) \right] dE . \nonumber  \\
\label{e.Jsleft}
\end{eqnarray}

By subtracting (\ref{e.Jsright}) and (\ref{e.Jsleft}) the
following expression for the heat current is obtained
\cite{sivan86}

\begin{equation}
U_1 = T J_{1S} = \frac{1}{h} \int T_{12}(E) (E - \mu) \left[f_1 -
f_2 \right] dE, \label{e.SivanImry}
\end{equation}
where $T$ and $\mu$ are the average temperature and chemical
potential.

In the linear transport regime it is convenient to cast
Eqs.\,(\ref{e.SivanImry}) and (\ref{e.twoterminal current}) in

matrix form,

\begin{equation}
\left( \begin{array}{c} I_1 \\ U_1 \end{array} \right) = \left(
\begin{array}{cc} L_0 & \frac{1}{T} L_1 \\ L_1 & \frac{1}{T} L_2
\end{array} \right) \left( \begin{array}{c} \mu_1 - \mu_2 \\ T_1 -
T_2 \end{array} \right),
\end{equation}
where
\begin{equation}
L_n = \frac{2e}{h} \int T_{12}(E) (E - \mu)^n
\left(-\frac{\partial f}{\partial E} \right) dE.
\end{equation}
An important property which can be determined experimentally is
the thermoelectric power or thermopower, defined by

\begin{equation}
S(\mu,T) = \frac{1}{eT} \frac{L_1}{L_0}.
\label{eq.thermopower-channels}
\end{equation}

Thermoelectrical properties of 2DEG quantum point contacts have
been studied experimentally by van Houten \ea \cite{vanhouten92}.
The experimental results on the thermopower in atomic contacts
will be discussed in Sect.~\ref{sss.thermopower}.

\subsubsection{Density of states and energetics within the scattering
approach} \label{sss.dos-scattering}

The scattering matrix is not only related to transport properties.
If the energy dependence of $\hat{s}_{\alpha,\beta}$ is known it
is also possible to relate this quantity to the density of states
of the mesoscopic sample. This type of relations were first
derived in the context of nuclear scattering theory
\cite{dashen69} and establish a connection between the phase
accumulated in the scattering region and the charge within this
region. For a mesoscopic conductor one can use the approximate
expression \cite{akkermans91,gasparian96}

\begin{equation}
\rho(E) = \frac{1}{2\pi i} \sum_{\alpha,\beta} \mbox{Tr} \left[
\hat{s}_{\alpha\beta}^{\dagger} \frac{\partial
\hat{s}_{\alpha\beta}}{\partial E} - \frac{\partial
\hat{s}^{\dagger}_{\alpha\beta}}{\partial E} \hat{s}_{\alpha\beta}
\right] . \label{dos-scattering}
\end{equation}

It should be mentioned that Eq. (\ref{dos-scattering}) is valid as
far as one can neglect the variation in the density of states on
the leads due to the presence of the sample \cite{levy00} which is
usually the case when the geometry is smooth and the WKB
approximation can be applied \cite{buttiker02}.

On the other hand, provided that Eq. (\ref{dos-scattering}) holds,
one can relate the scattering matrix with the total energy of the
system through the density of states. Of course, this gives only
the one-electron contribution to the total energy because
electron-electron interactions are not included within the
scattering approach. These relations provide a way to analyze the
free-electron contributions to the mechanical properties of atomic
contacts, which will be discussed in
Sect.\,\ref{sss.quantum-effects}.

\subsubsection{Limitations of the scattering approach}
\label{sss.theory_limitations}

In spite of its great success in describing many properties of
mesoscopic systems, the scattering approach is far from being a
complete theory of quantum transport. The scattering approach is
mainly a phenomenological theory whose inputs are the scattering
properties of the sample, contained in the $\hat{s}$ matrix. No
hints on how these properties should be obtained from a specific
microscopic model are given within this approach. Moreover, the
scattering picture is a one-electron theory which is valid only as
long as inelastic scattering processes can be neglected. A strong
assumption lies in considering electron propagation through the
sample as a fully quantum coherent process. According to normal
Fermi-liquid theory, such a description would be strictly valid at
zero temperature and only for electrons at the Fermi energy. Under
a finite bias voltage and at finite temperatures deviations from
this simple description might occur.

Between all possible limitations we shall concentrate below in
analyzing the following three: 1) lack of self-consistency in the
electrostatic potential, 2) inelastic scattering within the
sample, 3) electron correlation effects.

\

\underline{Self-consistency:} Although the expression
(\ref{e.twoterminal current}) for the current in a two-terminal
geometry is in principle valid for an arbitrary applied voltage
one should take into account that the transmission coefficients
can be both energy and voltage dependent. This dependence is in
turn determined by the precise shape of the electrostatic
potential profile developing on the sample, which should in
principle be calculated self-consistently. There are only few
studies of non-linear transport in mesoscopic devices which
include a self-consistent determination of the potential profile.
Between these studies we mention the one by Pernas \ea
\cite{pernas90} in which the current through a finite linear chain
modeled by a tight-binding Hamiltonian is calculated by means of
non-equilibrium Green functions. More recently Todorov
\cite{todorov00} calculated by similar methods the non-linear
conductivity for disordered wires.  Brandbyge \ea
\cite{brandbyge02} developed a method based on density functional
theory to calculate the high-bias conductance for atomic-sized
wires.

\

\underline{Inelastic scattering:} At finite bias coherent
propagation of electrons through the sample may be limited by
inelastic scattering processes due to electron-phonon and
electron-electron collisions. B\"uttiker \cite{buttiker86} has
proposed a phenomenological description of these processes within
the scattering approach. In this description the inelastic
scattering events are simulated by the addition of voltage probes
distributed over the sample. The chemical potential on these
probes is fixed by imposing the condition of no net current flow
through them. Thus, although the presence of the probes does not
change the total current through the sample they introduce a
randomization of the phase which tends to destroy phase coherence.
The current in such a structure will contain a coherent component,
corresponding to those electrons which go directly from one lead
to the other, and an inelastic component, corresponding to those
electrons which enter into at least one of the voltage probes in
their travel between the leads. A specific realization of this
construction using a tight-binding model has been presented by
D'Amato and Pastawski \cite{damato90}. On the other hand, Datta
\cite{datta90} and others \cite{hershfield91} have demonstrated
the equivalence between this phenomenological approach and the
case where there is an interaction between electrons and localized
phonon modes distributed over the sample treated in the
self-consistent Born approximation. We should point out that the
restricted problem of one electron interacting with phonons in a
one-dimensional conductor can be mapped into a multichannel
scattering problem \cite{bonca97,ness99}.

\

\underline{Electron correlation effects:} The presence of strong
Coulomb interactions may alter completely the description of
transport given by the scattering approach. This occurs, for
instance, when the sample is a small conducting region weakly
coupled to the external leads. In this case electronic transport
is dominated by charging effects and could be completely
suppressed by the Coulomb blockade effect \cite{vanhouten91}.
Although there have been some proposals to find the equivalent to
the Landauer formula for the case of strongly interacting
electrons \cite{meir92} it is in general not possible to account
for electron correlation effects within the simple scattering
approach presented above. For this case more sophisticated methods
based on Green function techniques are needed. The interested
reader is referred to chapter 8 in Ref. \cite{datta97a} for an
introduction to these techniques.

\subsection{Relation to other formulations: Kubo formula and Green
function techniques} \label{ss.Kubo}

The more traditional approach to transport properties of solids is
based on linear response theory in which the conductivity tensor
is given by the well known Kubo formula \cite{mahan90}. In this
section we shall discuss the connection between this approach and
the scattering picture of transport presented above. We shall
first give a short derivation of the Kubo formula for bulk
materials and then analyze the relation between conductance and
non-local conductivity for a mesoscopic sample connected to
macroscopic leads. This relation permits establishing a link
between linear response theory and the scattering approach. We
shall finally discuss how to express transport properties in terms
of one-electron Green functions.

\begin{figure}[!t]
\centerline{\includegraphics[width=8cm] {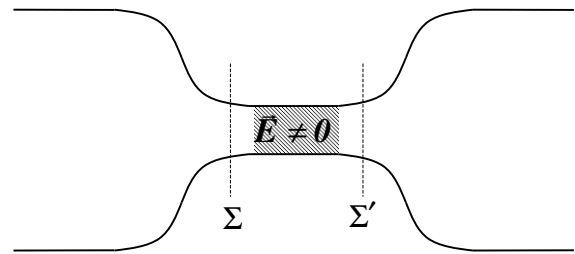}}
\caption{Schematic representation of a two-terminal conductance
system with the positions of the integration planes indicated.}
\label{f.Electric Field}
\end{figure}

When a finite electric field ${\bf E}$, oscillating with frequency
$\omega$, is applied to the sample, the relation between the
current density and the field is given by
\begin{equation}
{\bf J}({\bf r}) = \int d^3r^{\prime} \sigma({\bf r},{\bf
r}^{\prime},\omega) {\bf E}({\bf r}^{\prime}) ,
\end{equation}
where $\sigma({\bf r},{\bf r}^{\prime},\omega)$ is the non-local
conductivity tensor (for simplicity we shall ignore the tensorial
character of $\sigma$ and assume an isotropic response of the
electron system). We shall be interested in the zero frequency
limit. In this limit we may assume that the electric field is
uniform on the mesoscopic sample and vanishes on the leads
\cite{lee81,baranger89}. This situation is illustrated in Fig.
\ref{f.Electric Field}.

Let us first analyze the simpler case of an infinite homogeneous
conductor under an applied constant field. In this case the
relation between field and current density becomes local, i.e
${\bf J} = \sigma(\omega) {\bf E}$. One can obtain
$\mbox{Re}\sigma(\omega)$ by evaluating the energy absorption rate
$P$ due to electronic transitions induced by the applied field.
Taking the temporal average over one cycle one obtains

\begin{equation}
P = \langle  {\bf E} \cdot {\bf J} \rangle  = \mbox{Re}
\sigma(\omega) E^2. \label{e.Power}
\end{equation}

Notice that $\mbox{Re}\sigma(\omega)$ determines the dissipative
part of the current and that $\mbox{Re}\sigma(\omega) \rightarrow
\sigma(0)$ for $\omega \rightarrow 0$. Within lowest order
perturbation theory, the energy absorption rate is given by

\begin{equation}
P = \sum_{\alpha \beta} W_{\alpha \beta} \left(f_{\alpha} -
f_{\beta} \right) \left(\epsilon_{\alpha} -
\epsilon_{\beta}\right), \label{e.FGR}
\end{equation}
where
\begin{equation}
W_{\alpha\beta} = \frac{2\pi}{\hbar} |\langle
\alpha|\hat{V}|\beta\rangle |^2 \delta(\epsilon_{\alpha} + \hbar
\omega - \epsilon_{\beta}).
\end{equation}
In this expression $\alpha$ and $\beta$ denote different electron
eigenstates of the system, $f_{\alpha}$ and $f_{\beta}$ being the
corresponding Fermi factors. The perturbation term $\hat{V}$
corresponds to a uniform electric field in the $x$ direction, i.e.
\begin{equation}
\langle \alpha|\hat{V}|\beta\rangle  = - e E \langle
\alpha|\hat{X}|\beta\rangle .  \label{eq.pert-term}
\end{equation}
Combining Eqs.\,(\ref{e.Power}--\ref{eq.pert-term}) one obtains
for the conductivity
\begin{eqnarray}
\mbox{Re} \sigma(\omega) &=& \frac{2\pi e^2}{\hbar} \sum_{\alpha
\beta} |\langle \alpha|\hat{X}|\beta\rangle |^2
\left(\epsilon_{\alpha} - \epsilon_{\beta}\right) \nonumber
\\
& &\times  \left(f_{\alpha} - f_{\beta} \right)
\delta(\epsilon_{\alpha} + \hbar \omega - \epsilon_{\beta}).
\end{eqnarray}
This expression can be rewritten in terms of the momentum operator
by using the property
\begin{equation}
\langle \alpha|\hat{p}_x|\beta\rangle  = \frac{im}{\hbar} \langle
\alpha|\hat{X}|\beta\rangle \left(\epsilon_{\alpha} -
\epsilon_{\beta} \right),
\end{equation}
where $m$ is the electron mass. Thus, in the limit $\omega
\rightarrow 0$ the conductivity can be written as

\begin{eqnarray}
\sigma(0) &=& \frac{2 \pi e^2 \hbar}{m^2} \int dE \sum_{\alpha
\beta} |\langle \alpha|\hat{p}_x|\beta\rangle |^2
\left(-\frac{\partial f}{\partial E} \right) \nonumber
\\
& & \times \delta(E-\epsilon_{\alpha}) \delta(E -
\epsilon_{\beta}) . \label{e.Kubo}
\end{eqnarray}

Eq.\,(\ref{e.Kubo}) is known as the Kubo formula and relates the
conductivity to the equilibrium current fluctuations for a
homogeneous conductor. In order to analyze the conductance of a
finite sample it is necessary to go back to the non-local
conductivity. This quantity is obtained from (\ref{e.Kubo}) by
replacing $\langle \alpha|\hat{p}_x|\beta\rangle $ by $m\langle
\alpha|\hat{\jmath}_x({\bf r})|\beta\rangle /e$, where
$\hat{\jmath}_x({\bf r})$ is the current operator at position
${\bf r}$, yielding

\begin{eqnarray}
\sigma({\bf r},{\bf r}^{\prime},0) = 2 \pi \hbar \int dE
\sum_{\alpha \beta} \langle \alpha|\hat{\jmath}_x({\bf
r})|\beta\rangle \langle \beta|\hat{\jmath}_x({\bf
r}^{\prime})|\alpha\rangle \nonumber \\
\times \left(-\frac{\partial f}{\partial E} \right)
\delta(E-\epsilon_{\alpha}) \delta(E-\epsilon_{\beta}) .
\label{e.non-local Kubo}
\end{eqnarray}

One can obtain the total current through the sample $I$ by
integrating the current density on an arbitrary transversal
cross-section $\Sigma$ oriented perpendicularly to the current
direction $x$ (see Fig. \ref{f.Electric Field}), i.e.

\begin{equation}
I = \int_{\Sigma} ds J_x({\bf r})  = \int_{\Sigma} ds
\int_{\Omega} d^3r^{\prime} \sigma({\bf r},{\bf r}^{\prime},0)
E_x({\bf r}^{\prime}), \label{e.total current 1}
\end{equation}

Now, the integration over the volume of the sample $\Omega$ in
(\ref{e.total current 1}) can be divided into integration along
the $x$ direction and over a transversal cross-section
$\Sigma^{\prime}$. We thus have

\begin{equation}
I = \int_{\Sigma} ds  \int dx^{\prime} \int_{\Sigma^{\prime}}
ds^{\prime} \sigma({\bf r},{\bf r}^{\prime},0) E_x({\bf
r}^{\prime}), \label{e.total current 2}
\end{equation}

This expression can be simplified further by noticing that the
integration over the transversal cross-section $\Sigma^{\prime}$
does not depend on its position on the $x$ axis. This property
arises from current conservation which implies
\begin{equation}
\int_{\Sigma} ds \langle \alpha|\hat{\jmath}_x({\bf
r})|\beta\rangle  = \int_{\Sigma^{\prime}} ds^{\prime} \langle
\alpha|\hat{\jmath}_x({\bf r}^{\prime})|\beta\rangle .
\end{equation}
One can thus interchange the integration over $x^{\prime}$ and
integration over the transversal cross-section $\Sigma^{\prime}$,
to obtain
\begin{equation}
I = \int_{\Sigma} ds \int_{\Sigma^{\prime}} ds^{\prime}
\sigma({\bf r},{\bf r}^{\prime},0) \int dx^{\prime} E_x({\bf
r}^{\prime}).
\end{equation}
As the integration of the electric field along the $x$ axis gives
the voltage drop along the sample $V$, this expression has the
form $I = G V$, where the conductance $G$ is given by

\begin{equation}
G = \int_{\Sigma} ds \int_{\Sigma^{\prime}} ds^{\prime}
\sigma({\bf r},{\bf r}^{\prime},0). \label{e.conductance-Kubo}
\end{equation}

Notice that, due to current conservation, this expression does not
depend on the position of the transversal cross-sections $\Sigma$
and $\Sigma^{\prime}$ along the $x$ axis. For convenience these
surfaces can be taken well inside the leads, i.e. far away from
the region where the voltage drop is located. In this asymptotic
region the electron states become plane waves along the $x$
direction.

\subsubsection{The conductance in terms of Green functions} \label{sss.Green}

The Kubo formula for the conductivity is usually expressed in
terms of Green functions \cite{mahan90,economou83}. It is then
possible to find an expression for the conductance in terms of
Green functions through (\ref{e.conductance-Kubo}) and establish a
connection between the scattering picture of transport and linear
response theory \cite{lee81,baranger89}.

Let us first introduce the retarded and advanced Green functions
$G^{r,a}({\bf r},{\bf r}^{\prime},E)$ using an eigenstates
representation
\begin{equation}
G^{r,a}({\bf r},{\bf r}^{\prime},E) = \lim_{\eta \rightarrow 0}
\sum_{\alpha} \frac{\psi^*_{\alpha}({\bf r}) \psi_{\alpha}({\bf
r}^{\prime})}{E - \epsilon_{\alpha} \pm i \eta},
\end{equation}
which have the property
\begin{eqnarray}
G^a({\bf r},{\bf r}^{\prime},E) &-& G^r({\bf r},{\bf
r}^{\prime},E)
= \nonumber \\
& & 2 \pi i \sum_{\alpha} \psi^*_{\alpha}({\bf r})
\psi_{\alpha}({\bf r}^{\prime}) \delta(E - \epsilon_{\alpha}).
\end{eqnarray}
Then, Eqs.\,(\ref{e.non-local Kubo}) and
(\ref{e.conductance-Kubo}) allow us to write the conductance $G$
in the form

\begin{eqnarray}
G & = & \frac{e^2 \hbar^3}{8 \pi m^2} \int dE
\left(-\frac{\partial f}{\partial E} \right) \int_{\Sigma} ds
\int_{\Sigma^{\prime}} ds^{\prime} \nonumber \\
& &\left[ \frac{\partial}{\partial x} \frac{\partial}{\partial
x^{\prime}} \left(G^a - G^r\right) \left(\tilde{G}^a - \tilde{G}^r
\right)
\right. \nonumber \\
& & + \left(G^a - G^r\right) \frac{\partial}{\partial x}
\frac{\partial}{\partial x^{\prime}} \left(\tilde{G}^a -
\tilde{G}^r \right) \nonumber \\
& & - \frac{\partial}{\partial x} \left(G^a - G^r\right)
\frac{\partial}{\partial x^{\prime}} \left(\tilde{G}^a -
\tilde{G}^r
\right) \nonumber \\
& & - \left. \frac{\partial}{\partial x^{\prime}} \left(G^a -
G^r\right) \frac{\partial}{\partial x} \left(\tilde{G}^a -
\tilde{G}^r \right) \right], \label{e.GF1}
\end{eqnarray}
where $G^a - G^r = G^a({\bf r},{\bf r}^{\prime},E) - G^r({\bf
r},{\bf r}^{\prime},E)$ and $\tilde{G}^a - \tilde{G}^r = G^a({\bf
r}^{\prime},{\bf r},E) - G^r({\bf r}^{\prime},{\bf r},E)$.

We can simplify this expression considerably by taking the
surfaces $\Sigma$ and $\Sigma^{\prime}$ well inside the left and
the right lead respectively. When ${\bf r}$ and ${\bf r}^{\prime}$
correspond to points well inside the left and right leads the
eigenstates are simple combinations of plane waves. Each plane
wave corresponds to a lead mode $n$ with wave number $k_n$ in the
$x$ direction. In terms of the modes wavefunctions, $\chi_n$, the
Green functions can be expanded as
\begin{equation}
G^{r,a}({\bf r},{\bf r}^{\prime},E) = \sum_{n m}
\chi_m(\vec{\rho}) \chi^*_n(\vec{\rho}^{\prime}) G^{r,a}_{n
m}(x,x^{\prime}),
\end{equation}
where the indexes $n$ and $m$ refer to the modes on the left and
right leads respectively, and $\vec{\rho}$ and
$\vec{\rho}^{\prime}$ indicate the position on the transversal
surfaces. For $x \rightarrow -\infty$ and $x^{\prime} \rightarrow
+\infty$ the Green functions components
$G^{a,r}_{nm}(x,x^{\prime})$ behave as
\begin{equation}
G^{a,r}_{nm}(x,x^{\prime}) \sim e^{\mp i k_n x} e^{\pm i k_m
x^{\prime}} ,
\end{equation}
and in this way the derivatives in (\ref{e.GF1}) can be easily
computed. As a final result one obtains

\begin{equation}
G = \frac{e^2 \hbar^3}{\pi m^2} \int dE \left(-\frac{\partial
f}{\partial E} \right) \sum_{nm} k_n k_m
|G^{a}_{nm}(x,x^{\prime},E)|^2 .
\end{equation}

By comparing with the Landauer formula (\ref{e.Landauer}) the
transmission coefficient in terms of Green functions is the just
given by \cite{lee81}

\begin{equation}
T_{12}(E) = \hbar^2 \sum_{nm} v_n v_m
|G^{a}_{nm}(x,x^{\prime},E)|^2 ,
\end{equation}
where $v_n = \hbar k_n/m$ is the velocity on channel $n$. We
should point out that the above expression is useful to obtain the
transmission for realistic microscopic models. We shall come back
to this point in Sect.\,\ref{s.models}.

\section{Theory for current transport in superconducting point contacts}
\label{s.sctransporttheory}

When one or both electrodes in a point contact are
superconducting, transport properties may be dramatically
affected. For instance, the conductance in
normal-metal--superconductor (N-S) or
superconductor--superconductor (S-S) contacts exhibits very
peculiar non-linear behavior which is associated with the presence
of an energy gap in the excitation spectrum of the superconductor.
In addition, in S-S point contacts one can observe manifestations
of the ground state phase-coherence like the Josephson effect.

In this section we shall review the basic theoretical developments
which allow to understand the transport properties of
superconducting point contacts. It will provide one of the most
powerful approaches for experimentally determining the number of
conductance channels in atomic-sized contacts.

\subsection{The Bogoliubov de Gennes equation and the concept
of Andreev reflection} \label{ss.BdeG}

In a tunnel junction, where the coupling between the electrodes is
exponentially small, one can calculate the current-voltage
characteristic starting from the electron states of the isolated
electrodes and then using first order perturbation theory in the
coupling \cite{bardeen61}. Such calculations predict that the
current should vanish at zero temperature in a N-S junction when
the bias voltage $eV$ is smaller than the superconducting energy
gap $\Delta$. In the same way, they would predict that the current
vanishes for $eV < 2 \Delta$ in a symmetric S-S junction at zero
temperature (see for instance \cite{barone82}).

These type of calculations are, however, not suitable for point
contacts with finite normal conductance. In this case it is in
general necessary to adopt a non-perturbative approach. The
Bogoliubov-de Gennes (BdeG) equations \cite{degennes66},
describing the quasi-particle excitations in non-uniform
superconductors, provide a useful starting point for this case.

In a superconductor the quasi-particle excitations consist of a
mixture of electron-like and hole-like states. The BdeG equations
are two coupled linear differential equations from which the
amplitudes $u(r,E)$ and $v(r,E)$ of an excitation of energy $E$ on
the electron and hole states can be obtained. These equations can
be written as \cite{degennes66}

\begin{eqnarray}
E u(r,E)  =  \left[ -\frac{\hbar^2\nabla^2}{2m} + U(r) - E_F
\right] u(r,E) & &
\nonumber \\  + \Delta(r) v(r,E) & & \nonumber \\
E v(r,E) =  -\left[ -\frac{\hbar^2\nabla^2}{2m} + U(r) - E_F
\right] v(r,E) & & \nonumber \\  + \Delta^*(r) u(r,E), & &
\label{BdeGequations}
\end{eqnarray}
where $U(r)$ and $\Delta(r)$ are effective potentials to be
determined self-consistently \cite{degennes66}. In the case of a
uniform superconductor the BdeG equations can be solved trivially
in terms of plane waves to obtain the well known dispersion law
for quasi-particles in the superconductor
\begin{equation}
E_k = \sqrt{(\epsilon_k-E_F)^2 + \Delta^2},
\end{equation}
where $\epsilon_k = (\hbar k)^2/2m$, and the coefficients

\begin{equation}
u^2_k = \frac{1}{2} \left( 1 + \frac{\epsilon_k-E_F}{E_k}\right)
\hspace{1cm}  v^2_k = \frac{1}{2} \left( 1 -
\frac{\epsilon_k-E_F}{E_k}\right). \label{ukvk}
\end{equation}

The dispersion relation can be inverted to obtain the wavector
modulus in terms of the energy of the excitation
\begin{equation}
\hbar k^{\pm} = \sqrt{2m \left[E_F \pm \sqrt{E^2 -
\Delta^2}\right].}
\end{equation}
The excitations with $k^+$ arise from states above the Fermi
surface in the normal case and are called `quasi-electrons' and
the ones with $k^-$ arise from states below the Fermi surface and
are called `quasi-holes'.

Let us now consider the case of a point-contact between a normal
metal and a superconductor. For simplicity we shall assume that
there is a single channel connecting both electrodes and call $x$
a coordinate along the point contact. We shall also assume that
the pairing potential $\Delta(x)$ exhibits a step-like behavior
from zero to a constant value $\Delta$ at the N-S interface.

Let us consider an incident electron with energy $E$. The wave
functions at each side of the interface can be written as
\begin{eqnarray}
\Psi(x,E) & = & \left( \begin{array}{c} 1 \\ 0 \end{array} \right)
e^{iq^+ x} + r_{eh} \left( \begin{array}{c} 0 \\ 1 \end{array}
\right) e^{iq^- x}
\nonumber \\
& & + r_{ee} \left( \begin{array}{c} 1 \\ 0
\end{array} \right) e^{-iq^+ x} \;\; \mbox{for $x < 0$}
\nonumber \\
\Psi(x,E) &=& t_{ee} \left( \begin{array}{c} u(E) \\ v(E)
\end{array} \right) e^{ik^+ x}
\nonumber \\
& & + t_{eh} \left(\begin{array}{c} v(E) \\ u(E)  \end{array}
\right) e^{-ik^-x}\;\; \mbox{for $x > 0$},
\end{eqnarray}
where $\hbar q^{\pm} = \sqrt{2m (E_F \pm E)}$ and the two elements
in each column vector represent the electron and hole components
of the excitation. The four coefficients $r_{ee}$, $t_{ee}$,
$r_{eh}$ and $t_{eh}$ can be determined by imposing the conditions
of continuity of the wave function and its derivative at the
interface. They describe the four possible processes that can take
place for an incident electron on the N-S interface, i.e.
reflection as an electron, transmission as a quasi-electron,
reflection as a hole and transmission as a quasi-hole. An
additional approximation is to assume that for $E \sim \Delta$ one
has $q^+ \simeq q^- \simeq k^+ \simeq k^-$, which is valid as long
as $\Delta/E_F \ll 1$. This assumption is usually called the
Andreev approximation.

\begin{figure}
\centerline{\includegraphics[width= 6cm] {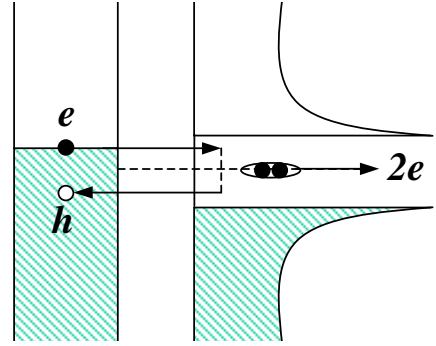}}
\caption{Schematic representation of an Andreev reflection
process.} \label{Andreevrefsch}
\end{figure}

In the simplest case of a perfect N-S interface with no mismatch
in the electrostatic potential one obtains $r_{ee} = t_{eh} = 0$
and
\begin{equation}
r_{eh}(E) =  v(E)/u(E),
\end{equation}
which gives the probability amplitude for reflection as a hole or
{\it Andreev reflection}. Taking into account the expressions for
the coefficients given by (\ref{ukvk}) it is easy to show that for
$E < \Delta$, $r_{eh}$ is just a phase factor given by
\begin{equation}
r_{eh}(E) = \exp{\left[-i \arccos\left(E/\Delta\right)\right]} ,
\label{reh}
\end{equation}
while for $E> \Delta$, $r_{eh}$ decays exponentially. The
probability of Andreev reflection is then equal to 1 for incident
electrons with energy inside the superconducting gap. It should be
noticed that in the whole process two electrons are injected from
the normal electrode, which finally gives rise to a new Cooper
pair in the superconductor as illustrated in
Fig.\,\ref{Andreevrefsch}. This is the basic mechanism for
converting a normal current into a supercurrent at a N-S
interface, first discussed by Andreev in 1964 \cite{andreev64}.

\begin{figure}
\centerline{\includegraphics[width= 8cm] {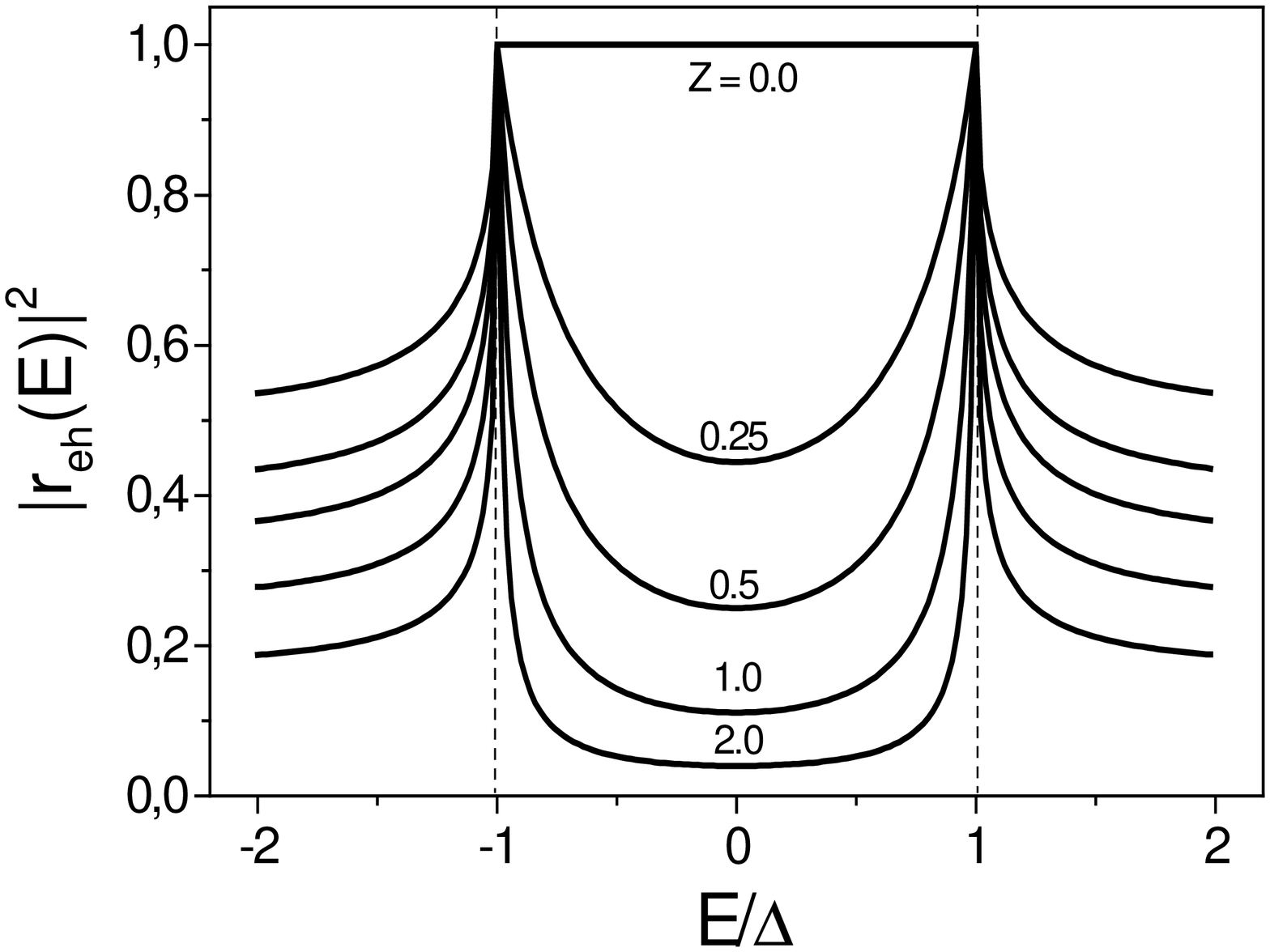}}
\caption{Andreev reflection probability in the model by Blonder,
Tinkham and Klapwijk \protect\cite{blonder82}.}
\label{Andreevreflection}
\end{figure}

In order to analyze the case of an imperfect interface with
arbitrary normal transmission one can use the model introduced by
Blonder, Tinkham and Klapwijk \cite{blonder82}. Within this model
the electron potential at the interface is represented by a delta
function, i.e $U(x) = H \delta(x)$. The boundary conditions on the
wavefunctions should now take into account the presence of this
delta-like potential. A straightforward calculation then yields
for $E<\Delta$ \cite{blonder82}

\begin{equation} |r_{eh}(E)|^2 = \frac{\Delta^2}{E^2 + (\Delta^2 - E^2)(1 + 2 Z)^2},
\end{equation}
where $Z = k_F H/2 E_F$ is a dimensionless parameter controlling
the strength of the delta-like potential. The energy dependence of
the Andreev reflection amplitude for increasing values of $Z$ is
represented in Fig.\,\ref{Andreevreflection}. As can be observed,
for $Z=0$ one recovers the case of a perfect interface with
$|r_{eh}(E)|^2=1$ inside the gap, while for $Z \rightarrow
\infty$, $r_{eh}(E) \rightarrow 0$ inside the gap, as expected for
a tunnel junction.

These results can be used to obtain the linear conductance of a
N-S quantum point. For this purpose, the scattering approach
described in Sect.~\ref{s.transport_theory} has to be extended in
order to incorporate Andreev reflection processes (see for
instance \cite{beenakker92}). At zero temperature and for a single
conduction channel one obtains
\begin{equation}
G_{NS} = \frac{4e^2}{h} |r_{eh}(0)|^2,
\end{equation}
where the factor 4 instead of the usual 2 in the Landauer formula
reflects the fact that two electrons are transmitted in an Andreev
reflection process. This expression has been generalized to the
multi-channel case by Beenakker \cite{beenakker92} who obtained
\begin{equation}
G_{NS} = \frac{4e^2}{h} \sum_n \frac{\tau^2_n}{(2 - \tau_n)^2},
\label{GNSmultichannel}
\end{equation}
where $\tau_n$ denotes the transmission probability on the $n$-th
conduction channel. Eq. (\ref{GNSmultichannel}) shows that the
normal conduction channels are not mixed by Andreev reflection
processes, i.e. an electron incident on a given conduction channel
is reflected as a hole on the {\it same} channel. This is
reasonable since the energy scale for superconducting correlations
is much smaller than the energy between the bottoms of the
conductance channels. We shall come back to this property when
analyzing the case of biased SNS contacts (Sect.\,\ref{ss.SNS2}).

\subsection{SNS contacts at zero bias}
\label{ss.SNS1}

In a similar way as in the previous subsection one can use the
BdeG equations to analyze the excitation spectrum in a SNS
structure. Let us consider the case of two superconducting
electrodes connected by a single normal channel of length $L$. For
simplicity let us first consider the case of perfect matching at
both interfaces. An electron in the normal region with energy $E <
\Delta$ will be Andreev reflected into a hole when reaching one of
the interfaces and, conversely, this hole will be converted into
an electron with the same energy at the other interface. The
combination of these processes gives rise to an infinite series of
Andreev reflections. The resonance condition can be easily
established by taking into account the amplitude for an individual
Andreev reflection given by Eq. (\ref{reh}) and the phase
accumulated by electrons and holes when propagating through the
normal region, which yields

\begin{equation}
E = \Delta \cos{\left\{\frac{\varphi}{2} - \frac{k_FL}{2}(E/E_F) -
n\pi \right\}}, \label{boundstateI}
\end{equation}
where $\varphi$ is the phase difference between the
superconducting electrodes (notice that the superconducting phase
enters into the Andreev reflection amplitude (\ref{reh}) simply as
an additional phase factor). This equation admits multiple
solutions which correspond to bound states inside the
superconducting gap, usually  called Andreev states or
Andreev-Kulik states \cite{kulik69}. In the limit $L \ll \xi_0$,
$\xi_0 = \pi v_F/\Delta$ being the superconducting coherence
length, only two bound states are found at energies
\cite{furusaki90}
\begin{equation}
E = \pm \Delta \cos{\left\{\frac{\varphi}{2}\right\}}.
\end{equation}

For finite phase difference these states carry a Josephson current
with opposite directions that can be computed using the
thermodynamic relation $I \sim dF/d\varphi$, where $F$ is the free
energy of the system. This relation yields $I \sim
\sin{\varphi/2}$ for perfect interfaces, a result which was first
derived by Kulik and Omelyanchuk \cite{kulik78} in 1978. For a
quantum point contact with an smooth (or adiabatic) geometry
accommodating $N$ conduction channels the current-phase relation
is given by \cite{beenakker91a}

\begin{equation}
I(\varphi) = \frac{Ne\Delta}{\hbar}
\sin{\left(\frac{\varphi}{2}\right)} \tanh{\left(\frac{E}{2 k_{\rm
B}T}\right)}.
\end{equation}

The case of a SNS structure with arbitrary normal transmission can
also be analyzed using the BdeG equations. For a single conduction
channel in the limit $L \ll \xi_0$ one obtains two bound states at
energies \cite{beenakker91a}

\begin{equation}
E = \pm \Delta \sqrt{1 - \tau
\sin^2{\left(\frac{\varphi}{2}\right)}}, \label{andreevstates}
\end{equation}
where $\tau$ is the normal transmission. From this expression the
current-phase relation turns out to be
\begin{equation}
I(\varphi) = \frac{e \tau \Delta}{2 \hbar}\frac{\sin(\varphi)}
{\sqrt{1 - \tau \sin^2{\left(\frac{\varphi}{2}\right)}}}
\tanh{\left(\frac{E}{2 k_{\rm B}T}\right)} . \label{ivsphi}
\end{equation}

This result has been obtained by several authors using different
techniques \cite{beenakker92,haberkorn78,arnold85,martinrodero94}.
It interpolates between the tunnel limit where the behavior $I
\sim \sin(\varphi)$ first predicted by Josephson
\cite{josephson62} is recovered, to the perfect transmission limit
where the $\sin(\varphi/2)$ behavior of Kulik-Omelyanchuk is
recovered. The maximum supercurrent at zero temperature as a
function of the contact transmission is given by \cite{levy95}
\begin{equation} I_{max} = \frac{e\Delta}{\hbar}\left[ 1- \sqrt{1- \tau}\right] .
\end{equation}

In superconducting quantum point-contacts the effect of thermal
fluctuations on the supercurrent can be rather large. Thermal
noise in a phase polarized contact has been calculated Averin and
Imam \cite{averin96} for the case of perfect transmission and by
Mart\'{\i}n-Rodero \ea \cite{martinrodero96} for the case of
arbitrary transmission.

\subsection{SNS contacts at finite bias voltage}
\label{ss.SNS2}

When applying a finite voltage the phase difference in a SNS
contact increases linearly with time according to the Josephson
relation, i.e. $\varphi = \omega_0 t$, where $\omega_0 =
2eV/\hbar$ is the Josephson frequency. The response of the system
cannot in general be described by Eq. (\ref{ivsphi}) in an
adiabatic approximation due to the fact that excitations of
quasiparticles come into play and can give the main contribution.
The total current through the contact contains all the harmonics
of the Josephson frequency and can be written as
\cite{averin95,cuevas96}
\begin{equation}
I(V,t) = \sum_n I_n(V) e^{in\omega_0t}.
\end{equation}
In this decomposition one can identify a dissipative part, which
is an odd function of $V$, and a non-dissipative part, which is an
even function of $V$. These two parts are given by \cite{cuevas96}
\begin{equation}
I_D = I_0 + 2 \sum_{m>0} \mbox{Re}(I_m) \cos{(m\omega_0t)},
\end{equation}
and
\begin{equation}
I_S = -2 \sum_{m>0} \mbox{Im}(I_m) \sin{(m\omega_0t)}.
\end{equation}
The quantity which is more directly accesible by experiments is
the dc component $I_0$, which we will discuss in more detail
below.

\begin{figure}
\centerline{\includegraphics[width= 8cm] {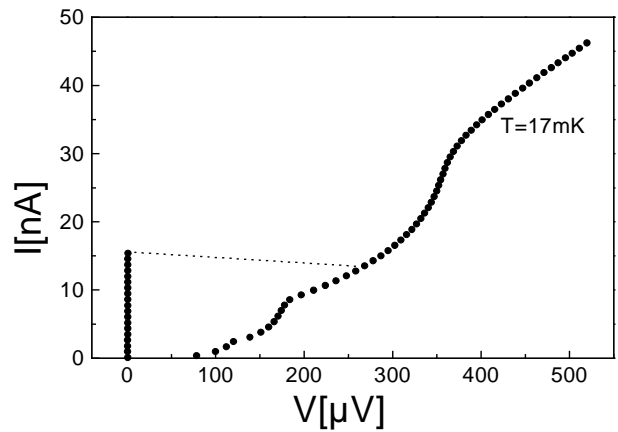}}
\caption{Current-voltage characteristic for an Al atomic contact
at 17\,mK. As the current is ramped up the contact is seen to
switch from the supercurrent branch near zero voltage to the
finite-voltage dissipative branch, which is highly non-linear.
Reprinted with permission from \protect\cite{goffman00}.
\copyright 2000 American Physical Society.} \label{f.expsgs}
\end{figure}

As an example, the measured current voltage characteristic of an
Al atomic contact at 17\,mK is shown in Fig.\,\ref{f.expsgs}, and
the experiment will be discussed more extensively in
Sect.~\ref{s.superconductors}. In this curve one can clearly
distinguish the supercurrent branch at zero voltage. On the other
hand, the dissipative branch exhibits a very peculiar structure
for bias voltages smaller than $2 \Delta/e$, which is called the
{\it subgap structure} (SGS). The SGS consists of a series of more
or less pronounced current jumps located at $eV = 2\Delta/n$ This
structure cannot be understood in terms of single quasi-particle
processes.

The first observations of the SGS were made in tunnel junctions by
Taylor and Burstein \cite{taylor63} and Adkins \cite{adkins63} in
1963. Since then, there have been several attempts to explain this
behavior theoretically. In the early 60's Schrieffer and Wilkins
\cite{schrieffer63} proposed a perturbative approach which is
known as multi-particle tunneling theory. In this approach the
current is calculated up to second order perturbation theory in
the tunneling Hamiltonian. Although it explains the appearance of
a current step at $eV = \Delta$ the theory is pathological as the
current becomes divergent at $eV = 2\Delta$. Another theory,
proposed by Werthamer in 1966, suggested that the SGS is due to
the coupling of the tunneling electrons with the Josephson
radiation \cite{werthamer66}. This theory predicted a different
behavior for even and odd $n$, which is not observed
experimentally.

\begin{figure}
\centerline{\includegraphics[width= 8cm] {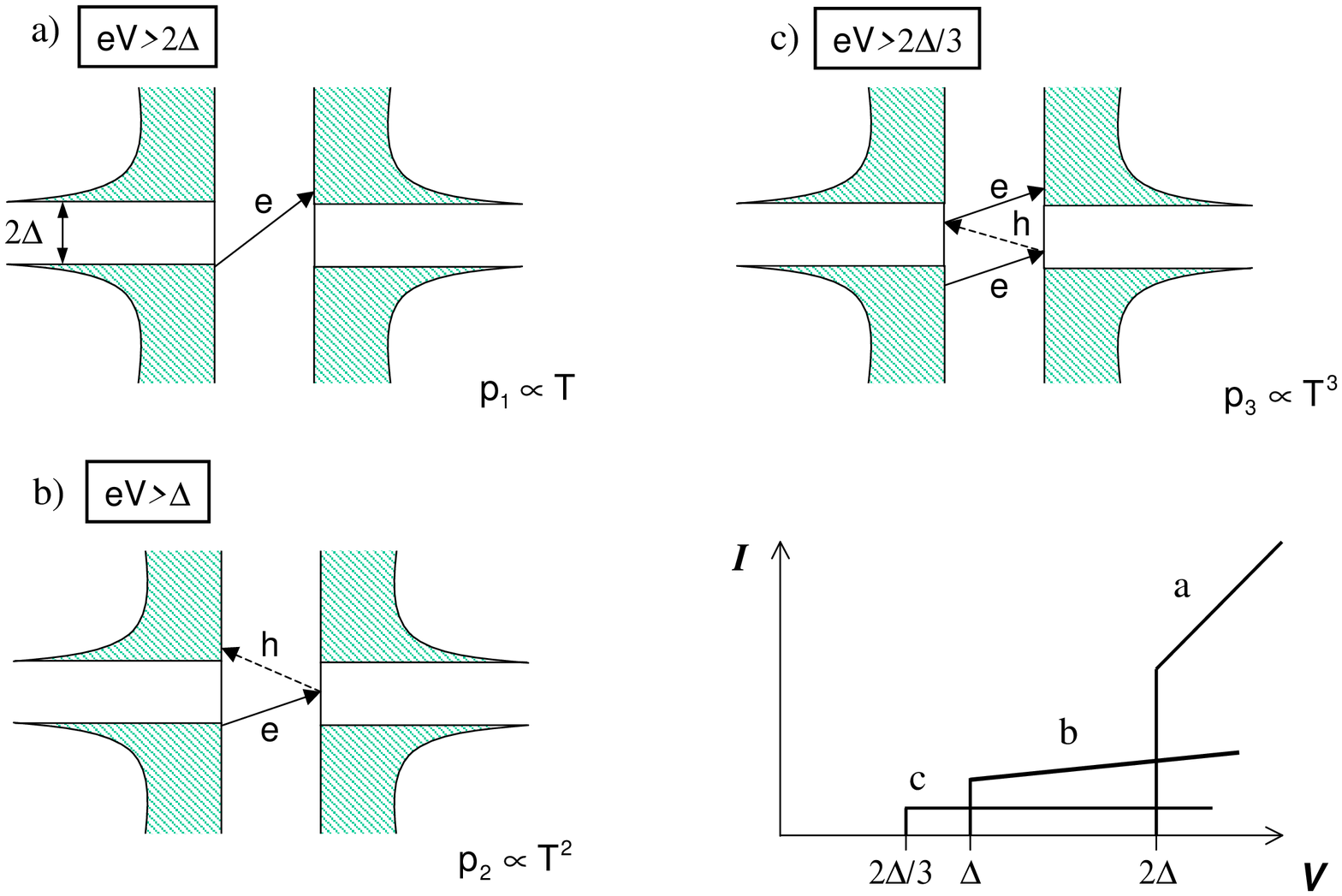}}
\caption{Schematic explanation of the subgap structure in
superconducting contacts.} \label{explanationSGS}
\end{figure}

There is nowadays clear evidence that the SGS can be understood in
terms of multiple Andreev reflections (MAR), as first proposed by
Klapwijk, Blonder and Tinkham \cite{klapwijk82}. A qualitative
explanation of the SGS in superconducting contacts in terms of MAR
is illustrated in Fig.\,\ref{explanationSGS}. Single
quasi-particle processes as those illustrated in
Fig.\,\ref{explanationSGS}(a) are only possible when $eV >
2\Delta$ in order for the transferred quasi-particle to reach the
available states on the right superconductor. The probability of
these type of processes is proportional to the contact
transmission and they give a contribution to the current-voltage
relation as indicated schematically in the lower right panel in
Fig.\,\ref{explanationSGS}. At lower bias voltages current is
possible due to Andreev processes. The simplest Andreev process is
depicted in Fig.\,\ref{explanationSGS}(b), where two
quasi-particles are transmitted with a probability proportional to
the square of the normal transmission, creating a Cooper pair on
the right side. These processes give a contribution to the $IV$
with a threshold at $eV = \Delta$. At even lower biases higher
order Andreev processes can give a contribution. In general, a
$n$-th order process in which $n$ quasiparticles are transmitted
gives a contribution proportional to the $n$-th power of the
transmission and with a threshold at $eV = 2\Delta/n$.

The first semi-quantitative determination of the SGS in terms of
MAR was presented by Klapwijk, Blonder and Tinkham
\cite{klapwijk82} for the case of a one-dimensional SNS structure
with perfect interfaces and generalized by Octavio \ea
\cite{octavio83} for the case of arbitrary transparency. These
last calculations were based on a Boltzmann kinetic equation
corresponding to a semi-classical description of transport. Fully
microscopic calculations using Green function techniques were
first presented by Arnold in 1987 \cite{arnold87}.

A deeper and more quantitative insight into the SGS has been
obtained in the last decade by analyzing the case of single-mode
contacts \cite{averin95,cuevas96,bratus95}. In 1995 Bratus \ea
\cite{bratus95} and Averin and Bardas \cite{averin95} calculated
the current in a voltage biased superconducting contact by
matching solutions of the time-dependent BdeG equations with
adequate boundary conditions (scattering approach). While Bratus
\ea analyzed the low transmission regime, Averin and Bardas
calculated the $IV$ curves for arbitrary transmission. On the
other hand, Cuevas \ea \cite{cuevas96} used a Hamiltonian approach
together with non-equilibrium Green function techniques to obtain
the current in a one channel contact with arbitrary transparency.
The results of the scattering and the Hamiltonian approaches have
been shown to be equivalent \cite{cuevas96}. Moreover, the
approach presented in \cite{cuevas96} demonstrates that the
pathologies of the multi-particle tunneling theory of Schrieffer
and Wilkins disappear when calculations are performed up to
infinite order in the tunneling Hamiltonian.

\begin{figure}
\centerline{\includegraphics[width= 8cm] {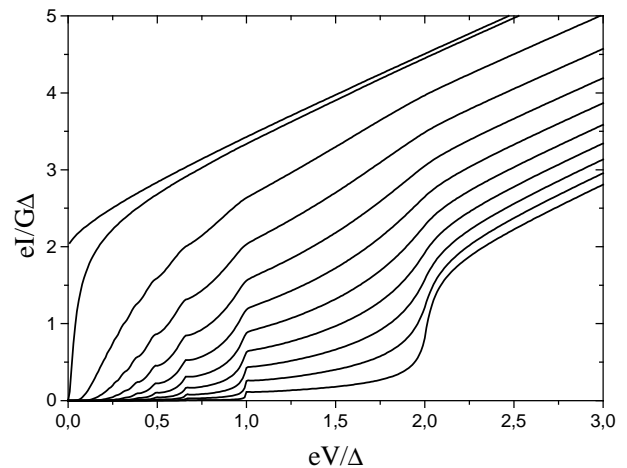}}
\caption{The zero-temperature dc component of the current in a
single mode superconducting contact, where the values of the
normal transmission increase from $\tau = 0.1$ in the lower curve
by increments of 0.1. The upper two curves correspond to
$\tau=0.99$ and $\tau = 1$. Reprinted with permission from
\protect\cite{cuevas96}. \copyright 1996 American Physical
Society.} \label{sgsteo}
\end{figure}

The theoretical results for the dc component of the current in a
single-mode contact for a number of transmission values are
presented in Fig.\,\ref{sgsteo}. The two relevant features of
these curves are the SGS for $eV < 2\Delta$ and the excess current
with respect to the normal case which is found for $eV \gg
\Delta$. As can be observed, the SGS is very pronounced at low
transmission and becomes smoother when the transmission is close
to 1. At perfect transmission the structure in the $IV$ curve
completely disappears and the excess current reaches the value $e
\Delta/\hbar$.

In the limit of small bias voltages the transport properties of
the contact can be understood in terms of the non-equilibrium
population of the Andreev states given by (\ref{andreevstates})
\cite{averin95,cuevas96}. The theory predicts a crossover from
supercurrent to dissipative current which takes place for $eV \sim
\Delta \sqrt{1-\tau}$. At larger biases the two Andreev states can
be connected by Landau-Zener transitions \cite{averin95}.

More recently shot noise in a single channel superconducting
contact has been studied \cite{cuevas99,naveh99}. It has been
demonstrated that the effective charge, defined as the ratio
between noise and current, i.e. $q^* = S/2I$, is quantized in the
limit of low transmission and increases as $1 + {\rm
Int}(2\Delta/V)$ for decreasing bias voltage. This prediction is
consistent with the fact that the current is due to transfer of
multiple quasi-particles mediated by Andreev processes as
schematically depicted in Fig.\,\ref{explanationSGS}.

For the purpose of this review the most important result is the
fact that the theory produces current-voltage curves that have a
very characteristic shape, which exclusively depends on the
transmission probability $\tau$. The results obtained above can be
easily generalized to the multichannel case. As in the N-S case,
one may assume that the conduction channels are the same as in the
normal state, i.e. that Andreev reflection processes do not mix
the normal channels. This is true as far as one can neglect the
energy dependence of the normal scattering matrix within the
superconducting gap region. The validity of this approximation
will be discussed in Sect.~\ref{s.exp_modes}. Further, the theory
is exact and can therefore be exploited to extract the
transmission probabilities and the number of relevant conduction
channels from the experiment. The analysis of experimental
$IV$-curves for superconducting atomic-sized contacts will be
presented in Sect.~\ref{s.exp_modes}.

\subsection{Current biased contacts}
\label{ss.RSJ}

In practice it is not possible to impose a constant voltage at
very low bias ($eV \ll \Delta$). In this limit the effective
impedance of the contact tends to zero due to the presence of the
supercurrent branch and the impedance of the circuit in which the
contact is embedded starts to play an important role. A way to
study this limit in detail is to impose a dc current bias through
the contact shunted with an external resistance $R$. This is the
equivalent to the `Resistively Shunted Junction' (RSJ) model well
studied in connection with superconducting tunnel junctions. As in
that case, provided that the contact capacitance is large enough,
the dynamics of the phase in this configuration is analogous to
the motion of a massless particle in a tilted `wash-board'
potential, governed by a Langevin equation. However, in the case
of an atomic contact the potential is not the usual sinusoid but
has a more general form, which depends on the occupation of the
Andreev states, given by \cite{averin96}

\begin{equation}
U_{p}=-I_{b}\,\varphi
+\sum\nolimits_{i=1}^{N}(n_{i+}-n_{i-})E(\varphi ,\tau _{i})
\end{equation}
where $I_{b}$ is the imposed current bias,
$E(\varphi,\tau)=-\Delta \sqrt{1 - \tau \sin^2(\varphi)}$ gives
the lower Andreev level energy in a quantum channel of
transmission $\tau$ and $n_{i\pm}$ denote the occupation of the
upper and lower Andreev states in the $i$-th quantum channel. The
Langevin equation which determines the phase evolution is given by

\begin{equation}
\frac{d\varphi}{dt} = -\frac{\partial U_p}{\partial \varphi} +
\frac{2eR}{\hbar} i_n(t)
\end{equation}
where $i_n(t)$ is a fluctuating current arising from thermal noise
in the shunting resistor. The noise intensity should satisfy the
fluctuation-dissipation theorem.

As a final ingredient one has to impose thermal equilibrium
population of the Andreev states at $\varphi = 2 n \pi$ as
boundary conditions. The corresponding Langevin equations can be
solved either by direct numerical simulation or by mapping the
problem onto a Fokker-Planck equation, following a procedure
introduced by Ambegaokar and Halperin \cite{ambegaokar69} for the
traditional RSJ model. The results of this model and the
comparison with experimental measurements have been presented by
Goffman \ea \cite{goffman00} and will be discussed in
Sect.~\ref{s.superconductors}.

\section{The conductance of atomic-sized metallic contacts: experiment} \label{s.conductance}

As we have seen above, in a conductor for which the dimensions are
much smaller than the phase coherence length the linear
conductance, $G$, is given in terms of the Landauer expression
\begin{equation}
G ={2e^2\over h} {\sum_{n} \tau_n} , \label{eq.landauer}
\end{equation}
where the $\tau_n$ describe the transmission probabilities for
each of the eigenmodes of the  conductor and the sum runs over all
occupied modes.  We have taken the expression in the limit of
$T\rightarrow 0$, which is a good approximation for metals since
the subband splitting is typically of the order of an
electronvolt. If we can contrive our experiment in such a way that
the $\tau_n$ are equal to 1, up to a mode number $N$, and equal to
zero for all other modes, then the conductance assumes values
which are an integer times the quantum unit of conductance,
$G_0=2e^2/h$.  It turns out to be possible to fabricate conductors
which have precisely this property, using a two-dimensional
electron gas (2DEG) semiconductor device, as was beautifully
demonstrated in the seminal experiments by van Wees \ea\
\cite{wees88} and Wharam \ea\ \cite{wharam88}.

For atomic-sized metallic contacts the Landauer expression is also
applicable and the number of channels involved is expected to be
small. This can be judged from the fact that the Fermi wavelength
in metals is of the order of the atomic diameter, $\lambda_F\simeq
5$~\AA\ .  This implies that, while the 2DEG experiments require
cooling to helium temperatures in order to be able to resolve the
splitting of $\sim$1~meV between the quantum modes, in metals the
mode splitting is ${\pi^2\hbar^2}/2 m \lambda_F^2\sim$1~eV, which
is sufficiently high to allow the observation of quantum effects
at room temperature. The number of relevant conductance channels
in a single-atom contact can be estimated as $N=({k_F a/2})^2$.
For copper we have $(k_Fa/2)^2=0.83$, which is close to 1,
suggesting that a single atom corresponds to a single conductance
channel. This implies that the atomic granularity will limit our
ability to smoothly reduce the contact size, in order to directly
observe quantum effects in the conductance. Moreover, a priori we
cannot expect the conductance for metallic atomic-sized contacts
to be given by simple multiples of the conductance quantum. The
wave function of the electrons inside the contact will resemble
the atomic wave functions, and the matching of these to the wave
functions in the leads will critically influence the transmission
probabilities for the quantum modes. The atomic structure of the
contact and the composition of the electronic quantum modes will
be interwoven, and we will now discuss the experimental
observations and what we can conclude from them.

\subsection{Contact making and breaking}\label{ss.steps&plateaus}
\begin{figure}[!t]
\centerline{\includegraphics[width=8cm]{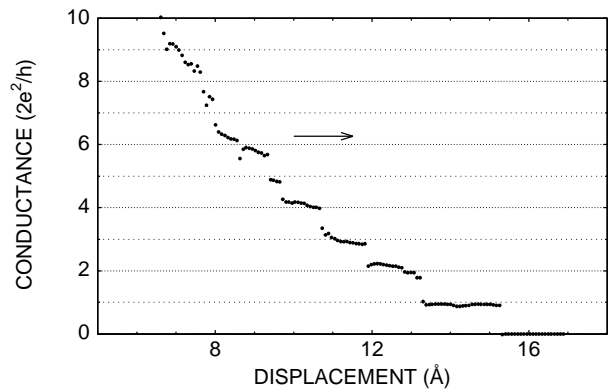}}
\caption{Conductance curve for a gold contact measured at room
temperature in UHV by pressing and STM tip into a clean gold
surface and recording the conductance while retracting the tip.
The time for recording the curve was approximately 20~ms. Both the
current and voltage were measured since both varied during the
experiment, but the voltage was smaller than 3~mV until entering
the tunneling regime. Reprinted with permission from
\protect\cite{brandbyge95}. \copyright 1995 American Physical
Society.} \label{f.goldstepsRT}
\end{figure}

\begin{figure}
\centerline{\includegraphics[width=8cm]{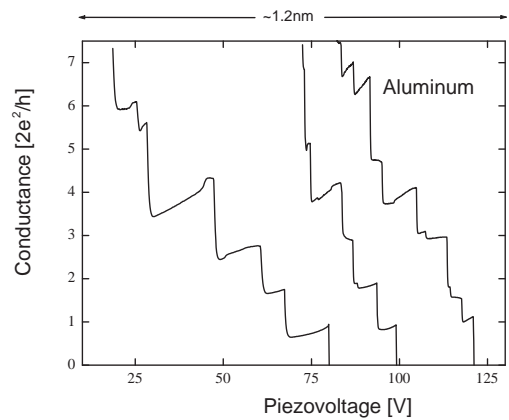}}
\caption{Three examples of conductance curves for aluminum
contacts measured at 4.2\,K as a function of the piezovoltage. In
most cases the last ``plateau'' before contact is lost has an
anomalous positive slope. An approximate scale for the range of
the displacement is indicated at the top of the graph. Reprinted
with permission from \protect\cite{krans93}. \copyright 1993
American Physical Society.}
\label{f.aluminum_curve}
\end{figure}

The experimental techniques most frequently employed in the study
of conductance of atomic-sized contacts involve mechanically
driven breaking and making cycles of a contact between two metal
electrodes. Examples of conductance curves for gold at low
temperatures were already shown in Sect.\,\ref{s.fabrication},
Fig.\,\ref{f.STM_curve_example}, when we discussed the various
experimental techniques. A further example of a conductance curve
measured on a gold sample at room temperature is shown in
Fig.\,\ref{f.goldstepsRT}.  The curves are recorded while breaking
the contact and, although the conditions for the two experiments
are widely different, the main features in the curves are similar.
We recognize a series of plateaus in the conductance, which are
nearly horizontal for the lowest conductance values, but have a
negative slope for larger contact size. At the end of a plateau a
sharp jump is observed, at which the conductance usually decreases
by an amount of order of the quantum unit, \g0. Although one is
inclined to see a coincidence of the plateaus with multiples of
\g0\  this coincidence is far from perfect, and in many cases
clearly absent. A marked exception to this rule is the position of
the last conductance plateau before contact is lost. For gold this
plateau is reproducibly found very near 1\g0. The latter property
is generally observed for monovalent metals (Cu, Ag and Au, and
the alkali metals Li, Na and K). For $sp$ and $sd$-metals the
plateaus are generally less regularly spaced, and the last plateau
before tunneling can be a factor of two or more away from 1\g0. An
example for the $sp$-metal aluminum is shown in
Fig.\,\ref{f.aluminum_curve} and for the $sd$-metal platinum in
Fig.\,\ref{f.MullerPt}. The behavior is much less regular than for
gold and in many regions the conductance is even seen to rise
while pulling the contact. Similar anomalous slopes are also seen
for other metals, an explanation for which will be presented in
Sect.\,\ref{ss.strain}.

Let us now discuss the problem of the interpretation of the sharp
steps that are observed in the conductance
(Figs.~\ref{f.STM_curve_example}, \ref{f.MullerPt},
\ref{f.goldstepsRT}, \ref{f.aluminum_curve}). As was pointed out
in some of the early papers \cite{muller92,todorov93a}, the
dynamics of the contact conductance around the steps strongly
favors an interpretation in terms of atomic rearrangements, which
result in a stepwise variation of the contact diameter. This
interpretation has been corroborated by a number of recent
experiments and classical molecular dynamics simulations have
helped to visualize the atomic rearrangements involved. The
molecular dynamics calculations of mechanics, and model
calculations of force and conductance will be discussed in
Sect.\,\ref{s.models}. The most direct experimental evidence comes
from simultaneous measurements of the conductance and the force in
the contact, which is to be discussed in
Sect.\,\ref{s.mechanical}. However, already from the dynamic
behavior of the conductance around the steps we can obtain very
strong evidence for this interpretation, and we will first review
this experimental evidence.

\begin{figure}
\centerline{\includegraphics[width=8cm]{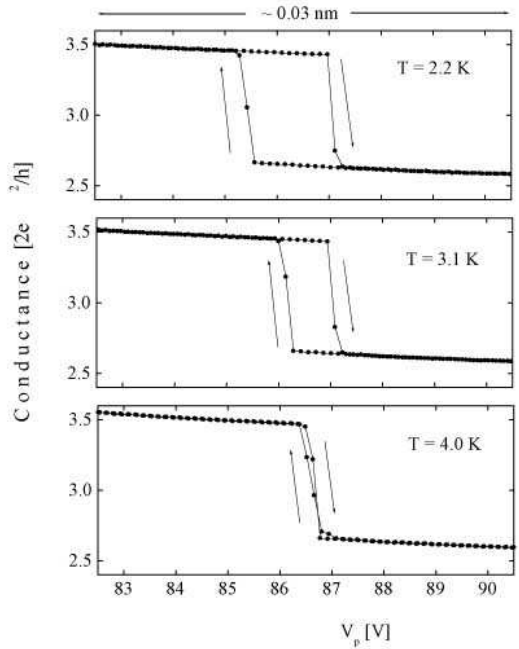}}
\caption{Measurement of the conductance for an atomic-sized Au
contact while sweeping the piezo voltage forward and backward
around a single step. Clear hysteresis of the order of 0.1~\AA\ is
observed at 2.2~K. At 3.1~K the width of the hysteresis is
reduced, and at 4.0~K it is entirely suppressed. Courtesy M. Krans
\protect\cite{krans96b}. }
\label{f.hysteresis}
\end{figure}

The first observation is that the steps are so steep that the
slopes cannot be resolved in the experiment on a time scale of the
order of 1~ms, suggesting that they correspond to very fast jumps
in the conductance. Further, the conductance traces are different
in each measurement. This is to be expected, since we do not
control the structural arrangement of the atoms in the contact,
and each time the contact is pressed together and slowly separated
again, the shape of the contact evolves through a different
sequence of structures. However, in the low temperature
experiments reproducible cycles can be obtained after `training'
the contact by repeatedly sweeping the piezovoltage, $V_p$, over a
limited range \cite{agrait94,brom98}. In such cycles, the steps
often show hysteresis in the position at which they occur for the
forward and backward sweeps. The hysteresis is sensitive to the
bath temperature of the experiment, and could in some cases be
removed by raising the temperature by only a few Kelvins
\cite{brom98,krans96b} (see Fig.\,\ref{f.hysteresis}).
For other steps, such hysteresis was not observed, but instead the
conductance showed spontaneous fluctuations of a two-level type,
between the values before and after the step. This phenomenon is
observed only in a very narrow range of  $V_p$ around the steps;
at the plateaus the conductance assumes stable values. These
two-level fluctuations (TLF) are discussed in more detail in
Sect.\,\ref{sss.heating}, where it is shown that the properties of
the system are consistent with two configurations of the contact
being nearly equal in energy, and the energy barrier separating
the two states can be overcome by thermal activation. When the
barrier is too high for thermal activation to be observable, we
observe hysteresis of the conductance steps, which can be viewed
as being the result of tilting of the potential of the two-level
system by the strain applied to the contact. For gold van den Brom
\ea \cite{brom98} investigated about one hundred individual steps,
all of which showed either hysteresis or TLF, and the steps were
always steep, with the slope limited by the experimental
resolution.

All these observations are clearly not consistent with a smoothly
varying contact radius and steps resulting directly from the
quantization of the conductance. A natural interpretation, rather,
is formed by a model that describes the contact as having a stable
atomic geometry over the length of a plateau in the conductance,
where the total energy finds itself in a local minimum. At the
jumps in the conductance, the local energy minimum for a new
geometry drops below that of the present state as a result of the
stress applied to the contact. When the energy barrier between the
old and the new minimum is large compared to the thermal energy,
we will observe hysteresis, and when it is small enough thermally
activated fluctuations between the two states will be observed.

The lengths of the plateaus are quite irregular in the
low-temperature experiments. For Au at room temperature the
elongation of the contact appears to be often more regular and
Marszalek \ea \cite{marszalek00} have shown that the lengths of
the plateaus can be classified into three groups of about 0.175,
0.335 and 0.535\,nm. They propose that that this result can be
interpreted in terms of an elongation process of a series of slip
events along \{111\} planes. The mechanical deformation properties
of atomic contacts will be further discussed in
Sect.\,\ref{s.mechanical}.

The height of the steps is of the order of the conductance
quantum, but no systematic correlation between the position of the
plateaus and quantized values can be discerned, except for the
last two or three conductance plateaus before the jump to the
tunneling regime, and only for simple metals. Again, this is
consistent with the picture of atomic rearrangements, where the
contact changes size by approximately the area of one atom, and as
we have seen, one atom contributes $\sim$1~\g0 to the conductance,
even in the semiclassical approximation.

This should be different for semimetals, such as antimony and
bismuth, where the density of electrons is more than three orders
of magnitude smaller than in ordinary metals. From the electron
density in Sb one estimates a Fermi wavelength of 55\,\AA, which
is much larger than the atomic diameter. Experiments on point
contacts for this semimetal \cite{krans94} show indeed that the
jump from vacuum tunneling to contact is found at
$\sim$1\,M$\Omega$, or $\sim$0.01\g0. Continuing to increase the
contact, steps in the conductance were again observed, with a step
height that is also of order 0.01\g0. This is consistent with the
notion of steps resulting from atomic rearrangements and with a
semi-classical estimate for the conduction of one atom of Sb. An
interpretation of the steps in terms of conductance quantization
is definitely ruled out.

We would like to stress again that there is an important
difference in experiments performed at room temperature compared
to those performed at low temperatures. The detailed study of
hysteresis and TLF behavior for individual steps is only possible
at low temperatures, where the drift of the contact is
sufficiently low. Further, it is important to guarantee a clean
metallic contact. At low temperatures this is a simple procedure.
Either by employing the MCBJ technique or by repeated indentation
of an STM tip into the sample \cite{agrait96} clean contact
surfaces can be exposed, which remain clean for very long times
due to the cryogenic vacuum conditions. Clean contacts are seen to
form a neck during stretching of the contact. In contrast, when a
surface is contaminated and the tip is pushed into it, the surface
is indented and contact is only formed upon deep indentation. Even
under UHV conditions the later is frequently observed, as was
already noted by Gimzewski \ea \cite{gimzewski87}. A conductance
well below 1\,\g0 often observed for atomic-sized metallic
contacts in air should in our view be attributed to contamination.
Even Au is very reactive for low coordination numbers of the
atoms, and the claim of the observation of localization in Au
nanowires \cite{pascual95} should be judged in this context.

Many authors have attempted to argue that the conductance values
at the plateaus should be strictly quantized. Deviations from
perfect quantization are then primarily attributed to back
scattering on defects near the contact. An ingenious argument was
presented by de Heer \ea \cite{deheer97} attempting to show that
the conductance for gold contacts can be described as effectively
being due to two quantized conductors in series. However, in our
view the interpretation hinges on the assumption that the numbers
of channels into which an electron scatters are well-defined
integer numbers, which is not expected. An analysis of defect
scattering near the contact is presented in
Sect.\,\ref{s.defect_scattering}.

Kassubek \ea \cite{kassubek01} even consider the possibility that
the dynamic behavior of the contact around the steps can be
explained by considering the sensitivity to shape distortions of
the total energy for a free-electron system confined to a
cylindrical wire. The model will be discussed further in
Sect.\,\ref{ss.force_models}. Although it is very interesting to
investigate how far one may go with this `minimal model', it seems
more natural to take the constraints to the contact size imposed
by the atomic granularity as the fundamental cause for the jumps
in the conductance. The fact that the dynamical behavior of the
steps, showing hysteresis and TLF, is very similar for monovalent
metals, which may be favorable for observing quantum induced force
jumps, as well as for $p$- and $d$-metals (where this is excluded
as we will see) argues strongly in favor of this point of view.

It is instructive to single out a specific jump in the
conductance, namely the one between contact and tunneling. This
will help to illustrate the points discussed above, and we will
show how we can distinguish the contact regime from the tunneling
regime of a junction in the experiments.

\subsection{Jump to contact}
\label{ss.jump-to-contact}

At sufficiently large distances between two metal surfaces one
observes a tunnel current that decreases exponentially with the
distance and that depends on the workfunction of the metal
according to Eq.~(\ref{eq.exponential_tunneling}). As the two
metal surfaces are brought closer together at some point a jump to
contact occurs, as first observed by Gimzewski and M\"oller
\cite{gimzewski87}. Once in contact this is followed by a
staircase-like conductance curve due to the atomic structure, as
described above. Thus, we will take this characteristic
distinction of smooth exponential distance dependence versus
staircase-like structure as the signatures for tunneling and
contact regimes, respectively. In some cases we will see that the
distinction is not as clear-cut. For clean metals the first
contact is always of the order of 10\, k$\Omega$. Only clean
contacts form this adhesive jump after which a connective neck is
pulled when separating the electrodes. After only 6 hours in UHV
the tip was seen to be sufficiently contaminated that indenting it
in a silver surface caused a dip in the surface rather than
pulling a neck, due to a repulsive interaction caused by
adsorbates \cite{gimzewski87}.

\begin{figure}[!t]
\begin{center}
\includegraphics[width=80mm]{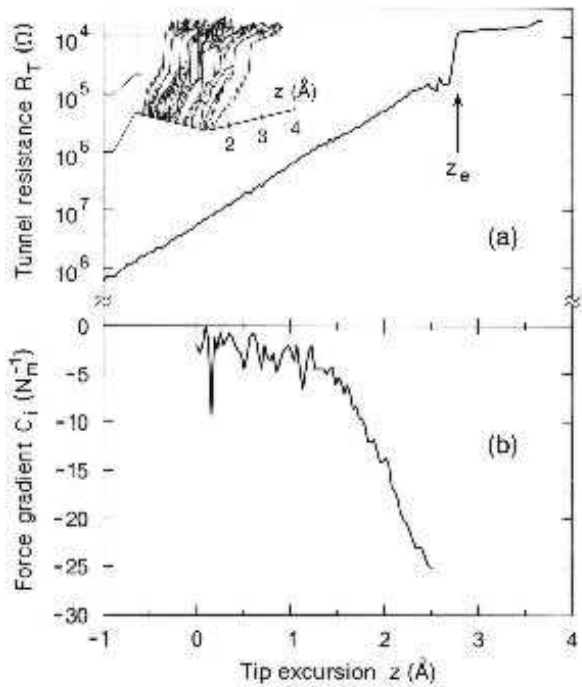}
\end{center}
\caption{(a) Tunnel resistance vs. tip excursion $z$ measured on
an Ir surface using an Ir tip at room temperature in UHV, averaged
over 64 cycles. Positive $z$ corresponds to decreasing the
tunneling-gap width. The transition to contact is indicated by the
arrow at $z_e$. Inset: Degree of reproducibility at the
transition. (b) Interaction force gradient vs. tip excursion
measured simultaneously with the tunnel resistance. Reprinted with
permission from \protect\cite{durig90}. \copyright 1990 American
Physical Society.} \label{f.force_in_tunneling}
\end{figure}

There are several mechanisms that may lead to deviations from
strict exponential tunneling. We do not want to digress too far
into the field of vacuum tunneling, and limit the discussion to
the two main influences. The first is the image potential that a
tunneling electron experiences from the two metallic surfaces,
which modifies the barrier. However, the apparent barrier, given
by the slope in a logarithmic plot of tunnel resistance versus
distance, is only affected at very small distances
\cite{binnig84,lang88,coombs88}. Second, since tunneling of
electrons is a result of a weak overlap of the electronic wave
function of the two metals in the vacuum space between them, the
overlap also results in a bonding force. For metals the attractive
interaction pulls the surface atoms closer to the other electrode,
effectively reducing the tunneling distance. This attractive force
in the tunneling regime was directly measured by D\"urig \ea
\cite{durig90} using a combination of an STM and a sample mounted
on a cantilever beam, Fig.\,\ref{f.force_in_tunneling}. The shift
in the resonance frequency of the cantilever beam served as a
measure for the force gradient. A semilogarithmic plot of the
tunnel resistance as a function of the distance was found to be
linear over the entire range of about 4 decades in resistance.
This was confirmed by Olesen \ea \cite{olesen96}, who argue that
the effect of the attractive force is almost exactly canceled by
the decrease of the tunnel barrier due to image potential
corrections. Deviations from exponential tunneling at close
distance to contact observed in low temperature MCBJ experiments
\cite{krans93,voets96} were attributed to the attractive force and
the observation of this effect is probably allowed by the higher
stability of the instrument. Understanding this force is important
for quantitative analysis of the surface corrugation measured in
STM \cite{ciraci89,wintterlin89,ciraci90,ciraci90a,clarke96}.

From continuum models and atomistic simulations Pethica and Sutton
\cite{pethica88} predicted that the attractive force between two
clean metal surfaces should lead to an intrinsic instability at a
distance of 1--3 \AA. This `avalanche in adhesion' \cite{smith89}
is expected to cause the surfaces to snap together on a time scale
of the order of the time it takes a sound wave to travel an
inter-planar spacing ($\sim 100$ fs, see
Sect.\,\ref{sss.simplemetals}). Originally it was believed that
one should be able to bring surfaces together in a continuous
fashion because atoms are more strongly bound to their neighboring
atoms than to the opposite layer. However, the elastic response of
many atomic layers produces an effective spring constant by which
the surface atoms are held. When the gradient of the force pulling
the surface atoms across the vacuum gap is greater than this
spring constant the surfaces snap together. When this elastic
medium is absent, as for a single atom held between two rigid
surfaces, a perfectly smooth transition takes place, as can be
shown by simple model calculations \cite{ciraci92}.

\begin{figure}[!b]
\begin{center}
\includegraphics[width=90mm]{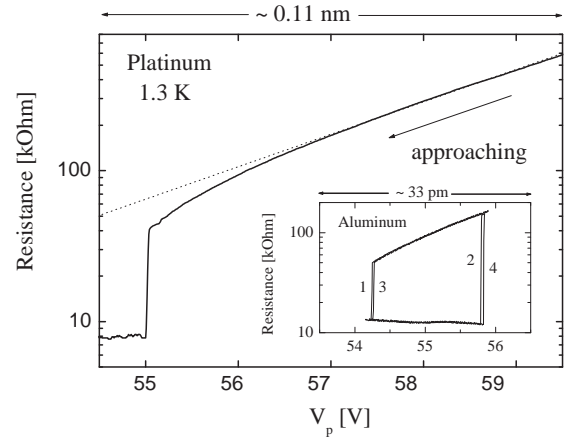}
\end{center}
\caption{Semi-logarithmic plot of an ac-resistance measurement of
a Pt MCBJ sample at 1.3\,K as a function of piezovoltage, $V_p$.
The curve is recorded starting a the high-resistance tunneling
side (right). The dotted line is an extrapolation from larger
$V_p$.  Between $\sim$500~k$\Omega$ and $\sim$40~k$\Omega$ a
downward deviation from exponential behavior is observed, followed
by a jump to a stable value. The inset shows two successive cycles
of an Al break junction at 4.2\,K, illustrating the hysteresis of
the jumps. The numbers indicate the sequential order of the jumps,
first up (1), then down (2) and up and down again (3 and 4).
Reprinted with permission from \protect\cite{krans93}. \copyright
1993 American Physical Society.} \label{f.jump-to-contact}
\end{figure}

It is this jump-to-contact that was observed in STM and MCBJ
experiments \cite{gimzewski87,agrait93,krans93}
(Figs.~\ref{f.force_in_tunneling},~\ref{f.jump-to-contact}). The
jump is associated with hysteresis, as expected, where the forward
and reverse jumps are separated by distances of the order of
1\,\AA, but can be as small as 0.2\,\AA. The tunneling-resistance
from which the jump-to-contact occurs is typically between 30 and
150\,k$\Omega$. The well-defined switching of the resistance
between values differing by an order of magnitude has been
proposed for applications as a quantum-switch device
\cite{smith95}. For some systems the jump-to-contact was found to
be absent \cite{good96,cross98}, although these appear to be
exceptions to the general rule. A possible explanation may be that
in these cases the force gradient never exceeds the effective
spring constant of the elasticity of the metal, but more work is
needed to investigate this.

At higher temperatures a jump-to-contact of quite different nature
has been observed by Kuipers and Frenken \cite{kuipers93}. When
studying lead surfaces between room temperature and the surface
melting temperature it was found that the tip jumps into contact
with the surface already from a distance of at least 10\,\AA. In
order to break this contact it was necessary to rapidly retract
the tip. From the relation between retraction speed and the
distance at which the contacts were broken the authors were able
to demonstrate that a neck spontaneously grows by surface
diffusion of Pb atoms. This is the result of the tendency of
surface atoms to diffuse towards points of highest concave
curvature. Later, Gai \ea \cite{gai98} showed that there is a
competition between the concave curvature {\it along} the neck and
the convex curvature around it. This leads to a spontaneous growth
of the cross section for short necks, but long thin necks tend to
diffuse out of contact. This forms a likely explanation for the
observation of spontaneous contact breaking for clean gold
contacts at room temperature \cite{muller96}. This also implies
that the protrusion that is left behind on the surface after
breaking of the neck diffuses away at room temperature, which
significantly increases the hysteresis cycle for returning to
contact. In a molecular dynamics simulation for contact formation
at room temperature S{\o}rensen \ea \cite{sorensen96a} show that a
sequence of atom hops is involved in the contacting process, which
they refer to as `diffusion-to-contact'. The same mechanism has
been invoked to describe the formation of a neck of silicon atoms
between a tip and a clean Si(111) surface \cite{hasunuma97}.

\subsection{Single-atom contacts}\label{ss.one-atom}

After indenting the STM tip into the sample surface, or pressing
the electrodes together in an MCBJ experiment, and subsequently
pulling the contact apart again, the above described staircase
structure in the conductance is observed. The conductance at the
last plateau prior to the jump back into the tunneling regime is
usually fairly reproducible and is believed to be associated with
a contact of a single atom. In Sects.~\ref{s.mechanical},
\ref{s.models}, and \ref{s.exp_modes} we will encounter many
arguments that support this interpretation, based on measurements
of the force, on model calculations, and on the experimentally
determined number of conductance channels, respectively. However,
already at this point we can see that this is a reasonable point
of view. In particular for gold, Fig.\,\ref{f.force_in_tunneling},
the last contact conductance is very reproducible, and close to
1\g0, suggesting a single atom contact provides a single
conductance channel. If a single atom were not enough to open a
conductance channel we would expect to find many contacts with
smaller last conductance values. Moreover, the hysteresis between
contact and tunneling for many metals is of the order of one tenth
of the size of an atom. If several atoms were involved in the
contact configuration one would expect to see much larger
hysteresis since there is more room for reconfiguration.

There have been very few direct experimental tests of single-atom
contacts. A very controlled experiment has been performed by
Yazdani \ea \cite{yazdani96} using a stable low temperature STM. A
clean nickel surface was prepared in UHV and a low concentration
of xenon adatoms was deposited on the surface. Making contact to a
Xe atom with a bare W tip and with a Xe-atom terminated tip
allowed measuring the resistance of a Xe atom contact and a
two-Xe-atom chain. The results were compared to model calculations
for this system and the agreement was very satisfactory.
Unfortunately, Xe is not a metal, which is reflected in the high
resistance for the `contact' of about 80~k$\Omega$ and
10~M$\Omega$ for the single and two-atom chain, respectively.
Also, no jump-to-contact is observed, consistent with the weak
adsorption potential for Xe.

Rather direct evidence comes from the observation of the atomic
structure of a gold contact in a room temperature UHV
high-resolution transmission electron microscope \cite{ohnishi98}.
For contacts that have a single atom in cross section the
conductance is found to be close to 12.9~k$\Omega$, or half this
value. Ohnishi \ea argue from the intensity profile that the
higher conductance in the later case is associated with a double
row of atoms along the line of view. For the lower-intensity
profiles the conductance agrees with a value of 1\,\g0.

Single-atom contacts can be held stable for very long times at low
temperatures. At room temperature, on the other hand, clean gold
contacts drift away from their starting value on a time scale of
milliseconds due to drift and thermal diffusion of the atoms
making up the contact, as discussed above. Although several
authors report a ``surprising stability'' over time scales of
order of an hour of atomic-sized gold contacts
\cite{pascual95,dremov95,abellan97,abellan98,correia99}, Hansen
\ea \cite{hansen00} have demonstrated that this enhanced stability
is very likely associated with adsorbates or other contaminants.
Since the presence of contaminants is seen to be associated with
non-linear IV characteristics we will return to this point in
Sect.\,\ref{ss.nonlinear}. An influence of adsorbates as capillary
forces mainly by moisture stabilizing the contacts was also
suggested by Abell{\'a}n \ea \cite{abellan98}.

Although the various results are not very explicit or sometimes
contradicting, it appears that any conductance value can be held
stable for gold contacts in air, not just those with a conductance
near a multiple of \g0 \cite{abellan98}. However, the experiments
in high vacuum at room temperature seem to suggest that multiples
of \g0 are preferentially formed when the contact is allowed to
choose its own size by diffusion of atoms \cite{muller96}. At low
temperatures diffusion is too slow, and in air adsorbates inhibit
the surface diffusion of atoms. Additional evidence is given by Li
\ea \cite{li99} who use an electrochemical fabrication technique
with feed-back (see Sect.\,\ref{ss.nanofab contacts}) to stabilize
the conductance. They find that non-integer multiples of \g0
cannot be held stable for very long times in contrast to integer
multiples. Similar observations were made by Junno \ea using AFM
manipulation of nanoparticles \cite{junno98}. These observation
may be related to a stabilization of the contact by the filling of
the quantum modes, as will be discussed in
Sect.\,\ref{ss.force_models}. It may also be related to the
observation by Ohnishi \ea \cite{ohnishi98} of the formation of
multiple parallel strands of single-atom gold wires in their HRTEM
images. This aspect certainly deserves further study.

\subsection{Conductance histograms}\label{ss.histograms}

From the discussion in the previous paragraphs it is clear that
contacts fabricated by these methods show a wide variety of
behavior, where the atomic structure of the contact plays a very
important role. The methods used, with the exception of very few
experiments \cite{ohnishi98,yazdani96}, do not permit detailed
knowledge over the atomic-scale built-up of the contact.
Nevertheless, one would like to investigate possible quantization
effects in the conductance and some experiments that we have
discussed seem to suggest that these exist. However, it is very
difficult to objectively separate the stepwise conductance
behavior as a function of contact size resulting from the atomic
structure and the stepwise pattern due to quantization of the
conductance.

In order to perform an objective analysis of the data it has
become customary to construct conductance histograms from large
sets of individual conductance versus displacement curves. The
method exploits the large variability in the data resulting from
the many possible atomic-scale contact configurations, and assumes
that all possible configurations, or effective contact sizes, are
equally likely to be formed. Under this assumption, one expects to
find peaks in the histogram corresponding to conductance values
that are preferred by the {\em electronic} system. A smooth,
adiabatic shape of the contact for a free-electron gas would then
lead to peaks at integer multiples of the quantum unit \g0. We
shall see that the assumption underlying this method is generally
not fulfilled, but with some caution one may still extract
evidence for quantization effects in the conductance from the
data.

\begin{figure}
\centerline{\includegraphics[width= 8cm]{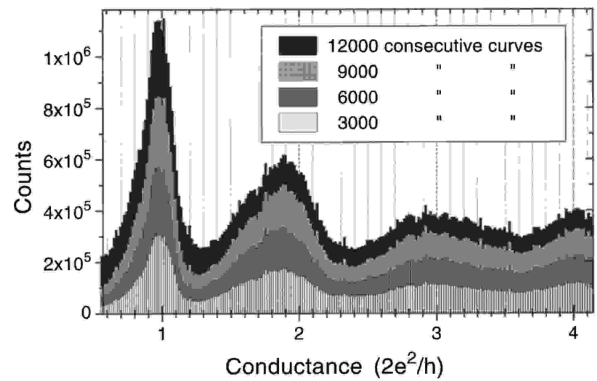}}
\caption{Conductance histogram for gold contacts recorded in air
at room temperature, using a dc voltage bias of 90.4\, mV. The
numbers of curves used for the construction of the histogram is
given in the inset, showing the gradual evolution of the peaks in
the histogram. The data have been corrected for an effective
series resistance of 490\,$\Omega$. Reprinted with permission from
\protect\cite{costa97}. \copyright 1997 American Physical
Society.} \label{f.HistoCostaK}
\end{figure}

An example of a conductance histogram for gold recorded at room
temperature in air is given in Fig.\,\ref{f.HistoCostaK}. The
conductance axis is usually divided into a number of bins, and the
conductance values falling within the range of each bin are
collected from a large number of individual scans of contact
breaking. The resolution is obviously limited by the width of the
bins but also by the digital resolution of the analog-to-digital
converter (ADC), and the mismatch of these two may produce
artificial periodic structures or `noise' in the histograms.
Modern 16-bit ADC's have ample resolution by which such artifacts
can be avoided. Alternatively, one may employ the method of
constructing a distribution curve introduced by Brandbyge \ea
\cite{brandbyge95}. Here, all the conductance values are collected
in a single file and sorted in increasing order of the
conductance. The height of the distribution $H$ at a conductance
measurement point $G_i$ is then calculated as $H(G_i)=
(2w/N)/(G_{i+w}- G_{i-w})$, where $w$ is a conveniently chosen
interval, determining the width of the smearing of the
distribution. The factor $1/N$ normalizes the area under the
distribution to 1.

In order to improve statistics one would like to average over as
many curves as required to resolve the smaller peaks. Since it
turns out that several thousands of curves are usually needed, the
contact breaking is done in a short time, of order of a second or
shorter. Also, when studying the breaking of macroscopic contacts,
as in relays, the final breaking process of the last few atoms
takes place on a time scale of the order of 10\,$\mu$s. For the
recording of these fast processes one employs fast digital
oscilloscopes equipped with flash-ADC's. The latter have usually a
limited resolution (8 or 10 bits) and a significant differential
non-linearity, as was pointed out by Hansen \ea \cite{hansen97},
meaning that the effective width of the digitization intervals has
a large spread. This results in artificial `noise' in the
histograms, which can be avoided \cite{hansen97} by dividing the
measured data by a normalizing data set recorded for a linear ramp
using the same digital recorder settings.

Below, we will present an overview of histograms recorded for
various metals. Before we start we would like to point out that
the results have been obtained under widely different
circumstances, some of which are known to influence the results,
and some influences have not yet been systematically investigated.
Clearly the environment should be considered, whether working in
air or in UHV. The purity of the sample material should be
specified, but when working in UHV the surface preparation
technique is even more important. In STM techniques one often
employs a tip of a hard metal, such as tungsten or
platinum/iridium, and indents this tip into a soft metal surface
such as gold. There is good evidence that the soft metal wets the
tip surface, so that after a few indentations, one has a
homogeneous contact of the sample metal only. However, this
depends on the wetting properties of the combination of the tip
and surface material, and one should be aware of possible
contaminations of the contacts by the tip material. Further, the
temperature has an effect on the results, but also the measurement
current, which may locally heat the contact, or modify the
electronic structure due to the finite electric field present in
the contact. The shape of the connective neck that is formed
during elongation of the contact may be influenced by these
factors, but also by the depth of indentation of the two
electrodes and the speed of retraction. The sample material may be
in the form of a polycrystalline wire, a single crystal or a thin
film deposited on a substrate and this has an effect on the
mechanical response of the metal but also on the electron mean
free path. Even with all these factors taken into account, and for
one and the same sample, some details of the histograms do not
always reproduce. Here, the local crystalline orientation of the
electrodes forming the contact may play a role as proposed in
Ref.\,\cite{rodrigues00}. Nevertheless, it turns out that there
are many features that are very robust, and reproduce under most
circumstances, as we will see below.

\subsubsection{The archetypal metal: gold}\label{sss.gold}

\begin{figure}
\centerline{\includegraphics[width=6cm,angle=270]{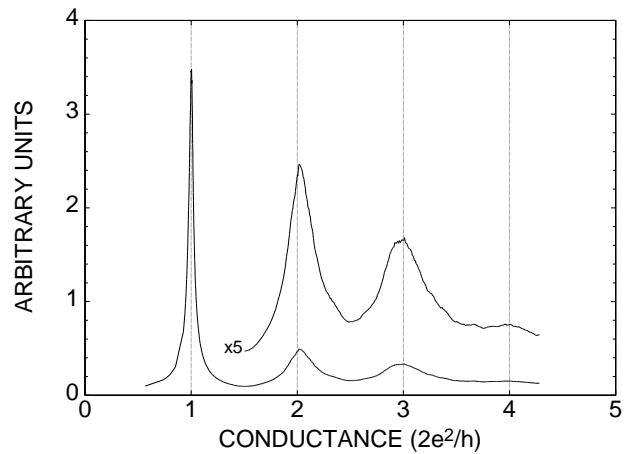}}
\caption{Histogram for gold contacts measured at room temperature
in UHV by pressing an STM tip into a clean gold surface and
recording the conductance while retracting the tip. The data have
been corrected for an effective series resistance of
150\,$\Omega$. Reprinted with permission from
\protect\cite{brandbyge95}. \copyright 1995 American Physical
Society. }
\label{f.goldhistoRT} 
\end{figure}

Gold contacts are the most widely investigated by the histogram
method
\cite{gai96,sirvent96,hansen97,costa97b,yasuda97,krans96b,deheer97,costa97,brandbyge97,costa97a,li98a,jarvis99,jian99,ludoph99,itakura99,shu00}.
The first results were presented by Brandbyge \ea
\cite{brandbyge95}, Fig.\,\ref{f.goldhistoRT}, using a
room-temperature STM under UHV. Many features of the histograms
recorded under widely different circumstances seem to reproduce
(cf. Fig.\,\ref{f.HistoCostaK}). This is presumably due to the low
reactivity of the gold surface and the fact that it is easily
cleaned. One generally observes peaks near 1, 2 and 3\,\g0, which
are shifted to somewhat lower values compared to exact multiples
of \g0. This shift has been attributed to backscattering of the
electrons on defects near the contact, as will be discussed in
Sect.\,\ref{s.defect_scattering}. Sometimes a broad feature at
4--5\,\g0 can be resolved, and there is one report of individual
peaks near 4 and 5\,\g0 \cite{costa97a}. The shape of the peaks
and their relative amplitudes are not reproducible in detail, but
the first peak is always very pronounced, and much higher than the
others. The other peaks tend to decrease in height with increasing
conductance. The first peak is very robust against the applied
bias voltage \cite{yasuda97,yuki01} and survives up to about 2\,V,
corresponding to formidable current densities of the order of
$2\cdot 10^{15}$\,A/m$^2$. Even the other structure in the
histogram is unchanged up to 500\,mV. Also the atmosphere does not
have a dramatic influence, and only by intentionally increasing
the concentration of reactive molecules in the atmosphere some
changes are observed. Li \ea \cite{li98} report that the peaks at
higher conductance values are gradually suppressed in going from
air to vapors of ethanol, pyridine and 4-hydroxyl thiophenol, in
increasing order of adsorption strength. However, the peak near
1\g0 survives under all circumstances.

Between helium- and room-temperatures no qualitative changes in
the histograms are observed \cite{sirvent96}, but the relative
height of the first peak grows toward lower temperatures. Although
no influence of the retraction speed was observed in the range
from 30 to 4000\,nm/s \cite{costa97}, the experiment by Muller \ea
\cite{muller96} seems to suggest that in the extremely slow limit,
where surface diffusion of gold atoms at room temperature is
expected, the shape of the histogram changes considerably. In the
latter experiment the peaks at higher conductance are more
pronounced.

Two reports give evidence for peaks near half-integer multiples of
\g0, notably near 0.5 and 1.5\,\g0. Both use a thermally
evaporated gold film on a mica substrate. One was measured with an
STM at room temperature under UHV \cite{jarvis99}. The other
\cite{shu00} is special, in that the histogram was recorded with
an STM tip immersed in an electrochemical cell, containing a
0.1\,M NaClO$_4$ or HClO$_4$. In this case the histogram had the
regular appearance, but below a threshold voltage in the
electrochemical potential in the cell the half-integer peaks
appeared. The authors discuss various possible explanations, but
discard all except the one proposed by de Heer \ea
\cite{deheer97}. However, as argued above, we believe also this
explanation is not viable, and the problem remains to be solved.

\begin{figure}[!t]
\centerline{\includegraphics[width= 8cm]{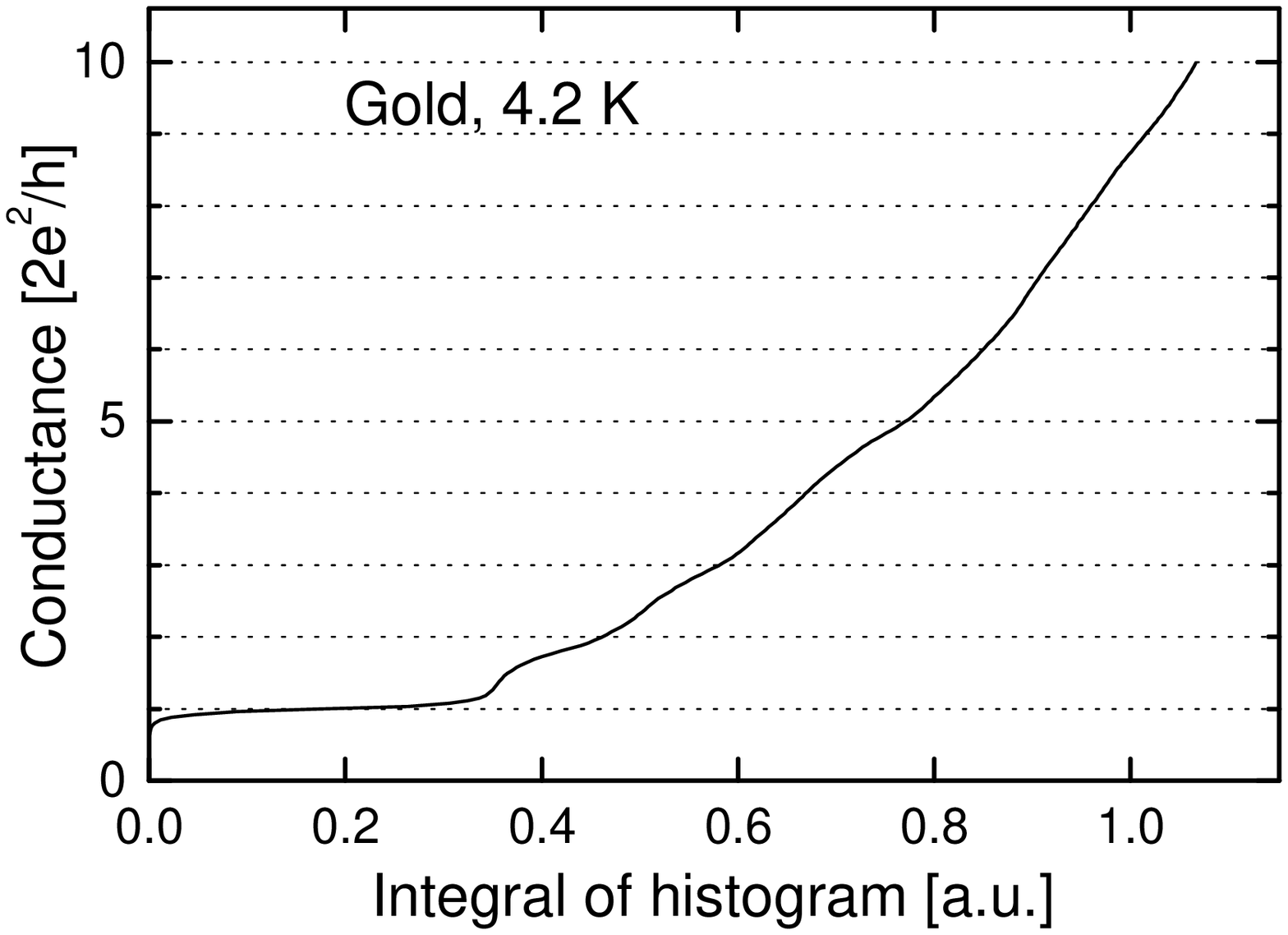}}
\caption{Curve obtained by integrating a gold histogram similar to
Fig.\,\protect\ref{f.goldhistoRT}. Courtesy A.I. Yanson
\protect\cite{yanson01}. } \label{f.integrated}
\end{figure}

When we assume that the contact breaking process produces any
effective contact diameter with equal probability, then the
histograms represent a derivative of the conductance with respect
to the effective diameter of the contact. It is instructive to
calculate the integral of the histogram, as was first done by Gai
\ea \cite{gai96}. Fig.\,\ref{f.integrated} shows such a curve,
obtained from a gold histogram similar to the one shown in
Fig.\,\ref{f.goldhistoRT}. This curve is to be compared to
conductance traces obtained for 2DEG semiconductor devices
\cite{wees88,wharam88}, for which the width of the contact can
directly and continuously be adjusted by the gate electrostatic
potential. Compared to the latter, the conductance steps in
Fig.\,\ref{f.integrated} are poorly defined, with the exception of
the first conductance quantum. Moreover, the first quantum feature
results from the fact that our assumption mentioned above is not
valid. The effective diameters produced during contact breaking
are strongly influenced by the possible atomic configurations. As
we will argue Sect.\,\ref{s.chains}, the step at 1\g0,
corresponding to the strong peak in the histogram for gold,
results from the formation of a chain of gold atoms during the
last stages of contact breaking. Disregarding the first level in
Fig.\,\ref{f.integrated}, we find that the conductance is not
strictly quantized, as the probability of finding a contact with a
conductance of, e.g., 2\g0 is only twice that of finding 1.5\g0.
However, the conductance is still determined by the quantum
states, as described in Sect.\,\ref{s.transport_theory} and is
carried by a limited number of modes. We will show in
Sects.~\ref{s.exp_modes} and \ref{s.defect_scattering} how the
quantum nature for monovalent metals is revealed by a tendency for
the modes to open one-by-one as the contact becomes larger.

The well-defined and robust features in the histograms for gold
are rather unique, and for other materials it is usually necessary
to work under clean UHV or cryogenic conditions to obtain
reproducible results.

\subsubsection{Free electron metals: Li, Na and K }\label{sss.alkali}

\begin{figure}[!t]
\centerline{\includegraphics[width= 8cm]{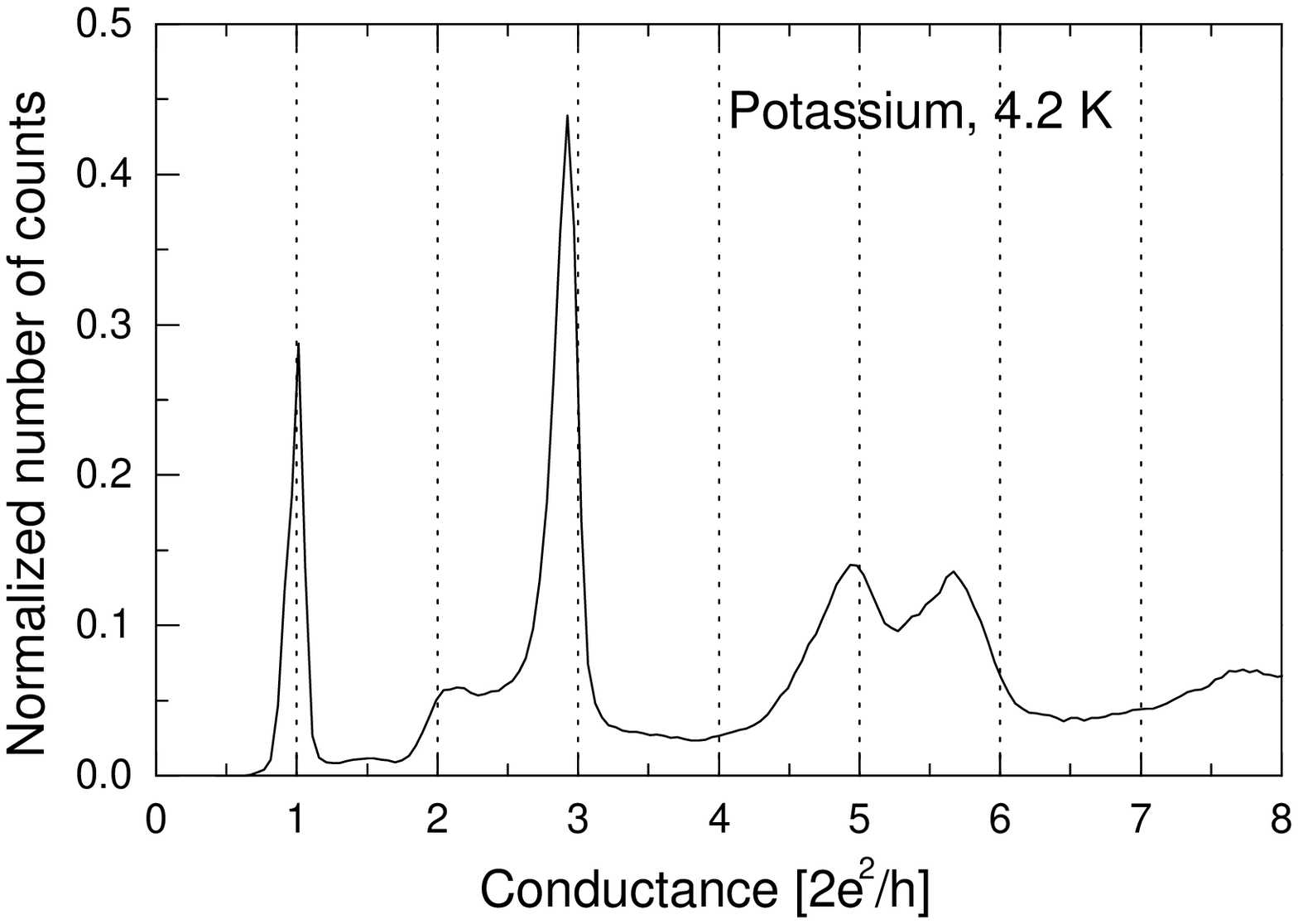}} \caption{
Histogram of conductance values, constructed from $G(V_{\rm
p})$-curves measured for potassium  at 4.2\,K with an MCBJ device,
involving several thousand individual measurements. The
measurements were done with a constant bias voltage of 10\,mV. The
characteristic sequence of peaks ($G= 1,3,5,6$) is regarded as a
signature for conductance quantization. Courtesy A.I. Yanson
\protect\cite{yanson01}.}
\label{f.histK}%
\end{figure}

For sodium in a low temperature experiment using the MCBJ
technique, a histogram with peaks near 1, 3, 5 and 6\g0 was
observed \cite{krans95}. Similar results have been obtained for
potassium (Fig.\,\ref{f.histK}), lithium
\cite{yanson99,yanson01,ludoph00a}, and cesium \cite{yansonTBP}.
Notice the sharpness of peaks, the absence of peaks at 2 and 4\g0
(disregarding the little shoulder below 3\g0) and the low count in
between the peaks. For lithium, the histogram looks qualitatively
similar, but the first two peaks are much smaller and the shift
below integer multiples of \g0, attributed to an effective series
resistance due to scattering on defects
(Sect.\,\ref{s.defect_scattering}), is bigger \cite{yanson01}.

The characteristic series 1-3-5-6, and the fact that peaks near 2
and 4\g0 are nearly absent points at an interpretation in terms of
a smooth, near-perfect cylindrical symmetry of the sodium
contacts. Sodium indeed forms a very good approximation to a free
electron system, and the weakly bound $s$-electrons strongly
reduce surface corrugation. As will be discussed in
Sect.\,\ref{sss.free_electrons}, for a model smooth, cylindrically
symmetric contact with continuously adjustable contact diameter
\cite{bogachek90,torres94}, the conductance increases from zero to
1~$G_0$ as soon as the diameter is large enough so that the first
conductance mode becomes occupied. When increasing the diameter
further, the conductance increases by two units because the second
and third modes are degenerate. The modes are described by Bessel
functions (assuming a hard wall boundary potential) and the first
mode is given by the $m=0$ Bessel function, which is not
degenerate. The second and third modes are the degenerate $m=\pm
1$ modes, followed by $m=\pm 2$ for further increasing contact
diameter. The next mode that will be occupied corresponds to the
second zero of the $m=0$ Bessel function, and is again {\it not}
degenerate. Thus the conductance for such a contact should
increase by 1, 2, 2 and 1 units, producing just the series of
conductance values observed in the sodium experiment.

\begin{figure}[!t]
\centerline{\includegraphics[width= 8cm]{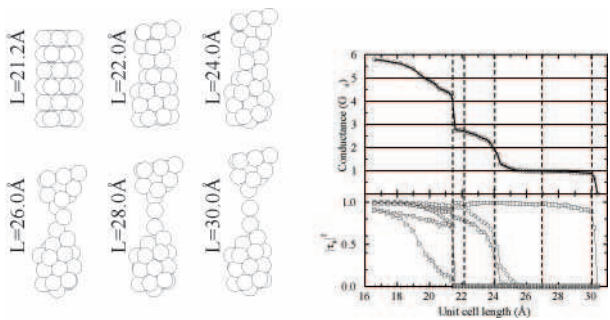}}
\caption{Conductance for the configurations of sodium atoms
illustrated at the left for several elongation stages of the unit
cell. The authors used a self-consistent local-density
approximation potential. The lower panel on the right shows the
evolution of the individual transmission probabilities for each
conductance channel in the system. Reprinted with permission from
\protect\cite{nakamura99}. \copyright 1999 American Physical
Society.} \label{f.nakamura99}
\end{figure}

Note that the conductance steps observed in the individual
conductance traces are still due to atomic reconfigurations.
Indeed, the sharpness of the steps is limited by the experimental
resolution, and they are associated with hysteresis and two-level
fluctuations as for any other metal. This picture can be
reconciled with the notion of conductance quantization in a
free-electron gas by considering the model calculation by Nakamura
\ea \cite{nakamura99}, Fig.\,\ref{f.nakamura99}. Without going
into the details of the calculations, which will be addressed in
Sect.\,\ref{s.models}, we find from the model that a single-atom
contact corresponds to a single mode, which is nearly perfectly
transmitted. At a unit-cell length of 22\,\AA\  three modes are
transmitted adding-up to a total conductance close to 3\g0, and
the narrowest cross section is made up of three atoms. The
potential that the electrons experience in this contact is nearly
cylindrical due to the fact that the electrons in Na are so weakly
bound. The calculation does not reproduce the abrupt transition
from 3 to 1\g0, probably due to the limited size of the model
system. The small shoulder at the lower end of the peak at 3\g0 in
Fig.\,\ref{f.histK} could then be due to the occasional formation
of a contact with two atoms in cross section, which is too small
for the three channels to be fully transmitted.

This picture of a strict separation of the effect of the atomic
structure in determining the geometry and the conduction modes
setting the conductance as dictated by this geometry, may need to
be refined by taking into account that the occupation of the
quantum modes may energetically favor specific atomic
configurations, as will be discussed in
Sects.~\ref{ss.force_models} and \ref{s.shells}.

\subsubsection{The noble metals}\label{sss.noble}

Similar as for gold, conductance histograms for copper and silver
have a dominant peak at or just below 1\g0
\cite{krans93,krans95,li98,hansen97,muller96,ludoph00a,costa97d,ono99,shu00}.
Above this peak one generally finds two additional peaks. At room
temperature in air these peaks are broad and form more or less a
single broad peak having two maxima \cite{hansen97,costa97d}.
Attempts to interpret these observations in terms of conductance
quantization peaks shifted by a series resistance are not quite
satisfactory since the series resistance required is rather large,
of the order of 500\,$\Omega$, and the correction does not shift
all peaks simultaneously to the desired positions. At low
temperatures the second and third peaks are more clearly separated
\cite{krans95,ludoph00a}. In contrast to gold where the height of
the peaks decreases systematically for increasing conductance, for
copper the third peak is often much more pronounced than the
second one. For silver the difference is less extreme, but a
reverse peak intensity has also been observed. Furthermore, the
third peak is much closer to 3\g0 compared to the distance of the
second peak below 2\g0. This suggests that Cu has an appearance
intermediate between that found for Au and for Na and K. The
second peak would then be mainly due to the atomic structure of a
two-atom contact, while the third peak represents all
configurations that admit three near-fully transmitted modes. Some
indirect evidence for this interpretation was given in
\cite{ludoph99,ludoph00a}. As was observed for gold, using slow
scans at room temperature more peaks can be seen and the peaks
tend to be closer to integer multiples of \g0
\cite{li98,muller96,shu00}, which we propose to attribute to the
effect of surface diffusion of atoms.

Three studies fall outside this picture. Rodrigues \ea
\cite{rodrigues02} have used a MCBJ under UHV at room temperature
and found a conductance histogram for Ag that, apart from the peak
just below 1\g0, has a rather strong peak at 2.4\g0 and a broader
feature just above 4\g0. They attribute the peak at 2.4\g0 to a
stable nanowire geometry along the [110] crystallographic
direction, which they have identified in HR-TEM images. Although
this is plausible, the difference with the commonly observed
histograms remains to be explained. Ono \ea \cite{ono99} use a
home-built relay-type set up which switches the contact formed
between the apex of a thin Cu wire and a thin Cu film evaporated
onto a glass substrate, under ambient conditions. The histogram
shows only two peaks, but sharp and centered at 1 and 2\g0. It is
not clear what distinguishes the technique used here from the
other studies, but it should be noted that the histograms are
built from a very limited number of curves ($\sim$20) and that the
curves were selected to ``have at least one plateau''. It is also
remarkable, as we shall discuss below, that they find several
sharp peaks for Ni. It is likely that these peaks are the result
of stable contact cycles, which result after training of the
contact as discussed in Sect.\,\ref{ss.steps&plateaus}, and
therefore reflect recurring contact configurations. Finally, Li
\ea \cite{li00} use an STM to study the influence of adsorbate
molecules on Cu atomic-sized contacts and find that after the
addition of an organic molecule, 2,2'-bipyridine, the conductance
histogram shows additional peaks near the half-integers 0.5 and
1.5\g0. These results resemble those for Au obtained by the same
group \cite{shu00}, suggesting that also the latter may be
attributed to the effect of adsorbates. A full explanation for the
effect is still lacking, but Li \ea have proposed to exploit this
sensitivity of the atomic conductance to the presence of certain
molecules as a chemical sensor.

\subsubsection{Transition metals}\label{sss.transition}

\begin{figure}[!t]
\centerline{\includegraphics[height=6cm,angle=270]{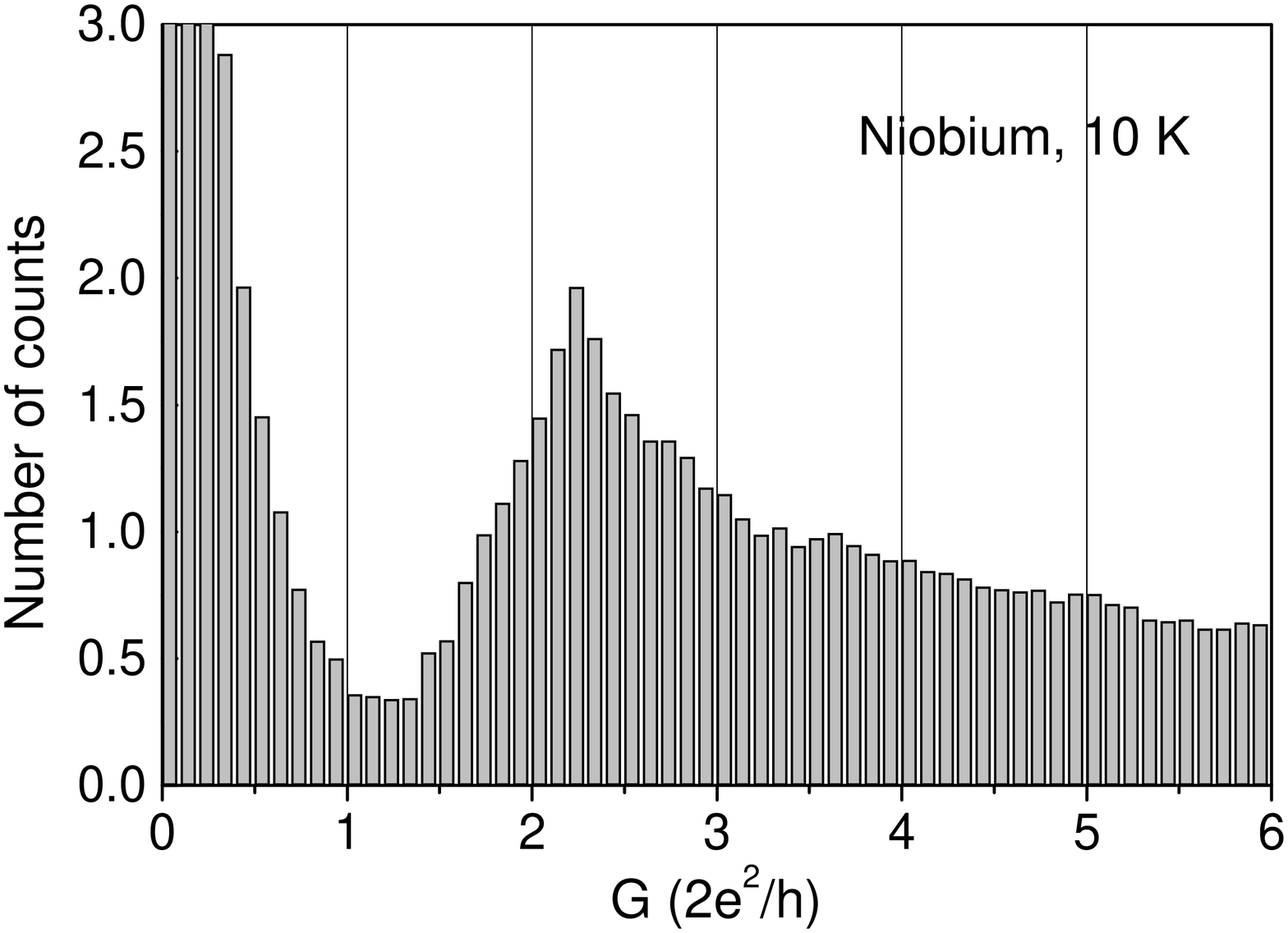}}
\caption{Histogram constructed from 2400 individual conductance
curves for a niobium sample. Each curve was recorded while
stretching the contact to break, using the MCBJ technique at a
temperature of 10~K, which is just above the superconducting
transition temperature. The conductance was measured using a DC
voltage bias of 20~mV. (Reprinted with permission from
\protect\cite{ludoph00}. \copyright 2000 American Physical
Society.}
\label{f.histoNb}
\end{figure}

For the non-magnetic transition metals (we will discuss the
ferromagnetic ones below) the histograms show generally very few
features. In MCBJ experiments at low temperatures the transition
metals with partially filled $d$-shells, as far as they have been
studied, show a single broad peak centered well above 1\g0. This
peak can generally not be identified with an integer value of the
conductance; for example niobium shows a wide peak centered near
2.3--2.5\g0 (Fig.\,\ref{f.histoNb}). The peak at zero conductance
arises from the fact that there was no low-conductance cut-off
applied in the data and the conductance measured in the tunneling
regime causes an accumulation of points at low conductance. The
jump between contact and tunneling is relatively small for Nb, and
the tunnel current can rise even somewhat above 1\g0 just before
the jump to contact, leading to a nearly continuous cross-over
from the data points obtained in tunneling and those obtained in
contact. Results for vanadium are comparable  to those for Nb
\cite{yanson01}.

Similarly, for Rh, Pd, Ir and Pt at low temperatures histograms
with a single peak, centered in the range from 1.5 to 2.5\g0 have
been obtained \cite{krans93,sirvent96a,yanson01,smit01}, with some
weaker features at higher conductances. At room temperature, both
in UHV \cite{olesen95} and in air \cite{costa97} the first peak
for Pt is surprisingly found near 1\g0. Recent work suggests that
this peak near 1\,\g0 may arise from contamination of the metal
surface with hydrogen \cite{smit02}. On the other hand, histograms
for Ru, Rh, and Pt obtained by a very fast relay technique under
ambient conditions but at high voltage-bias show features that are
somewhat similar to the low-temperature data
\cite{itakura00,yuki01a}.

Contamination of the contacts is most likely responsible for the
fact that most experiments on transition metals at room
temperature do not show any reproducible structure. When
concentrating therefore on the data obtained under cryogenic
vacuum, these seem to point at a general interpretation of the
first peak in terms of the characteristic conductance of a
single-atom contact. This is in excellent agreement with the
expected conductance, see Sects.\,\ref{ss.TBcalculation} and
\ref{s.exp_modes}, where it is argued that a single-atom contact
for a transition metal with a partially filled $d$-band has five
conductance channels available, which are only partially open.

For Zn, having a completed $d$-band, there is a first dominant
peak well below 1, at about 0.7\g0, which resembles the results
for aluminum, and a second smaller peak near 2\g0
\cite{yanson01,scheerTBP}.

\subsubsection{Ferromagnetic metals}\label{sss.ferromagnets}

The ferromagnetic metals have attracted special interest as a
result of speculations that the strong exchange splitting of the
electron bands may lift the spin-degeneracy of the conductance
modes, which would give rise to half-integer ($e^2/h$ instead of
$2e^2/h$) conductance steps as obtained for simple free-electron
models. However, as mentioned above, the number of channels for a
single atom is expected to be five, and all modes are only
partially open so that the total conductance is in the range of
1.5--3\g0. This is consistent with what is observed at low
temperatures in MCBJ experiments on Fe \cite{ludoph00a}, where the
conductance histogram shows a single peak at 2.2\g0 very similar
to that observed for Nb in Fig.\,\ref{f.histoNb}. Low-temperature
STM experiments similarly show that the last contact value for Ni
is typically 1.6\g0 \cite{sirvent96a}.  At room temperature in air
the histograms for Fe, Co and Ni are entirely featureless
\cite{hansen97,costa97}, at least in some of the early
experiments. This is attributed to a contamination of the contacts
by adsorbates from the atmosphere, which washes out all regular
metallic conductance features.

More recently there have been a few reports of sharp features for
the ferromagnets \cite{ono99,ott98,oshima98,komori99}. Ott \ea
\cite{ott98} obtain three sharp peaks near 1, 2 and 3\g0 for Fe,
using a relay-type technique under ambient conditions, while for
Ni they find the more familiar featureless histogram. The authors
suggest that the state of magnetization may be of importance,
since they report having saturated the magnetization state of
their wires. Note that the histogram was constructed from a rather
limited set of 80 conductance traces. The data have some
resemblance with those of Ref.~\cite{komori99} recorded at 4.2\,K.
In this work the number of accumulated scans is also limited, but
in addition the contact was first `trained' and only very shallow
indentation cycles to a depth of about 4\g0 were used. This
implies that the histogram is not obtained by averaging over many
contact configurations, but rather represents a reproducible
contact-configuration cycle. The histogram can then not be
interpreted in terms of intrinsic conductance properties.

Ono \ea \cite{ono99} showed data for Ni comparable to those for Fe
by Ott \ea\ The technique is not very different from the method by
Ott \ea, except that they contact the Ni wire to a thin Ni film
evaporated onto a glass substrate, as also used for Cu, see above.
However, in addition they report that the `integer' peaks near 1
and 2\g0 that are observed without magnetic field are joined by
additional peaks near 0.5 and 1.5\g0 for fields above 5\,mT. The
number of curves in each histogram is only 20, and the curves have
been selected to have at least one plateau. The latter suggests
that many individual curves are featureless.

\begin{figure}[!t]
\centerline{\includegraphics[width= 8cm]{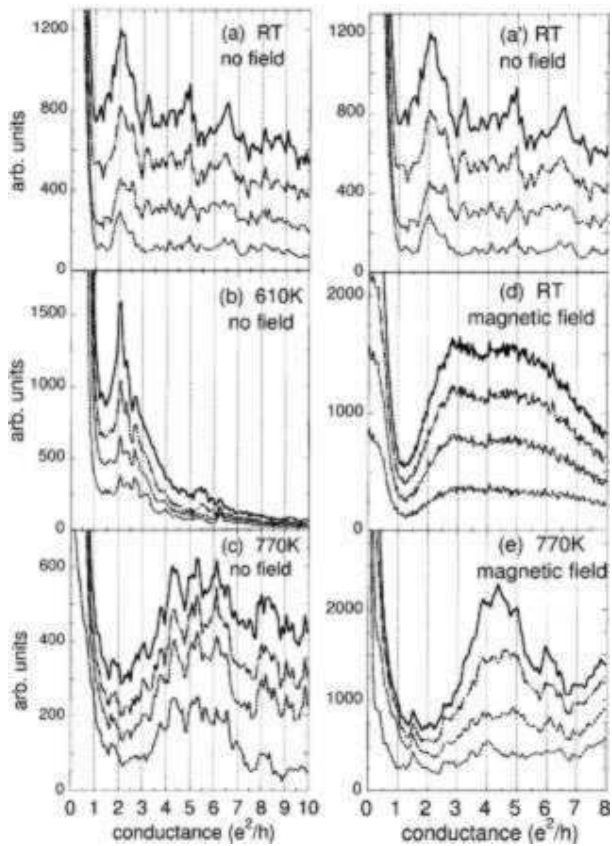}}
\caption{Conductance histogram constructed from conductance curves
for Ni. The left column of panels shows the evolution of the
histograms with temperature: (a)room temperature, (b) 610K, (c)
770K. The right column shows the effect of applied magnetic field:
(a') same as (a), (d) in a field of 0.12\,T at room temperature
and (e) the same field at 770\,K. The histograms are presented in
a cumulative fashion for 40, 140, 240 and 340 traces in (a),(a')
and for 100, 200, 300 and 400 traces in (b)--(e). Reprinted with
permission from \protect\cite{oshima98}. \copyright 1998 American
Institute of Physics. } \label{f.Oshima}
\end{figure}

Oshima and Miyano \cite{oshima98} constructed a relay between a Ni
wire and a Ni coil that could be heated by a current well above
room temperature. The set-up was placed in a vacuum chamber with a
pressure in the range 0.5--$1\cdot 10^{-4} $\,Pa and a magnetic
field up to 0.12\,T. At room temperature a peak at 1\g0 is
observed, which survives up to 610\,K (Fig.\,\ref{f.Oshima}).
However, above the Curie temperature for Ni ($T_{\rm C}=$631\,K)
the histogram changes into a broad peak centered around 2.7\g0.
Also the application of a magnetic field removes the sharp feature
at 1\g0. The results are remarkable, but the explanation offered
is qualitative and not entirely convincing. Note the similarity of
the high-T and high-field data to the low-temperature results for
the $d$-metals. The authors mention that the application of the
magnetic field changes the mechanics of the contact breaking,
because the repulsive magnetic force between the like-oriented
electrodes results in a contact breaking on microsecond time
scales.

Finally, Garc{\'\i}a \ea \cite{garcia99} have produced stable
atomic-sized Ni contacts at room temperature by embedding the
contact between two Ni wires in a resin. The magnetization state
of the two wires was switched by field coils wound around the
wires. The contact resistance was seen to switch between $\sim 3$
and $\sim 10$\,k$\Omega$ upon reversing the applied field of
2\,mT. Reference experiments on Cu-Cu and Cu-Ni contacts of
similar conductance did not show any significant response to the
magnetic field. An explanation of these results has been proposed
in terms of scattering on the domain wall trapped inside the
constriction \cite{tatara99}.

Clearly, the results on the ferromagnets are not all consistent,
but some evidence exists that the magnetization state modifies the
conductance and the histogram. However, it is not yet clear what
are the conditions to go from featureless histograms to histograms
showing peaks near integer and even half-integer values of \g0.
More work is needed to clarify the experimental situation.

\subsubsection{Aluminum and other $sp$-metals}\label{sss.aluminum}

\begin{figure}[!b]
\centerline{\includegraphics[width=7cm]{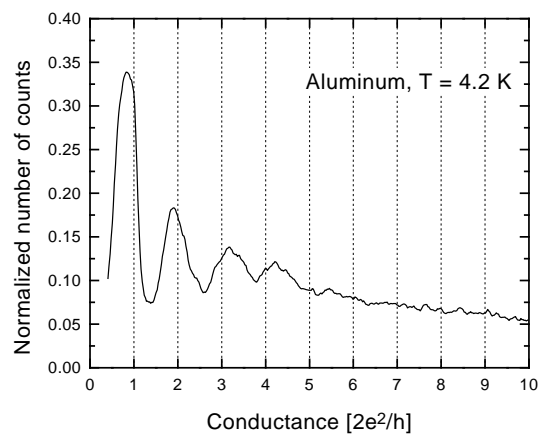}}
\caption{Histogram constructed from 30\,000 individual conductance
curves for two different samples of aluminum, using the MCBJ
technique at 4.2\,K at a sample bias voltage of 10\,mV. Reprinted
with permission from \protect\cite{yanson97}. \copyright 1997
American Physical Society. } \label{f.AluminumHisto}
\end{figure}

Figure~\ref{f.AluminumHisto} shows a histogram for aluminum
obtained at 4.2\,K \cite{ludoph00a,yanson97}. In \cite{yanson97}
these data were taken as evidence that structure in the histogram
cannot exclusively be interpreted as arising from quantum
structure in the conductance modes. Indeed, the strong first peak
lies even somewhat below 1, at about 0.8\g0, while a single-atom
contact for the $sp$-metals is believed to be associated with
three partially transmitting conductance channels (see
Sect.\,\ref{s.exp_modes}). Also, a series resistance
interpretation of the shift of the peaks is inconsistent with the
fact that the first two peaks lie somewhat below integer values,
while the next two weaker features lie above 3 and 4\g0,
respectively. As will be discussed below, the total conductance
for the three channels in a one-atom contact are expected to add
up to about 1\g0, which suggests that the peak is due to an {\it
atomic configuration} of a single atom in the contact. It further
suggests that also the higher conductance peaks are due to
preferred atomic configurations.

This interpretation is strongly supported by calculations by Hasmy
\ea \cite{hasmy01}. They obtained a histogram of the effective
contact cross sections deduced from a large series of molecular
dynamics simulations for the stretching of Al nanowires. At low
temperatures (4\,K) clear peaks are found at the positions
corresponding to 1, 2, 3 and 4 atoms in cross section. Although
the conductance could not be obtained from the same calculation,
it demonstrates that these specific contact areas contribute with
more than average weight in the histograms. Taking a typical
conductance of 0.92\,\g0 per atom one obtains a fair agreement
with the experimental conductance histogram, without taking any
quantum effects of the electron gas into account. Further, the
authors observe that the peaks in the cross section histogram are
not equally well pronounced for different crystalline orientations
of the wire, which may explain the small variations in appearance
of the conductance histograms.

Histograms for Pb and Sn measured at low temperatures have a
single dominant peak at about 1.7\g0 that is more than 1.5\g0
wide, with some weaker features at higher conductances
\cite{yanson01}. The width of the peak is consistent with the
gradual variation of the conductance over the last plateau,
stretching often from $\sim$3\g0 to 1\g0 (see
Refs.~\cite{agrait94,cuevas98} and Sect.\,\ref{ss.strain}). The
interpretation fits with the conductance expected for a
single-atom contact geometry.

Lewis \ea \cite{lewis99} reported conductance histograms measured
on 5N-purity Ga samples, using a variable-temperature STM. The
histograms, constructed from 500--1000 curves, while featureless
at room temperature, gradually develop a peak at (or slightly
above) 1\g0 plus a wider feature near 2\g0.

\subsubsection{Semimetals and semiconductors}\label{sss.semimetals}

As mentioned above, for antimony the jump-to-contact was observed
at about 1\,M$\Omega$ \cite{krans94}, a value much higher than for
the regular metals. This is consistent with the fact that electron
density for Sb is three orders of magnitude lower than for common
metals, giving a corresponding Fermi wavelength of about 55\,\AA.
Although the bulk bandstructure for such materials may not be
applicable at the atomic scale, one can still infer that a
one-atom contact will not be sufficiently large to transmit a full
channel. The conductance is then reduced by orders of magnitude
compared to the conductance quantum, since only tunneling
contributes to the current transport. At still larger contact
diameters, no evidence was found for quantization around the unit
values of conductance \cite{krans94}. The conductance is observed
to increase in a stepwise fashion, similar to metallic contacts,
but the step heights are much smaller that a quantum unit. This
behavior is naturally attributed to the atom-by-atom decrease of
the contact size with elongation. Similar steps at high-resistance
contacts have been observed for the semiconductor Si
\cite{hasunuma98}. The Si contacts are found to show Schottky-like
rectification characteristics that is influenced by the type of
doping of the Si material.

Costa-Kr\"amer \ea \cite{costa97c} measured a conductance
histogram for Bi using an STM at helium temperatures. The
histogram collected from 3000 curves shows a broad peak with a
maximum at $\sim$2\g0, and a shoulder near 1\g0. Although the
authors make the connection with quantized conductance, the data
are insufficient to support such conclusion. One of the further
complications is the fact that the data are recorded for a voltage
bias of 90\,mV, whereas the Fermi energy, $E_F$, for Bi is only
25\,meV. At $eV \gg E_F$ the simple models that produce quantized
conductance cross over to a regime where half-integer conductance
values dominate \cite{glazman89,pascual97,bogachek97a}.

More recently, Rodrigo \ea \cite{rodrigo02} recorded histograms
for Bi at 4\,K and 77\,K, using a voltage bias of 10--20\,mV. They
observed a strikingly different behavior for the two temperatures.
At 4\,K they observed sub-quantum conductance steps similar to Sb.
The curves between the conductance jumps are typically short, of
order 0.1 to 0.2\,nm, and often curve strongly upward upon
stretching. At 77\,K on the other hand the plateaus are much more
regular and flat, and stretch over several nanometers. Most
significant is the observation that nearly all curves have a
well-defined plateau near 1\,\g0, which results in a pronounced
peak in the conductance histogram. A smaller peak near 2\,\g0 can
also be seen. Rodrigo \ea propose an interesting explanation for
these observations based on the bandstructure for Bi. Besides the
band with a light effective electron mass, that is responsible for
the usual Fermi surface properties of Bi, they identified a
low-lying heavy-electron band. Under conditions of lateral quantum
confinement the quantum level in the light band, which has a
strong dispersion, is pushed above the lowest level in the
heavy-electron band. At low temperatures the heavy electron band
determines the conductance in the atomic contacts. However, at
77\,K the light electrons can be observed to determine the
conductance for larger-size contacts, when the confinement is less
severe. The authors present a simplified model for the evolution
of the contact size that appears to explain the data. Despite the
complications of the bandstructure for Bi and the anomalous
temperature dependence this appears to be the most clear-cut
evidence for conductance quantization, that does not suffer from
the discreteness of the atomic structure.

\subsubsection{Metallic alloys and compounds}\label{sss.alloys}

The number of experiments on compounds and alloys is very limited.
The additional complication here is that the interpretation of the
data requires knowledge of the atomic structure formed at the
contact. The composition of the material at the atomic scale may
be very different from the bulk composition due to surface
segregation and the mechanical work done on the contact.

Many alloys may have properties similar to those for elemental
metals, as indicated by experiments on Au with approximately 5\%
Co \cite{hansen97}. The first experiment shows that the addition
of a few percent Co does not significantly modify the histogram
compared to pure Au. For Au and Ag one can form random alloys in
the entire concentration range and one finds a gradual cross-over
from the Au to the Ag histograms \cite{enomoto02}. In studies for
alloys of Cu, Ag and Au with transition metals it is found that
the peak at 1\,\g0, that is characteristic for the noble metals,
survives for transition metal concentrations well over 50\%
\cite{enomoto02,bakker02,heemskerk02}. The interpretation for this
observation requires further study. There is evidence for
segregation of the noble metals away from the contact under the
application of a high bias current \cite{heemskerk02}.

Volkov \ea \cite{volkov95} used a low-temperature STM to study the
contact between a Pt tip and the narrow-gap semiconductor Pb$_{\rm
1-x}$Sn$_{\rm x}$Se. They find a jump-to-contact sometimes to a
conductance near 1\g0, and sometimes to plateaus with a much
smaller conductance. The latter can be explained by the
semiconducting nature of the sample. The histogram they present
shows three rather broad peaks, near 1, 2 and 4\g0 for which the
authors do not attempt to offer an explanation. Ott \ea
\cite{ott98} apply the same technique as for Fe described above to
the ferromagnetic perovskite La$_{\rm 0.75}$Sr$_{\rm
0.25}$MnO$_{\rm 3}$ and show a histogram with seven rather sharp
peaks. However, the number of scans used is limited to 60 and they
mention that ``\ldots it is possible that some of the features
could shade off if much larger data sets were considered.
Unfortunately these ceramic crystals are unsuitable for high rate
measurements because of the difficulty of establishing reliable
contacts\ldots ''. The latter agrees with experience from MCBJ
experiments on heavy fermion metals, high-temperature
superconductors and organic conductors at low temperatures
\cite{ludoph99b}, where it was seen that it is usually impossible
to identify a clear jump-to-contact, and the mechanical properties
of atomic-sized contacts for these unconventional materials differ
fundamentally from the plastic behavior observed for the elemental
metals.

A rather exceptional material studied using conductance histograms
is that composed of multiwalled nanotubes \cite{frank98}. It is
beyond the scope of this review to discuss the rapid developments
in the study of these interesting materials for which one may
consult a recent review \cite{dekker99} and references therein.
Single-walled carbon nanotubes come in various modes of chirality,
most of which are semiconductors while some are metallic. The
metallic nanotubes are predicted to have two conductance channels.
Direct measurements of this number of channels has been difficult,
since the conductance is severely modified by the contact barriers
to the leads and by defects along the nanowires, but recent work
agrees with two conductance channels per nanowire \cite{kong01}.
For multiwalled nanotubes no general predictions can be made. In
experiments on a multiwalled carbon nanotube attached to the gold
tip of an STM and measured at room temperature by immersing it
into liquid metals Frank \ea \cite{frank98} found a conductance
histogram with two very sharp peaks. The best results were
obtained using liquid Hg and the peaks were found very close to 1
and 2\g0. Part of the explanation may be that only a single carbon
wall, possibly the outer one, contributes to the conductance, but
that still leaves a factor of 2 to be explained.

\subsection{Non-linear conductance}\label{ss.nonlinear}

The discussions above have mostly been limited to the linear
conductance for small bias. We defer discussion of non-linear
contributions to the conductance at low temperatures due to
superconductivity and due to scattering on defects to
Sects.~\ref{s.exp_modes} and \ref{s.defect_scattering},
respectively. However, even at room temperature many authors have
reported non-linear contributions to the conductance for fairly
large bias voltages, in the range from 0.1 to 1\,V
\cite{pascual95,dremov95,costa97b,abellan97,srikanth92}. Most of
these experiments were performed for gold under ambient conditions
and the current-voltage ($IV$) relation generally has a
significant cubic term, $I = g_0 V + g_3 V^3$, with $g_3 > 0$ so
that the current at high bias lies above the extrapolation from
the zero-bias conductance. As was pointed out by Hansen \ea
\cite{hansen00} this contradicts the observations of Sakai and
coworkers \cite{yasuda97,itakura99} that the peaks in the
conductance histogram for gold remain at their initial positions
up to bias voltages larger than 0.5\,V, as discussed in
Sect.\,\ref{sss.gold} above.

Hansen \ea made a careful study of this problem and developed a
technique that allows them to record accurate $IV$-curves within
about 10\,$\mu$s \cite{hansen00,hansen00a}. They show convincingly
that under clean UHV conditions the $IV$-curves for gold at room
temperature are nearly linear up to at least 0.5\,V. Only when the
surface is intentionally contaminated do they observe
non-linearities of the magnitude reported before. In addition,
they observe that the clean contacts tend to break spontaneously
on a time scale of milliseconds, while the contaminated contacts
can be held stable at a conductance near the quantum unit for
hours. They accompany their observations with a calculation based
on a tight binding model for the atomic-sized contact under the
application of a finite bias voltage (see also
\cite{brandbyge99}). The calculations reproduce the experimentally
observed nearly linear $IV$-characteristics, and only above about
1\,V a slight curvature was obtained with a sign opposite to that
observed in the room temperature experiments. The authors propose
that contaminated contacts contain a tunnel barrier composed of
adsorbates, which naturally explains the observed curvature in the
$IV$-curves. At the same time, the mechanical contact area would
be much larger than that for a metallic contact with the same
conductance, explaining the enhanced stability. The proposed
effect of impurities has been confirmed by {\it ab initio}
calculations that self-consistently include the applied bias
voltage \cite{mehrez02}. The linear $IV$ dependence for clean Au
is modified to have a pronounced $g_3$-term when a sulfur impurity
is inserted in a gold atomic contact.

For metals other than gold there are not many results on the
$IV$-characteristics available yet. However, from the work on gold
we learn that great care is needed to ensure clean contacts, since
most other metals will be more sensitive to adsorbates than gold.
Recent work by Nielsen \ea shows that clean Pt contacts have a
much stronger non-linear voltage dependence, which was shown to
agree with first-principles calculations \cite{nielsen02}.

\section{Mechanical properties of atomic-sized point contacts}
\label{s.mechanical}

How do the mechanical properties of matter change as size is
reduced down to the atomic scale? This question is of fundamental
interest, not only theoretically but also from an applied point of
view since contact in macroscopic bodies typically occurs at
numerous asperities of small size, whose mechanical properties
determine those of the contact. This explains the interest of
investigating small size contacts for many technologically
important problems like adhesion, friction, wear, lubrication,
fracture and machining
\cite{landman90,sutton90,bhushan95,landman96}. The appearance of
proximal probes like the STM and related techniques, together with
computational techniques for simulating tip-surface interactions
with atomic detail have contributed to the growth of the new field
of nanotribology.

In metallic contacts mechanical and electrical properties  are
intimately related. Experiments in which mechanical and electrical
measurements are combined are essential for understanding the
physics of these systems. Not many of this kind of experiments
have been done due to the technical difficulties involved. Note
that in most of the experiments reviewed in the foregoing sections

 the geometry of the contact and its evolution during an
experiment are result of the stresses acting on them. Hence the
evolution of the conductance in these experiments reflects not
only the changes in size of the constriction but also the
mechanical processes taking place at the constriction itself.

In this section,  we first review some of the basic concepts of
the mechanical properties of metals, specifically elastic and
plastic deformation, fracture, and contact mechanics. The elastic
properties of a metal are not expected to change much as size is
decreased to nanometer dimensions, since they reflect the
resistance of atomic bonds to stretching. However for one-atom
contacts or atomic chains things could be otherwise.
Nanometer-size specimens are also stronger since the strength
(i.e., the resistance to plastic deformation) depends on the
presence of dislocations which could be absent in small specimens.
We will discuss some simple models of metallic constrictions based
on contact mechanics. These are based on macroscopic continuum
theory, which is applicable, in principle, for distances large
compared with the distances between the atoms. Hence such models
cannot be expected to describe atomic-sized systems accurately,
but they can serve as a starting point for interpreting the
experimental results. We will see in Sect.~\ref{s.models} that
many of the phenomena displayed by these continuum models can be
recognized in the microscopic, atomistic models. In
Sect.~\ref{s.shape}, we use these models combined with the
experimental results to deduce, approximately, the shape of
mechanically drawn contacts. The experiments in which the
conductance and forces of contacts down to one atom are
simultaneously measured are discussed in
Sect.~\ref{s.experiments}. Experiments on the mechanical
properties of atomic chains will be described in
Sect.~\ref{s.chains}. Theoretical work on the mechanical and
electrical properties of atomic contacts using molecular dynamics
simulations will be reviewed in  Sect.~\ref{ss.MD}.

\subsection{ Mechanical properties of metals}
\subsubsection{Elastic deformations}

When a solid is subjected to a load it undergoes a change in
shape. For small loads this deformation is elastic and the
specimen recovers its  original dimensions as the load is removed.
For isotropic materials, the stress tensor $\sigma$ and the strain
tensor $\epsilon$  are linearly related \cite{Landau-ela,Courtney}
\begin{equation}
\sigma_{ik}=\frac{E}{1+\nu}\left[\epsilon_{ik}+\frac{\nu}{1-2\nu}\left(\sum_{l}
\epsilon_{ll}\right)\delta_{ik}\right],
\end{equation}
where $E$ is the modulus of elasticity or Young's modulus, and
$\nu$ is Poisson's ratio. This law, valid for small deformations,
is called Hooke's law.

\begin{figure}[!t]
\begin{center}
\includegraphics[width=80mm]{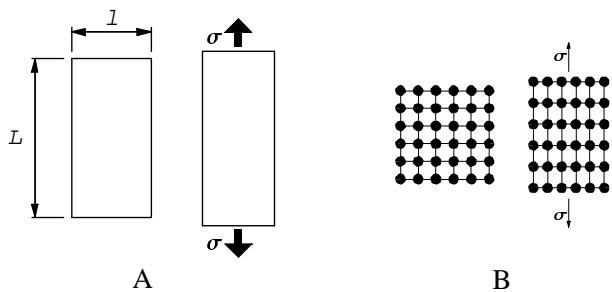}
\end{center}
\caption{\label{f.elastic} (A) Elastic deformation of a rod under
uniaxial stress. (B) Atomistic view of elastic uniaxial
deformation.}
\end{figure}

For homogeneous deformations, in which case the strain and stress
are constant in all the solid, the relation between strain and
stress is particularly simple. For example (see Fig.\
\ref{f.elastic}(A)), the uniaxial extension or compression of a
rod of length $L$ and lateral dimension $l$, whose axis is in the
$z$-direction, due to a force $F$, is given by
\begin{equation}
\epsilon_{zz}=\frac{\sigma}{E}  \mbox{\hspace{1cm} or
\hspace{1cm}} \frac{\delta L}{L}=\frac{1}{E}\frac{F}{l^{2}},
\end{equation}
and the lateral deformation by
\begin{equation}
\epsilon_{xx}=\epsilon_{yy}= -\nu \epsilon_{zz}.
\end{equation}
For metals, values of Poisson's  ratio range between 0.25 and 0.4,
which implies that the volume is not conserved during elastic
deformation. We may write the response of the bar as an effective
spring constant $k_{\mbox{\scriptsize{eff}}}=E l^{2}/L$.

Single crystals are not isotropic and in order to specify the
elastic properties several elastic constants are needed, their
number depending on the symmetry of the crystal. The least number
of non-zero constants is three, for cubic symmetry. In this case,
the Young's modulus depends on the direction of the applied stress
relative to the crystal axes \cite{Landau-ela,Courtney}. For
instance, in the case of Au, the maximum value of the Young's
modulus is $E_{\langle 111 \rangle}=117$ GPa, and the minimum is
$E_{\langle 100 \rangle}=43$ GPa. Macroscopic polycrystalline
samples are in practice isotropic, since they are composed of
crystallites with random orientations. The Young's modulus for
polycrystalline gold is taken to be 80 GPa.

On an atomic scale,  elastic strain consists in small changes in
the inter-atomic spacing, that is, in the stretching of
inter-atomic bonds, as illustrated in Fig.\
\ref{f.elastic}(B).Consequently the modulus of elasticity $E$ is a
measure of the resistance of the inter-atomic bonds to
deformation.

\subsubsection{Plastic deformations\label{sss.plastic}}

For most metallic materials elastic deformation  is possible only
for strains smaller than about 0.005 (or 0.5\%). As the material
is deformed beyond this point, permanent, nonrecoverable, or {\em
plastic deformation} occurs. In macroscopic metal specimens this
transition from elastic to plastic behavior, or yielding, occurs
gradually, and it is difficult to assess accurately the lower
limiting stress below which no plastic deformation is found.
Conventionally the {\em yield strength} $\sigma_{\mbox{\scriptsize
y}}$ is defined as the stress necessary to produce a plastic
strain of 0.002 under uniaxial stress. The yield strength of a
metal is very sensitive to any prior deformation, to the presence
of impurities and to heat treatment, in contrast to the modulus of
elasticity which is insensitive to these factors.

When the material is in a complex state of stress,  as in the case
of a point-contact, the load at which plastic yield begins is
related to the yield strength $\sigma_{\mbox{\scriptsize y}}$
through an appropriate yield criterion, which is written in terms
of the principal stresses or eigenvalues of the stress tensor,
$\sigma_{1},\sigma_{2}$, and $\sigma_{3}$ \footnote{The stress
tensor is symmetric and consequently can be diagonalized at any
point.}. The simplest criterion is due to Tresca
\begin{equation}
\max
\{|\sigma_{1}-\sigma_{2}|,|\sigma_{2}-\sigma_{3}|,|\sigma_{3}-\sigma_{1}|\}
=\sigma_{\mbox{\scriptsize y}},
\end{equation}
and a somewhat more accurate criterion is von Mises' criterion
\begin{equation}
\frac{1}{\sqrt{2}}[(\sigma_{1}-\sigma_{2})^{2}+(\sigma_{2}-\sigma_{3})^{2}+(\sigma_{3}-\sigma_{1})^{2})]^{1/2}=\sigma_{\mbox{\scriptsize
y}}.
\end{equation}
These criterions \cite{Johnson}, which in practice are equivalent,
are in fact based on the idea that plastic flow is caused by shear
stresses. It can be shown that the greatest shear stress, that is,
the maximum value of the off-diagonal elements of the stress
tensor, is given by $\tau_{\mbox{\scriptsize
max}}=(\sigma_{\mbox{\scriptsize max}}-\sigma_{\mbox{\scriptsize
min}})/2$, where $\sigma_{\mbox{\scriptsize max}}$, and
$\sigma_{\mbox{\scriptsize min}}$ are the largest and smallest
eigenvalue, respectively.

\begin{figure}[!t]
\begin{center}
\includegraphics[width=80mm]{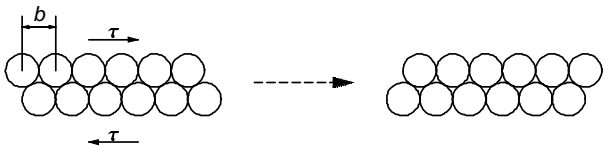}
\end{center}
\caption{\label{f.frenkel} Slip in a perfect crystal under shear
stress.}
\end{figure}

On an atomic scale, plastic deformation corresponds to the
breaking of bonds between neighboring atoms and the reforming of
bonds with the new neighbors. The atoms change positions, that is,
they change their configuration. Upon removal of the stress they
do not return to their original positions and there is a permanent
change in the shape of the body. The simplest model of plastic
deformation of a `perfect' crystal, that is, one with no defects,
considers the sliding of two compact planes with respect to each
other. Frenkel \cite{frenkel26} calculated the maximum shear
stress required for this process to occur. He considered two
neighboring planes in a crystal with a repeat distance $b$ in the
direction of shear and spacing $h$. These planes are assumed to be
undistorted as a shear stress $\tau$ is applied, as illustrated in
Fig.\ \ref{f.frenkel}. It is then assumed that $\tau$ varies with
shear displacement $x$ as
\begin{equation}
\tau=\frac{Gb}{2\pi h} \sin\frac{2\pi x}{b},
\end{equation}
where $G$ is the shear modulus \footnote{$G=E/2(1+\nu)$ in
polycrystalline materials.}. The maximum value of $\tau$ is then
\begin{equation}
\tau_{\mbox{\scriptsize max}}=\frac{Gb}{2\pi h}.
\end{equation}
For the \{111\} planes of a face-centered cubic metal we take
$b=a/\sqrt 6$, and $h=a/\sqrt 3$, where $a$ is the lattice
parameter, thus $\tau_{\mbox{\scriptsize max}}\approx G/9$. A more
extensive discussion \cite{kelly86} gives $\tau_{\mbox{\scriptsize
max}}\approx G/30$.

These values of the shear stress are much larger than those
observed in macroscopic metal specimens. This discrepancy is
explained by the presence of dislocations, which can glide at low
stress values \cite{Courtney,kelly86}. The atomic distortions
accompanying the motion of dislocations, which are linear defects,
are considerably less than those happening during ``perfect
slip'', which requires glide of atomic planes in a correlated
manner. High values of the shear stress, close to the theoretical
prediction, are observed in experiments with dislocation-free
specimens like whiskers in bending or uniaxial tension
\cite{kelly86}, and are also to be expected for nanometer volumes
of metals, since dislocations are unstable and are quickly
expelled from small-volume samples.

In macroscopic specimens plastic flow involves the motion of large
numbers of dislocations in response to stress. Dislocation motion
occurs through a process termed {\em slip}, which occurs in a
preferred crystallographic plane (slip plane) and along a specific
direction (slip direction). The combination of the slip plane and
slip direction (called the slip system) is such that the atomic
distortion that accompanies the motion of a dislocation is
minimal. The slip plane is the plane of the most dense atomic
packing and the slip direction corresponds to the direction, in
this plane, having the highest linear density. For the fcc
structure the slip planes are the $\{ 111\}$ planes and the slip
directions are of the $\langle 110\rangle$-type, and there are 12
slip systems. The sliding of atomic planes in a perfect crystal
will also occur in the slip plane and along the slip direction,
because this is the least energetic process. In some crystal
structures, the unit dislocation dissociates into partial
dislocations, the so-called {\em Shockley partials}. For example,
in fcc crystals atomic glide on a $\{ 111\}$ plane is somewhat
easier if the motion is  divided in two partial slip steps. Such a
partial slip leads to a disruption of the characteristic ABC
stacking of the fcc structure, producing a stacking fault.

In a single crystal, the various slip systems are oriented
differently, and plastic flow will initiate in the slip system
which has the greatest resolved shear stress, i.e. the greatest
stress acting on the slip plane and in the slip direction. The
critical value of the resolved shear stress
$\tau_{\mbox{\scriptsize CRSS}}$, for which plastic flow is
initiated when the single crystal is subjected to a tensile (or
compressive) stress, depends on the orientation of the crystal
with respect to the tensile axis and is characteristic of the
material and smaller than the yield strength
$\sigma_{\mbox{\scriptsize y}}$. The value of
$\tau_{\mbox{\scriptsize CRSS}}$ depends on temperature and strain
rate. For high temperatures ($T \ge 0.7 T_{m}$, with $T_{m}$ being
the material's melting temperature in Kelvin),
$\tau_{\mbox{\scriptsize CRSS}}$ decreases rapidly with increasing
temperature and decreasing strain rate as a result of the
important role played by diffusive processes. At these
temperatures plastic deformation can be effected over a period of
time  at a stress level well below the materials yield strength.
This time-dependent deformation is called {\em creep}. At
intermediate temperatures ( $0.25 T_{m} \le T \le 0.7 T_{m}$,
$\tau_{\mbox{\scriptsize CRSS}}$ is essentially constant. For $T
\le 0.25 T_{m}$, $\tau_{\mbox{\scriptsize CRSS}}$ is again a
function of temperature and strain rate, increasing with
decreasing temperature and increasing strain rate. This is a
result of the resistance to dislocation motion presented by
short-range barriers \cite{Courtney}.

Although dislocation glide is the dominant mechanism in plastic
deformation in macroscopic crystals, permanent shape changes can
be effected by the mechanism of {\em twinning} \cite{Courtney}.
Twinning is more likely to be observed in bcc materials than in
fcc metals.

A peculiarity of nanometer-scale specimens is that they have a
high surface-to-volume ratio, and as a consequence surface energy
effects may be of importance in plastic deformation. Let us
consider a cylindrical specimen of radius $a$, which for
simplicity will be assumed perfectly plastic with yield strength
$\sigma_{\mbox{\scriptsize y}}$, and which is elongated a small
distance $\Delta l$. The work $E_{v}$ required to deform its
volume may be written as
\begin{equation}
E_{v}=\pi a^2 \sigma_{\mbox{\scriptsize y}} \Delta l,
\end{equation}
and the work $E_{s}$ required to extend its surface, while
conserving volume, in the form
\begin{equation}
E_{s}=\gamma \pi a \Delta l,
\end{equation}
where $\gamma$ is the surface energy. Using  the values of
$\gamma$ and $\sigma_{\mbox{\scriptsize y}}$ for typical metals,
we find that $E_{s}$ becomes larger than $E_{v}$ for $a\sim$ 1--10
nm. Hence, surface energy effects which are negligible for larger
specimens are important in atomic-sized contacts. One may expect a
liquid-like behavior in the final stages of rupture where the
contact necks down to atomic dimensions.

\subsubsection{Fracture}

Fracture is the separation of a body in two pieces in response to
an  imposed stress at temperatures below the melting temperature
of the material. A given material may fracture in a variety of
ways, depending on temperature, stress state and its time
variation, and environmental conditions \cite{Courtney}. Tensile
fracture occurs by the stress-assisted separation of atomic bonds
across the plane of fracture. As with plastic deformation, at
relatively high temperatures, it is aided by diffusion. Fatigue
fracture is associated with cyclically applied strains or
stresses. Static fatigue and embrittlement are associated with
hostile or corrosive environments.

\begin{figure}[!b]
\begin{center}
\includegraphics[width=80mm]{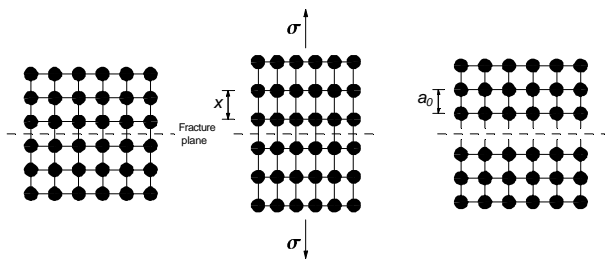}
\end{center}
\caption{\label{f.fracture} Atomistic model of theoretical tensile
fracture. The equilibrium structure (A) is altered due to the
application of stress $\sigma$ (B). After fracture (C), two new
surfaces form and the atoms return to their equilibrium positions.
After \cite{Courtney}.}
\end{figure}

Let us consider the ideal theoretical strength of a solid in the
same spirit as for plastic deformation. For a perfect crystal,
fracture would take place by the simultaneous rupture of all
atomic bonds across the fracture plane as depicted in
Fig.~\ref{f.fracture}. A simple estimate of the expected tensile
stress or cleavage stress \cite{kelly86,Courtney} is obtained by
considering that the the variation of the force to pull apart two
adjacent atomic planes, separated by a distance $a_{0}$, is of the
form
\begin{equation}
\sigma=\frac{E}{\pi}\left(\frac{a}{a_{0}}\right)
\sin\frac{\pi}{a}(x-a_{0}), \label{theo_cleavage}
\end{equation}
where $a$ is a measure of the range of the inter-atomic  forces.
The work done to separate the two atomic planes should be related
to the surface energy $\gamma$ for the two newly exposed surfaces,
\begin{equation}
\int_{a_{0}}^{a_{0}+a} \sigma\mbox{ d}x=2\gamma,
\end{equation}
whence $ a^{2}=\pi^{2}\gamma a_{0}/E$. The theoretical cleavage
stress $\sigma_{\mbox{\scriptsize th}}$ is given by the maximum
value of $\sigma$ in Eq.\,(\ref{theo_cleavage}),
\begin{equation}
\sigma_{\mbox{\scriptsize th}}=\sqrt\frac{E\gamma}{a_{0}}.
\end{equation}
The values obtained from this estimate are much larger than those
found experimentally. This is due to the existence of interior or
surface cracks that catalyze fracture. Cracks may also be
introduced in the material by plastic deformation. For a
completely brittle solid the theoretical strength will be attained
at the tip of the crack.

\begin{figure}[!t]
\begin{center}
\includegraphics[width=80mm]{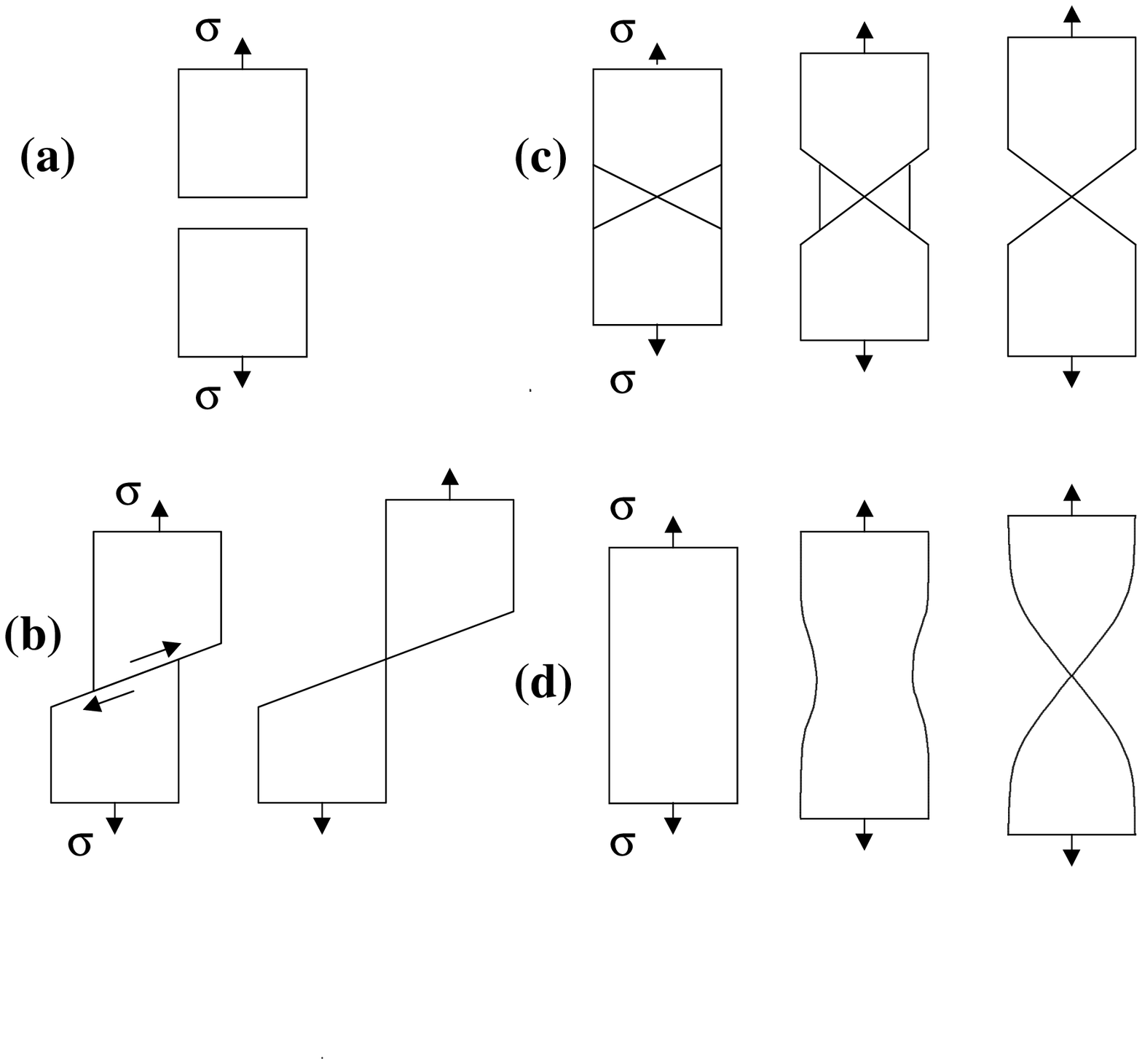}
\end{center}
\caption{\label{f.tensile_fracture} Schematic illustration of the
various modes of fracture. (a) Brittle fracture. (b) Fracture
occurring by glide on a single slip plane. (c) Rupture fracture in
a single crystal. (d) Rupture fracture in a polycrystal. After
\cite{Courtney}.}
\end{figure}

When the sample does not contain preexisting cracks tensile
fracture may be preceded by varying degrees of plastic
deformation. When fracture takes place prior to any plastic
deformation it is termed {\em brittle} (Fig.
\ref{f.tensile_fracture}a). The fracture mechanism is called {\em
cleavage} if fracture occurs within grains (or in a single
crystal) or {\em brittle intergranular fracture} if it progresses
along grain boundaries in a polycrystal. {\em Ductile fracture} is
preceded by varying degrees of plastic deformation. In a single
crystal fracture may occur by gliding on a slip plane, in which
case the fracture will be atomically flat (Fig.
\ref{f.tensile_fracture}b). {\em Rupture fracture} corresponds to
a 100 \% reduction of the minimal cross section of the specimen by
plastic deformation. In single crystals this is effected by
multiple slips (Fig. \ref{f.tensile_fracture}c), whereas in a
polycrystal is associated with necking (Fig.
\ref{f.tensile_fracture}d). Rupture fracture is the extreme case
of ductile fracture. In some materials there is a gradual
transition from brittle to ductile as temperature is increased.

In ductile metals plastic yield is related to shear stresses
because the maximum shear stress is smaller than the theoretical
cleavage stress, and consequently plastic deformation takes place
by shearing not by cleaving. Even, if we consider perfect metal
crystals the ideal maximum shear stress $\tau_{\mbox{\scriptsize
max}}$ will be reached much before $\sigma_{\mbox{\scriptsize
th}}$, indicating that the metal will prefer to flow by shear
rather than to cleave. For instance, for gold in the $ \langle 111
\rangle $ direction $\sigma_{\mbox{\scriptsize th}}$ is 27 GPa,
while $\tau_{\mbox{\scriptsize max}}$ is 0.74 GPa.

\subsubsection{Contact mechanics}

Bodies whose surfaces are non-conforming \footnote{Two surfaces
are said to be conforming if they fit together without
deformation.} when brought into contact, touch first at a point or
along a line and, even under load, the dimensions of the contact
area are generally small compared to the dimensions of the bodies
themselves. In these circumstances the contact stresses are highly
concentrated  and decrease rapidly away from the point of contact.
The shape of the bodies is not important and the stresses can be
calculated assuming that each body is an elastic half-space
\cite{Landau-ela,Johnson}. For high loads, when the elastic limit
is exceeded, only the region of the contact will deform
plastically.

The response of the elastic half space to a concentrated load can
be calculated in terms of the pressure distribution. For a load
acting on a circular region of radius $a$, solutions can be found
in closed form for pressures of the form  $p=p_{0}
(1-r^{2}/a^{2})^{n}$ \cite{Johnson}. Defining an effective elastic
constant $k_{\mbox{\scriptsize eff}}$ as the ratio of the load
\begin{equation}
F=2\pi \int_{0}^{a} p(r) r \mbox{\scriptsize d}r
\end{equation}
to the average displacement of the surface
\begin{equation}
\langle u_{z} \rangle=2\pi \int_{0}^{a} u_{z} r \mbox{d}r,
\end{equation}
we obtain
\begin{equation}
k_{\mbox{\scriptsize eff}}= \frac{CEa}{(1-\nu^{2})}.
\end{equation}
The constant $C$ equals $3\pi^{2}/16$, for $n=0$, the uniform
pressure distribution. For $n=1/2$, the so-called Hertzian
pressure, which is the pressure distribution resulting from the
contact of two spheres, or Hertzian contact, we have $C=16/9$.
For $n=-1/2$, we have $C=2$ and the pressure distribution
corresponds to a uniform displacement of the contact area, as in
the case of an ideally rigid punch indenting a softer surface.
Note that the effective elastic constant $k_{\mbox{\scriptsize
eff}}$ is not very sensitive to the detailed pressure
distribution.

Plastic deformation of the elastic half space will start when the
yield condition is satisfied, that is, when the maximum shear
stress $\tau_{\mbox{\scriptsize max}}$ reaches the value
$\sigma_{\mbox{\scriptsize y}}/2$ anywhere in the solid. For the
Hertzian pressure distribution, we have $\tau_{\mbox{\scriptsize
max}}=0.44 F/\pi a^{2}$ and for the uniform pressure distribution
$\tau_{\mbox{\scriptsize max}}=0.31 F/\pi a^{2}$. Both values
above are computed for $\nu=0.4$ which applies for gold. These
maximal values are reached within the solid directly below the
center of the contact, at a depth of $z=0.51a$ and $z=0.67a$ for
the Hertzian and uniform pressures, respectively.

The standard model contact considered in textbooks
\cite{Landau-ela,Johnson} is the so-called Hertzian contact, first
considered by H. Hertz (\cite{Timoshenko}, p. 409). In a Hertzian
contact the contacting bodies are assumed to have spherical
surfaces at the point of contact. In this case, the contact radius
varies with load as the surfaces deform elastically. This model
has been used to interpret the results of nanoindentation
experiments \cite{tangyunyong93}, and also of friction experiments
using AFM \cite{carpick96,carpick96a} but it is not adequate for
metallic contacts since the effective radius of curvature is too
small and plastic deformation takes place before the area could
vary due to elastic deformation.

\begin{figure}[!t]
\begin{center}
\includegraphics[width=50mm]{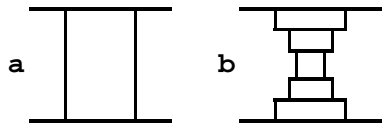}
\end{center}
\caption{Model geometry for metallic nanocontacts. (a) Simple
cylindrical shape model for a constriction. (b) Slab model.
\label{f.contact}}
\end{figure}

As the simplest model for metallic nanocontacts, we can consider a
short cylinder of radius $a$ and length $L$ between two
semi-infinite half planes (the electrodes) as depicted in
Fig.\,\ref{f.contact}(a). This is similar to the model we have
used for the transport properties. The response of this system to
the applied strain is linear, in contrast to the response of
Hertzian contacts. The elastic constant of this constriction is
given by $k=(1/k_{\mbox{\scriptsize cyl}}+2/k_{\mbox{\scriptsize
ele}})^{-1}$, where $k_{\mbox{\scriptsize cyl}}$ and $k_{ele}$ are
the elastic constants of the cylinder and electrodes,
respectively. As seen above, for the cylinder we have
$k_{cyl}=E\pi a^{2}/L$, and for the electrodes (the half spaces),
$k_{\mbox{\scriptsize ele}}=B Ea/(1-\nu^{2})$, with $B\approx 2$.
Note that the elasticity of the  electrodes can be more important
than that of the constriction itself if this is short, namely
$L<\pi a$. In this model the contact radius is almost constant,
only decreasing (increasing) slightly during elongation
(contraction) due to Poisson's ratio.

A somewhat more elaborate model of the constriction consists of
slabs varying cross-section \cite{torres96,untiedt97} (see
Fig.~\ref{f.contact}(b)). In this case, the elasticity is given by
all the slabs acting like a series of spring $1/k=\sum_{i}
1/k_{i}$. This model is valid only if the diameters of two
adjacent slabs are not too different, that is, if the changes in
cross section of the constriction are not very abrupt.

Note that for a contact composed of slabs the stress is
concentrated in the narrowest cross-section which, consequently,
will govern the yield condition. The maximum load
$F_{\mbox{\scriptsize max}}$ that a contact can sustain before
plastic deformation either for long or short constrictions,  is
given by $ F_{\mbox{\scriptsize max}}=\pi a^{2}
\sigma_{\mbox{\scriptsize y}}$.

This completes our brief summary of the relevant concepts in
continuum mechanics. We shall see that many of these familiar
properties of solids can be observed with slight modifications
down to the atomic scale, but quantum effects may lead to dramatic
modifications, as we will see most clearly in
Sects.\,\ref{s.chains} and \ref{s.shells}.

\subsection{Simultaneous measurement of conductance and force}
\label{s.experiments}

\begin{figure}[!t]
\begin{center}
\includegraphics[width=60mm]{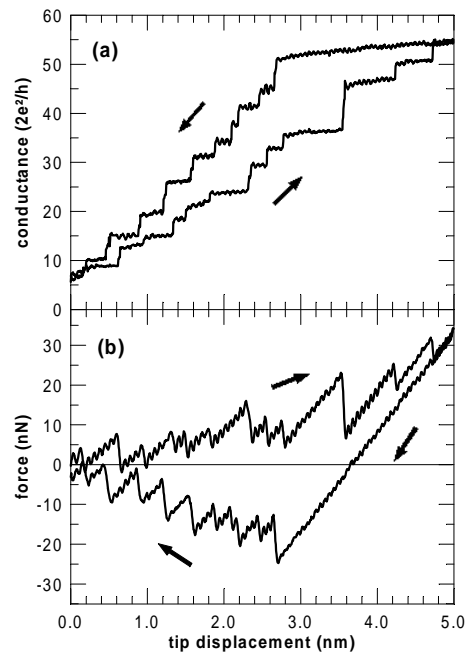}
\end{center}
\caption{\label{f.rubio1996_fig2} Simultaneous recording of
conductance (a) and force (b) during a cycle of contraction and
elongation of a constriction without breaking the contact.
Reprinted with permission from \protect\cite{rubio96}. \copyright
1996 American Physical Society. }
\end{figure}

In experiments measuring the conductance and force simultaneously
in atomic-sized contacts pressures consistent with the ideal
strength of the metals were found, much larger than those for
macroscopic contacts. In experiments for Pb \cite{agrait94} at 4.2
K, using an STM supplemented by a force sensor, it was found that
for contact radii between 3 nm and 13 nm  the pressure in the
contact during plastic deformation was  approximately 1\,Gpa.,
Improved resolution in subsequent experiments, on Au contacts
between 2 and 6 nm in diameter, at 4.2 K \cite{agrait95} and at
room temperature \cite{agrait96}, and for contacts down to a
single atom at room temperature \cite{rubio96}, show that the
deformation process, either for contraction or elongation,
proceeds in alternating elastic and yielding stages (see
Fig.~\ref{f.rubio1996_fig2}). These elastic stages are linear and
the elastic constant obtained from the slope is consistent with
the model of a short constriction, using the bulk  Young's
modulus. Yield takes place for pressures of 2--4 GPa, consistent
with the ideal strength, increasing up to about 13 GPa before
rupture. In these experiments the mechanical relaxations are
perfectly correlated to the jumps in the conductance, showing
beyond doubt that they have a mechanical origin. Somewhat larger
values for the pressure have been reported for experiments on Au
at room temperature \cite{stalder96}. However, in this  latter
experiment the resolution was not enough to resolve the
conductance plateaus or the elastic stages in the force.

High strengths have also been  observed in microindentation
experiments, where contacts were much larger. These experiments
where pioneered by Gane and Bowden \cite{gane68} who, using a 600
nm diameter flat punch on a specimen of annealed gold, measured a
yield stress of 1 Gpa. This is close to the shear strength
$\tau_{\mbox{\scriptsize Y}}=0.74$ GPa calculated for the slip of
an ideal Au lattice on \{111\} planes \cite{kelly86}, and much
higher than the typical value of 0.2 GPa for bulk polycrystalline
gold. The experiment was performed inside of a scanning electron
microscope which permitted to perform microindentation tests on
regions of the specimen free from dislocations. More recently,
similar results were obtained in nano-scale contacts by Michalske
and his collaborators using a variant of AFM (the so-called
interfacial force microscope). These experiments were performed on
gold thin films of 200 nm thickness \cite{tangyunyong93,thomas93},
and the curvature of the W tip  was $\sim 400$ nm. The surface of
gold was passivated by a self-assembling monolayer film to avoid
the adhesive interaction between probe and substrate, and the
results could be described by the Hertzian contact theory. They
found that for small contacts (radii of the order of  30 nm) the
yield stress was $\sigma_{\mbox{\scriptsize Y}}\sim 1$ GPa.

Catastrophic fracture-like yielding in Au nanocontacts (of
diameter larger than 5 nm) has been reported in experiments
performed at room temperature \cite{stalder96}. In these
catastrophic events the neck cross-section changes by one order of
magnitude. This observation contrasts with the abovementioned
experiments \cite{agrait95,agrait96} in contacts of similar sizes,
at low and room temperatures, and can be traced to the different
elastic constants of the sensor used (160 N/m in \cite{stalder96}
vs 705 and 380 N/m in \cite{agrait95} and \cite{agrait96},
respectively). A relatively soft sensor causes elastic energy
accumulation that is suddenly released in a catastrophic
avalanche.

Gold nanocontacts typically deform plastically down to the last
atom contact before fracture takes place, as demonstrated by the
numerous experiments in Au contacts showing a well-defined value
of the conductance of 1 $G_{0}$. This is the extreme mode of
ductile fracture (rupture). The force necessary to break this
one-atom contact is found to be also quite well defined with a
value of $1.5\pm 0.1$\,nN \cite{rubio96}, as shown in
Fig.~\ref{f.rubio1996_fig1}. Similar results have been found for
the mechanical forces during the elongation and rupture of an
atomic chain of gold atoms \cite{rubio01} which will be reviewed
in the Sect.~\ref{s.chains}.

The stiffness (or the effective spring constant) of the contact
has been measured directly using an ac method in UHV
\cite{jarvis99}. In this experiment an Pt-coated tip was used to
form contacts on a gold sample, and the first conductance step was
observed at approximately half integer values of the quantum of
conductance.

The variations of length of the plateaus for gold contacts at room
temperature  \cite{marszalek00}, which was found to be quantized,
has been related to the underlying processes during plastic
deformation.

The  effect of temperature on the mechanical properties of
metallic atomic contacts has not been studied systematically. For
Au contacts no significant difference is observed between the
mechanical properties at room temperature and those at 4.2 K
\cite{rubio96,agrait96,rubio01}, except for the much larger
stability of the contacts at low temperatures, which is in part
due to the slowing down of diffusion processes and in part due to
the higher stability of the experimental setup.

In a contact, since stresses are concentrated in the minimal cross
section, it is natural to assume that plastic deformation will
involve mostly the narrowest part of the constriction, which also
controls the conductance. Using this assumption in conjunction
with volume conservation it is possible to estimate the length
involved in plastic deformation from the curves of conductance vs
elongation \cite{untiedt97,stalder96} (see the next section). For
the smallest contacts it is found to be independent of contact
diameter and involves 5 to 6 atomic layers \cite{stalder96}, while
for larger contacts it varies.

From the experimental results for nanometer-scale contacts of Au
we can conclude that atomic constrictions go through a sequence of
discrete atomic configurations.  Each of these configurations
deforms elastically (and reversibly) until it yields, changing to
a new configuration in order to relax the stress. This new
configuration is of the order of one atomic spacing shorter (for
contraction) or longer (for elongation) than the previous one. For
each configuration the conductance is approximately constant, with
a variation that can be accounted for by the elastic changes in
the constriction due to Poisson's coefficient. The measured slope
during the elastic stages is proportional to $k_{c}+k_{s}$ where
$k_{s}$ is the finite stiffness of the force sensor, and $k_{c}$
is the effective spring constant of the contact, including the
electrodes. The measured effective spring constants are consistent
with the macroscopic Young's modulus for Au, taking into
consideration the uncertainty in the exact geometry of the
constriction \cite{rubio96,agrait95}.  The yield point for each
atomic configuration presents very large values consistent with
the  ideal strength in the absence of dislocations. The yield
point for each atomic configuration is independent of the sensor
stiffness but if the sensor is too soft an avalanche will occur
during elongation and some of the configurations will not be
accessible. Ideally the force sensor should have zero compliance,
in order not to affect the dynamics of the measurement.

\subsection{The shape of mechanically drawn metallic contacts}
\label{s.shape}

The scanning capabilities  of STM have been used to obtain an
estimate of the contact geometry. In the first work on metallic
contacts, Gimzewski  \ea \cite{gimzewski87}, studied the local
modifications induced by point contacts of a Pt-Ir tip and an Ag
substrate in UHV, by imaging the area where the tip has touched
the sample surface. They found that for clean metal-metal
contacts, after gentle indentation of the substrate, the
topography shows a pronounced protrusion of nanometer dimensions,
consistent with formation, stretching and breaking of necks of
atomic dimensions. The same approach has been used for contacts of
different sizes \cite{pascual93,agrait95,agrait96,gimzewski87a}
and  serves at best to give an estimate of the maximum dimensions
attained by the contact during the indentation by measuring the
extent of the plastically distorted area in the substrate but
gives no information of the relation between shape and electrical
or mechanical properties.

\begin{figure}[!t]
\begin{center}
\includegraphics[width=80mm]{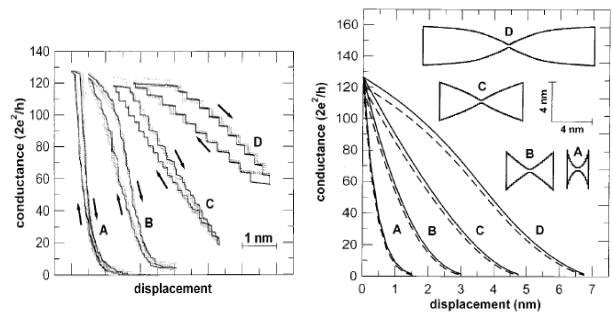}
\end{center}
\caption{\label{f.untiedt1997} Experimental conductance curves for
four different Au contacts at low temperature (left panel).
Fitting of the experimental curves gives the estimated shapes just
before breaking depicted in the right panel. All the necks have
the same initial cross section. Reprinted with permission from
\protect\cite{untiedt97}. \copyright 1997 American Physical
Society.}
\end{figure}

The shape of the constriction produced by plastic deformation can
be estimated from the conductance vs displacement curves ($I$-$z$
curves) \cite{untiedt97}, using the fact that the conductance of a
ballistic constriction is to a good approximation proportional to
the contact area, Eq.\,\ref{eq.sharvin} with a small perimeter
correction which depends on the shape (Sect.\,\ref{sss.sharvin}).
In a constriction submitted to a tensile force, stresses are
largest in the narrowest part and, consequently, we may safely
assume that plastic deformation occurs mostly in this area, in a
zone of extent $\lambda$, leaving the rest of the neck unmodified.
Assume that the contact at any point of its evolution can be
represented by the slab model of the previous section, which can
be considered symmetrical with respect to the center of the
contact for simplicity. Under tensile force the whole constriction
will deform elastically until the yield stress is reached in the
narrowest slab whose cross sectional area is $A_{i}$. We assume
that only a central portion of this slab of length $\lambda_{i}$
deforms plastically generating a longer and narrower slab in order
to relax stress. The cross sectional area  of the new slab
$A_{i+1}$ is given by volume conservation
\begin{equation}
A_{i+1}=\frac{A_{i}\lambda_{i}}{(\lambda_{i}+\Delta l)},
\end{equation}
where $\lambda_{i}+\Delta l$ is the length of the new slab. Since
only the central portion of the narrowest slab is modified, the
shape of the constriction after a number of this plastic
deformation stages can be deduced form the sequence of values of
$A_{i}$ and $\lambda_{i}$.

The plastic deformation length, $\lambda$, defined above, is
related to the portion of the constriction that participates in
the plastic deformation process, and in general it will depend on
the cross-section, length, and history of the constriction. It can
be calculated from the experimental $I(z)$ curve by noting that in
the limit $\Delta l \rightarrow 0$, $\lambda=-(d\ln A/dl)^{-1}$,
where $A$ is the cross-section of the narrowest portion of the
contact \cite{untiedt97}. For small contacts $\lambda$ is
typically constant ranging from 0.2 to 1 nm and corresponding to
an exponential behavior of $A(z)$. These results imply that for
small contacts only a few atomic layers participate in the plastic
deformation process \cite{stalder96}. In contrast experiments on
larger contacts show that $\lambda$ depends on the cross section
$A$ and also on the deformation history of the contact. As shown
in Fig.\,\ref{f.untiedt1997} contacts of similar cross sections
can have very different shapes. Typically, constrictions that have
been submitted to {\em training} by repeatedly compressing and
elongating before breaking are longer and have $\lambda\propto
A^{1/2}$. Thus, in this case the plastic deformation length is
proportional to the radius of the contact.  Au necks formed at low
temperature are not very different from those at room temperature.
For the latter long necks are somewhat easier to form and they are
typically about 30\% longer. This small difference is expected,
since room temperature is still much lower than the melting
temperature for gold.

\begin{figure}[!t]
\begin{center}
\includegraphics[width=70mm]{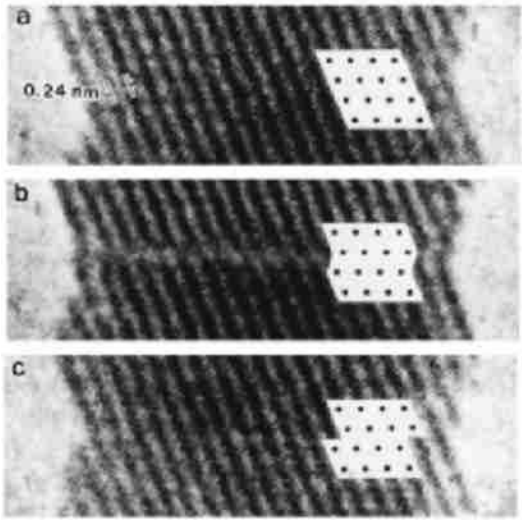}
\end{center}
\caption{\label{f.kizuka1998} Time-sequence series of
high-resolution images of elemental processes of slip during shear
deformation of a Au contact. Reprinted with permission from
\protect\cite{kizuka98a}. \copyright 1998 American Physical
Society. }
\end{figure}

The  effects of surface tension and surface diffusion have been
included in a continuum model calculation for the shape evolution
of a constriction in order to explain the experimental results in
Pb contacts at room temperature \cite{gai98}. These results show
that surface diffusion can either lead to the growth of the neck
or to its thinning and breaking depending on its curvature.
Note that at room temperature, surface diffusion effects are
expected to be more important in Pb than in Au due to the much
lower melting temperature of Pb.

The shape and even direct atomistic visualization of the process
of mechanical deformation in gold contacts has been possible using
high resolution transmission electron microscopy (HRTEM). A
simplified STM-like setup (piezo-driven specimen holder without
tunneling current  feedback control) was mounted inside a 200 keV
HRTEM, and a time resolution of 1/60 s and space resolution of 0.2
nm were achieved \cite{kizuka97}. Compression, tensile and shear
deformation experiments were performed in nanometer-sized gold
contacts \cite{kizuka98,kizuka98a}, showing that deformation
proceeds by slip and twinning in dislocation-free contacts of
about 4 nm width (see Fig. \ref{f.kizuka1998}). Pillar-like
structures were observed during retraction in 2 nm wide contacts
\cite{kizuka98}. For these smaller contacts, it was not clear
whether lattice slips proceed by a dislocation mechanism ({\em
i.e.,} by introduction of a dislocation or dislocation-like
localized strain) or by simultaneous displacement of lattice
planes, whereas in the 4nm wide necks the introduction of rapidly
disappearing partial dislocations was observed during the
deformation \cite{kizuka98a}. At room temperature surface
diffusion contributes to neck growth in addition to compressive
deformation for the smaller contacts \cite{kizuka98}.

\begin{figure}[!t]
\begin{center}
\includegraphics[width=70mm]{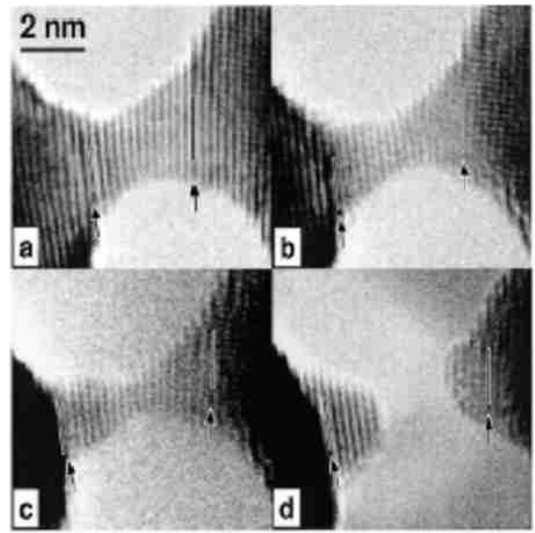}
\end{center}
\caption{\label{f.rodrigues2000} Nanowire evolution when stretched
along the [111] direction for gold. After the rupture the two
apexes reorganize and retract. Reprinted with permission from
\protect\cite{rodrigues00}. \copyright 2000 American Physical
Society. }
\end{figure}

Another possibility for the observation of the structure of
nanowires using HRTEM is to generate the nanowires {\em in situ}
by focusing the electron beam with a large current density  on
different sites of a self-supported metal thin film, which
produces holes that are allowed to grow until a nanometric neck is
formed \cite{ohnishi98,kondo97,rodrigues01,rodrigues00}. The beam
intensity is then reduced to perform image acquisition. Kondo \ea
\cite{kondo97} studied the structure of stable nanowires ranging
from 0.8 to 2 nm in thickness, and from 5 to 10 nm in length,
formed in a 3 nm thick Au(001) film. These nanowires, which are
remarkably straight and of uniform thickness along their axes, are
probably stabilized by their hexagonal surface reconstructions.
Rodrigues \cite{rodrigues00} formed nanowires in a polycrystalline
film of 5 nm thickness.
 The apexes forming the contact appeared to move
spontaneously with respect to each other, probably due to thermal
expansion of the whole film, leading to a slow elongation of the
nanowires, see Fig.\,\ref{f.rodrigues2000}. They observed that
just before rupture the gold nanowires are crystalline  and
display only three atomic configurations where either [100], [110]
or [111] directions lie approximately parallel to the elongation
direction. Mechanical behavior was brittle or ductile depending on
orientation. Single atom wires have also been resolved
\cite{kondo97,rodrigues01,takai01,kizuka01} but discussion of this
aspect will be deferred to Sect.\,\ref{s.chains}.

\section{Model calculations for atomic-sized contacts}
\label{s.models}

A full description of the transport and mechanical properties of
atomic-sized contacts requires a quantum-mechanical treatment of
both their nuclear and electronic degrees of freedom. This is the
idea behind {\em ab-initio} molecular dynamics simulations. The
problem is that the computational requirements are  so high that
systems that can be studied this way are limited to a small number
of atoms (see Sect.\,\ref{ss.abinitio}). Different approaches can
be followed to simplify this problem, dealing with different
aspects separately. In classical molecular dynamics (MD)
simulations (see Sect.\,\ref{ss.MD}), the dynamics and energetics
of the system are calculated assuming that the atoms respond to
adequately parameterized forces. The electrons, which in fact give
rise to these interatomic forces, are not taken into account
explicitly. The conductance can be obtained from the calculated
structure using different models but neglecting its effect on the
dynamics and energetics. On the other hand, it is possible to
concentrate on the electronic effects. The atomic structure can be
assumed fixed like in tight-binding (TB) (see
Sect.\,\ref{ss.TBcalculation}) and {\em ab initio} models or
completely ignored like in free electron (FE) (see
Sect.\,\ref{ss.force_models}) models.   These partial approaches
have been very illuminating, clarifying important aspects of the
problem.

\subsection{Molecular dynamics simulations of contact evolution}
\label{ss.MD}

The first molecular dynamics (MD) simulations of the formation
and fracture of metallic nanocontacts were carried out by Landman
{\em et al.} \cite{landman90}  and by Sutton and Pethica
\cite{sutton90}. These simulations modeled the interaction between
a tip and a substrate showing that contact formation is associated
with  an atomic-scale instability which leads to the
jump-to-contact phenomenon and involves the inelastic motion of
atoms in the vicinity of the interfacial region. The process of
elongation of the atomically thin neck formed during pull-off
proceeds via structural  atomic rearrangements, which occur when
the constriction becomes mechanically unstable. In between this
structural transformations the constriction deforms elastically.

The  conductance of a metallic contact during the process of
contact formation and fracture using MD was first calculated by
Todorov and Sutton \cite{todorov93} using a tight-binding (TB)
scheme. They found that the abrupt variations of the conductance
observed experimentally \cite{muller92} are related to the sudden
structural atomic rearrangements of the atoms in the contact. The
possible relation of the conductance steps with conductance
quantization in metallic contacts and the contradictory
experimental evidence
\cite{muller92,agrait93,pascual93,olesen94,brandbyge95}, led to
more MD simulations using either TB \cite{bratkovsky95} or
free-electron (FE) \cite{olesen94,brandbyge95,bratkovsky95}
schemes for the calculation of the conductance. In particular, the
question was whether the steps were due to abrupt decreases in the
contact cross section or whether they were due to the abrupt
pinching off of conductance channels during a smooth decrease in
the cross section. The correlation of the changes in conductance
with the changes in cross sectional area and the effect on the
conductance of scattering from the surface rugosity  of the
constriction \cite{brandbyge95,bratkovsky95} and internal defects
\cite{bratkovsky95} was also investigated. In these early
simulations not much attention was paid to the total force on the
contact, since the experimental results were not widely available.

\begin{figure}[!t]
\begin{center}
\includegraphics[width=70mm]{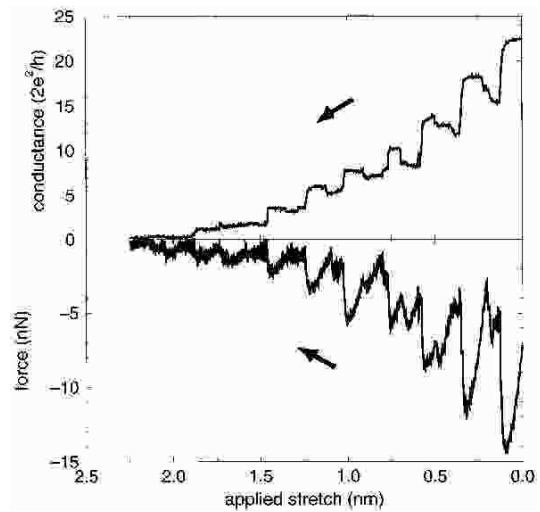}
\end{center}
\caption{\label{f.todorov-force} The force and conductance
throughout a dynamic simulation of the pull off of a Au contact at
1 K, with a pull-off rate of 4.08 m/s.
 Reprinted with permission
from \protect\cite{todorov96}. \copyright 1996 American Physical
Society.}
\end{figure}

Later, Todorov and  Sutton \cite{todorov96} found a correlation
between the force applied to the contact and the conductance
jumps, as shown in Fig.\,\ref{f.todorov-force}. This agrees with
the simultaneous jumps in force and the conductance found in the
experiments of Fig.\,\ref{f.rubio1996_fig1}. Brandbyge \ea
\cite{brandbyge97} analyzed the conductance of realistic contacts
in terms of transmission eigenchannels. They found that, except
for the smallest contacts (1--3 $G_{0}$), there are several
partially open channels due to scattering in the constriction.
That is, conductance quantization is lost above about 3 $G_{0}$.
The effect of crystalline orientation, and temperature was
considered by Mehrez \ea \cite{mehrez97,mehrez97a}.

Landman  \ea \cite{landman96a,landman96} studied the atomistic
mechanisms of deformation in relatively large constrictions, their
results explain the `reversibility' of plastic  deformation
observed experimentally \cite{agrait93,untiedt97}. S{\o}rensen \ea
\cite{sorensen98} also studied larger contacts and considering the
effect of crystalline orientation on the deformation and fracture
processes.

Barnett and Landman \cite{barnett97} and Nakamura \ea
\cite{nakamura99} used {\em ab initio} MD to simulate the breaking
of a sodium wire. For Al atomic contacts the jump-to-contact has
been simulated by \cite{heinemann97}. The mechanisms of formation,
evolution and breaking of atomically thin gold nanowires have been
recently investigated using classical  MD simulations by
Rubio-Bollinger \ea \cite{rubio01} and tight-binding MD
simulations by da Silva \ea \cite{dasilva01}.

\subsubsection{Principles of MD simulations}

Molecular dynamics (MD) simulations  consist of the modeling  of
the motion of the individual atoms or molecules within a system of
interacting species. The energetics and dynamics of the atoms are
obtained from interaction potentials from which the Newtonian
forces experienced by these atoms are derived. In  {\em ab initio}
or first principles methods the evaluation of the potential energy
is obtained from a quantum-mechanical description of the system,
and is limited, at present, to systems with a small number of
atoms and relatively short simulation times due to computational
demands. In contrast, the empirical and semi-empirical potentials
used in conventional MD simulations make it possible to simulate
much larger systems for much longer times. In between first
principles and empirical methods is the tight-binding molecular
dynamics (TBMD) method \cite{mehl96}, which is more accurate than
empirical potential methods  because it explicitly includes the
electronic structure and much faster than first principles
methods.

Conventional MD simulations use phenomenological inter-atomic
potentials to model the energetics and dynamics of the system.
Although simple pair potentials of the Lennard-Jones type have
been used \cite{sutton90}, an accurate description of metallic
systems requires more complex potentials that include many-body
interactions. These potentials contain the physics of the model
systems and their functional form is selected on the basis of
theoretical considerations and are typically fitted to a number of
experimental or theoretically calculated data.

The embedded atom method (EAM) \cite{foiles86} and effective
medium theory (EMT) \cite{jacobsen87} potentials derived from
density functional theory (DFT) in its quasi-atom \cite{stott80}
or effective medium \cite{norskov80} versions are often used to
model metallic systems. In these models the potential energy of
the system is written as a sum of a short-range pair-interaction
repulsion, and an embedding energy for placing an atom in the
electron density of all the other atoms:
\begin{equation}
E_{\mbox{\scriptsize pot}}=\sum_{i}    F_{i} [\rho_{h,i}]
+\frac{1}{2}\sum_{i} \sum_{j \ne i} V_{ij}(r_{ij}), \label{eq.EAM}
\end{equation}
where $V_{ij}(r_{ij})$ is  a two-body potential which depends on
the distance $r_{ij}$ between atoms $i$ and $j$, and
$F_{i}[\rho_{h,i}]$  is the {\em embedding energy} for placing an
atom at position $i$, where the host electron density  due to the
rest of the atoms  in the system is $\rho_{h,i}$. The latter is
given by
\begin{equation}
\rho_{h,i}=\sum_{j \ne i} \rho (r_{ij}),
\end{equation}
$\rho (r_{ij})$ being the `atomic density' function. The  first
term in Eq.\,(\ref{eq.EAM}) represents, in an approximate manner,
the many-body interactions in the system. These potentials
provide a computationally efficient approximate description of
bonding in metallic systems, and  have been used with significant
success in different studies \cite{rafii-tabar00}.

Closely related are the Finnis-Sinclair (FS) potentials, which
have a particularly simple form
\begin{equation}
E^{\mbox{\scriptsize FS}}_{\mbox{\scriptsize pot}}=\epsilon \left[
\frac{1}{2} \sum_{i} \sum_{j \ne i} V(r_{ij})-c\sum_{i}
\sqrt{\rho_{i}} \right],
\end{equation}
with $V(r_{ij})=(a/r_{ij})^{n}$, and $\rho_{i}= \sum_{j \ne
i}(a/r_{ij}^{m})$, where $a$ is normally taken to be the
equilibrium lattice constant, $m$ and $n$ are positive integers
with $n>m$, and $\epsilon$ is a parameter with the dimensions of
energy. For a particular  metal the potential is completely
specified by the values of $m$ and $n$, since the equilibrium
lattice condition fixes the value of $c$. The square root form of
the second term, which represents the cohesive many-body
contribution to the energy, was motivated by an analogy with the
second moment approximation to the tight binding model
\cite{rafii-tabar00}.

It must be emphasized that the  applicability and predictive power
of MD simulations using empirical potentials is  limited in
circumstances where the system evolves into regions of
configuration space not covered by the fitted data. That is,
potentials fitted to bulk properties may not be adequate to
describe low-coordinated systems such as surfaces or clusters or,
in particular, atomic-sized contacts. An improvement to this
situation consists in expanding the database used for the fitting
to include a set of atomic configurations calculated by {\em ab
initio} methods. This set may include not only three-dimensional
crystals with different lattice parameters but also slabs, layers
and atomic chains \cite{mishin99}.

\subsubsection{Implementation of MD simulations}

An MD simulation proceeds by constructing a finite  portion of an
infinite model system with any desired configuration in a primary
computational cell. The cell is replicated generating periodic
images of the system. This periodic boundary condition is
introduced to remove the undesirable effects  of the artificial
surfaces associated with the finite size of the simulated system.
In MD simulations of atomic-sized contacts the periodic boundary
conditions have been applied either along all three sides  of the
computational cell \cite{sutton90}, or only parallel to the
contact, keeping static several atomic layers on the top and
bottom of the computational cell  and using them as grips
\cite{landman90}. In the first case, all atoms are treated
dynamically, but cells above and below the contact are in
mechanical contact. In the second case, only a two-dimensionally
periodic array of contacts is modeled, but two artificial
interfaces in each cell between dynamic and static atoms are
introduced.

The equations of motion are integrated via the velocity Verlet
algorithm or predictor-corrector algorithms, with time steps
varying from 1 fs \cite{dasilva01} to 100 fs \cite{mehrez97}.
Constant temperature is imposed by controlling the average
temperature by means of an adequate thermostat, which can be
applied to all the atoms in the cell or, in the case where there
are static layers, just to the deepest dynamic layer.

Three different  geometries have been used in MD simulations of
nanocontacts. A tip that is lowered into contact with a slab and
then is pulled off
 \cite{landman90,sutton90,pascual95,todorov93,brandbyge95,landman91,sutton92,sutton94,pascual95a,todorov96}, a neck  that is either pulled off
 \cite{brandbyge95,bratkovsky95,sorensen98}, or elongated and
compressed \cite{landman96}, and a nanowire that is pulled off
\cite{olesen94,hasmy01,dasilva01}.

The displacement  of the tip with respect to the substrate, or
elongation and contraction of the constriction, is simulated by
either varying uniformly the height of the periodic cell or
rigidly displacing the grip layers. The speed of increment or
decrement of the height of the computational cell ranged from
0.004 m/s \cite{mehrez97,mehrez97a} to 60 m/s \cite{todorov93}.
These speeds,  although well bellow the speed of sound in the
material (of the order of $10^{3}$ m/s), are several orders of
magnitude higher than the speeds attained in STM and MCBJ
experiments of contact formation and fracture which  are in the
range of $10^{-10}$ and $10^{-7}$ m/s, and may be too fast to take
into account diffusive, thermally activated motion (see below).

The validity of the MD simulations for  the interpretation of real
life experiments must be carefully evaluated due to the enormous
difference in time scales \cite{todorov96,sutton96}. In an
experiment there can be different relaxation mechanisms, spanning
a wide range of time scales. As the strain rate is decreased, or
the temperature raised, slower relaxation processes, like
collective relaxation processes, come into play resulting in
differences in the mechanical evolution of the contact as put in
evidence by performing the same simulation at different rates
\cite{todorov96,mehrez97}. This could be particularly important in
the evolution of the shape of the neck in the final stages of the
pull off.

\subsubsection{Calculation of conductance in atomistic MD models}
\label{sss.MDconductance}

The calculation of the  conductance in a MD simulation of atomic
contacts is essential for comparison  with the experimental
results, since in most of the experiments only the conductance is
measured. One possibility is to use a tight-binding  (TB) model
(see Sect.\,\ref{ss.TBcalculation}) to calculate the conductance
of a given atomic configuration using the atomic coordinates
generated by the classical MD simulations. This method has been
used to study the formation and fracture of Ir \cite{todorov93}
and Au \cite{todorov96} contacts, and of the elongation and
fracture of a Ni constriction \cite{bratkovsky95}. These
calculations only utilize one atomic orbital (1$s$) and
consequently do not represent the electronic structure of the
metals accurately, however, the method enables a sufficiently
rapid calculation of the conductance to be made, in which the
positions of all the atoms within the contact are taken into
account explicitly.

Another possibility to  calculate the conductance in an MD
simulation is to use a free electron (FE) model
(Sects.~\ref{sss.free_electrons}). In the FE methods the internal
structure  of the constriction is replaced by a free electron
jellium in a hard-wall potential defined by the positions of the
atoms obtained from classical MD. This leaves out the effects of
the ionic disorder which may be present in real contacts. The
constriction potential profile is constructed by putting solid
spheres with the Wigner-Seitz radius at the atomic positions
obtained from the MD simulations \cite{bratkovsky95} or by
overlapping the free atom electron densities and calculating the
effective one-electron potential in the local density
approximation (LDA) \cite{brandbyge95}. Once the hard-wall
boundary is defined the free-electron cross section along the
constriction is determined, and a smoother axisymmetric
\cite{bratkovsky95} or rectangular \cite{brandbyge95} equivalent
profile is defined. The Schr\"{o}dinger equation in this profile
is solved exactly including interchannel scattering
\cite{bratkovsky95} or approximately neglecting interchannel
scattering \cite{brandbyge95} to obtain the transmission
probabilities of the different modes. The boundary roughness can
be treated perturbatively \cite{bratkovsky95}. An earlier, rather
too simplified approach was adopted by Olesen \ea \cite{olesen94}
who assumed that the constriction was adiabatic and, consequently,
transmission channels would be either totally open or closed.

Comparison between these two types of conductance calculations for
the same atomic configurations \cite{bratkovsky95}, shows a
similar conductance in both cases but a different detailed
structure. In this work the FE model is determined solely by the
profile of the contact, while the TB model depends on the precise
atomic structure of the contact. This difference becomes
particularly important in situations where the contact develops
structural defects \cite{bratkovsky95}.

Scattering due to internal disorder can be also taken into account
by taking a step further FE models. Brandbyge \ea
\cite{brandbyge97a} and S{\o}rensen \ea \cite{sorensen98} rather
than using a hard-wall potential for the constriction, considered
a one-electron potential generated from the atomic coordinates by
constructing the electronic density as a sum of free-atom electron
densities. The potential is then generated from the density using
the local density approximation (LDA). The macroscopic electrodes
are described by a free-electron model and join the contact in a
smooth manner. This potential is not self-consistent but gives a
good description  of the corrugation near the boundary. The
quantum transmission of electrons through the three-dimensional
potential is calculated using a numerical exact, recursive
multichannel method \cite{brandbyge97}. A similar approach is
followed by Mehrez \ea  \cite{mehrez97,mehrez97a}. However, these
authors axisymmetrize the potential and use the transfer matrix
method to calculate the conductance.

In the case of larger contacts an estimate of  the conductance of
a contact can be obtained from the contact radius using a
semiclassical modification of Sharvin's expression \cite{torres94}
(see Sect.\,\ref{sss.sharvin}).

\subsubsection{Results for simple metals}
\label{sss.simplemetals}

Several questions have  been elucidated by the classical MD
simulations of atomic scale contacts, namely, the mechanisms of
contact formation and fracture for homogeneous and heterogeneous
contacts; the mechanism of plastic deformation for small and
relatively large contacts; and the conductance of realistic atomic
contacts and its relation with the plastic deformation processes.
We will discuss neither one-atom contacts nor atomic chains since
classical MD simulations are not reliable in situations were
metallic atoms have low coordination.

Contact formation between the approaching surfaces of  tip and
sample is associated with an atomic-scale instability which causes
the atoms of the interfacial region to irreversibly jump to
contact at a distance of a few angstroms, in a short time span of
$\sim 1$ ps (compare to the experimental results in
Sect.\,\ref{ss.jump-to-contact}). Further advance of the tip
results in the onset of plastic deformation.  Separating the tip
and sample leads to ductile deformation of the contact, producing
an atomic-sized constriction or neck, which eventually fractures.
The mechanism of elongation of this constriction consists of a
sequence of brief atomic structural rearrangements during which
the constriction disorders and re-orders with the introduction of
a new atomic layer. In between these rearrangements, which occur
when the contact becomes mechanically unstable due to stress
accumulation, the contact deforms elastically. This mechanism of
plastic deformation seems to be a feature of atomic-sized metallic
contacts. This was shown in the MD simulations of the Au/Ni system
by Landman {\em et al.} \cite{landman90,landman91} using EAM
potentials. Qualitatively similar results were obtained by Sutton
and Pethica \cite{sutton90} using a Lennard-Jones  pair potential.

In the case of contacts  between different metals most of the
plastic deformation takes place in the softest metal
\cite{landman90,rafii-tabar00,landman91}. In the process of
contact formation, most of the jump-to-contact is due to the atoms
of the softer metal, irrespective of their being part of the tip
or substrate. As the tip continues advancing beyond this point, a
hard tip will indent a soft substrate, whereas a soft tip will
flatten against a hard substrate. If the softest metal wets the
hardest metal, the atomic size constriction that forms upon
retraction consists solely of the softest metal atoms. After
fracture a patch of the softest metal atoms remains on the hardest
metal surface.

Atomic scale contacts  show very high yield strengths. For Au
contacts, Landman and collaborators \cite{landman90,landman91}
obtained maximum pressures of the order of 10 GPa both under
tensile and compressive stress,  which implies shear strengths
even larger than the theoretical value for bulk Au in the absence
of dislocations \cite{kelly86}.

\begin{figure}[!t]
\begin{center}
\includegraphics[width=70mm]{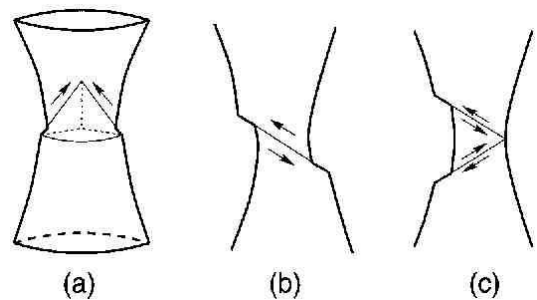}
\end{center}
\caption{\label{f.slip} Schematic illustration of selected slip
processes.
 (a) A three plane slip. (b) A simple single-plane slip.
 (c) A slip in two nonparallel slip planes. Reprinted with permission from \protect\cite{sorensen98}.
\copyright 1998 American Physical Society.}
\end{figure}

New mechanisms of deformation are revealed in simulations  with
larger constrictions. In these simulations, the initial state of
the simulation is a constriction cut from a perfect crystal in a
given orientation. Initially, the atomic positions are fully
relaxed for a certain interval of time. As the contact is
stretched,  the constriction deforms via a succession of
alternating stress accumulation and relief stages, during which it
undergoes inelastic structural transformations. In the case of fcc
metals, like Au,  and Ni, these transformations consist
preferentially of slip in one or several closed-packed $\{ 111 \}$
planes \cite{landman96,bratkovsky95,sorensen98}. When slip occurs
in several nonparallel planes, defects and some local disorder can
be introduced. When disordered regions are present, the subsequent
deformation mechanisms tend to involve the atoms in these regions,
thereby changing the atomic structure and, in general, reducing
the amount of disorder. In this way disorder often anneals out
during the elongation process. The zone involved in the structural
transformations, i.e. the {\em active zone,} extends over many
atomic layers in the contact, and is not limited to the narrowest
cross section \cite{sorensen98}. Two distinct mechanisms for slip
have been identified. For relatively large constrictions, slip
occurs via glide of a dislocation nucleated at the surface of the
contact \cite{landman96,bratkovsky95,todorov96,sorensen98}.
Typically, the dislocation is dissociated into Shockley partials,
which glide completely across the constriction in a structural
rearrangement, producing a stacking fault. The second partial
could glide through the same path in a subsequent rearrangement
and remove the stacking fault. For thinner constrictions slip is a
homogeneous shear of one plane of atoms over another plane of
atoms \cite{sorensen98}. In this case the slip is also dissociated
into partials. S{\o}rensen \ea find that in their simulations the
crossover between these two mechanisms takes place at contact
diameters  around 15 \AA\ \cite{sorensen98}. However, dislocation
glide has also been described by Bratkovsky \ea
\cite{bratkovsky95} in smaller diameters. Probably  not only the
diameter but also the aspect ratio of the constriction and the
type of metal play a role.

Deformation of  constrictions with different crystalline
orientations differ substantially due to the different orientation
of the slip planes \cite{mehrez97,sorensen98}. In the fcc
structure, there are four sets of $\{ 111 \}$ planes, which lie
parallel to the sides of a regular tetrahedron. For a constriction
oriented in the [111] direction, one set of these planes is
perpendicular to the constriction axis. The other three are
inclined with respect to this axis, and they are the active planes
in which slip can occur to relieve stress. These three active slip
planes are equivalent and a three-plane slip is possible  (see
Fig.\,\ref{f.slip}). For a constriction oriented in the [110]
direction, two sets of close-packed $\{ 111 \}$ planes lie
parallel to the axis and are, consequently, inactive. In contrast,
for a constriction oriented in the [100] direction all four sets
of slip planes are active. Formation of an atomic chain before
rupture also seems to depend on the crystalline orientation, being
more likely, in the case of Au, in contacts oriented in the [100]
direction \cite{sorensen98}.

Due to the possibility of having a large number of active glide
planes in various regions, the separation between force
relaxations may be quite irregular \cite{landman96,sorensen98}. In
contrast, as the wire thins down stresses concentrate on the
narrowest region and the number of active glide planes decreases,
resulting in a more regular pattern \cite{mehrez97}.

This slip mechanism explains the reversibility of plastic
deformation \cite{landman96}, observed experimentally for
elongation-contraction cycles in relatively large contacts. In the
experiments the contact could go over and over again through the
same sequence of conductance steps, the traces of the successive
cycles superposing almost exactly \cite{agrait93,untiedt97}. In
the simulation by Landman \ea \cite{landman96}, similar stress
accumulation and relief mechanisms, and atomic structural
rearrangement processes, including glide, occur during both
extension and compression of the contact. Overall mechanical and
structural reversibility is observed, but equivalent
configurations may differ in the position of some atoms.

\begin{figure}[!t]
\begin{center}
\includegraphics[width=80mm]{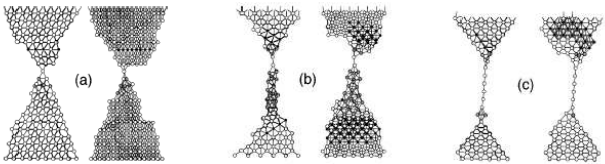}
\end{center}
\caption{\label{f.lastcontact} Snapshots of atomic configurations
 from MD simulations of Au contacts. The pictures show, from two different
 viewpoints, the final configuration just before rupture of three different
  contacts; (a) Au[111], (b) Au[110], and (c) Au[100]. Reprinted with permission from \protect\cite{sorensen98}.
\copyright 1998 American Physical Society.}
\end{figure}

In the last stages of elongation of a large constriction, as it
becomes relatively thin, the nature of the deformation changes.
The process in this regime involves localized atomic
rearrangements similar to those described for small constrictions
\cite{landman96,bratkovsky95,mehrez97,sorensen98}. Depending on
their crystalline orientation, contacts may become permanently
disordered in the narrowest region for the last part of the
elongation until rupture of the contact (see
Fig.\,\ref{f.lastcontact}).

As in the case of small constrictions the maximum axial stress
before yield  (the  yield strength) is very high.  It depends on
the precise atomic configuration, and for Au it is in the range
3--6 GPa \cite{landman96a,sorensen98}. This value, as mentioned
above, is of the order of magnitude expected  for bulk Au  in the
absence of dislocations. The mechanical ideal nature of the
nanowires can be related to their characteristic small dimensions
and the inability to support dislocation sources.

MD simulations show qualitative differences in the fracture
behavior for different materials \cite{sorensen98}. Contacts of Au
tend to be longer and smoother in the central region than  Ni
contacts. Ni contacts often break at a cross section of two atoms
or more which is seldom observed in Au contacts. Quantitatively,
differences are those to be expected from the difference in
macroscopic properties. The yield strength for Ni is found to be
10--20 GPa, while for Au it is 4--6 GPa.

The effect of the  compliance of the setup has been considered by
Brandbyge \ea \cite{brandbyge97a} and S{\o}rensen \ea
\cite{sorensen98}. A high compliance can arise from a soft
cantilever or the compliance of the macroscopic wires (say, a
sharpened tip). As a result new mechanical instabilities appear,
which prevent some atomic configurations to be probed during the
experiment. Conductance plateaus become flatter and the jumps in
the conductance or force are more pronounced. Well-ordered contact
structures, which are relatively strong, will tend to be probed
rather than weaker disordered configurations.

MD simulations show qualitative differences in the fracture
behavior for different materials \cite{sorensen98}. Contacts of Au
tend to be longer and smoother in the central region than  Ni
contacts. Ni contacts often break at a cross section of two atoms
or more which is seldom observed in Au contacts. Quantitatively,
differences are those to be expected from the difference in
macroscopic properties. The yield strength for Ni is found to be
10--20 GPa, while for Au it is 4--6 GPa.

As we will see below  in  Sect.\,\ref{sss.free_electrons},
transport through an adiabatic constriction (one that varies
sufficiently smoothly) for free and independent electrons is
quantized, in the sense that transmissions eigenchannels are
either open or closed. Mechanically drawn metallic constrictions
simulated using MD appear to be, in general, non-adiabatic and,
although quite crystalline, can have a number of defects, like
surface rugosity, stacking faults, vacancies, and local disorder.
These defects cause backscattering that can easily perturb the
conductance quantization
\cite{brandbyge95,bratkovsky95,brandbyge97a}. Brandbyge \ea
\cite{brandbyge97a} analyzed the conductance of Au contacts
resulting from MD simulations in terms of transmission
eigenchannels, and found that, except for the smallest contacts
(one or two atoms), several eigenchannels were partially open, and
the conductance had steps at non-integer values of $2e^{2}/h$. Due
to the irregularities in the contact the conductance is not always
controlled by the narrowest cross section, as would be the case
for smooth contacts. Higher temperatures
\cite{bratkovsky95,mehrez97,sorensen98} and lower deformation
rates \cite{todorov96,mehrez97} favor a higher degree of
crystallinity (see Sect.\,\ref{s.shape} for the structure of
contacts formed at room temperature) during the process of
elongation  and more regular constrictions \cite{bratkovsky95}
and, consequently, the observation of a quantized conductance.
Disordered contacts are found to be not only weaker than
well-ordered contacts of similar thickness but can also be pulled
longer and thinner than ordered contacts in the last part of the
elongation. As a consequence, lower temperature and higher pulling
rate favors longer contacts. The effect  of the temperature on
conductance histograms was considered by Hasmy {\em et al.}
\cite{hasmy01} by simulating cross-section histograms at 4, 300
and 450 K.

Note that these pulling rates are still at least 5 and 3 orders of
magnitude faster than the experiments. This indicates that the
real structure could be more ordered than those observed in the
simulations. As remarked above, the observed phenomena in the MD
simulations must be interpreted with some caution, since phenomena
on a longer time scale, inaccessible to the simulation, could play
a role in the deformation of atomic contacts. For example, the
contact could disorder in an atomic rearrangement, and have
insufficient time to reorder before the next rearrangement. For a
given volume of disordered material the time taken to reorder will
decrease with increasing temperature up to the melting point.
However, simulations performed at different temperatures offer, at
best, only a  qualitative idea of the effect of temperature. A
process that occurs in an experiment at a given temperature might
not take place in a MD simulation at the same temperature as, for
instance, diffusion processes may be inaccessible.

It is now well-established  from the results of numerous MD
simulations with realistic parameters that the abrupt jumps in the
conductance are the result of   sudden structural atomic
rearrangements leading to stress relaxation
\cite{brandbyge95,todorov93,bratkovsky95,todorov96,%
mehrez97,sorensen98,brandbyge97a}. As a consequence, jumps in the
force and conductance generally coincide with each other
\cite{todorov96,sorensen98,brandbyge97a}. The variations of the
narrowest cross section do not always coincide with the
conductance jumps \cite{bratkovsky95}, but this is not surprising
since the conductance does not depend exclusively on the narrowest
cross section of the contact. Occasionally atomic  rearrangements
occur far from the narrowest part, specially when the constriction
is long, which cause a mechanical relaxation and only a small
change in the conductance.

\subsection{Free-electron gas conductance and force models}
\label{ss.force_models}

In this section we will consider the free-electron (FE) models of
atomic contacts. The free electron approximation has proven its
value for the calculation of various properties of metals. It
works best for simple, monovalent metals, in particular the alkali
metals, for which the Fermi surface is nearly spherical. In FE
models of atomic contacts electrons move freely in a confining
potential,  which consists of wide (bulk) regions separated by a
more or less abrupt constriction whose dimension is of the order
of the wavelength of the electrons. The electrons are considered
independent, that is, the electron-electron interaction is
neglected and the only effect of the positive ions is to create a
uniform positive charge background that confines the electrons.
(In this section we will also several {\em jellium} calculations
beyond the FE approximation.) One is interested in the electronic
properties of this system as the shape of the constriction is
varied smoothly in an arbitrary but prescribed way. The lateral
electronic confinement leads to the quantization of the electronic
spectrum with the formation of subbands, and under certain
circumstances to the quantization of the conductance. Electronic
cohesion, thermoelectric effects, noise properties and magnetic
effects  have also been studied in the FE approximation. These FE
models are useful and instructive as they permit the  study  of
electronic effects without unnecessary complications. The virtue
of this approach is that both conductance and force are calculated
using the same physical laws. Nevertheless, it has a fundamental
deficiency: the atomic nature of metals is completely ignored.
This is particularly relevant when the mechanical properties of
the material are considered. Indeed, the mechanical response of
materials involves structural changes through displacement and
discrete rearrangement of the atoms, as we have seen in the
previous section. As a consequence, one has to be cautious with
the interpretation of  experimental results exclusively in terms
of FE models. Note that, in the experiments, the variation in the
shape of a metallic contact is, in fact, a mechanical process, and
since the atomic dimensions in metals are of the order of the
Fermi wavelength, variations in the shape of the constriction
cannot be smooth.

\subsubsection{Conductance calculations: conditions for the quantization  of the conductance}
\label{sss.free_electrons} As a first simplified model of a
contact or  constriction we may consider a uniform cylindrical
wire of radius $R$ and length $L$, with free and  independent
electrons connected to two bulk reservoirs \cite{bogachek90}. We
have to solve Schr\"{o}dinger's equation,
\begin{equation}
-\frac{\hbar^{2}}{2m^{*}}\nabla^{2}\psi ({\bf r})=E \psi ({\bf
r}),
\end{equation}
with the boundary condition $\psi[r=R]=0$. In this coordinate
system  and for  these boundary conditions the Schr\"{o}dinger
equation is separable, and the eigenstates are given by
\begin{equation}
\psi_{mn}(r,\phi,z)=J_{m}(\gamma_{mn}r/R)e^{im\phi} e^{ikz},
\end{equation}
where the $z$  coordinate is taken along the cylinder axis;
$m=0,\pm 1, \pm 2,\pm 3, \dots $, $n=1,2,3,\dots$, are the quantum
numbers; and $\gamma_{mn}$ is the  $n$-th zero of the Bessel
function of order $m$, $J_{m}$. The energies of the eigenstates
are
\begin{equation}
E_{mn}(k)=\frac{\hbar^{2}k^{2}}{2m^{*}}+E_{mn}^{c} ,
\end{equation}
where
\begin{equation}
E_{mn}^{c} = \frac{\hbar^{2}}{2m^{*}}\frac{\gamma_{mn}^{2}}{R^{2}}
. \label{eq.bandbottom}
\end{equation}
Since $J_{-m}(r)=(-1)^{m}J_{m}(r)$, the states $m$ and $-m$ are
degenerate. The electron states are divided into a set of
parabolic one-dimensional subbands. The bottom of each subband is
located at an energy $E_{mn}^{c}$. The separation of these
subbands increases as the radius of the cylinder decreases.
Neglecting scattering from the connections to the reservoirs, the
conductance of the wire is simply given by the number of subbands
that cross the Fermi level, $E_{F}$, with each subband
contributing $2e^{2}/h$ to the conductance (see
Sect.\,\ref{ss.scattering_approach}).  The conductance as a
function of the diameter of the contact will show perfectly sharp
steps, increasing by one unit at each zero of $J_{0}$ and by two
units at each zero of any of the other Bessel functions
\cite{bogachek90}. In the limit of low voltage and low
temperature, that is, $k_{B}T$ and $eV$ much smaller than the
subband splitting, the conductance will be {\em quantized} showing
steps at $G_{0}, 3G_{0}, 5G_{0}, 6G_{0},\dots$

\begin{figure}[!t]
\begin{center}
\includegraphics[width=60mm]{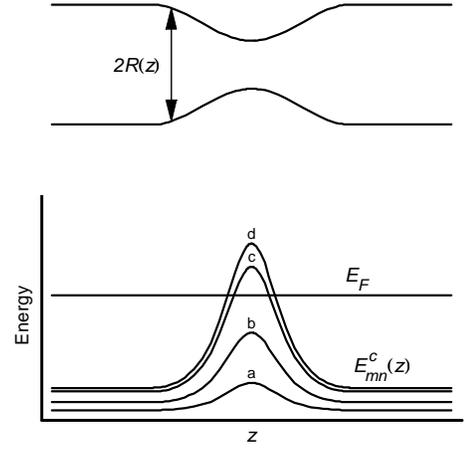}
\end{center}
\caption{\label{f.adiabatic-constriction} Quantization in an
adiabatic constriction. In a constriction with a slowly varying
profile the electrons moving in the $z$ direction see an effective
potential energy $E_{mn}^{c}(z)$. The curves labeled a, b, c, and
d correspond to the profiles of  $E_{mn}^{c}(z)$ for $(m,n)$ equal
$(0,2)$, $(2,1)$, $(1,1)$, and $(0,1)$, respectively. Only modes
with a barrier less than $E_F$ contribute to the conductance. }
\end{figure}

A more realistic model for an atomic contact  should take into
consideration explicitly the connections with the electrodes.
Consider a narrow axisymmetric constriction in a much larger wire
whose axis is in the $z$-direction. The profile of the
constriction is given by $R(z)$. Now we have to solve
Schr\"{o}dinger equation with the boundary condition
$\psi[r=R(z)]=0$. We find the solution by separating the lateral
and the longitudinal motion of the electron
\cite{brandbyge95,bratkovsky96},
\begin{equation}
\psi(r,\phi,z)=\sum_{mn} \chi_{mn}(z)u_{mn}^{z}(r,\phi),
\end{equation}
where $u_{mn}^{z}(r,\phi)$ is a solution of radial motion,
\begin{eqnarray}
-\frac{\hbar^{2}}{2m^{*}} \left( \frac{\partial^{2}}{\partial
r^{2}}+\frac{1}{r}\frac{\partial}{\partial r
}+\frac{1}{r^{2}}\frac{\partial^{2}}{\partial \phi^{2}} \right)
u_{mn}^{z} (r,\phi)= \nonumber \\
E_{mn}^{c}(z) u_{mn}^{z}
(r,\phi),
\end{eqnarray}
with the boundary condition $u_{mn}^{z}[r=R(z)]=0$, and has the
following form:
\begin{eqnarray}
u_{mn}^{z}(r,\phi)&=& \frac{1}{\sqrt\pi R(z) J_{m+1}(\gamma_{mn})}
J_{m}(\gamma_{mn}r/R(z))e^{im\phi}, \mbox{\hspace{5mm}}\nonumber \\
& &  m=0,\pm 1, \ldots.
\end{eqnarray}
Substituting this solution in Schr\"{o}dinger equation,
multiplying it by $u^{*}_{pq}(z)$, and integrating over $r$ and
$\phi$ at a given z, we obtain the following system of equations:
\begin{eqnarray}
\left( -\frac{\hbar^{2}}{2m^{*}}
\frac{d^{2}}{dz^{2}}+E_{mn}^{c}(z) \right) \chi_{mn}(z)+ \sum_{pq}
F_{mnpq} \chi_{pq}(z)\nonumber
\\ = E \chi_{mn}(z),
\nonumber
\\ \label{eq.adiabatic}
\end{eqnarray}
where
\begin{equation}
E_{mn}^{c}(z)=\frac{\hbar^{2}}{2m^{*}}\left(\frac{\gamma_{mn}}{R(z)}\right)^{2}.
\end{equation}
The operators $F_{mnpq}$ which  depend on $dR/dz$ and
$d^{2}R/dz^{2}$ couple the different solutions $\chi_{mn}(z)$.
The system of differential  equations  in
Eq.\,(\ref{eq.adiabatic}) can be solved exactly
\cite{bratkovsky96}. However, if the variation of $R$ with $z$ is
sufficiently  slow, the coupling between the different modes {\em
(mode mixing)} can be neglected, and we are left with a
one-dimensional Schr\"{o}dinger equation for each pair of quantum
numbers $(m, n)$ with an effective potential barrier
$E_{mn}^{c}(z)$ as depicted in
Fig.\,\ref{f.adiabatic-constriction}. This is called the {\em
adiabatic} approximation. The pair of quantum numbers $(m,n)$,
describe the transverse motion, and define the individual {\em
conductance channels}, which in this approximation are also
eigenmodes of the wire without the constriction. In contrast, for
contacts of arbitrary shape, the eigenchannels are given by linear
combinations of the eigenmodes of the system without the
constriction. The constriction {\em mixes} the different modes
(mode mixing) as described in Sect.\,\ref{sss.eigenchannels}.

In the adiabatic  approximation, modes for which the maximum of
the potential barrier $E_{mn}^{c}(z_{0})$ falls below the Fermi
energy $E_{F}$, will be perfectly transmitted and those for which
it is above will be reflected. However, when $E_{F}$ falls just
below the maximum of the barrier part of the mode will be
transmitted by tunneling through the barrier. When $E_{F}$ falls
just above the maximum of the barrier, part of the mode will be
reflected due to the variation of the potential height
(over-the-barrier scattering). Thus, the conductance, which is
controlled by the minimal radius of the contact $a_{0}$, will show
steps, as in the case of the cylindrical constriction, but the
steps will be somewhat smeared due to tunneling and
over-the-barrier scattering. Only in the case of very long and
slowly varying constrictions is this effect negligible.

\begin{figure}[!t]
\begin{center}
\includegraphics[width=70mm]{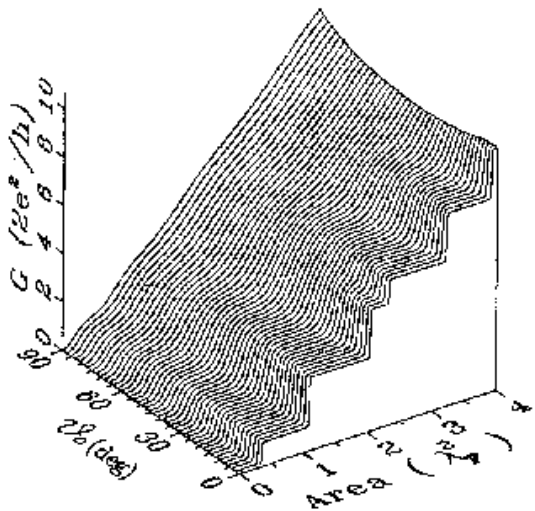}
\end{center}
\caption{\label{f.torres94} Conductance of a  quantum constriction
modeled as a hyperbola of revolution, as a function of the area
$A$ of the narrowest section and as a function of the opening
angle $\theta$ of the constriction. Reprinted with permission from
\protect\cite{torres94}. \copyright 1994 American Physical
Society.}
\end{figure}

The transition  from adiabatic to non-adiabatic contact is
elegantly illustrated in the calculation by Torres {\em et al.}
\cite{torres94}. These authors modeled the constriction as a
hyperbola of revolution and calculated the conductance exactly
using spheroidal oblate coordinates, taking advantage of the fact
that in these  coordinates Schr\"{o}dinger's equation is
separable. For small opening angles of the hyperbola,
$\theta_{0}$, the constriction has an elongated shape that tends
to a cylinder, whereas for $\theta_{0}=\pi/2$, the model
represents an infinitely thin barrier separating the two
electrodes, pierced by a circular hole. Their results are shown in
Fig.\,\ref{f.torres94}. The sharp steps obtained for a long
cylindrical wire ($\theta_{0}=0$) gradually smear as $\theta_{0}$
approaches $\pi/2$ as a consequence of tunneling and
over-the-barrier scattering which become important as adiabaticity
is lost. This shows that only in the case of adiabatic
constrictions the conductance is strictly quantized, that is,
transmission channels are either open or closed. For general
geometries, several evanescent channels with transmissions less
than one contribute to the conductance \cite{lopezciudad99}.

Contacts with cylindrical symmetry always have an orbital
degeneracy of the wave functions, which results in steps of height
$2G_{0}$ when a degenerate channel opens. This degeneracy is
lifted when the constriction is not exactly symmetrical, and only
conductance steps of height $G_{0}$ will be observable
\cite{bratkovsky96}. Modeling a constriction using a saddle point
potential may result in steps with higher degeneracies which are
also partially broken for non axisymmetric contacts
\cite{scherbakov96}.

Boundary roughness tends to suppress conductance quantization.
This question has been addressed by Bratkovsky and Rashkeev
\cite{bratkovsky96} for axisymmetric contacts using a perturbative
approach, and by Brandbyge \ea \cite{brandbyge97a} for contacts
with arbitrary shape using the recursion transfer-matrix  method
\cite{hirose95}. They found that the conductance steps shift
downward, and for strong corrugations, resonances start to build
up. In contrast, the effect of a localized scatterer within the
constriction is to smear or close single steps selectively
\cite{brandbyge97a}. The reason is that the modes that have a node
at the position of the scatterer will suffer less scattering.
These factors suppress conductance quantization since there will
be partially open modes. Nevertheless the stepped structure might
still be observable though not at integer values of the
conductance. The effects of disorder in the constriction and its
vicinity are further discussed in
Sect.\,\ref{s.defect_scattering}.

As we mentioned in Sect.\,\ref{sss.MDconductance}  the shape of
the contacts obtained from MD simulations can be used to define a
FE constriction and their evolution with plastic deformation
\cite{brandbyge95,bratkovsky95}. These calculations show plateaus
in the conductance but not always at integer values of the
conductance. The correlation observed in the MD simulations
\cite{brandbyge95,bratkovsky95} between the changes in conductance
and the atomic rearrangements in the contact show that the
conductance plateaus are related to mechanically stable atomic
configurations.

Now, we can summarize the circumstances  under which conductance
quantization can be observed within a FE model. Perfect
quantization in a contact is only obtained when the contact, in
addition to being ballistic, i.e. the length and diameter of the
contact being shorter then the electron mean free path, is
adiabatic, i.e. its cross section is a smooth function of the
longitudinal coordinate. Surface roughness and internal defects
cause backscattering that destroy conductance quantization.
Additionally, all energies $eV$ and $k_{B}T$ must be smaller than
the subband splitting.

\subsubsection{The relation between cross section
and conductance: Corrections to Sharvin's formula}
\label{sss.sharvin}

According to the semi-classical formula of Sharvin the conductance
of a small contact is proportional to the area of the contact.
This result is based on the proportionality of the number of
modes, with energies smaller than a given value in a 2-dimensional
system, to the area. However, in a finite system it is necessary
to add corrections. Consider a 2-dimensional rectangular box,
whose sides are $a$ and $b$, and assume that the modes have zero
amplitude at the boundary.  The  number of modes $N$ with energies
smaller than $E_{F}=\hbar^2 k_{F}^{2}/2m^{*}$ is  given by the
so-called Weyl expansion
\cite{hoppler98,weyl11,Morse&Feshbach,kac66}
\begin{equation}
N=\frac{k_{F}^{2}}{4\pi} ab -\frac{k_{F}}{2\pi} (a+b)
+\frac{1}{4},
\end{equation}
where the second term arises because the modes with $k_{x}$ or
$k_{y}$ equal to zero are not to be included. The third term
appears because the mode $k_{x}=0$ and $k_{y}=0$ appears twice in
the second term. For a general connected boundary without sharp
corners having an area $A$ and a perimeter $P$, the number of
modes is given by  \cite{hoppler98}
\begin{equation}
N=\frac{k_{F}^{2}}{4\pi} A -\frac{k_{F}}{4\pi} P +\frac{1}{6},
\label{eq.sharvinmodes}
\end{equation}
Note that the third term changes from $1/4$ to $1/6$. The
semi-classical  conductance, in the case of a uniform constriction
with cross section $A$ and perimeter $P$ will be simply given by
\begin{equation}
G=\frac{2e^2}{h}N.
\end{equation}
For an adiabatic constriction $A$ and $P$ correspond to the
narrowest cross section. The quantum oscillations are superimposed
to this smooth Sharvin conductance.

In the context of quantum point contacts this result was first
noted by S\'{a}enz and coworkers who found from their exact
quantum computation mentioned above \cite{torres94}, that the
conductance of a circular point contact deviates from Sharvin's
result. They observed that the perimeter correction (the second
term in Eq.\,(\ref{eq.sharvinmodes})) depends on the opening angle
since for large opening angle the constriction is no longer
adiabatic. Thus we may write
\begin{equation}
G \approx \frac{k_{F}^{2}}{4\pi} A -\alpha \frac{k_{F}}{2\pi}P.
\end{equation}
In the limit of a circular hole (opening angle
$\theta_{0}=\pi/2$), $\alpha=1/4$, while for an adiabatic
constriction (opening angle $\theta_{0}=0$) $\alpha=1/2$.

Another correction to the Sharvin conductance comes from the fact
that, in a real metal, electrons are not so strongly confined
\cite{garciamartin96,bogachek97}. There will be a certain spill
out of the wave functions that depends  on the work function of
the metal. The main influence of the `soft wall' boundary
condition is to increase the effective radius  of the constriction
relative to the `hard wall' case. This spill out can be taken into
account by using an effective radius $a_{0}+\delta a_{0}$ for the
constriction. For metals $k_{F}\delta a_{0}=$ 0.72--1.04 and
$\alpha=$ 0.13--0.025 \cite{garciamartin96}.

\subsubsection{Effect of magnetic fields}

Scherbakov \ea \cite{scherbakov96} used a saddle-point potential
to study the effect of a longitudinal magnetic field on the
conductance. They found that for relatively weak magnetic fields
(diameter of the constriction much smaller than the cyclotron
radius), the conductance exhibits Aharonov-Bohm-type oscillations.
This behavior transforms, in the strong field limit, into
Shubnikov-de Haas oscillations with an Aharonov-Bohm fine
structure.  The magnetic fields necessary to observe these effects
in small constrictions are quite high  since they depend on the
magnetic flux embraced by the contact. To observe the
Aharonov-Bohm oscillations or the splitting of the thermopower
peaks the magnetic flux through the contact area should be of the
order of the flux quantum ($hc/e$), and much higher for the
Shubnikov-de Haas oscillations. A reasonably high magnetic field
is of the order of $10$\,T. This means that the area of the
contact should be at least 100\,nm$^{2}$, which for a metal gives
a conductance of the order of 1000\,$G_{0}$. In the case of a
semimetal the situation is much better since the Fermi wavelength
is much larger. For example, in bismuth such a large contact would
have of the order of 2 channels.

Field effects have also been calculated for the thermoelectric
properties \cite{bogachek96} and for shot noise
\cite{bratkovsky96,scherbakov98}, in the framework of FE models.

\subsubsection{Nonlinear effects in the conductance}

Analyzing the nonlinear regime within the scattering approach
requires in principle a self-consistent determination of the
potential profile within the sample (see
Sect.\,\ref{sss.theory_limitations}). As a very crude
approximation one can neglect the voltage dependence of the
transmission coefficients and calculate the current as
\begin{equation}
I=\frac{2e}{h}\int_{E_{F}-eV/2}^{E_{F}+eV/2} \sum_{n}
\tau_{n}(\epsilon) d\epsilon,
\end{equation}
where $V$ is the bias voltage and $\tau_{n}(\epsilon)$ is the
transmission  probability of an electron with energy $\epsilon$.
Thus, the opening of new transmission channels will be gradual,
i.e. a new channel will be opening while the bottom of the band
traverses an interval $eV$ around the Fermi energy, and the sharp
steps in $G=I/V$ will be smeared \cite{garciamartin00}.

For a symmetric contact, the differential  conductance $g=dI/dV$
may be approximated by \cite{pascual97,bogachek97,garciamartin00}
\begin{eqnarray}
g(V)\approx \frac{2e^{2}}{h} \left( \frac{1}{2}\sum_{n}
\tau_{n}\left( E_{F}+\frac{eV}{2} \right)+ \right.
\nonumber \\
\left. \frac{1}{2}\sum_{n} \tau_{n}\left( E_{F}-\frac{eV}{2}
\right) \right).
\end{eqnarray}
Thus, $g$ can be regarded as the average of two zero voltage
conductances at different effective Fermi energies. As a
consequence the plateaus at integer values of $2e^{2}/h$ evolve
into  half-integer values
\cite{pascual97,bogachek97,garciamartin00}. The sequence of values
for these plateaus depends on the geometry of the contact. For a
non-axisymmetric contact at voltages of the order of $0.2
E_{F}/e$, there will be plateaus at half-integer and integer
values of $G_{0}$, while for contacts with cylindrical geometry
the sequence is
$0.5G_{0},1G_{0},2G_{0},3G_{0},4G_{0},4.5G_{0},5.5G_{0},6G_{0},\ldots$.
For abrupt constrictions, nonadiabatic effects manifest themselves
as an overall decrease of the conductance towards saturation at
very high voltages \cite{pascual97}.

The combined effect of a magnetic field  and high bias voltage has
been discussed by Bogachek \ea \cite{bogachek97}.

\subsubsection{Simulation of conductance histograms}

Conductance histograms are useful for presenting the  experimental
results on the evolution of the conductance of metallic contacts
with size, due to the variability of the particular features in
each experimental curve (see Sect.\,\ref{ss.histograms}).  In
order to approach the experimental situation S\'{a}enz and
coworkers \cite{lopezciudad99,garciamartin00} have introduced the
use of histograms in model calculations. They assume a certain
evolution of the contact area $A$ and opening angle $\theta$ as a
function of elongation $d$. The dependence of $A$ on $d$ is taken
to be almost exponential, as observed for the last part of the
elongation in the experiments \cite{untiedt97,stalder96}. The
qualitative results will not depend on the exact dependence,
however. The actual evolution will result in a discrete set of
points on a conductance curve $G(A(d),\theta(d))$. Assuming that
the contact can take any cross section along this curve, a
conductance histogram is obtained. The histogram shows peaks at
$1G_{0},2 G_{0},3 G_{0},4 G_{0},5 G_{0},\ldots$ for
non-axisymmetric contacts and $1G_{0},3 G_{0},5 G_{0},6
G_{0}\ldots$ for axisymmetric contacts. The peaks in the histogram
become more smeared as the conductance increases, reflecting the
fact the conductance does not show sharp steps for larger values
of the opening angle. Thus, the quantum nature of transport
manifests itself in the peaks of the histogram, but the
conductance is not quantized as any value is possible.

In an earlier work \cite{torres96} the  evolution of the
conductance was obtained from a mechanical slab model similar to
that described in Sect.\,\ref{s.shape}. Histograms have been also
used \cite{garciamartin00} to compare the nonlinear effects due to
a large bias voltage with experimental results
\cite{yasuda97,itakura99}. The effects of disorder on the
conductance histograms is discussed in
Sect.\,\ref{s.defect_scattering}.

\subsubsection{Quantum effects in the force}
\label{sss.quantum-effects}

Quantum-size effects on the mechanical properties of metallic
systems have previously been observed in metal clusters, which
exhibit enhanced stability for certain magic numbers of atoms.
These magic numbers have been rather well explained in terms of a
shell model based on the jellium approximation (see
Sect.\,\ref{ss.clusters}). In a metallic constriction of nanometer
dimensions lateral confinement of the electrons leads to a
discrete number of subbands that act as delocalized chemical
bonds. As an adiabatic constriction is stretched to the breaking
point, the force resulting from this electronic cohesion shows
force oscillations synchronized with the quantized jumps in the
conductance. Consider the simple model of
Sect.\,\ref{sss.free_electrons}, namely, the uniform cylindrical
wire \cite{ruitenbeek97,blom98}. In this case, the density of
states (DOS) $D(E)$ is given by
\begin{equation}
D(E)=L \sum_{mn}^{+} \sqrt{\frac{2m}{\hbar^{2} \pi^{2}}}
\frac{1}{\sqrt{E-E_{mn}^{c}}},
\end{equation}
where the + on the summation indicates that $m$ and $n$ run to the
maximum value for which the argument of the square root is
positive.  The number of electrons $N$ in the wire is obtained by
integration,
\begin{equation}
N=2L \sum_{mn}^{+} \sqrt{\frac{2m}{\hbar^{2} \pi^{2}}}
\sqrt{E_{F}-E_{mn}^{c}},
\end{equation}
and the total (kinetic) energy $E_{\mbox{\scriptsize tot}}$ can be
found by integrating over the product of the DOS and the energy,
\begin{equation}
E_{\mbox{\scriptsize tot}}=NE_{F}-\frac{4L}{3}\sum_{mn}^{+}
\sqrt{\frac{2m}{\hbar^{2}\pi^{2}}} (E_{F}-E_{mn}^{c})^{3/2}.
\end{equation}

The longitudinal force resulting when the wire is elongated at
constant volume, can be obtained from the derivative of the
thermodynamic potential $\Omega=E_{\mbox{\scriptsize tot}}-E_{F}N$
(assuming that the chemical potential is fixed):
\begin{eqnarray}
F=-\frac{d\Omega}{dL}= \sum_{mn}^{+} \sqrt{
\frac{2m}{\hbar^{2}\pi^{2}} }  \left\{
\frac{4}{3}(E_{F}-E_{mn}^{c})^{3/2} - \right.
\nonumber \\
\left. 2(E_{F}-E_{mn}^{c})^{1/2}E_{mn}^{c} \right\} .
\label{eq.force1}
\end{eqnarray}

More realistic constrictions with nonuniform profiles can be
treated within the scattering approach using the relation between
the scattering matrix and the density of states discussed in
Sect.\,\ref{sss.dos-scattering} \cite{stafford97,kassubek99}. This
makes possible to study transport and mechanical properties of
the nanowire using the same formalism.

\begin{figure}[!t]
\begin{center}
\includegraphics[width=70mm]{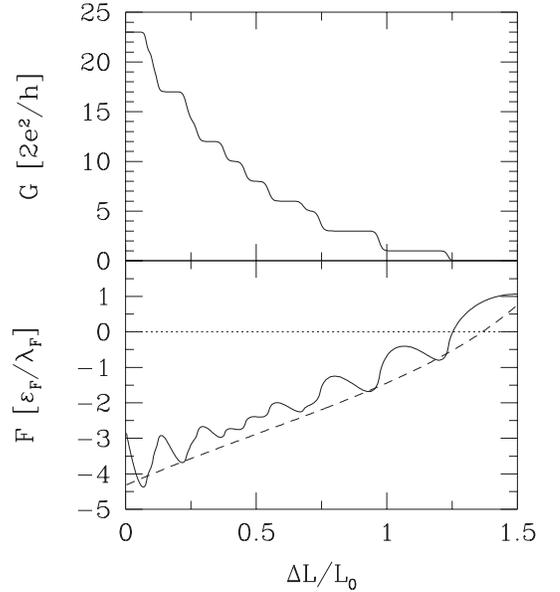}
\end{center}
\caption{\label{f.force-stafford} Calculation of the electrical
conductance $G$ and tensile force $F$ for a jellium model of an
adiabatic constriction in a cylindrical quantum wire versus
elongation. The dashed line indicates the contribution to the
force  due to the macroscopic surface tension. Reprinted with
permission from \protect\cite{stafford97}. \copyright 1997
American Physical Society.}
\end{figure}

Fig.\,\ref{f.force-stafford} shows the evolution of the electrical
conductance and force as an adiabatic constriction is elongated
\cite{stafford97}. The total force can be divided into a smooth
component, due to the increase of surface area, and an oscillating
component, which is of the order of $E_{F}/\lambda_{F}$ {\em
(universal force fluctuations)}. This oscillating correction to
surface tension is correlated with the conductance, presenting
local minima (taking maximal elongational force) as the
conductance approaches a step-drop and local maxima (minimal
elongational force) as the conductance channel closes.
Qualitatively, closing a conductance channel whose contribution to
the conductance is $2e^{2}/h$ requires a force of the order of
$E_{F}/\lambda_{F}$ independent of the total number of conducting
channels. Similar results (with sharper local maxima) are obtained
for a uniform cylindrical wire \cite{blom98}, and a non-adiabatic
wide-narrow-wide geometry \cite{kassubek99}. disorder in the
constriction has a strong effect on cohesion, which is sensitive
to the specific impurity distribution at the center of the
constriction \cite{stafford98,burki99,stafford99}.

The magnitude of these force oscillations ($E_{F}/\lambda_{F}$=1.7
nN) is similar to the magnitude of the forces observed
experimentally in Au for the smallest contacts \cite{rubio96} (see
Fig.\,\ref{f.rubio1996_fig1}) and it has been suggested
\cite{stafford97} that the abrupt atomic rearrangements observed
during deformation of metallic contacts may be caused by the
breaking of extended metallic bonds formed by each conductance
channel. However, there are two aspects of the experimental curves
that cannot be explained within this FE model: (a) The magnitude
of the force oscillations  observed experimentally depends on the
total conductance (being larger for larger conductances). For Au
contacts with conductances in the range of 10 and 60 quantum
units, force oscillations range between 6 and 15 nN \cite{rubio96}
(see Fig.\,\ref{f.rubio1996_fig2}), while for contacts ranging
from 100 to 300 quantum units, the force oscillations are in the
range of 20 to 50 nN \cite{agrait95}. (b) Metals, and solids in
general, can stand compressive stresses. The experimental evidence
is that this holds also true for Au contacts with conductances
larger than 10--20 quantum units (see Fig.\,\ref{f.rubio1996_fig2}
a compressive force (positive) is needed to contract the contacts.
Only for smaller contacts the force is tensile (negative) during
compression.

Both these aspects are related to the same deficiency of the
model: the ionic background is incompressible but cannot oppose to
shear deformations, that is, it is {\em liquid-like} and the shear
strength of the material would be zero. Shear strength appears
naturally in crystals as a resistance to sliding of atomic planes
with respect to each other as discussed in
Sect.\,\ref{sss.plastic}. However, as the constrictions become
smaller, the surface-to-volume ratio increases and we can expect
that for the smallest  contacts the behavior will be dominated by
surface tension (see the end of Sect.\,\ref{sss.plastic}).

The free  electron model assumes that the chemical potential is
fixed by the electrodes and is constant as the wires change
configuration. This leads to charge fluctuations
\cite{ruitenbeek97}, which are of the order of $e$, the charge of
the electron \cite{kassubek99}. However, in metals, screening is
very effective down to a scale $\sim 1$ \AA. One would expect
screening of the charge oscillations to occur in nanowires. As a
first approximation, one can impose a charge neutrality
constraint, determining the electrostatic potential
self-consistently to enforce global charge neutrality. Note that,
in this case, it is the electrochemical potential that matches the
Fermi energy. The resulting fluctuations in the electrostatic
potential would be observable as fluctuations in the work function
\cite{ruitenbeek97,pogosov01}.  However, screening will be poorer
for the smallest wires, since the charging energy needed to
establish the electrostatic potential woud be too large.

The effect of screening on the force oscillations was
overestimated in Ref.\,\cite{ruitenbeek97}, because the
interaction of the positive jellium background with the
self-consistent potential was neglected. Kassubek \ea
\cite{kassubek99} showed that this effect is small, justifying the
FE approximation. In fact, the force oscillations obtained from
more elaborate calculations that go beyond the free electron
approximation are very similar to those obtained in the FE model,
both for a uniform cylindrical wire \cite{yannouleas97,zabala99}
and an adiabatic constriction \cite{yannouleas98}. Yannouleas \ea
\cite{yannouleas98} used the so-called shell-correction  method,
which uses non-selfconsistent electronic structures from an
extended Thomas-Fermi theory, but takes shell structure into
account as a correction. This method is not selfconsistent but has
been shown, in clusters, to yield results in excellent agreement
with experiments and selfconsistent calculations.  Zabala {\em et
al.} \cite{zabala99} use the stabilized jellium model within
spin-dependent density-functional theory in order to obtain a more
realistic description which takes into account not only screening
but also electron exchange and correlation.

\subsection{Tight-binding models for the conductance}
\label{ss.TBcalculation}

Tight-binding models provide a simple description of electron
states in solids that can be considered as complementary to the FE
models. Instead of using plane waves as a starting point, TB
models are based on localized orbitals. TB models were originally
developed to describe the bands arising from tightly bound $d$ or
$f$ valence orbitals in transition metals. There exist, however,
several parameterization schemes which allow describing the bands
in $sp$-like metals rather accurately using these type of models.
TB models can also be regarded as a discretization of the
Schr\"odinger equation which allows  studying the electron states
in  non-homogeneous systems of arbitrary geometry. There exist
also a large experience with these type of models coming from
localization studies in disordered conductors (see, for instance,
\cite{mackinnon81,lee81a}).

In its simplest version, the TB model uses an orthogonal basis
$\left\{ |i>\right\}$ corresponding to a spherically symmetric
local orbital at each atomic site in the system. Within this
basis, the model Hamiltonian adopts the form

\begin{equation}
\hat{H} = \sum_i \epsilon_i |i><i| + \sum_{i\neq j} t_{ij} |i><j|,
\end{equation}
where the $\epsilon_i$ correspond to the site energies and
$t_{ij}$ denote the hopping elements between sites $i$ and $j$,
which are usually assumed to be non-zero only between nearest
neighbors.

Within these models, electronic and transport properties are
conveniently analyzed in terms of Green function techniques. The
retarded and advanced Green operators are formally defined as
\begin{equation}
\hat{G}^{r,a}(\omega) = \lim_{\eta \rightarrow 0} \left[ \omega
\pm i \eta - \hat{H} \right]^{-1}.
\end{equation}
The matrix elements of $\hat{G}^{r,a}(\omega)$ are directly
related to the local densities of states (LDOS) by
\begin{equation}
\rho_i(\omega) = \mp \frac{1}{\pi} \mbox{Im} <i|
\hat{G}^{r,a}(\omega) |i>.
\end{equation}

\begin{figure}[!t]
\begin{center}
\includegraphics[height=70mm,angle=270]{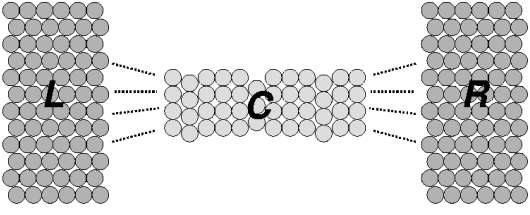}
\end{center}
\caption{\label{l-c-r.geometry} Typical geometry for conductance
calculations of atomic-sized conductors using TB models}
\end{figure}

In order to study the conductance of a finite system described by
a tight-binding Hamiltonian one may consider the geometry depicted
in Fig.\,\ref{l-c-r.geometry} in which a central region of atomic
dimensions is connected to two semi-infinite systems playing the
role of current leads. The total Hamiltonian for this system can
be decomposed as $\hat{H} = \hat{H}_L + \hat{H}_R + \hat{H}_C +
(\hat{V}_{L,C} + \hat{V}_{C,R} + h.c. )$, where $\hat{H}_{L,R}$
describe the electronic states in the uncoupled left and right
leads, $\hat{H}_C$ corresponds to the central region and the
$\hat{V}_{\alpha,\beta}$ terms describe the coupling between the
central region and the left and right leads. The zero-temperature
linear conductance is given in terms of Greens functions by the
following expression
\begin{equation}
G = \frac{8e^2}{h} \mbox{Tr} \left[ \mbox{Im}
\hat{\Sigma}^a_L(E_F) \; \hat{G}^a_{C}(E_F) \; \mbox{Im}
\hat{\Sigma}^a_R(E_F) \; \hat{G}^r_{C}(E_F) \right],
\label{conductanceTB}
\end{equation}
where $\hat{\Sigma}^{r,a}_{\alpha} = \lim_{\eta \rightarrow 0} \,
\hat{V}_{\alpha,C} \left[\omega \pm i\eta -
\hat{H}_{\alpha}\right]^{-1} \hat{V}_{C,\alpha} \,$ (with $\alpha
= L,R$) are self-energy operators introducing the effects  on the
dynamics of the electrons in the central region due to the
coupling with the leads, and $\hat{G}^{r,a}_{C}$ are the Green
operators projected on the central region.

Different versions of Eq.\,(\ref{conductanceTB}) have been derived
by many different authors using different approaches
\cite{todorov93,caroli71,levy92}. This expression is in fact
similar to the one derived using linear response theory in
Sect.\,\ref{sss.Green}. However, it does not rely on the
assumption of ideal leads used in that case. For more technical
details the reader may consult Ref. \cite{datta97a}. By using the
cyclic property of the trace, the expression (\ref{conductanceTB})
can be written in the usual form $G= (2e^2/h) \mbox{Tr}
[\hat{t}(E_F) \hat{t}^{\dagger}(E_F)]$, where \cite{cuevas98a}

\[ \hat{t}(E) = 2 \left[ \mbox{Im} \hat{\Sigma}^a_L(E_F) \right]^{1/2}
G^r_C(E) \left[ \mbox{Im} \hat{\Sigma}^a_R(E_F) \right]^{1/2}. \]

The existence of $\left[ \mbox{Im} \hat{\Sigma}^a_{L,R}(E_F)
\right]^{1/2}$ is warranted by $\mbox{Im}
\hat{\Sigma}^a_{L,R}(E_F)$ being positive definite matrices. The
knowledge of the $\hat{t} \hat{t}^{\dagger}$ matrix in terms of
Green functions allows determining the conduction channels for a
given contact geometry, as will be discussed below.

\subsubsection{Results for simple model geometries}
\label{sss.modelgeometries}

TB models allow obtaining some analytical results for the
conductance in certain limiting cases. Mart\'{\i}n-Rodero \ea
\cite{ferrer88,martinrodero88} developed a model for the
transition from tunneling to contact between a STM tip and
metallic sample for which the conductance is given by

\begin{equation}
G = \frac{8e^2}{h} \frac{\pi^2 T^2 \rho^0_{1} \rho^0_{2}}{|1 +
\pi^2 T^2 \rho^0_{1} \rho^0_{2}|^2}, \label{alvaro}
\end{equation}
where $T$ is the hopping element between the tip apex atom and the
nearest atom on the sample surface, and $\rho^0_{1,2}$ is the
unperturbed local densities of states on these two sites. This
expression predicts that the conductance quantum is reached when
$\pi^2 T^2 \rho^0_0 \rho^0_1 \simeq 1$, a condition that is
approximately fulfilled when using parameters appropriate for
simple metals.

Todorov \ea \cite{todorov93} analyzed the conductance for a single
atom between two semi-infinite fcc crystals cut along their (111)
planes using a tight-binding model with one $1s$ orbital per site.
They found that the conductance reaches the maximum value $2e^2/h$
for half-filled band when the atom is connected to the three first
neighbors on each surface with the same hopping element as in the
bulk crystal.

The conductance for model geometries of single-atom contacts have
also been analyzed using more realistic TB models including
several orbitals per atom in order to account for $s$, $p$ and $d$
bands in simple and transition metals. Sirvent \ea
\cite{sirvent96} studied in this way the influence of $d$ orbitals
in the conductance of Au, Ni and Pt. They used two types of model
geometries: one in which a single atom is at the center of a small
cluster having a threefold or a fourfold symmetry; and one in
which they replaced the central atom by two atoms along the
symmetry axis of the cluster. In both cases the semi-infinite
leads were simulated by Bethe lattices attached to the outermost
sites of the cluster. They showed that Ni and Pt clusters exhibit
a larger conductance than the corresponding Au clusters for all
the geometries considered. While in the case of Ni and Pt the
obtained conductance values ranged from $G_0$ to 3$G_0$, in the
case of Au these values were in general smaller than $G_0$. These
differences have been attributed to the contribution of $d$
orbitals in transition metals which provide additional channels
for conduction \cite{sirvent96}.

\subsubsection{Electron-electron interactions and the charge neutrality condition}

One question which was still open at that time was the strong
tendency to quantized values in first steps of the experimental
conductance traces for monovalent metals like Au, Ag or Cu. In FE
models, one can adjust the contact cross section and the electron
density according to the type of atom in order to get nearly
perfect quantization for simple metals
\cite{brandbyge95,torres96}. However, these results depend
critically on the ratio between the Fermi wavelength and the
contact diameter. In the case of TB models, conductance
quantization is harder to obtain using realistic models for the
contact atomic structure \cite{bratkovsky95}. The issue of the
apparent robustness of conductance quantization in experiments on
atomic contacts of monovalent metals was addressed in
\cite{levy97} using TB models in combination with model geometries
for the atomic structure. A fundamental difference with previous
TB or FE calculations was the inclusion of the charge neutrality
condition which provides a simple way to account for
electron-electron interactions in the neck region.

The assumption of local charge neutrality is reasonable for
metallic systems with a screening length comparable to interatomic
distances \cite{pernas90}, as was also pointed out in the context
of FE models in \ref{sss.quantum-effects}. Within a TB model this
is achieved by a self-consistent variation the diagonal elements
of the Hamiltonian at each site in the contact region
\cite{levy97}. The charge neutrality condition provides thus a
certain consistency between the Hamiltonian parameters and the
contact geometry.

In the first part of \cite{levy97} it was shown that sharp tips
modeled by pyramids grown along different crystallographic axis on
a fcc lattice exhibit narrow resonances in their LDOS. They showed
that these resonances become very sharp for tips grown along the
(111) axis while broader resonances are obtained for the (100)
direction. When forming a contact by connecting two of these
pyramids by a central common atom the resonances still survive and
tend to be pinned at the Fermi level due to the charge neutrality
condition. The conductance for such a situation can be
approximated by the expression $G \simeq \frac{8e^2}{h}
\frac{x}{(1+x)^2}$, where $x = \mbox{Im} [\Sigma_R(E_F)]/
\mbox{Im} [\Sigma_L(E_F)]$ measures the asymmetry between the left
and right sides of the contact. The conductance quantization due
to this resonance mechanism is thus very robust to fluctuations in
the atomic positions: even with $x \sim 2$ one still obtains $G
\simeq 1.8 e^2/h$, i.e. $90 \%$ of the quantum unit.

Recently, Kirchner \ea \cite{kirchner01} have proposed a sum rule
that would allow to include electron-electron interactions in a
parameterized Hamiltonian for an atomic contact beyond the charge
neutrality condition.

\subsubsection{Eigenchannels analysis}
\label{sss.TBeigenchannels}

In \cite{cuevas98a}, Cuevas \ea analyzed the conductance channels
in atomic contacts of $sp$ metals, like Al and Pb, using a TB
model. This work was motivated by the experimental results of
Scheer \ea \cite{scheer97}, discussed in Sect.\,\ref{ss.sgs_exp},
indicating that three channels contribute to the conductance of Al
one-atom contacts. This result was rather surprising in view of
the fact that the conductance on the first plateau of Al is
usually smaller than $1 G_0$.

For describing the $sp$ bands in Al Cuevas \ea \cite{cuevas98a}
used a TB model with parameters obtained from fits to ab-initio
calculations for bulk metals \cite{papaconstantopoulos86}. For
self-consistency between the atomic geometry and the electronic
structure they imposed the charge neutrality condition as in
\cite{levy97}. They showed that for the case of Al one-atom
contacts with an ideal geometry the conductance arises from the
contribution of three channels: a well transmitted channel having
$sp_z$ character (here $z$ indicates the axis along the contact)
and two degenerate poorly transmitted channels having  $p_x p_y$
character. These results were shown to be robust with respect to
disorder in the leads surrounding the central atom. It should be
pointed out that, within these type of models, the maximum number
of channels for one atom is fixed by the number of valence
orbitals having a significant contribution to the bands at the
Fermi energy, which would yield 4 in the case of $sp$ metals.
However, as these authors pointed out, the anti-symmetric
combination of $s$ and $p_z$ orbitals do not contribute to the
conductance due to a destructive interference effect which holds
even for large deviations from the ideal geometry
\cite{cuevas98a}, so that one-atom contacts for $sp$ metals are
expected to have three conductance channels.

Ref. \cite{scheer98} discusses the results of a combined
theoretical and experimental effort designed to test the
predictions of TB models for the conduction channels of one-atom
contacts for a large variety of materials ranging from $sp$
metals, like Al and Pb; transition metals, like Nb; and simple
metals like Au. The theoretical calculations predicted 3 channels
for $sp$ metals, 5 for transition metals and only one for
monovalent metals. The experimental evidence supporting these
predictions will be discussed in Sect.\,\ref{ss.sgs_exp}.

\begin{figure}[!t]
\begin{center}
\includegraphics[width=70mm]{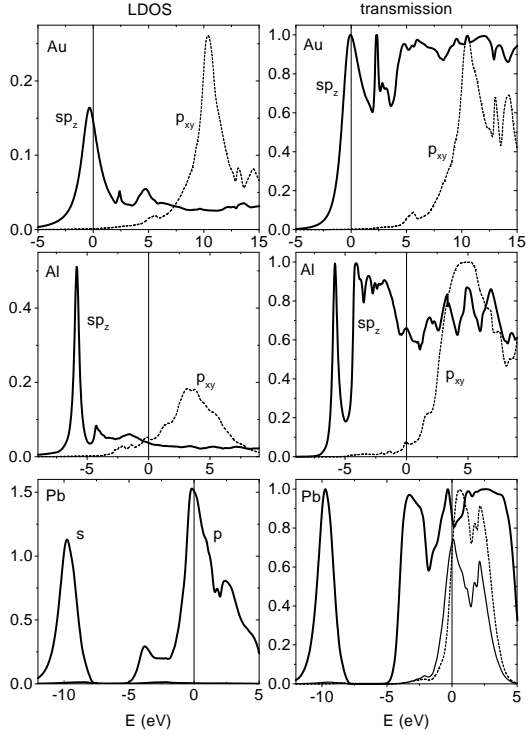}
\end{center}

\caption{LDOS at the central atom and transmissions vs energy for
ideal geometries representing Au, Al and Pb one-atom contacts. The
vertical line indicates the position of the Fermi level. Reprinted
with permission from \protect\cite{cuevas98}. \copyright 1998
American Physical Society.} \label{f.ldos+channels}
\end{figure}

TB models provide a direct microscopic insight on the conduction
channels of  atomic-sized contacts. This is illustrated in
Fig.\,\ref{f.ldos+channels} where both the LDOS at the central
atom and the transmissions as a function of energy for each
individual channel are shown for Au, Al and Pb one-atom contacts.
In the case of Au the calculation predicts the presence of a
single relevant channel at the Fermi energy. This channel arises
mainly from the contribution of the $6s$ orbitals. As can be
observed in the top panel of Fig.\,\ref{f.ldos+channels}, the LDOS
exhibits a resonance around the Fermi energy. The charge
neutrality condition pins the corresponding transmission resonance
to the Fermi energy and provides, as discussed in \cite{levy97}, a
strong mechanism for the almost perfect conductance quantization
for Au at the first plateau. For the case of Al (middle panel in
Fig.\,\ref{f.ldos+channels}) both $3s$ and $3p$ orbitals have an
important weight at the Fermi energy. Although the relative
position and shape of the $s$ and $p$ bands is similar to the case
of Au, the Fermi level lies closer to the $p$ bands. One thus
finds three channels with non-negligible transmission: the $sp_z$
channel with transmission $\sim 0.6$, and the $p_{x,y}$ channels
with transmission of the order of 0.1.  Finally, in the case of
Pb, with an extra valence electron as compared to Al, the Fermi
level moves to a region where both $sp_z$ and $p_{x,y}$ are widely
open. the calculated conductance for the ideal geometry is $\sim
2.8 G_0$.

More recently, Brandbyge \ea \cite{brandbyge99} have used a
non-orthogonal TB model to study the conduction channels of Au
one-atom contacts at finite bias voltage. The parameters for a
non-orthogonal basis (consisting of $s$, $p$ and $d$ orbitals) are
obtained from fits to ab-initio band calculations
\cite{papaconstantopoulos86}. These fits are in general more
accurate than the ones obtained using an orthogonal basis. On the
other hand, Brandbyge \ea imposed the charge neutrality condition
in order to obtain self-consistently the voltage drop along the
contact in a voltage bias situation. They considered chains of 3
or 6 gold atoms attached to layers of 4 and 9 atoms in both ends,
which again are connected to the (100) faces of perfect
semi-infinite electrodes. They found that the current is due to
the contribution of a single channel with nearly perfect
transmission up to bias voltages of the order of 1 $V$. They also
analyzed model geometries with (111) oriented electrodes, for
which they found a rather narrow resonance with perfect
transmission around the Fermi energy in agreement with
\cite{levy97} and \cite{cuevas98a}. The voltage drop along the
contact was found to have no symmetry despite the left-right
symmetry of the model geometry. This was attributed to the absence
of electron-hole symmetry in the LDOS.

\subsection{Ab-initio calculations}
\label{ss.abinitio}

In classical  MD simulations and TB models of metallic
point-contacts of the previous sections the interaction potentials
and hopping parameters are given as parameterized functional
forms. However, an accurate description of the energetics and
dynamics of a material would require a first-principles
calculation. Within this approach the total energy of a system of
ions and valence electrons can be written as \cite{barnett93}
\begin{eqnarray}
E_{\mbox{\scriptsize total}} (\{{\bf r}_{I}\}, \{ \dot{{ \bf
r}}_{I} \})=\sum_{I} \frac{1}{2} m_{I} |\dot{{\bf r}}_{I}|^{2} +
\nonumber \\
\sum_{I>J} \frac{Z_{I}Z_{J}}{|{\bf r}_{I}-{\bf
r}_{J}|} + E_{\mbox{\scriptsize elect}}(\{{\bf r}_{I}\}),
\label{eq.abinitio}
\end{eqnarray}
where ${\bf r}_{I}$, $m_{I}$,  and $Z_{I}$ are the position, mass,
and charge  of the $I$-th ion, and $E_{\mbox{\scriptsize
elect}}(\{{\bf r}_{I}\})$ is the ground-state energy of the
valence electrons evaluated for the ionic configuration $\{{\bf
r}_{I}\}$. The first two terms in this equation  correspond to the
ionic kinetic and interionic interaction energies, respectively.
In this equation we have considered that the electrons follow the
instantaneous configuration of the ions (Born-Oppenheimer
approximation).

The major task in  first principles or {\em ab initio} methods is
to calculate the ground-state electronic energy  which is
typically done via the Kohn-Sham (KS) formulation \cite{kohn65} of
the density functional theory (DFT) of many-electron systems
within the local density (LDA) or local-spin-density (LSD)
approximation. The ground-state energy of the valence electrons,
which according to the Hohenberg-Kohn theorem \cite{hohenberg64}
depends only on the electronic density, is given by
\begin{equation}
E_{\mbox{\scriptsize elect}}=T_{e}+E_{eI}+E_{ee},
\end{equation}
where $T_{e}$ is the  kinetic energy; $E_{eI}$ is the electron-ion
interaction energy, with the interaction between ions and valence
electrons typically described by pseudopotentials; and $E_{ee}$ is
the electron-electron interaction energy, which consists of
Hartree and  exchange-correlation parts.

In the Kohn-Sham (KS)  method the many-body problem for the
ground-state electronic density  $n({\bf r})$ of an inhomogeneous
system of $N$ electrons in a static external potential (due to the
positive ions) is reduced to solving self-consistently the
independent-particle Sch\"{o}dinger equation
\begin{equation}
[-\nabla^{2}+v_{\mbox{\scriptsize eff}}(n)]
\psi_{j}=\epsilon_{j}\psi_{j}, \label{eq.KS1}
\end{equation}
with the electronic density given by
\begin{equation}
n=\sum_{j} f_{j}|\psi_{j}|^2, \label{eq.KS2}
\end{equation}
where $f_{j}$ are the occupation of the $j$th (orthonormal)
orbital. The KS effective  potential $v_{\mbox{\scriptsize eff}}$
is given by the functional derivative of $E_{\mbox{\scriptsize
elec}}$
\begin{equation}
v_{\mbox{\scriptsize eff}}=\frac{\delta E_{\mbox{\scriptsize
elec}}}{\delta n},
\end{equation}
and since it depends on the electronic density it must be obtained
self-consistently.

{\em Ab initio} calculations  of the  structure and properties of
metallic point-contacts have been performed using different
degrees of approximation. Self-consistent electronic  structure,
field and current calculation have been performed by Lang and
collaborators \cite{lang87,lang95,lang97,lang97a,lang98,lang00}
and Kobayashi and collaborators
\cite{kobayashi99,kobayashi99a,kobayashi00}, for one-atom contacts
\cite{lang87,lang97} and for atomic chains of Al
\cite{lang95,kobayashi99,kobayashi99a,kobayashi00}, Na
\cite{lang97a,kobayashi99,kobayashi99a,kobayashi00}, and C
\cite{lang98,lang00}. In these calculations the metal electrodes
are described using the uniform-background (jellium) model and the
atomic cores, whose positions are pre-assigned, using a
pseudopotential. Since the electrostatic potential is also
self-consistently calculated, it is possible to study systems
under finite voltages \cite{lang00}. This type of calculations are
very demanding computationally and the number of atoms that can be
included is very limited. Simplified (non-selfconsistent)
approaches to the calculation of the conductance have been
followed by several authors to study Al wires
\cite{mehrez97,wan97} and Na wires \cite{sim01}.

First-principle  molecular dynamics simulations have been used to
simulate the breaking of an atomic-sized sodium wire at high (190
K) \cite{barnett97} and low temperature \cite{nakamura99}. In
these calculations the electronic structure, total energy and
forces on the ions are calculated self-consistently, while the
current is obtained from the KS orbitals using  the
linear-response Kubo formula \cite{barnett97} or by calculating
the transmission probability through the self-consistent potential
\cite{nakamura99}. Barnett and Landman \cite{barnett97} find that,
at high temperature, as the nanowires are stretched to just a few
atoms, the structure of the neck can be described in terms of the
configurations observed in sodium clusters. However, at this
temperature the structure is undergoing perpetual thermally
excited configuration changes and it is not clear  whether the
cluster-derived structures have a significant weight in the
time-averaged structure of the contact.  Nakamura \ea
\cite{nakamura99} study the interplay between conductance modes
and structural deformation.

In {\em ab initio} studies  of structural properties the atomic
configuration are allowed to relax under a given constraint until
the total forces on the atoms is negligible. The atomic and
electronic structure and stability of atomic chains of Au have
been studied extensively
\cite{sanchez99,okamoto99,torres99,hakkinen99,hakkinen00,demaria00},
triggered by the experimental evidence on the formation of atomic
gold chains. Atomic chains of Al \cite{taraschi98,sen01}, C
\cite{lang98}, Ca, Cu, and K \cite{sanchez01} have also been
investigated. The bond strength and breaking forces in Au
\cite{rubio01}, and Ni, Pd, Pt, Cu, Ag, and Au \cite{bahn01} have
been studied, in conjunction with EMT MD to gain insight on the
formation mechanisms of atomic chains. Okamoto \ea
\cite{okamoto99} also calculated the conductance of Au chains of
different lengths using a non-selfconsistent
recursion-transfer-matrix method. Taraschi {\em et al.}
\cite{taraschi98} investigated the structural properties of Al
nanowires with cross sections from one to a few atoms, studying
the crossover from the atomic wire behavior.

Recently, Mehrez \ea \cite{mehrez02} and Brandbyge \ea
\cite{brandbyge02} presented fully self-consistent DFT
calculations of the conductance of atomic contacts in which the
atomic structure of the whole system (the contact and the
electrodes)  is considered on the same footing. The effect of the
finite potential is also taken into account using non-equilibrium
Green's functions. The effect of the detailed atomistic
description of the electrodes in the conductance within {\em ab
initio} calculations has been discussed by Palacios \ea
\cite{palacios02}. They find that, in contrast to the case of Au,
the conductance in Al atomic contacts is very sensitive to the
precise geometry of the whole system.

Most  of the results of first-principles calculations are related
to atomic chains, since in this case the size of system is
sufficiently small to be handled. These results will be reviewed
in Sect.\,\ref{s.chains}.

\section{The character of the conductance modes in a single atom}
\label{s.exp_modes}

Conductance measurements, discussed in Sect.\,\ref{s.conductance},
reveal that the last plateau before the contact breaks is of the
order of $G_0$ for most metallic elements. A closer inspection
shows that there are important differences between different
materials. Thus, while noble metals like Au exhibit typically a
rather constant last plateau very close to $G_0$, in other metals
like Al and Pb the last plateau usually exhibits a clear slope
when elongating the contact, which is positive for Al and negative
for Pb.

These typical behaviors are certainly reflecting important
differences in the electronic structure of all these metals, which
give rise to differences in the structure of its conductance modes
in an atomic-sized contact. However, conductance measurements by
themselves do not yield much information on these modes. They only
give us the sum of the corresponding transmission coefficients and
its variation upon elongation or compression of the contact.
Although we can be sure that more than a single channel
contributes to the conductance when $G$ is larger than $G_0$ we
cannot claim that only one mode is contributing when $G \le G_0$.
It is possible that several poorly transmitted channels add to
give a total transmission smaller than one.

To obtain further information on the conductance modes requires
the measurement of additional independent transport properties.
What we need is a quantity which should be a non-linear function
of the transmission coefficients, so that it provides information
independent of the conductance. The aim of this section is to
present different experimental techniques that have been proposed
in the last few years to extract information on the conduction
channels. We shall also analyze the results for several metallic
elements like Al, Pb, Nb and Au. It will be shown that these
results are consistent with the simple picture of conduction
channels arising from the atomic orbital structure of each
element, which was presented in Sect.\,\ref{ss.TBcalculation}.

\subsection{Experiments on the superconducting subgap structure}
\label{ss.sgs_exp}

Metallic elements like Al, Pb, Nb, etc. become superconducting at
temperatures routinely attainable in the laboratory. The
combination of STM or MCBJ techniques and cooling to low
temperatures have permitted to explore the superconducting
properties of atomic-sized contacts of different metallic elements
\cite{scheer97,scheer98}. The highly non-linear $IV$
characteristic of a superconducting contact, discussed
theoretically in Sect.\,\ref{s.sctransporttheory}, has revealed to
be a powerful tool to extract information on the conductance
modes. This technique will be discussed below.

\subsubsection{First experiments: the tunneling regime}
\label{sss.subgaptunneling}

\begin{figure}[!t]
\centerline{\includegraphics[width= 8cm] {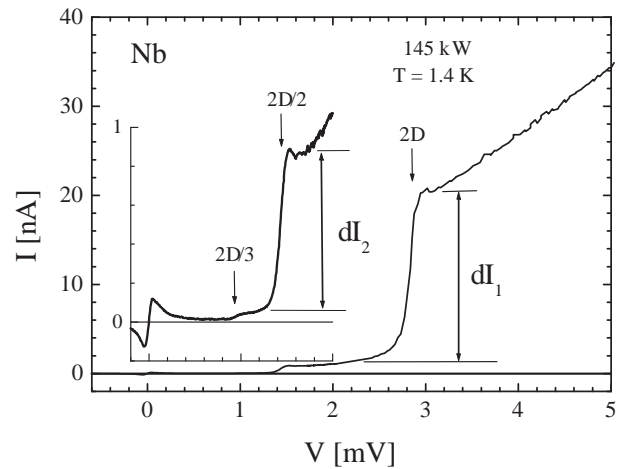}}
\caption{Current-voltage characteristics for a Nb vacuum tunnel
junction, using a MCBJ device at 1.4 K. The superconducting
transition temperature is $T_c = 9.1$\,K and the gap is $\Delta=
1.4$\,meV. In addition to the usual steep rise of the current at
$eV = 2\Delta$ there are smaller current onsets at $2\Delta/n$, as
is clear on the expanded scale for $n=2$ and 3. Reprinted with
permission from \protect\cite{post94}. \copyright 1994 American
Physical Society.} \label{f.Nbtunnelcurve}
\end{figure}

The first quantitative analysis of the subgap structure in
superconducting atomic contacts was presented by van der Post \ea
\cite{post94}. For these experiments and the ones described below
it is essential that all electrical wires to the sample space are
carefully filtered to block radio-frequency radiation onto the
atomic junction. Appropriate techniques for filtering are
described in Refs.\,\cite{vion95,glattli97}. Van der Post \ea
studied the $IV$ characteristic of Nb and Pb contacts in the
tunneling regime produced by the MCBJ technique
(Fig.\,\ref{f.Nbtunnelcurve}). They showed that the sizes of the
current steps in the subgap structure are proportional the
\nobreak{$n$-th} power of the transmission, where $n$ is the order
of the step, for $n=1, 2$, and 3. This decrease in step height is
expected, based on their explanation in terms of multiple Andreev
reflections, as discussed in Sect.\,\ref{s.sctransporttheory}.
The precise expression for the $n$-th current step at low
transmission $\tau$, first derived by Bratus \ea \cite{bratus95},
is given by

\begin{equation}
\delta I_{n} = \frac{e \Delta }{\hbar} \tau^n
\left(\frac{2n}{4^{2n-1}} \right) \left(n^n/n! \right)^2 .
\end{equation}

In fitting the experimental results of Ref.\,\cite{post94} a
single channel was assumed. This is a natural assumption for the
tunneling regime in an atomic-sized contact because the various
orbitals of the front-most atoms have different exponential decay
lengths into the vacuum and the slowest decaying mode will soon
dominate the conductance. Indeed, at closer separation (higher
tunnel conductance), contributions of more than one channel have
been detected \cite{ludoph00}. The study of the $IV$ curves in the
tunneling regime was an important first step towards the analysis
of the subgap structure in the contact regime.

\subsubsection{$sp$-metals: Al and Pb}

The first fits of the $IV$ curves of atomic contacts in the
superconducting state using the numerical results of the
microscopic theory were presented by Scheer \ea \cite{scheer97}.
They studied Al atomic contacts produced by the MCBJ technique in
combination with litographically defined samples, described in
Sect.\,\ref{sss.micro-MCBJ}.

\begin{figure}[!t]
\centerline{\includegraphics[width=7cm]{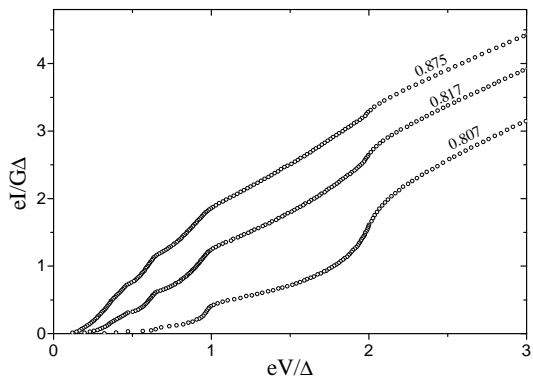}}
\caption{Typical superconducting $IV$ curves recorded on the last
conductance plateau for Al atomic contacts at 30\,mK, well below
the superconducting transition temperature $T_{\rm c}=1.17$K. The
voltage has been scaled to the superconducting gap value,
$\Delta$, and the current to the product of the conductance and
the gap, $G\Delta$. Three cases with similar normal conductance
(0.875, 0.817 and 0.807 $G_0$) show very different subgap
structure. Data taken from \protect\cite{scheer97}.}
\label{f.sgsAl}
\end{figure}

Typical $IV$ curves recorded on the last conductance plateau are
shown in Fig.\,\ref{f.sgsAl}. The normal conductance is given by
the differential conductance at bias voltages much larger than
$2\Delta$. One of the more surprising results found by these
authors is that contacts having a very similar value of the normal
conductance exhibit very different $IV$ curves in the
superconducting state, as can be observed in Fig.\,\ref{f.sgsAl}.
This result is already suggesting that the channel decomposition
of the conductance should differ from contact to contact.

\begin{figure}[!b]
\centerline{\includegraphics[width= 7cm]{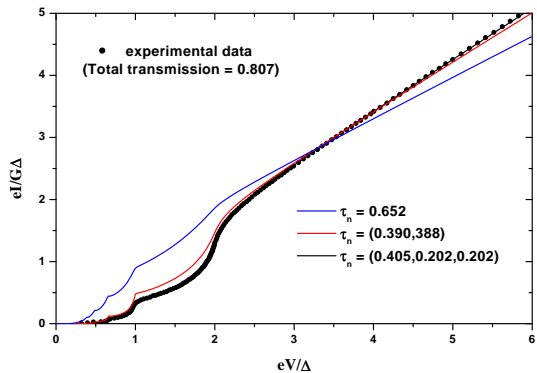}}
\caption{Comparison of one of the experimental curves of
Fig.\,\protect\ref{f.sgsAl} with theoretical fits using one
(blue), two (red) and three channels (black curve). The sets of
transmissions used were $0.652$, $(0.390,0.388)$ and
$(0.405,0.202,0.202)$. Data taken from \protect\cite{scheer97}.}
\label{f.sgsAl2}
\end{figure}

The $IV$ curves on the last plateau for Al cannot in general be
fitted satisfactorily using the single channel theory. The best
fit using a single channel for one of the curves in
Fig.\,\ref{f.sgsAl} is shown in Fig.\,\ref{f.sgsAl2}. In the
fitting procedure a set of 100 numerical $IV$ curves with
transmissions covering evenly the $(0,1)$ range and calculated
with the microscopic theory of Ref.\,\cite{cuevas96} were used. As
can be observed, although the single channel theory reproduces the
correct qualitative behavior it fails to reproduce the $IV$ curves
quantitatively. This is particularly evident for the excess
current at large bias voltages.

Scheer \ea proposed that the discrepancy with theory is due to the
fact that more than one conduction channel is contributing in the
Al contacts even though the total conductance $G$ can be smaller
than $G_0$. Assuming that the contribution of the different
channels to the superconducting $IV$ curves are independent the
total current can be written as

\begin{equation}
I(V) = \sum_n I_0(V,\tau_n) ,
\end{equation}
where ${\tau_n}$ is the set of transmission coefficients
characterizing the contact and $I_0(V,\tau)$ are the $IV$ curves
given by the single channel theory. The theoretical conditions for
the validity of the superposition hypothesis have been discussed
in Sect.\,\ref{ss.SNS2}.

When the number of conduction channels and the transmission for
each of them are used as fitting parameters the agreement with the
experimental results can be highly improved, as illustrated in
Fig.\,\ref{f.sgsAl2}. In general it is found that three channels
are enough to fit the $IV$ curves for Al contacts on the last
conductance plateau \cite{scheer97,scheer98}, with an accuracy
better than $1\%$ of the current above the gap. The quality of the
fit can be measured by the so called $\chi^2$ factor, given by the
sum of the square of the deviation of the measured current at all
points from the theoretical curve at the same bias voltage,
divided by the number of recorded points. Typical values of
$\chi^2$ are smaller than $10^{-4}$. When additional channels are
included in the fitting their transmissions are found to be
negligible (smaller than 0.01) and the value of $\chi^2$ is not
significantly reduced. In this way, the fitting procedure becomes
a very precise method to determine the channel decomposition (or
channel content) of a given contact. The obtained channel ensemble
described by the set of transmission values $\{\tau_n\}$ was shown
to be robust as a function of temperature, up to $T_{\rm c}$, and
as a function of magnetic field, up to the critical field
\cite{scheer00}. This also confirms that the conduction channels
in the normal and in the superconducting states are equivalent,
and the onset of superconductivity does not modify them.

\begin{figure}[!t]
\centerline{\includegraphics[width= 7cm]{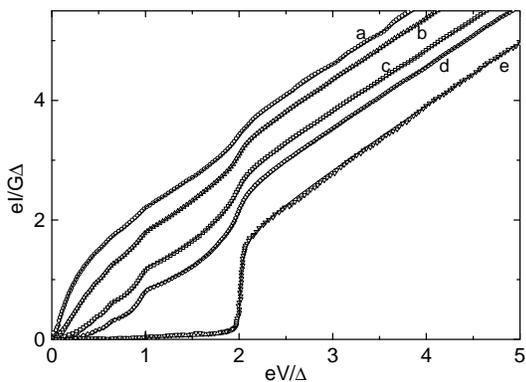}}
\caption{Superconducting $IV$ curves on the last plateau for Pb
atomic contacts fabricated using the STM technique (taken from
Ref.\,\cite{scheer98}). The channel transmissions obtained in the
fitting procedure were: for curve $a$: $\tau_1=0.955$,
$\tau_2=0.355$, $\tau_3=0.085$, $\tau_4=0.005$; curve $b$:
$\tau_1=0.89$, $\tau_2=0.36$, $\tau_3=0.145$, $\tau_4=0.005$;
curve $c$: $\tau_1=0.76$, $\tau_2=0.34$, $\tau_3=0.27$,
$\tau_4=0.02$; curve $d$: $\tau_1=0.65$, $\tau_2=0.34$,
$\tau_3=0.29$, $\tau_4=0.12$; curve $e$: $\tau_1=0.026$. Reprinted
with permission from Nature \protect\cite{scheer98}. \copyright
1998 Macmillan Publishers Ltd.} \label{f.sgsPb}
\end{figure}

Similar results are obtained in the case of Pb \cite{scheer98}.
The Pb atomic contacts studied in Ref.\,\cite{scheer98} were
produced using a low-temperature STM with a Pb tip and Pb
substrate. The superconducting $IV$ curves for Pb contacts on the
last conductance plateau at 1.5\,K, well below $T_{\rm c}=7.2$\,K,
are shown in Fig.\,\ref{f.sgsPb}. In this figure curves $a$, $b$,
$c$ and $d$ correspond to a very similar total conductance close
to 1.4\,$G_0$, and curve $e$ corresponds to the tunneling regime.
In cases $a$ to $d$ four channels were needed to fit the
experimental results, where transmission of the fourth channel was
usually very small, and smaller than 0.12 in all cases. The effect
of a magnetic field on the subgap structure in Pb atomic contacts
was studied in Refs. \cite{suderow00a,suderow00b}. In
\cite{suderow00a} the $IV$ curves were fitted using a modified MAR
theory which incorporates pair-breaking effects due to the
magnetic field. As in the case of Al, the set of transmission
values was found to be quite robust as a function of magnetic
field. In Ref.\,\cite{suderow00b} it is shown that a smeared
subgap structure can be observed even for magnetic fields larger
than the bulk critical field.

\subsubsection{Transition metals: Nb}

In Refs.\,\cite{scheer98} and \cite{ludoph00} the superconducting
$IV$ curves of Nb atomic contacts were analyzed both in the
contact and in the tunneling regime. These contacts were produced
by the MCBJ technique, and measured at a temperature of about
1.5\,K, while $T_{\rm c}=9.0$\,K,.

\begin{figure}[!t]
\centerline{\includegraphics[width= 7cm]{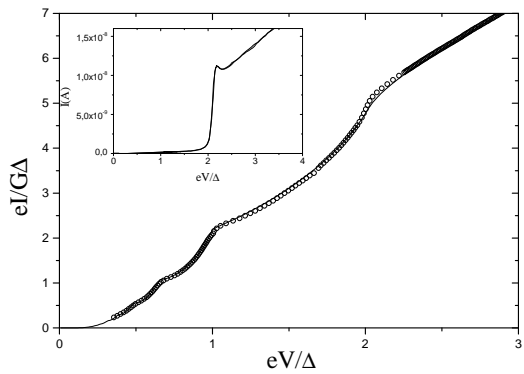}}
\caption{Superconducting $IV$ curve on the last plateau for a Nb
atomic contact fabricated using the MCBJ technique and best fit
using five conduction channels. The resulting channel
transmissions were $\tau_1=0.731$, $\tau_2=0.599$, $\tau_3=0.351$,
$\tau_4=0.195$ and $\tau_5=0.018$. The inset shows a typical $IV$
for Nb in the tunneling regime. Data taken from
\protect\cite{ludoph00}.} \label{f.sgsNb}
\end{figure}

The last conductance plateau in Nb is somewhat higher than in Al
and Pb, with typical values ranging between 1.5 and 2.5 $G_0$.
These values already indicate that several channels are
contributing to the conductance in a one-atom contact. An example
of the superconducting $IV$ curves recorded on the last plateau is
shown in Fig.\,\ref{f.sgsNb}. These curves are best fitted using
five conduction channels, with a typical $\chi^2$ factor of the
order of $10^{-3}$. This agreement with the theoretical curves,
although quite satisfactory, is not so remakable as in the case of
Al and Pb. The reason of this slight discrepancy becomes clear
when analyzing the $IV$ curves in the tunneling regime,
illustrated as an inset in Fig.\,\ref{f.sgsNb}. As can be
observed, the $IV$ curves exhibit a ``bump" around $V = 2 \Delta$,
a feature which is not present in the theoretical tunneling curves
corresponding to a BCS superconductor. Although there is at
present no clear explanation for this bump, a similar structure
appears in the case of tunneling from a thin film of a normal
metal on top of a superconductor, suggesting that the order
parameter could be depressed in the contact region.

As a second complication, the relatively large number of fitting
parameters cannot be determined to great accuracy anymore, since a
shift in one value can be largely compensated by a shift in all
the others. However, the number of channels, which is the most
significant parameter, is very well determined at five.

\subsubsection{$s$-metals: Au}

The subgap structure analysis cannot be applied directly to
non-superconducting metals like Ag or Au. However, a normal metal
in contact with a bulk superconductor acquires superconducting
properties due to the proximity effect \cite{degennes66}. This
effect has been exploited to induce superconductivity in Au
atomic-contacts \cite{scheer98,scheer01}. In these experiments a
thin layer of Au with a thickness of 20\,nm was deposited on top
of a lithographically defined thick Al layer. A photograph of the
device is shown in Fig.\,\ref{f.turtles}.

These samples exhibit less well-defined plateaus than contacts of
pure Au produced by STM or the MCBJ techniques. In particular, the
conductance in the smallest contacts is usually much smaller than
$G_0$, i.e. the typical value for pure Au one-atom contacts. These
differences may be due to structural disorder induced by the
fabrication process.

The superconducting $IV$ curves in the tunneling regime exhibit a
reduced energy gap and the characteristic bump around $V =
2\Delta$ denoting a non-BCS spectral density. These curves can be
reasonably well described by taking into account a reduced and
modified density of states in the gold layer near the contact due
to the proximity effect \cite{scheer01}. A full microscopic theory
of the proximity effect in nanoscale structures is, however, still
lacking.

In spite of these limitations, the $IV$ curves in the contact
regime can be fitted satisfactorily using the theory for BCS
superconductors. It is found that a single channel is sufficient
to describe the $IV$ curves for the smallest contacts with $G <
G_0$, indicating that a single channel carries the current for a
contact of single gold atom.

\subsubsection{Summary of results and discussion}

The analysis of the subgap structure has permitted to extract
information on the channel content for atomic contacts of a
variety of metallic elements in the periodic table. The results
are consistent with the predictions of a simple model based on
atomic orbitals, discussed in Sect.\,\ref{sss.TBeigenchannels}.
Thus, one atom contacts of monovalent metals like Au, in which the
density of states at the Fermi energy is dominated by
$s$-electrons, appear to sustain a single conduction mode.
One-atom contacts of metals in groups III and IV, like Al and Pb,
are characterized by a maximum of four channels, among which one
is usually negligible. Atomic contacts of transition metals like
Nb, where $d$ electrons play a dominant role, are better described
by five conduction channels.

Certainly, one should be critical about of these conclusions which
relies on the accuracy of the fitting procedure. As discussed
above, the quality of the fitting somewhat depends on the
material. For both Al and Pb the more reliable results are
obtained, corresponding to the fact that these metals are well
described by BCS theory. On the other hand, in the case of Nb and
Au contacts the accuracy of the fitting procedure is limited by
the non-BCS features in the spectral density. A detailed analysis
shows that in Nb the best transmitted channels can be determined
with an accuracy of nearly 10\%. However, the number of channels
with a significant transmission can be determined with high
accuracy studying the evolution of the $\chi^2$ factor when adding
more channels to the fit \cite{ludoph00}.

For the case of Al one-atom contacts, an alternative explanation
of the subgap structure assuming a single conduction channel has
been proposed \cite{bascones98}. This explanation requires the
presence of tunneling barriers close to the contact region, giving
rise to an energy dependent transmission. The agreement with the
experimental curves is, however, much poorer than what is obtained
with the multi-channel hypothesis.

Although the subgap structure is the most powerful of the methods
available to extract information on the conductance modes in
atomic-sized contacts, there are several independent experiments
that confirm the validity of the channel decomposition obtained
from the subgap structure analysis. In particular for the
monovalent metals, where the subgap analysis is complicated by the
proximity effect, shot noise experiments discussed in the next
section give unambiguous evidence for a single conductance
channel. Further information can be obtained from measurements of
thermopower and conductance fluctuations as a function of the
applied bias voltage, which will be discussed in
Sect.\,\ref{s.defect_scattering}. The values for the transmission
probabilities obtained from the superconducting subgap structure
can be further tested by measuring the supercurrent, which can be
quantitatively predicted from these values, as will be discussed
in Sect.\,\ref{s.superconductors}.

\subsection{Shot noise: saturation of channel transmission}
\label{ss.exp_shotnoise}

For a perfect ballistic point contact, in the absence of
back-scattering, i.e., all channel transmission probabilities are
either 1 or 0, shot noise is expected to vanish. This can be
understood from the wave nature of the electrons, since the wave
function extends from the left bank to the right bank of the
contact without interruption. When the state on the left is
occupied for an incoming electron, it is occupied on the right as
well and there are no fluctuations in this occupation number. In
other words, the incoming electron is not given the choice of
being transmitted or not, it is always transmitted when it enters
an open mode. In order to have noise, the electron must be given
the choice of being reflected at the contact. This will be the
case when the transmission probability is smaller than 1 and
larger than 0. In single-channel quantum point contacts, shot
noise is predicted to be suppressed by a factor proportional to
$\tau(1-\tau)$, as was derived in
Sect.\,\ref{sss.theory_shot_noise}. This quantum suppression was
first observed in point contact devices in a 2-dimensional
electron gas \cite{reznikov95,kumar96}. Since the general
expression for shot noise in a multi-channel contact,
Eq.\,(\ref{eq.shotnoise}), depends on the sum over the second
power of the transmission coefficients, this quantity is
independent of the conductance, $G=G_0\sum \tau_n$, and
simultaneous measurement of these two quantities should give
information about the channel distribution.

Shot noise in atomic scale contacts was measured using the MCBJ
technique \cite{brom99}, where its high degree of stability was
further improved by careful shielding from external
electromagnetic, mechanical and acoustic vibrations. By measuring
at low temperatures the thermal noise is reduced. However, the
noise level of the pre-amplifiers in general exceeds the shot
noise to be measured. Using two sets of pre-amplifiers in parallel
and measuring the cross-correlation, this undesired noise is
reduced. By subtracting the zero-bias thermal noise from the
current-biased noise measurements, the pre-amplifier noise,
present in both, is further eliminated. For currents up to
1\,$\mu$A the shot noise level was found to have the expected
linear dependence on current. For further details on the
measurement technique, we refer to \cite{brom99}.

\begin{figure}[!t]
\centerline{
\includegraphics[width=8cm]{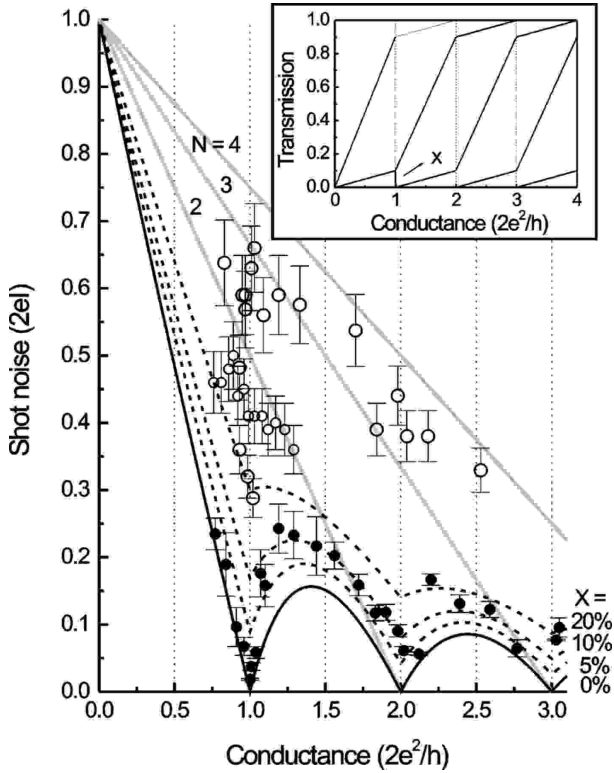}}
\caption{Measured shot noise values for gold (filled circles) and
aluminum contacts (open circles) at 4.2\,K with a bias current of
0.9\,$\mu$A. For gold, comparison is made with calculations
described in the text and in the inset (full and dashed black
curves). For aluminum, comparison is made with the maximum shot
noise that can be produced by $N$ modes (gray curves), as
explained in the text. The minimum shot noise is given by the full
black curve. Note that in the limit of zero conductance, the
theoretical curves all converge to full shot noise. {\em Inset: }
Model for visualizing the effect of contributions of different
modes to the conductance and shot noise. The model gives a measure
for the deviation from the ideal case of channels opening one by
one, by means of a fixed contribution
$(1-\tau_{n-1})+\tau_{n+1}=x$ of the two neighboring modes. As an
illustration the case of a $x=10\%$ contribution from neighboring
modes is shown. Reprinted with permission from
\protect\cite{brom00a}. \copyright 2000 Springer-Verlag.}
\label{f.SN(G)}
\end{figure}

First we discuss the results for the monovalent metal gold, for
which a single atom contact is expected to transmit a single
conductance mode. In Fig.\,\ref{f.SN(G)} the experimental results
for a number of conductance values are shown (filled circles),
where the measured shot noise is given relative to the classical
shot noise value $2eI$. All data are strongly suppressed compared
to the full shot noise value, with minima close to 1 and 2 times
the conductance quantum. We compare our data to a model that
assumes a certain evolution of the values of $\tau_n$ as a
function of the total conductance. In the simplest case, the
conductance is due to only fully transmitted modes ($\tau_{n}=1$)
plus a single partially transmitted mode (full curve). The model
illustrated in the inset gives a measure for the deviation from
this ideal case in terms of the contribution $x$ of other
partially open channels; the corresponding behavior of the shot
noise as a function of conductance is shown as the dashed curves
in Fig.\,\ref{f.SN(G)}. This model has no physical basis but
merely serves to illustrate the extent to which additional,
partially open channels are required to describe the measured shot
noise.

\begin{figure}[!b]
\centerline{
\includegraphics[width=8cm]{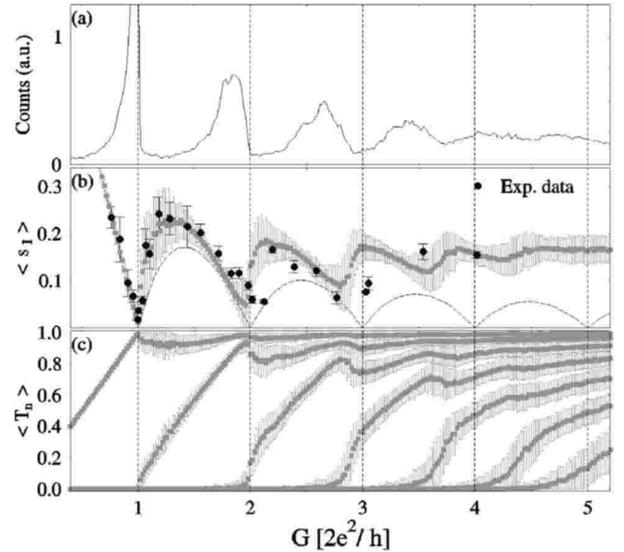}}
\caption{(a) Conductance histogram produced from an ensemble of
contact shapes and defect configurations for a 2DEG point contact.
(b) Mean values of the shot noise $\langle s_I \rangle$ in the
ensemble (grey squares) together with the experimental data for
gold from Fig.\,\protect\ref{f.SN(G)} (black dots). (c) Calculated
mean values for the transmission probabilities $\langle \tau_n
\rangle$. The error bars on the numerical results indicate the
standard deviations over the ensemble. Reprinted with permission
from \protect\cite{burki99}. \copyright 1994 American Physical
Society.} \label{f.calculated-noise}
\end{figure}

B{\"u}rki and Stafford calculated the conductance and noise
simultaneously from an ensemble of impurity configurations and
contact shapes for 2DEG quantum point contacts \cite{burki99}. The
parameters were chosen such as to reproduce the conductance
histogram for gold. The shot noise obtained for the same data set
closely reproduces the observed noise for Au, as shown in
Fig.\,\ref{f.calculated-noise}. Choosing a 2DEG avoids the problem
of degeneracies of the second and third conductance mode in a
cylindrical contact, but may not be fully realistic. However, the
agreement with the data is very satisfactory, suggesting that the
spread in the conductance values giving rise to the histogram is
consistent with the measured shot noise values.

For all points measured on the last conductance plateau before the
transition to tunneling, which has $G\le G_{0}$ and is expected to
consist of a single atom (or a chain of single atoms, see
Sect.\,\ref{s.chains}), the results are well-described by a single
conductance channel. This is in agreement with the fact that gold
has only a single valence orbital. On the other hand, for
$G_{0}<G<$\,2\,$G_{0}$ there is about 10\% admixture of a second
channel. For $G>$\,2\,$G_{0}$ the contribution of other partially
open channels continues to grow.

When the experiment is repeated for aluminum contacts, a different
behavior is observed. For contacts between 0.8\,$G_{0}$ and
2.5\,$G_{0}$ the shot noise values vary from 0.3 to 0.6\,(2$eI$),
which is much higher than for gold (see Fig.\,\ref{f.SN(G)}). A
systematic dependence of the shot noise power on the conductance
seems to be absent. From the two measured parameters, the
conductance, $G$, and the shot noise, $S_{I}$, one cannot
determine the full set of transmission probabilities. However, the
shot noise values found for aluminum, especially the ones at
conductance values close to $G_{0}$, agree with
Eq.\,(\ref{eq.shotnoise}) only if we assume that more than one
mode is transmitted. The maximum shot noise that can be generated
by two, three or four modes as a function of conductance is
plotted as the gray curves in Fig.\,\ref{f.SN(G)}; the minimum
shot noise in all cases is given by the full black curve. Hence,
for a contact with shot noise higher than indicated by the gray
$N$-mode maximum shot noise curve, at least $N+1$ modes are
contributing to the conductance. From this simple analysis we can
see that for a considerable number of contacts with a conductance
close to 1\,$G_{0}$, the number of contributing modes is at least
three. Again, this is consistent with the number of modes expected
based on the number of valence orbitals, and with results of the
subgap structure analysis. Note that the points below the line
labeled $N=2$ should not be interpreted as corresponding to two
channels: the noise level observed requires at least two channels,
but there may be three, or more.

\begin{figure}[!t]
\centerline{\includegraphics[width=8cm]{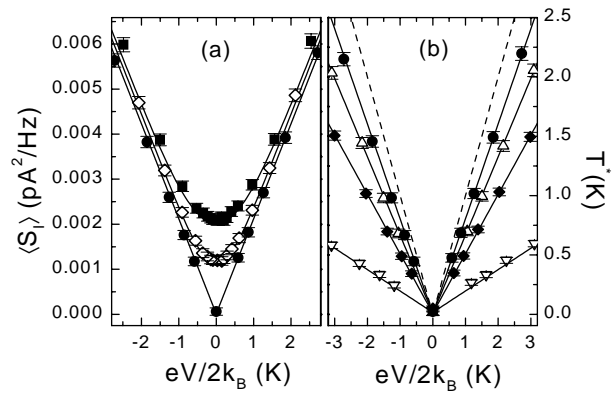}}
\caption{(a) Symbols: measured average current noise power density
$\langle S_I\rangle$ and noise temperature $T^*$, defined as
$T^*=S_I/4k_BG$, as a function of reduced voltage, for a contact
in the normal state at three different temperatures (from bottom
to top: 20, 428, 765 mK). The solid lines are the predictions of
Eq.\,(\ref{eq.shotnoise}) for the set of transmissions
\{0.21,0.20,0.20\} measured independently from the $IV$ in the
superconducting state. (b) Symbols: measured effective noise
temperature $T^*$ versus reduced voltage for four different
contacts in the normal state at $T = 20$ mK. The solid lines are
predictions of Eq. (\ref{eq.shotnoise}) for the corresponding set
of transmissions (from top to bottom: \{0.21,0.20,0.20\},
\{0.40,0.27,0.03\}, \{0.68,0.25,0.22\}, \{0.996,0.26\}. The dashed
line is the Poisson limit. Reprinted with permission from
\protect\cite{cron01}. \copyright 2001 American Physical Society.
\label{f.cronfig3}}
\end{figure}

More recently, shot noise measurements by Cron \ea \cite{cron01}
have provided a very stringent experimental test of the
multichannel character of the electrical conduction in Al. In
these experiments the set of transmissions ${\tau_n}$ are first
determined independently by the technique of fitting the subgap
structure in the superconducting state, discussed in
Sect.\,\ref{ss.sgs_exp}. In words of the authors of
Ref.\,\cite{cron01}, these coefficients constitute the `mesoscopic
PIN code' of a given contact. The knowledge of this code allows a
direct quantitative comparison of the experimental results on the
shot noise with the theoretical predictions of
Eq.\,(\ref{eq.shotnoise}). The experiments were done using Al
nanofabricated break junctions which exhibit a large mechanical
stability. The superconducting $IV$ curves for the smallest
contacts were measured below 1\,K and then a magnetic field of
50\,mT was applied in order to switch into the normal state. The
voltage noise spectrum was measured in a frequency window from 360
to 3560\,Hz. To extract the intrinsic current noise of the
contact, $S_I$, other sources of background noise arising from the
preamplifiers and from the thermal noise due to the bias resistor
were subtracted (for details see \cite{cron01}). The measured
voltage dependence of $S_I$ is shown in Fig.\,\ref{f.cronfig3}a
for a typical contact in the normal state at three different
temperatures, together with the predictions of
Eq.\,(\ref{eq.shotnoise}), using the mesoscopic pin code
${\tau_n}$ measured independently. The noise measured at the
lowest temperature for four contacts having different sets of
transmission coefficients is shown in Fig.\, \ref{f.cronfig3}b,
together with the predictions of the theory. This excellent
agreement between theory and experiments provides an unambiguous
demonstration of the presence of several conduction channels in
the smallest Al contacts and serve as a test of the accuracy that
can be obtain in the determination of the $\tau$'s from the subgap
structure in the superconducting $IV$ curve. The experimental
results of Ref.\,\cite{cron01} for the shot noise in the
superconducting state will be discussed in \ref{ss.shotnoise in
subgap}.

\subsection{Strain dependence of the conductance} \label{ss.strain}

As already commented, the behavior of the conductance on the last
plateaus before breaking the contact is a characteristic of each
metallic element. Krans \ea \cite{krans93} were the first who
pointed out the characteristic tendencies of several metals when
elongating or contracting the contact. They used the MCBJ to study
the case of Cu, Al and Pt, which exhibit different behaviors. In
Cu the plateaus show a slightly negative slope upon elongation but
the last plateaus are rather flat, specially the last one which is
rather constant around 1 $G_0$. The reduction of the conductance
with elongation coincides with the intuitively expected result.
However, in the case of Al and Pt the opposite behavior is found.

Within a free-electron model Torres and S\'aenz \cite{torres96}
have suggested that the different slopes on the last plateaus may
arise as a result of variations in the effective length of the
contact. They are able to reproduce different slopes for the
plateaus in their calculations, but the sign of the slopes is not
element specific: positive or negative slopes are found both for
Au and for Al. The variation in the slopes they observe is
probably related to resonances in the electron waves due to
scattering from the interfaces between the cylindrical slabs used
in their model calculation.

\begin{figure}[!t]
\centerline{\includegraphics[width=6cm]{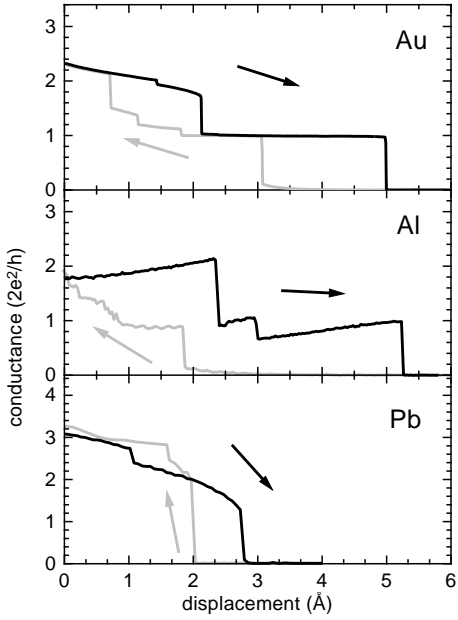}}
\caption{Typical behavior of the experimental conductance at the
last plateaus before breaking the contact for Au, Al and Pb. The
contacts were produced by the STM technique at low temperatures.
Reprinted with permission from \protect\cite{cuevas98}. \copyright
1998 American Physical Society.} \label{strain-dependence1}
\end{figure}

A different explanation of the characteristic tendencies has been
proposed by Cuevas \ea \cite{cuevas98}. They analyzed in
particular the case of Au, Al and Pb which exhibit three different
behaviors, as illustrated in Fig.\,\ref{strain-dependence1}. In
order to describe the electronic properties of these contacts,
Cuevas \ea used an atomic orbital basis and an idealized model
geometry following the proposal of Ref.\,\cite{cuevas98a}. Upon
elongation, the elastic deformation of a one-atom contact was
assumed to produce a slight increase in the distance between the
central atom and its nearest neighbors. For deformations not
larger than 5 or 10\% of the equilibrium distances, the variation
of the hopping integrals within the parameterization of Ref.
\cite{papaconstantopoulos86} can be described by power laws of the
form

\begin{equation}
t_{l,l^{\prime}} \sim \frac{1}{d^{l+l^{\prime}+1}} ,
\label{scaling-laws}
\end{equation}
where $d$ is the interatomic distance and $l$, $l^{\prime}$ denote
the corresponding angular momenta.

\begin{figure}[!t]
\centerline{\includegraphics[width=6cm]{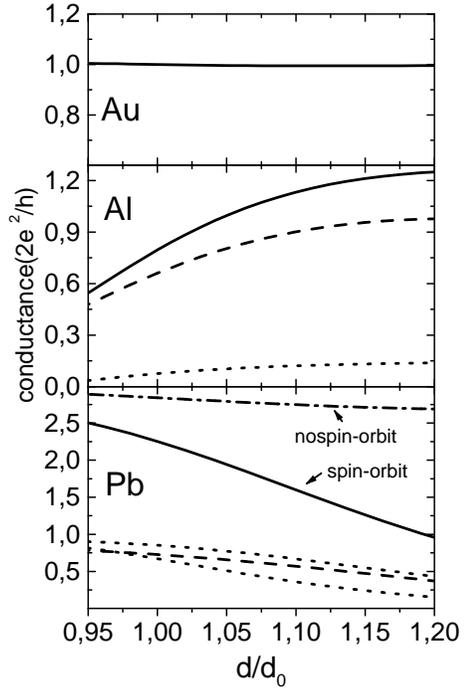}}
\caption{Theoretical results for the total conductance (full line)
and its channel decomposition for Au, Al and Pb one-atom contacts
as a function of the distance $d$ between the central atom and its
nearest neighbors, in units of the equilibrium distance $d_0$.
Reprinted with permission from \protect\cite{cuevas98}. \copyright
1998 American Physical Society.} \label{strain-dependence2}
\end{figure}

The theoretical results of Ref.\,\cite{cuevas98} for the evolution
of the conductance upon elastic deformation are shown in
Fig.\,\ref{strain-dependence2}. These results correspond to the
simplest model geometry of a single atom in between to
semi-infinite fcc crystals exposing (111) surfaces. As can be
observed, the results are in qualitative agreement with the
observed experimental tendencies. Moreover, these model
calculations allow understanding the microscopic origin of these
tendencies, as will be discussed below.

In the case of Au, the calculations predict a fully open single
conduction channel arising from the $6s$ orbital on the central
atom slightly hybridized with the $6p_z$ orbital. The LDOS on the
central atom exhibits a resonance located around the Fermi energy,
as can be observed in Fig.\,\ref{f.ldos+channels}. In this case,
the condition of charge neutrality plays a major role in pinning
this resonance at the Fermi energy when the contact is elongated,
which yields to a rather constant conductance of 1 $G_0$. These
results coincide with more sophisticated calculations for Au based
on LDA by Brandbyge \ea \cite{brandbyge97}.

The case of Al is qualitatively different, as both $3s$ and $3p$
orbitals give a significant contribution to the conductance. The
Fermi energy is in this case located somewhere in between the
center of the $s$ and the $p$ bands in order to accommodate three
electrons per atom. In the equilibrium situation there is a widely
open channel with $sp_z$ character and two degenerate poorly
transmitted channels of $p_x$ and $p_y$ character, as was
discussed in Sect\,\ref{ss.TBcalculation}. The elongation of the
contact induces a narrowing of the $sp_z$ and the $p_{x,y}$ bands,
which is more pronounced in this last case, as expected from the
scaling laws of the hopping elements. As a result, the Fermi
energy tends to be located around the $sp_z$ resonance and the
transmission on the $p_{x,y}$ modes tends to vanish, leaving a
single fully open channel. Again, more sophisticated
first-principles calculations have been performed by Kobayashi
\ea, which confirm that three channels participate in the
conductance for Al single-atom contacts \cite{kobayashi00}.
However, they propose a different explanation for the slope of the
plateau: the decreasing conductance results from straightening the
linear arrangement of Al atoms.

Also in Pb both $s$ and $p$ orbitals play an important role. As
already commented in Sect\,\ref{ss.TBcalculation}, this model
predicts the same number of channels for Al and Pb. However, there
is an extra valence electron in Pb which forces the Fermi energy
to a region where both the $sp_z$ and $p_{x,y}$ channels are
widely open, leading to a total conductance of 2.8 $G_0$ in
equilibrium. Using the parameterization of Ref.
\cite{papaconstantopoulos86} together with the scaling laws,
Eq.\,(\ref{scaling-laws}), a slight decrease of the conductance
upon elongation was found \cite{cuevas98}, which is less
pronounced than in the experimental results. As was pointed out,
the inclusion of spin-orbit coupling within the model leads to a
better agreement between the theoretical and experimental results.
This coupling is of the order of 1\,$eV$ in Pb and produces a
splitting of the $p$ resonance into a $p_{1/2}$ and a $p_{3/2}$
resonance, with the Fermi level in between. This splitting would
be responsible of the pronounced decrease of conductance upon
elongation.

\section{Corrections to the bare contact conductance} \label{s.defect_scattering}

The electronic transport properties of atomic-sized contacts are
predominantly determined by the properties of the narrowest part
of the contact, as we have seen above. With a single atom forming
the contact -- the configuration that we are especially interested
in -- the chemical nature of this atom determines the number of
conductance channels. The transmission probabilities for these
channels depend on connection of this atom to the neighboring
atoms in the metallic leads. We will define the atom plus its
direct environment as the `bare' contact. Only in the absence of
defects and surface corrugations in the leads close to the central
atom, and in the absence of excitations of other degrees of
freedom, would we measure the transmission probabilities for the
bare contact unaltered. In reality, any sudden variation in the
cross section of the conductor as we move away from the center
gives rise to partial reflection of the electron wave, as a result
of the mismatch of the waves at both sides of this jump in the
wire cross section. More generally, the electron wave can be
reflected off surface corrugations, defects and impurities. This
reflected partial wave only alters the current, and therefore the
conductance, to the extent that it passes back through the
contact. As we move further away from the center of the contact,
the probability for returning to the contact is reduced by the
solid angle at which the contact is seen from the scattering
object. Therefore, only scatterers close to the contact center
will have a significant effect on the conductance of the junction.

The scattering centers near the contact give rise to a number of
corrections, the most obvious of which is a reduction of the total
conductance.\footnote{In view of the fluctuations to be discussed
below, we take this conductance to be the ensemble-averaged
conductance, i.e. averaged over all possible defect configurations
for the same contact geometry.} All transmissions will be reduced
and the shift in the averaged sum of the transmission values is
often discussed in terms of an effective `series resistance' to
the contact. Quantum interference between different partial waves
in this scattering problem further gives rise to conductance
fluctuations. We will address the latter first
(Sect.\,\ref{ss.cond-fluct}) because it will provide a good basis
for the discussion of series resistance corrections.

Next we will discuss corrections to the conductance due to
electron-phonon scattering at finite bias, and the related problem
of heating in point contacts. When we include correlations between
the electrons additional effects arise that produce an anomalous
dip in the differential conductance centered at zero bias voltage.
Widely different mechanisms all lead to a similar kind of
zero-bias anomaly. When magnetic impurities are introduced into
the metal at sufficiently high concentrations to have a measurable
effect, a zero-bias conductance minimum is observed due to Kondo
scattering of the electrons (Sect.\,\ref{ss.Kondo}). The Kondo
scattering appears to be sensitive to the size of the contact.
Non-magnetic scattering centers have been observed to produce
similar zero-bias anomalies. This has been attributed to fast
two-level systems and it has been proposed that a 2-channel Kondo
model gives an accurate description of these observations
(Sect.\,\ref{ss.2-channel-Kondo}). Finally, Coulomb interactions
between the conduction electrons themselves give rise to a feature
similar to Coulomb blockade in tunnel junctions. In order to
observe this effect one needs to tailor the electromagnetic
environment of the junctions (Sect.\,\ref{ss.ECB}).

\subsection{Conductance fluctuations}\label{ss.cond-fluct}
Universal conductance fluctuations (UCF) are observed in a
mesoscopic conductor for which the phase coherence length is much
longer than the sample size, while, on the other hand, the elastic
scattering length is much smaller than the size of the system
\cite{lee85,altshuler85}. Electrons entering the system have many
possible trajectories for being scattered back to the lead from
which they entered and all partial waves sum up coherently. For
every new configuration of scattering centers the conductance is
slightly different and varies randomly. The root-mean-square
amplitude of the conductance fluctuations resulting from the
interference has a universal value, independent of sample size,
$\delta G_{\rm rms}\simeq e^2/h$. In practice it is not feasible
to measure an ensemble average by modifying the defect
distribution for one and the same sample. A practical way to
obtain an equivalent result is to measure the conductance as a
function of magnetic field. The field enclosed by the electron
trajectories modifies the phase, and this phase shift is different
for each different size of electron loop.

The universality of this fluctuation amplitude breaks down as soon
as the sample size falls below the average distance between
elastic scattering events. Therefore, it is not surprising that
the amplitude for conductance fluctuations in a ballistic point
contact is much smaller than $e^2/h$. Holweg \ea  \cite{holweg91}
studied the fluctuations for relatively large nanofabricated
contacts of the type described in Sect.\,\ref{ss.nanofab
contacts}, with a typical size corresponding to a resistance of
$\sim10\,\Omega$. The amplitude of the fluctuations in the
conductance as a function of magnetic field that they obtained was
reduced by two orders of magnitude compared to the universal
value. This is due to the geometrical factor that enters when we
take into account that it is unlikely for the electron to return
to the contact from a remote region. For the contributions from
interference of two partial waves scattering back to the contact
one typically obtains $\delta G_{\rm rms}\simeq (a/\ell)^2
(e^2/h)$ \cite{kozub94}, where $a$ is the radius of the contact
and $ \ell$ is the elastic mean free path. However, Kozub \ea
\cite{kozub94} noted that this gives an estimate that is far
smaller that the observed fluctuations. They proposed to repair
this discrepancy by assuming a dominant scatterer near the
contact. However, as explained below, the contact itself can be
regarded as a scatterer, which automatically resolves this
problem, and it turns out that $\delta G$ is reduced by only a
single factor $\sim (a/\ell)$.

\begin{figure}[!t]
\centerline{\includegraphics[height=7cm,angle=270]{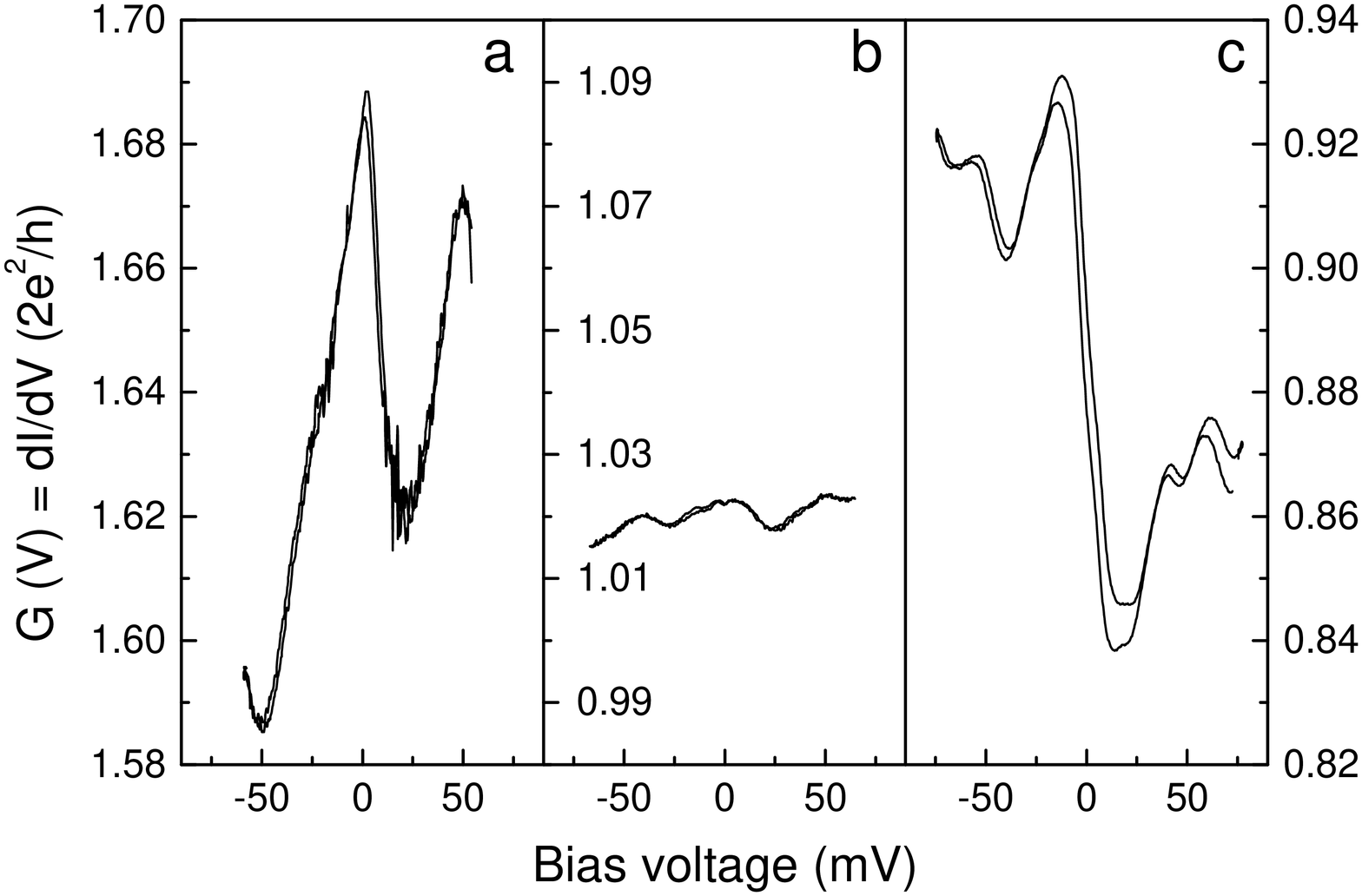}}
\caption{Differential conductance $dI/dV$ as a function of bias
voltage, measured with a modulation amplitude of less than
0.35\,mV, for three different gold contacts, with $G\simeq
1.65\,G_0$ (a), $\simeq 1.02\,G_0$ (b), and $\simeq 0.88\,G_0$
(c). For all three curves the vertical scale spans $0.12\,G_0$.
The curves have been recorded twice, once for decreasing bias
voltage, and back, to illustrate the reproducibility of the
features. Reprinted with permission from \protect\cite{ludoph99}.
\copyright 1999 American Physical Society.  }
\label{f.fluctuations}
\end{figure}

Nevertheless, when $a$ becomes as small as a single atom this
geometrical reduction factor is so severe that conductance
fluctuations as a function of magnetic field are extremely weak.
In this case the fluctuations are more sensitively probed by
following the differential conductance as a function of bias
voltage. Examples are shown in Fig.\,\ref{f.fluctuations}, which
illustrates the fact that a different pattern is obtained for each
new realization of the contact.

\subsubsection{Theory for defect scattering near a point contact}
\label{sss.theory-scattering} 
\begin{figure}[!t]
\centerline{\includegraphics[width=5cm]{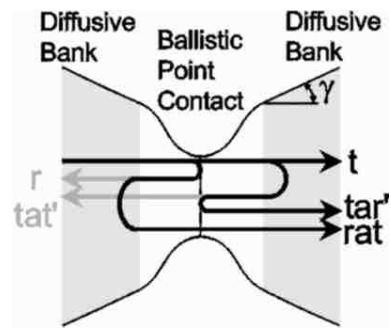}}
\caption{Diagram showing the bare contact (light) sandwiched
between diffusive regions (gray). The dark lines with arrows show
the paths that interfere with each other and contribute to the
conductance fluctuations in lowest order.  }
\label{f.trajectories}
\end{figure}

For an evaluation of the dominant
correction terms arising from defect scattering we will model the
contact as illustrated in Fig.\,\ref{f.trajectories}. It has a
ballistic central part (the `bare' contact), which can be
described by a set of transmission values for the conductance
modes. This is sandwiched between diffusive banks, where electrons
are scattered by defects characterized by an elastic scattering
length $\ell$. An electron wave of a given mode $n$ falling onto
the contact is transmitted with probability amplitude $t_n$ and
part of this wave is reflected back to the contact by the
diffusive medium, into the same mode, with probability amplitude
$a_n\ll 1$. This back-scattered wave is then {\em reflected} again
at the contact with probability amplitude\footnote{ The partial
wave amplitudes are related to the transmission probability of
this mode as $\tau_n = |t_n|^2= 1-|r_n|^2$.} $r_n$. The latter
wave interferes with the original transmitted wave. This
interference depends on the phase accumulated by the wave during
the passage through the diffusive medium. The probability
amplitude $a_n$ is a sum over all trajectories of scattering, and
the phase for such a trajectory of total length $L$ is simply
$kL$, where $k$ is the wave vector of the electron. The wave
vector can be influenced by increasing the voltage over the
contact, thus launching the electrons into the other electrode
with a higher speed. The interference of the waves changes as we
change the bias voltage, and therefore the total transmission
probability, or the conductance, changes as a function of $V$.
This describes the dominant contributions to the conductance
fluctuations, and from this description it is clear that the
fluctuations are expected to vanish either when $t_n=0$, or when
$r_n=0$. For those events only the much smaller higher order terms
involving two diffusive trajectories remain.

Elaborating this model Ludoph \ea  \cite{ludoph99,ludoph00a}
obtained the following analytical expression for the
root-mean-square value of the variation of the conductance with
voltage, $\sigma_{GV}=\langle dG/dV\rangle_{\rm rms}$, to lowest
order in the return amplitudes $a_n$:
\begin{equation}
\sigma_{GV} = \frac {2.71\ e\ G_0}{\hbar k_{\rm F} v_{\rm F}
\sqrt{1-\cos\gamma}} \Bigl(\frac {\hbar v_{\rm F}/\ell}{eV_{\rm
m}}\Bigr)^{3/4} \sqrt{\sum_n \tau_n^{2}(1-\tau_n)}\; .
\label{eq.sigmaGV}
\end{equation}
Here, $k_{\rm F}$ and $v_{\rm F}$ are the Fermi wave vector and
Fermi velocity, respectively, and $ \ell $ is the scattering
length. The shape of the contact is taken into account in the form
of the opening angle $\gamma$ (see Fig.\,\ref{f.trajectories}),
and $V_{\rm m}$ is the voltage modulation amplitude used in the
experiment. Here, it is assumed that $eV_{\rm m}\gg \max(k_{\rm
B}T, \hbar v_{\rm F}/L_{\varphi})$, with $L_{\varphi}$ the phase
coherence length. In deriving expression (\ref{eq.sigmaGV}) the
diffusive banks were treated semi-classically: the probability to
return to the contact after a time $t_{\rm s}$ was assumed to be
given by the classical diffusion expression. In adding over all
diffusion times $ t_{\rm s}$ a phase factor, $\exp(-iE t_{\rm
s}/\hbar)$ that an electron with energy $E$ accumulates during its
traversal of the diffusive region was taken into account.

The choice of where the boundary between the contact and the
diffusive banks is taken is somewhat arbitrary. It does not
explicitly appear in (\ref{eq.sigmaGV}), because a new choice for
the boundary is absorbed in modified values for the transmission
probabilities $\tau_n$. For any choice of boundaries we can find a
set of eigenchannels, but they will only slightly differ as long
as the distance $D$ between the contact center and the boundary is
large compared to the contact diameter. On the other hand, this
implies that the effects of surface corrugation very close to the
contact are represented by a reduction of the $\tau_n$'s, which
will then also have a distinct dependence on energy. The
fluctuations in the $\tau_n$'s will be visible on a large voltage
scale only, as long as $D$ is small enough, i.e. $D\ll\hbar v_{\rm
F}/eV$. For metallic contacts and on the voltage scales considered
here both conditions can be fulfilled reasonably well when we take
$D\sim 1$\,nm.

Apart from the fluctuations in the conductance there is a shift in
the total conductance of the contact. Including only the lowest
order correction the average total transmission probability is
given by $\sum_{n=1}^{N} \tau_{n}
(1-\sum_{m=1}^{N}\tau_{m}(\langle|a_{l_{mn}}|^{2}\rangle +
\langle|a_{r_{mn}}|^{2}\rangle))$. The last term, $ a_{r_{mn}}$,
describes the partial amplitude for an electron that is
transmitted through the contact in mode $n$, scattered back
towards the contact in the right diffusive bank, and then
transmitted through the contact a second time, in mode $m$, in the
opposite direction. The term $ a_{l_{mn}}$ describes a similar
trajectory for diffusion in the left lead. These processes will
lead to a smaller conductance than expected for the bare contact
conductance alone since part of the transmitted electrons are
scattered back, reducing the net forward current flow.

At higher conductance values, we expect a significant contribution
of higher order terms in the return probabilities $a_{l_{mn}}$ and
$a_{r_{mn}}$ to the conductance. Hence, the lowest order
correction used above will not suffice. Keeping track of higher
order terms, becomes very complicated for many channels. However,
using random matrix theory an expression for the correction to the
conductance of a quantum point contact connected to diffusive
leads has been derived \cite{mello91,beenakker94},
\begin{equation}
\langle G \rangle =  {2e^2 \over h} \left[
\frac{N}{1+\gamma_N}-\frac{1}{3}\left(
\frac{\gamma_N}{1+\gamma_N}\right)^{3} \right]. \label{eq.rcorr}
\end{equation}
\noindent Here, $\gamma_N=(N+1)G_0R_{\rm s}$ is roughly equal to
the ratio of the conductance of the bare contact to that of the
banks. The diffusive scattering in the banks is represented
through a sheet resistance $ R_{\rm s}$. In the theory all open
channels were assumed to be perfectly transmitting, $\tau_n=0$, or
1. To lowest order Eq.\,(\ref{eq.rcorr}) is consistent with the
correction to the average total transmission probability derived
from the backscattering above. The first term in (\ref{eq.rcorr})
is nearly equal to the expression for the classical addition of a
resistor $R$ and a conductor $G$, which would give  a conductance
$G/(1+GR)=NG_0/(1+N G_0 R)$. The second term in (\ref{eq.rcorr})
is a weak localization correction, and can usually be neglected
because it results from interference of two partial waves
scattered in the banks. Explicit calculations for model systems
show that the expression describes the shifts of the peaks in a
conductance histogram correctly
\cite{burki99a,garciamochales96,bascones98}. Note, however, that
Eq.\,(\ref{eq.rcorr}) was derived for a 2D electron system, with
leads of constant width. For the metallic point contacts
considered here a three dimensional analysis is applicable, with
leads that widen out to infinity far away from the contact, and we
anticipate significant deviations for larger contact size, as will
be further discussed in Sect.\,\ref{ss.series-resistance}.
Metallic contacts can be made arbitrarily large, in which case the
resistance decreases to zero, while Eq.\,(\ref{eq.rcorr}) is
bounded by the sheet resistance $R_s$ for $N\rightarrow\infty$.

Indeed, for metallic point contacts of larger size one has since
long made use of the expression due to Wexler \cite{wexler66},
\begin{equation}
G_{\rm W}=G_{\rm S} \left[ 1 +
\frac{3\pi}{8}\Gamma(\frac{\ell}{a})\frac{a}{\ell}\right]^{-1},
\label{eq.wexler}
\end{equation}
which interpolates between the Sharvin conductance $G_{\rm S}$,
Eq.\,(\ref{eq.sharvin}) and the Maxwell classical conductance
(\ref{eq.maxwell}) that applies for large contact radius $a$. The
function $\Gamma(K)$ is a slowly varying function, with
$\Gamma(0)=1$ and $\Gamma(\infty)=0.694$. We propose to replace
$G_{\rm S}$ by the bare quantum conductance of the contact, to
obtain an interpolation formula that will predict the conductance
of metallic contacts in the presence of disorder scattering.

\subsubsection{Experimental results}
\label{sss.exp-cond-fluct} We will mainly discuss results for
gold, which has been best characterized, and for which example
curves are presented in Fig.\,\ref{f.fluctuations}, while the
similarities and differences for other metals will be briefly
mentioned. Slight modifications of a metallic constriction,
induced by displacing the two banks over a small distance, can
have a dramatic effect on the interference pattern seen in the
voltage dependence of the conductance (see e.g.
Fig.\,\ref{f.fluctuations}), while the overall value of the
conductance remains almost constant \cite{ludoph00a,untiedt00}.
This agrees with the idea that the interference terms are
sensitive to changes in the electron path length on the scale of
$\lambda_{\rm F}$.

\begin{figure}[!t]
\centerline{\includegraphics[width=7cm]{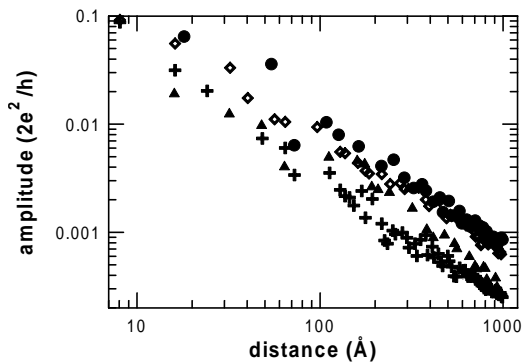}}
\caption{Dependence of the amplitude of the components of the
conductance fluctuations as a function of the distance traveled in
the banks, for several contacts for gold at 4.2K. The distance $d$
is obtained from the period of the oscillation, $\Delta V$, via
$d=\lambda_{\rm F}(E_{\rm F}/e\Delta V)$. The amplitude for the
components is obtained by taking the Fourier transform of the
differential conductance measured as a function of the bias
voltage. Reprinted with permission from \protect\cite{untiedt00}.
\copyright 2000 American Physical Society.   }
\label{f.one-over-d-dependence}
\end{figure}

The amplitudes of the spectral components of the conductance as a
function of bias voltage for a single contact decrease with their
frequency roughly as the inverse of the distance $d$ traveled by
the partial wave, where $d$ is deduced from this frequency, as
shown in Fig.\,\ref{f.one-over-d-dependence} \cite{untiedt00}.
This is what is to be expected when the scattering from the banks
is dominated by single-scattering events. Contributions of
scattering paths up to 100\,nm long are observable. For higher
temperatures the high frequency (long path) components are
gradually suppressed, reflecting the decrease in coherence length,
but even at room temperature conductance fluctuations remain
visible \cite{untiedt00}.

\begin{figure}[!t]
\centerline{\includegraphics[width=8cm]{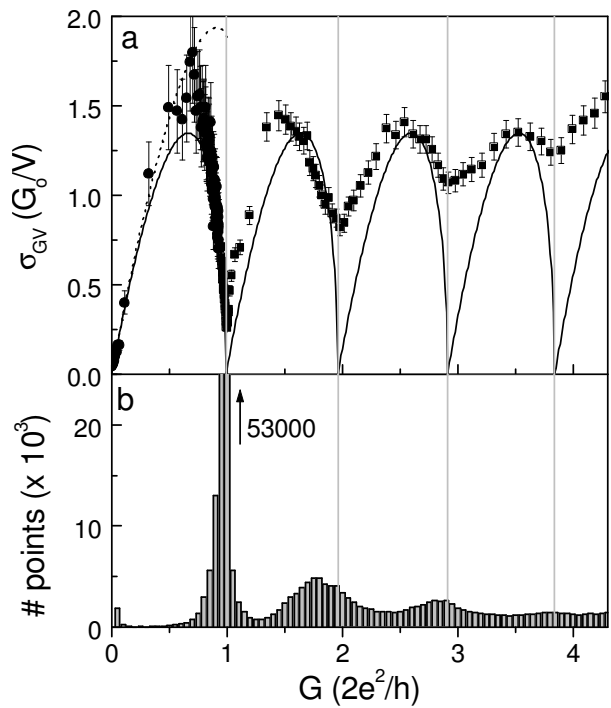}} \caption{(a)
Standard deviation of the voltage dependence of the conductance,
$\sigma_{GV}$, versus the conductance, $G$, obtained from 3500
contact-breaking cycles. The circles are the averages for 300
points, and the solid squares for 2500 points. The solid curves
depict the calculated behavior for a single partially-open
channel. The dashed curve is for a random distribution over two
channels. The vertical dotted lines are the corrected integer
conductance values (see text). The modulation voltage amplitude
was 20\,mV at a frequency of 48\,kHz. (b) Conductance histogram
obtained from the same data set. The peak in the conductance
histogram at $G_{0}$ extends to 53000 on the y-scale. Reprinted
with permission from \protect\cite{ludoph99}. \copyright 1999
American Physical Society.} \label{f.sigmaGV}
\end{figure}

In contrast to UCF studied in mesoscopic samples, for point
contacts one can make, to a very good approximation, a direct
ensemble average. This can be done by recording the conductance
and its derivative simultaneously and collecting the data for many
contact-breaking cycles \cite{ludoph99,ludoph00a}. In practice
this is done by simultaneously recording the first and second
harmonic of the modulation frequency with two lock-in amplifiers.
The values for the derivatives of the conductance from all these
contacts, having the same average conductance $\langle G \rangle$,
are used to calculate the standard deviation
$\sigma_{GV}=\sqrt{\langle(\partial G/\partial V)^2\rangle}$ for
each value of $\langle G \rangle$. Results obtained using the MCBJ
technique for gold at 4.2\,K are shown in Fig.\,\ref{f.sigmaGV}. A
fairly large modulation voltage was used in order to permit fast
data acquisition. This enhances the sensitivity for the
long-period components but in the theory for the ensemble average,
Eq.~(\ref{eq.sigmaGV}), the finite modulation amplitude has been
explicitly taken into account.

The full curves in Fig.\,\ref{f.sigmaGV}a are obtained from
Eq.\,(\ref{eq.sigmaGV}), assuming a single partially-open channel
at any point, i.e., assuming that in the interval $G/G_{0} \in
[0,1]$ there is a single channel contributing to the conductance
with $G= \tau_{1} G_{0}$, in the interval $[1,2]$ there are two
channels, one of which is fully open, $G= (1+\tau_{2}) G_{0}$,
etc. This is the same succession of channel openings as giving
rise to the full black curve for the shot noise in
Fig.\,\ref{f.SN(G)} The amplitude of the curves in
Fig.\,\ref{f.sigmaGV}a is adjusted to best fit the data, from
which a value for the mean free path $\ell=5\pm 1$~nm is obtained.
Similar experiments \cite{ludoph99,ludoph00a} for the monovalent
metals Cu and Ag and for Na also show the quantum suppression of
conductance fluctuations observed here for Au, while for the $sp$
metal Al and the $sd$ metals Nb and Fe this is not observed. The
absence of pronounced minima in $\sigma_{GV}$ for the non-$s$
metals agrees again with the results discussed in
Sect.\,\ref{s.exp_modes}, showing that for those metals typically
several partially transmitted modes are participating in a single
atom for any accessible value of the conductance.

The minimum observed at 1\,$G_{0}$ in Fig.\,\ref{f.sigmaGV} is
very sharp, close to the full suppression of fluctuations
predicted for the case of a single channel. In order to describe
the small deviation from zero, it is sufficient to assume that
there is a second channel which is weakly transmitted, $\tau_2\ll
1$, and $\tau_1 \simeq 1$ such that $\tau_1 + \tau_2 = 1$. For
this case it is easy to show that the value of $\sigma_{GV}$ at
the minimum is proportional to $\sqrt{\langle \tau_2\rangle}$,
from which $\langle \tau_2\rangle=0.005$ is obtained. This implies
that, on average, only 0.5\,\% of the current is carried by the
second channel. For the minima near 2, 3 and 4\,$G_0$ higher
values are obtained: 6, 10 and 15\,\%, respectively. The
well-developed structure observed in $\sigma_{GV}$ for gold in
Fig.\,\ref{f.sigmaGV}a, with a dependence which closely follows
the $\sqrt{\sum \tau_n^2(1-\tau_n)}$ behavior of
Eq.~(\ref{eq.sigmaGV}), agrees with the saturation of transmission
channels \cite{ludoph99} that was also seen in the shot noise
experiments in Sect.\,\ref{ss.exp_shotnoise}.

Note that the minima in Fig.\,\ref{f.sigmaGV}a are found slightly
below the integer values. A similar shift was obtained in
simulations of the shot noise for quantum point contacts by
B{\"u}rki and Stafford \cite{burki99}. The shift can be described
by taking a total return probability $\langle \sum_m
|a_{mn}|^2\rangle=0.005$, from which we derive a value for the
mean free path of $\ell=4\pm1$\,nm. This value agrees well with
the value obtained from the fluctuation amplitude. This gives
strong experimental support for the notion of a shift of the
average conductance by scattering on defects that can
approximately be described by a series resistance, which in this
case is about 130\,$\Omega$. We will return to discuss the
accuracy of this statement in Sect.\,\ref{ss.series-resistance}
below.

It is interesting to compare the positions of the maxima in the
conductance histogram and those for the minima in $\sigma_{GV}$ in
Fig.\,\ref{f.sigmaGV}. It appears that these positions do not all
coincide, which is most evident for the peak in the histogram at
about $G=1.8\,G_0$. As discussed in Sect.\,\ref{ss.histograms},
the histograms give preferential conductance values, which may
reflect a quantization effect in the conductance as a function of
contact diameter, but also a preference for forming contacts of
certain effective diameters. Such preferential contact diameters
may be expected based on the fact that the contact is only a few
atoms in cross section, which limits the freedom for choosing the
diameter. It appears that at least the peak at 1.8\,$G_0$ in the
histogram for gold arises from this atomic geometry effect.
Although the shot noise (Fig.\,\ref{f.SN(G)})and conductance
fluctuation experiments (Fig.\,\ref{f.sigmaGV}a) both show that
the conductance for gold contacts with $G\simeq 2\,G_0$ is carried
by two nearly perfectly transmitted modes, this conductance is not
preferred, as evidenced by the conductance histogram. This
interpretation agrees with {\it ab-initio} calculations for a
double strand of gold atoms for which H{\"a}kkinen \ea
\cite{hakkinen00} obtain a conductance of 1.79\,\g0 corresponding
to two channels, one of which is nearly fully open.

The validity of Eq.~(\ref{eq.sigmaGV}) has been tested by
measuring the amplitude of the conductance fluctuations in gold at
modulation voltages ranging from 10 to 80\,mV and the expected
dependence $ V_{\rm m}^{-3/4}$ was obtained \cite{ludoph00a}. The
agreement with the fluctuations in the thermopower described below
may serve as a further test on smaller energy scales.

\subsubsection{Thermopower fluctuations} \label{sss.thermopower}
\begin{figure}[!t]
\centerline{\includegraphics[width=7cm]{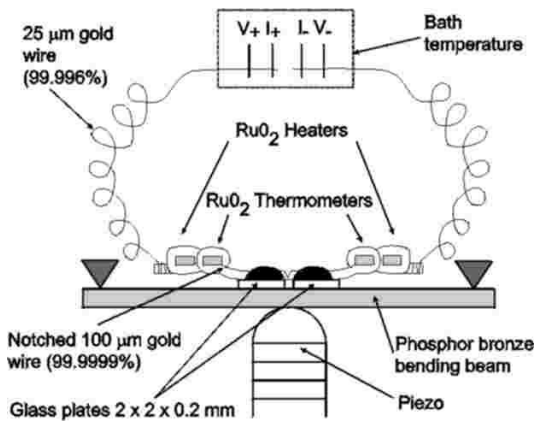}}
\caption{ Schematic diagram of the modified MCBJ configuration,
used for the simultaneous measurement of conductance and
thermopower. Reprinted with permission from
\protect\cite{ludoph99a}. \copyright 1999 American Physical
Society.} \label{f.thermopower}
\end{figure}

In a linear response approximation the thermal voltage induced by
a temperature difference $\Delta T$ over a contact is given by
\begin{equation}
V_{\rm tp} = S\cdot \Delta T = - {{\pi^2 k_{\rm B}^2 T}\over{3 e}}
{{\partial \ln G}\over{\partial \mu}} \Delta T
\label{eq.thermopower}
\end{equation}
This expression illustrates that the thermopower $S$ is a quantity
that is qualitatively similar to the voltage dependence of the
conductance. The role of voltage is now taken by the chemical
potential $\mu$ and the scale is set by the temperature difference
over the contact, $\Delta T$. It can be expressed in terms of the
transmission probabilities of the conductance channels with the
help of Eq.~(\ref{eq.thermopower-channels}). Based on a free
electron model of a quantum point contact Bogachek \ea
\cite{bogachek96} suggested that the thermopower should be
positive, having maxima midway between the contact widths for
which the conductance is at integer multiples of the conductance
quantum. In experiment however, the values measured for
atomic-sized gold contacts have both positive and negative values,
showing a random distribution centered around zero
\cite{ludoph99a}. The principle of the measurement is illustrated
in Fig.\,\ref{f.thermopower}.

By applying a constant temperature difference over the contacts,
the thermally induced potential could be measured simultaneously
with the conductance. Large thermopower values were obtained,
which jump to new values simultaneously with the jumps in the
conductance. The values are randomly distributed around zero with
a roughly bell-shaped distribution. The thermopower signal was
demonstrated to be dominantly of the same origin as the
conductance fluctuations discussed in the previous section. An
expression similar to (\ref{eq.sigmaGV}) can be derived to
describe the results \cite{ludoph00a,ludoph99a}. The experimental
results follow the law obtained from this defect-scattering model,
and quantum suppression of the thermopower at $G=1$\g0 was
observed. In fact, a scaling relation between the amplitude of the
fluctuations in the thermopower and that of the conductance can be
derived, free of any adjustable parameters \cite{ludoph00a}. The
two experimental techniques are very different, and the typical
energy scales of excitation are at least an order of magnitude
apart. This scale is set by the modulation voltage amplitude,
10--80\,mV, in one case and the temperature, $\sim$10\,K,
equivalent to $\simeq 1$\,mV, in the other. Therefore, these
results give strong support for the description and interpretation
presented above.

\subsection{The series resistance of a quantum point contact}
\label{ss.series-resistance} In the interpretation of conductance
histograms a phenomenological series resistance is often taken
into account in order to describe the shift of the peaks to lower
values \cite{krans95,brandbyge95,hansen97,costa97b,costa97a}, as
was discussed in Sects.\,\ref{ss.histograms} and
\ref{sss.free_electrons}. This practice was inspired on a similar
procedure commonly applied in point contact experiments for 2DEG
systems \cite{wees88}. However, as was pointed out at the end of
Sect.\,\ref{sss.theory-scattering} although this can be justified
for 2D systems, it is less accurate for metallic leads that widen
out to macroscopic size in three dimensions.

\begin{figure}[!t]
\centerline{\includegraphics[width=6cm]{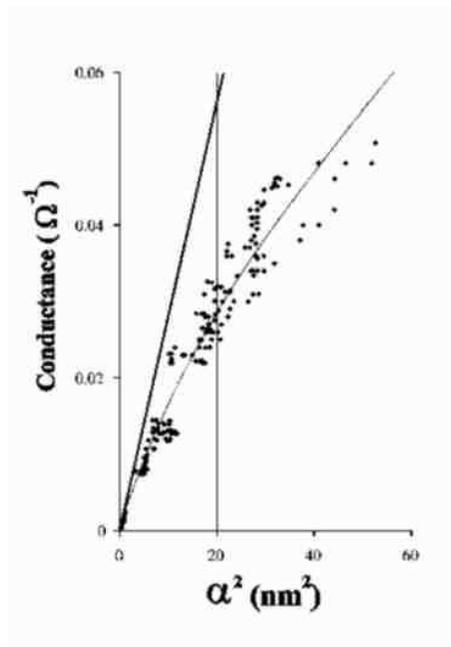}}
\caption{Measured conductance of gold point contacts as a function
of the square of the contact radius, as determined from
simultaneously recorded TEM images. The measurements were taken at
room temperature at a voltage bias of 10\,mV. The straight line
gives the Sharvin conductance, while the curve through the data
points is obtained from Wexler's interpolation formula, taking
$\Gamma=0.7$. The only adjustable parameter is the mean free path,
for which $\ell=3.8$\,nm was obtained. Reprinted with permission
from \protect\cite{erts00}. \copyright 2000 American Physical
Society.} \label{f.wexler}
\end{figure}

The dependence of the conductance of a contact on its  diameter
was tested in a direct measurement by Erts \ea\  \cite{erts00}.
They measured the size of contacts encountered in transmission
electron microscopy images of an STM tip contact at room
temperature in vacuum. The conductance measured simultaneously,
was then plotted against the contact area, see
Fig.\,\ref{f.wexler}. From a fit to the Wexler formula,
Eq.\,(\ref{eq.wexler}) they obtain a value for the mean free path
of $\ell=3.8$\,nm, which agrees closely with the value obtained
from the conductance fluctuation analysis for gold contacts
Sect.\,\ref{sss.exp-cond-fluct}). The agreement is perhaps better
than one should expect, in view of the difference in experimental
conditions, notably the temperature. The values for the mean free
path obtained are much shorter than what is normally found for
bulk samples, and can probably be attributed to scattering on
defects and surface roughness near the contact, introduced in the
process of mechanical contact formation. Assuming surface
scattering is indeed responsible, an important property of the
mean free path which has been neglected here is that $\ell$ will
not be a constant as a function of the conductance, but rather
increase as the contact diameter becomes larger. However, this
size dependence of the mean free path is not expected to be very
significant as long as the contact size is smaller than $\ell$.

It has been proposed in several places in the literature  that all
deviations from integer values in a conductance trace for the
noble metals can be attributed to backscattering. Although the
relatively short $\ell$ observed is responsible for a significant
shift in the ensemble averaged conductance, it is still too long
to hold backscattering responsible for the frequent measurement of
non-quantized values. Also, if scattering is held primarily
responsible for reducing the conductance from, e.g., a perfect
conductance of 2\,$G_{0}$ to 1.5\,$G_{0}$, then it is not
unreasonable to assume that contacts with a perfect conductance of
1\,$G_{0}$ are reduced to 0.5\,$G_{0}$ with a probability of the
same order of magnitude. This is not observed (at least in the
low-temperature experiments), as contacts with a conductance of
0.5\,$G_{0}$ occur more than 500 times less frequent for silver
and copper than contacts with a conductance of 1.5\,$G_{0}$. (The
formation of atomic chains, see Sect.\,\ref{s.chains}, reduces
this ratio to about 20 times in the case of gold, since the
conductance of the chains is quite sensitive to distortions making
contacts with a conductance of 0.5\,$G_{0}$ occur with an enhanced
frequency). If, on the other hand, one assumes that contributions
from tunneling, e.g. due to geometrical considerations, are more
important, the appearance of non-quantized values above 1\,$G_{0}$
finds a natural explanation. The formation of geometries with a
conductance smaller than 1\,$G_{0}$ is highly unlikely since the
smallest contact geometry is that of a single atom with
conductance 1\,$G_{0}$ and when the contact breaks, the banks
relax back in a jump to tunneling
(Sect.\,\ref{ss.jump-to-contact}). On the other hand, defect
scattering is clearly responsible for a shift of the peak
positions, as was demonstrated in a study of Cu-Ni random alloys
as a function of concentration \cite{bakker02}. The experiment
also showed that a straightforward application of a series
resistance correction does not work in the high-concentration
regime.

\subsection{Inelastic scattering}\label{ss.inelastic}

\subsubsection{Electron-phonon scattering}\label{sss.phonons}

When we say that a conductor is ballistic, we usually mean that
its characteristic length $L$ is much smaller than the mean
distance between scattering events. However, this does not imply
that scattering is entirely absent or unimportant. From the
sections above it is clear that elastic scattering plays a role in
atomic-sized contacts in reducing the average conductance and in
producing random conductance fluctuations as a function of the
applied bias voltage. In addition, at finite bias voltage the
electrons can undergo inelastic scattering events, which leads to
heating of the contact. This does not contradict the notion of a
ballistic contact: the contact is ballistic as long as the
electrons travel on average a distance much larger than the
contact size before scattering. When speaking about the most
common form of inelastic excitations, the phonons, every electron
that traverses the contact has a small, but finite, probability to
deposit some of its energy in the lattice vibrations inside the
contact itself.

Traditionally, electron-phonon spectroscopy in (large) metallic
contacts is described by considering the non-equilibrium electron
distribution near the contact that results from the applied bias
voltage, as illustrated in Fig.\,\ref{distribution}
\cite{khotkevich95,yanson74,jansen80}. Electrons that arrive in
the left electrode, coming from the right, are represented in a
Fermi surface picture by a cone with an angle corresponding to the
solid angle at which the contact is viewed from that position in
the metal. These electrons have $eV$ more energy that the other
Fermi surface electrons, and they can be scattered inelastically
to all other angles outside the cone. Only those that scatter back
into the contact will have a measurable effect on the current.

As the energy difference $eV$ increases this backscattering
increases due to the larger phonon density of states, which will
be observed as a decreasing conductance. Ignoring higher order
processes, the decrease of the conductance comes to an end for
energies higher than the top of the phonon spectrum, which is
typically 20--30\,meV. By taking the derivative of the conductance
with voltage one obtains a signal that directly measures the
strength of the electron-phonon coupling. An example for gold is
illustrated in Fig.\,\ref{f.PCS}. One can derive the following
expression for the spectrum, \cite{jansen80,zavaritskii73}
\begin{equation}
{{\rm d}^2 I \over {\rm d} V^2} = {4 \over 3 \pi} {e^3 m^2 v_{\rm
F} \over \hbar^4} a^3 \alpha^2 F_{\rm p}(eV),
\label{eq.e-p-spectrum}
\end{equation}
where $a$ is the contact radius, and the function $\alpha^2 F$ is
given by
\begin{eqnarray}
\alpha^2 F_{\rm p}(\epsilon) = {m^2 v_{\rm F} \over 4 \pi h^3}
\int {\rm d}^2 n \int {\rm d}^2 n' |g_{nn'}|^2
\nonumber \\
\times \eta(\theta({\bf n},{\bf n'})) \delta(\epsilon - \hbar
\omega_{nn'}). \label{eq.e-p-function}
\end{eqnarray}
Here, the integrals run over the unit vectors of incoming and
outgoing electron wave vectors (${\bf n}={\bf k}/|k|$), $ g_{nn'}$
is the matrix element for the electron-phonon interaction, and
$\eta$ is a function of the scattering angle that takes the
geometry into account, such that only backscattering through the
contact is effective, $\eta(\theta)=(1-\theta/\tan\theta)/2$. From
this expression, and by considering Fig.\,\ref{distribution}, it
is clear that the contribution of scattering events far away from
the contacts is suppressed by the effect of the geometric angle at
which the contact is seen from that point. The probability for an
electron to return to the contact decreases as $(a/d)^2$, with $a$
the contact radius and $d$ the distance from the contact. This
implies that the spectrum is dominantly sensitive to scattering
events within a volume of radius $a$ around the contact, thus the
{\em effective volume} for inelastic scattering in the case of a
clean opening (the contact) between two electrodes is proportional
to $a^{3}$. Clearly, this effective volume must depend on the
geometry of the contact. For a long cylindrical constriction, the
electrons scattered within the constriction will have larger
return probability, the effective volume, in this case, increases
linearly with the length \cite{yanson86}.

There is still very little theoretical work on phonon scattering
in the quantum-size limit of point contacts. In this case, we must
view the inelastic scattering process as mixing between the
different conductance eigenchannels. Furthermore, as the contact
becomes smaller, the signal will come from scattering on just a
few atoms surrounding the contact. The spectrum will no longer
measure the bulk phonons, but rather local vibration modes of the
contact atoms. This is also what leads to the interest in
measuring it. Bon{\v c}a and Trugman \cite{bonca95} have
introduced a formalism to calculate the interaction of tunneling
electrons with localized inelastic excitations. This method was
extended to the interaction of conduction electrons in a
single-mode quantum wire with local vibration modes
\cite{bonca97,ness99,emberly00}. The extension to quantum
conductors with many channels, allowing for transitions between
the channels, has not been made, to our knowledge.

\begin{figure}[!t]
\centerline{\includegraphics[width=7cm]{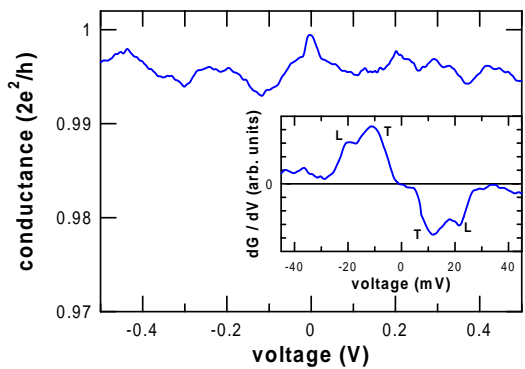}}
\caption{Differential conductance as a function of the applied
bias voltage for a one-atom Au contact at 4.2\,K. The contact was
tuned to have a conductance very close to 1\,\g0, which suppresses
the amplitude of the conductance fluctuations. This allows the
observation of a phonon signal, which is seen as a maximum at zero
bias. Inset: By taking the derivative of the conductance the
transverse (T) and longitudinal (L) acoustic branches can be
recognized symmetrically positioned around zero. Note the expanded
scale of the voltage axis in the inset. Reprinted with permission
from \protect\cite{untiedt00}. \copyright 2000 American Physical
Society.} \label{f.one-atom-phonons}
\end{figure}

In attempting to measure the phonon signal for small contact sizes
one encounters the problem that the phonon signal intensity
decreases, according to (\ref{eq.e-p-spectrum}), while the
amplitude of the conductance fluctuations remains roughly
constant, or slightly increases. The result is that the phonon
signal is drowned in the conductance fluctuations for the smallest
contacts. A solution to this problem is obtained for the special
and interesting event of a contact made up of a single channel
with nearly perfect transmission probability \cite{untiedt00}.
Indeed, for this situation according to (\ref{eq.sigmaGV}) the
conductance fluctuations are suppressed. Under these conditions
the features due to phonon scattering become clearly visible, as
illustrated in Fig.\,\ref{f.one-atom-phonons}. Surprisingly, one
observes a spectrum that still closely resembles the bulk phonon
spectrum, although the relative intensities of the features in the
spectrum are different. More interesting spectra are obtained for
a chain of metal atoms, where the one-dimensional features can be
clearly identified \cite{agrait02a}, as will be discussed in
Sect.\,\ref{s.chains}.

\subsubsection{Heating in atomic-sized contacts} \label{sss.heating}

From the presence of a phonon signal in the current-voltage
characteristics we deduce that there is a finite amount of heating
of the lattice by the current. Nevertheless it appears to be
possible to apply a voltage of up to nearly 2\,V over a contact
made up of a single gold atom without destroying it
\cite{untiedt00}. This gives an astonishingly high current density
of $2 \cdot 10^{9}$\,A/cm$^2$, which is more than five orders of
magnitude higher than for  macroscopic metallic wires. The reason
that this is possible is, of course, that almost all of the
electrical power $P=IV$ that is taken up by the junction is
converted into kinetic energy of the ballistic electrons. On
average, this excess kinetic energy is deposited into the phonon
system away from the contact at a distance equal to the inelastic
mean free path, $l_{\rm i}$. Since $l_{\rm i}$ can be as large as
1\,$\mu$m the thermal energy is strongly diluted in the banks on
either side of the contact.

Experimental evidence for an increase in the lattice temperature
of the contacts is obtained from the study of two-level
fluctuations (TLF). As outlined in Sect.\,\ref{ss.steps&plateaus},
when a contact is stretched one observes a sequence of steps in
the conductance that are usually associated with hysteresis, at
least when the contact has been `trained' a little
(Fig.\,\ref{f.hysteresis}).  For other steps, such hysteresis is
not observed, but in stead the conductance shows spontaneous
fluctuations of a two-level type, between the values before and
after the step, also known as random telegraph noise. This
phenomenon is observed only in a very narrow range of the piezo
voltage controlling the contact elongation; at the plateaus the
conductance assumes stable values. In some cases it is even
possible to tune the `duty cycle' of the TLF, i.e. the relative
portion of the time spent in the upper compared to the lower
state, by fine adjustment of the piezo voltage
\cite{ruitenbeek97a}.

\begin{figure}[!b]
\centerline{\includegraphics[width=7cm]{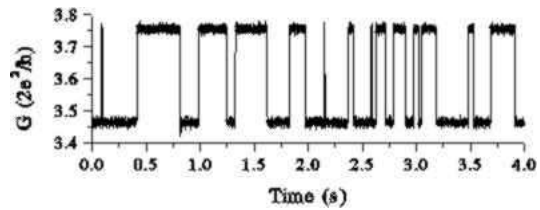}}
\caption{Two-level fluctuation observed for a copper point contact
measured with a bias current of $3\,\mu$A at $T=4.2$\,K. Courtesy
H.E. van den Brom \protect\cite{brom00}.} \label{f.TLF}
\end{figure}

As illustrated in Fig.\,\ref{f.hysteresis} the hysteresis becomes
smaller, and can often be suppressed, by increasing the bath
temperature. The same effect can be obtained by increasing the
current through the contact. Once the hysteresis is fully
suppressed, at still higher currents the system shows TLF,
fluctuating between the conductance values of the plateaus left
and right of the conductance step \cite{brom98}, as illustrated
for a Cu contact in Fig.\,\ref{f.TLF}. The fluctuation rate
increases very rapidly for larger currents.

The mechanism of these current-induced fluctuations can be
understood as an energy transfer of the non-equilibrium electrons
to the degrees of freedom of the atomic structure. The way atoms
rearrange during breaking or making of the contact can be
described in terms of a potential landscape in configuration
space. The total potential energy
$V(x_1,y_1,z_1;\ldots;x_N,y_N,z_N)$ of the contact is a function
of the position coordinates $(x_i,y_i,z_i)$ of all $N$ atoms
making up the contact. The actual positions of the atoms in a
given stable configuration correspond to a local minimum in this
space. By pulling or pushing the contact we impose a subset of the
atomic coordinates and the other coordinates rearrange to find a
new minimum.

Before atoms actually rearrange, the configuration has to be
lifted over an energy barrier $E_{\rm B}$. This lifting
corresponds to elastic deformation, while a jump over the barrier
results in an atomic rearrangement.  When we reverse the
piezo-voltage sweep direction immediately after a jump, the
contact can behave in two distinct ways: either the potential
landscape is such that the previous configuration is the most
favorable one, or a third nearby minimum presents itself. In the
first case the result will be a closed loop of hysteresis. In the
second case, the contact will take a different configuration,
which is observed as a non-retraceable step \cite{brom98}.  In the
latter case the contact will eventually search for the two lowest
minima in the neighborhood, when the contact is cycled over a
small range of the electrode displacement. After a few cycles the
contact will then be `trained' \cite{agrait94}.  If $E_{\rm B}$ is
low enough and the two available positions on both sides have
nearly equal energy, thermally activated jumps back and forth will
appear, manifesting themselves in the form of TLF, as in
Fig.\,\ref{f.TLF}.

An estimate for the typical values of the barrier $E_{\rm B}$ can
be obtained by considering the typical force jumps between two
configurations, which we have seen in Sect.\,\ref{s.mechanical}
are of the order of 1\,nN. Combining this number with the distance
over which the contacts need to be stretched between two jumps,
$\sim$0.1\,nm, we obtain $E_{\rm B} \sim 1$\,eV. From molecular
dynamics simulations for atomic-sized contacts estimates in the
range of 0.1 to 1 \,eV were obtained by S{\o}rensen \ea\
\cite{sorensen96a}. They developed a method to calculate the
lowest energy trajectory for an atomic structure in the transition
between two stable states. Although it is not clear whether the
calculated configurations are representative for a typical contact
jump, it appears that $E_{\rm B}\sim 0.1 - 1$\,eV is a reasonable
value.

The fluctuation rate of the observed TLF as function of the
current may be regarded as a local atomic-sized thermometer giving
information on the heating inside the contact. A study of the
temperature and current dependence of TLF was first done by Ralls,
Ralph and Buhrman for larger nanofabricated contacts of fixed
size, of the type described in Sect\,\ref{ss.nanofab contacts}
\cite{ralls88,ralls89,holweg92}. For these larger contacts, the
TLF are due to  unknown defects somewhere in or near the contact.
Most contacts show one or more TLF, but they have a distribution
in activation energies that cannot be controlled. It was argued
that a large collection of interacting TLF likely form a
microscopic mechanism for the ubiquitous 1/f-noise in macroscopic
conductors. Moreover, by studying the dependence of the
fluctuations on the polarity of the voltage bias it was possible
to show that the current exerts a net force on the defects. These
forces are responsible for electromigration of defects known to
occur in metallic systems that are subjected to long-term high
current densities.

In Refs.\,\cite{ralls88,ralls89,holweg92} the fluctuation rate of
individual TLF was measured, both as a function of temperature and
as a function of the bias voltage. By combining values for the
temperature and the bias voltage that give similar fluctuation
rates they observed that the effective temperature of the
two-level system could be described as  $ k_{\rm B}T_{\rm eff}=
\gamma eV$, for $eV \gg k_{\rm B}T$, with $\gamma \simeq 0.15$ .
Ralls \ea\  \cite{ralls89} proposed a model to describe these
results that allows for energy exchange between the
non-equilibrium electrons and the local fluctuator. They set up a
rate equation that takes heating and cooling by the electrons into
account, plus a parameter that allows for relaxation to the
lattice.

For the nanofabricated contacts the effective temperature of the
defect is higher than that of the lattice, because of the poor
relaxation coupling to the bulk phonons. For atomic-sized
contacts, however, the TLF presumably result from collective
rearrangements of all the atoms that make up the contacts, so that
this distinction does not exist. Todorov \cite{todorov98} has
proposed to describe heating in atomic-sized contacts by regarding
each atom as an independent oscillator, taking up energy from the
electrons. The thermal energy that assists the atomic
configurations responsible for TLF to cross the barrier is again
given by $\gamma eV$, with $\gamma = 5/16= 0.3125$ in a simple
free-electron approximation. However, relaxation of the vibrations
by thermal conduction through the lattice cannot be neglected. The
estimates Todorov makes for the local temperature including
lattice thermal conduction are about an order of magnitude
smaller, but the relevant parameters cannot be determined with
great confidence.

In the limit of a classical point contact (Maxwell limit,
Sect.\,\ref{ss.Maxwell}) the Joule heating produces an effective
temperature at the contact center given by \cite{holm67}
\begin{equation}
T_{\rm eff}^2 = T_{\rm b}^2 +  V^2/4L
\end{equation}
This simple relation was derived by assuming that the heat
conductivity, $\kappa$, is dominated by the electronic part, and
that this is related to the electrical conductivity, $\sigma$, by
the Wiedemann-Franz law, $\kappa/\sigma=LT$, where $L$ is the
Lorentz number. For low bath temperature $T_{\rm b}$, the
effective temperature in the contact is proportional to the bias
voltage $V$, and we obtain $k_{\rm B} T = \gamma e V$ with $\gamma
= 0.275$, which is not very different from the value of $\eta$ for
ballistic contacts quoted above.

Returning to the experiments in atomic-sized contacts, we note
that the transition rate for jumps over the barrier is expected to
be of Arrhenius form, $\nu = \nu_0 \exp(- E_{\rm B}/k_{\rm
B}T_{\rm eff})$. Here, $\nu_0$ is the attempt frequency, which is
of the order of phonon frequencies, $\sim$10$^{13}\, $s$^{-1}$;
the Boltzmann factor $\exp(- E_{\rm B}/k_{\rm B}T_{\rm eff})$
should be such that the TLF are observable on the laboratory time
scale, i.e.\  have a minimal frequency of order 0.1\,s$^{-1}$.
Hence in order to observe TLF at the temperature of the helium
bath, 4.2\,K, we must have a barrier $ E_{\rm B}$ smaller than
about 0.01\, eV. Since this is one or two orders of magnitude
smaller than our estimate for $E_{\rm B}$ this explains why we
observe mostly steps with hysteresis. For those events when
spontaneous TLF at 4.2\,K are observed a low-barrier fluctuating
system must be present. This could perhaps be a single atom at the
surface of the constriction with two equally favorable positions
separated by a small distance.

\begin{figure}[!t]
\centerline{\includegraphics[width=7cm]{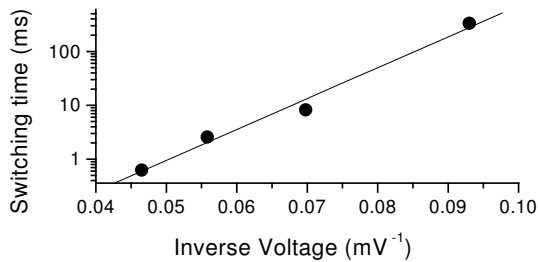}}
\caption{Semi-logarithmic plot of the switching time as a function
of the inverse bias voltage, for TLF in an atomic-sized Cu
contact, with a mean conductance of 3.6\,\g0 at a bath temperature
of 4.2\,K. Courtesy H.E. van den Brom \protect\cite{brom00}.}
\label{f.TLF-rate-vs-V}
\end{figure}

The bias dependence of the switching rate has been investigated
for TLF in Cu and Pt contacts in \cite{muller92,brom01}. For large
bias voltages, the rate $\nu$ increases as a function of bias
voltage and can be described by,
\begin{equation}
\nu=\nu_0\exp\left(-\frac{E_{\rm B}}{\alpha eV}\right),
\end{equation}
as is illustrated in Fig.\,\ref{f.TLF-rate-vs-V}. This suggests
that the effective temperature is indeed proportional to the bias
voltage. However, for some TLS a cross over was observed at low
voltages, where the rate becomes independent of the bias voltage.
Assuming that this occurs when the local temperature $\alpha eV$
becomes of the order of the bath temperature, one obtains
estimates for $\alpha$ of $\sim0.1 - 0.5$ \cite{brom00}. Adopting
this estimate for $\alpha$ we conclude that for experiments
performed at helium temperature, the lattice temperature at
100\,mV is of order 100--500\,K. Indeed, the fact that we can
convert a hysteretic loop into a TLF by increasing the current
implies that the thermal energy injected by the current into the
two-state system must be considerable. Single-atom gold contacts
have been found to survive bias voltages of nearly 2\,V, for which
the lattice temperature must approach the melting point. Note that
we explicitly distinguish the lattice temperature from the
electron temperature. At high bias the system is far from
equilibrium and a true temperature can probably not be properly
defined. Any effective values for the temparature of the lattice
and that of the electron gas are expected to be very different and
strongly position dependent.

In order to exploit TLF quantitatively as a local thermometer a
single two-level system for a contact needs to be measured as a
function of temperature {\em and} voltage. However, upon heating
the contact appears to undergo uncontrolled changes that were
attributed to thermal expansion of the substrate and the
electrodes \cite{brom01}, and attempts at measuring the lattice
temperature of the contact have not yet produced reliable results.

\subsection{Kondo scattering on magnetic impurities}
\label{ss.Kondo}

The Kondo effect is a prototypical electron correlation effect in
condensed matter physics. It arises in dilute magnetic alloys due
to the interaction between conduction electrons and the localized
magnetic moments. This interaction gives rise to a minimum in the
electrical resistivity as a function of temperature
\cite{wilson53}. If the concentration of local moments is
sufficiently small, the effect of interactions between the
impurities is negligible and it is then appropriate to consider
just a single impurity coupled to the conduction electrons.  This
problem, known as the `Kondo problem', is by now well understood
theoretically. The development of its solution started with the
work by Kondo \cite{kondo64}, continued with the scaling ideas of
Anderson and the renormalization group analysis by Wilson, until
the  finding of exact solutions based on the Bethe ansatz
\cite{hewson93}. However, a renewed interest in this problem has
emerged in more recent years associated with the possibility of
exploring `Kondo physics' in nanofabricated devices
\cite{ralph94,chenG91,chandrasekhar94,goldhaber98,madhavan98}.

The basic energy scale in Kondo physics is determined by the Kondo
temperature $T_{\rm K}$. This is related to the exchange coupling
constant $J$ between the conduction electrons and the local
magnetic moment, and to the density of states at the Fermi energy,
$N_{\rm F}$, through $T_{\rm K} = T_{\rm F} \exp({-1/JN_{\rm
F}})$, where $T_{\rm F}$ is the Fermi temperature
\cite{daybell73}.

The use of point contact spectroscopy to study metals with
magnetic impurities started almost two decades ago
\cite{lysykh80}. It revealed many features analogous to the
well-known phenomena in the electrical resistivity of bulk alloys,
the voltage playing the role of temperature in this analogy. With
the development of the MCBJ technique it has become possible to
study size effects in Kondo phenomena for contact diameters
decreasing from $\sim$50\,nm down to the atomic scale. Yanson \ea\
\cite{yanson95} reported a large broadening and increase of the
relative amplitude in the zero bias maximum of the differential
resistance for various noble metals (Cu, Au) doped with magnetic
impurities (Mn) as the contact diameter was decreased. They
interpreted their data as due to an enhancement of the exchange
coupling parameter $J$, leading to a surprisingly large increase
in the Kondo temperature $T_{\rm K}$ for small contact diameters.
In a subsequent paper, van der Post \ea\  \cite{post96a} using the
same technique studied the case of Fe impurities in Cu and found a
much weaker increase of $T_{\rm K}$ with decreasing contact
diameter. It was also argued that the large variation in $T_{\rm
K}$ observed in the first experiments \cite{yanson95} arises due
to the application of the standard weak-scattering result, which
breaks down for large $T_{\rm K}$. This leaves still a very strong
enhancement of $T_{\rm K}$. A possible explanation was proposed by
Zarand and Udvardi \cite{zarand96}, where the increase of $T_{\rm
K}$ was associated with fluctuations in the local density of
states in the contact region.

The enhancement of the Kondo scattering observed in these
experiments appears to contradict a suppression of the Kondo
resistivity observed for thin-films and microfabricated wires of
Kondo alloys \cite{chenG91,blachly95}. These results, in turn,
were disputed by Ref.\,\cite{chandrasekhar94}, where no size
effect was detected at all. An explanation for size effects in
terms of surface-induced anisotropies due to spin-orbit coupling
was proposed by Ujsaghy and Zawadowski
\cite{ujsaghy98a,ujsaghy98b}, and they present arguments that may
bring the various experiments into agreement. More recently, a
very pronounced size effect in the thermopower of mesoscopic AuFe
Kondo wires was discovered by Strunk \ea  \cite{strunk98}, that
appears to agree with the proposed spin-orbit induced anisotropy
near the surface of the wires.

Using a different experimental setup Ralph and Buhrman
\cite{ralph94} were able to observe for the first time
Kondo-assisted tunneling and simple resonant tunneling from a
single impurity. The devices used by Ralph and Buhrman were
nanofabricated Cu point contacts produced by the method described
in Sect.\,\ref{ss.nanofab contacts}. Although in most cases the
electron transport in such devices is through the metal filament,
on certain occasions they found in parallel a contribution due to
tunneling via charge traps in the silicon nitride adjacent to the
narrowest region of the Cu contact. Ralph and Buhrman identified
this system as an experimental realization of the Anderson model
out of equilibrium, a problem that has received considerable
attention from theory \cite{meir92,hershfield91a,ng93,levy93}. The
presence of a charge trap in the silicon nitride gives rise to a
very narrow peak in the differential conductance around $V=0$, in
agreement with the theoretical predictions for the Anderson model
\cite{meir92,hershfield91a,ng93,levy93}. The peak was shown to
exhibit Zeeman splitting, which is unambiguous evidence that is
was due to a magnetic defect. The fact that the signal shows up as
a peak in the conductance allowed to identify it as Kondo-assisted
tunneling through the silicon nitride rather than scattering from
a magnetic impurity within the Cu, which would produce a dip in
the conductance at $V=0$.

More recently, the use of scanning tunneling microscopes has
allowed to study the Kondo effect on a single magnetic atom on a
metal surface \cite{madhavan98,jli98,chen99,jamneala00}. One of
the first experiments of this kind by Madhavan \ea
\cite{madhavan98} was performed on a Au(111) surface after
deposition of 0.001 of a monolayer of Co. The $dI/dV$ spectra
taken in the vicinity of a single Co atom revealed the presence of
a narrow feature (of the order of a few mV) around $V=0$. Instead
of a simple lorentzian peak Madhavan and coworkers observed a dip
followed by a shoulder, a form which is characteristic of Fano
resonances \cite{fano61}. These type of resonances arise from the
interference between two possible channels for tunneling between
tip and sample. According to Madhavan \ea one channel would be
provided by the Kondo resonance associated with $d$ orbitals in
the Co atom and the other would be due to the surrounding
continuum of conduction band states. Ab-initio calculations show
in fact that the local density of states at the Co site exhibits a
narrow resonance at the Fermi energy with a $d$ character
\cite{weissmann99}. Similar experimental findings were reported by
Li \ea  \cite{jli98} for Ce atoms on Ag(111) surfaces. These
authors showed that the Fano type resonances were not observed for
nonmagnetic Ag adatoms. The systematic behavior of the $dI/dV$
spectrum for $3d$ transition metal adatoms was studied by Jamneala
\ea \cite{jamneala00}. They found very pronounced features around
$V=0$ only for Ni, Co and Ti. No traces of a Kondo type resonance
were found in the case of Fe, Mn, Cr or V, which can be explained
by $T_K$ being less than the experimental temperature of 6 K in
these experiments. For the `end' elements (Ni and Ti) the features
around $V=0$ could be due to a combination of the Kondo resonance
and the bare $d$ band resonances, which are closer to the Fermi
energy for these elements.

\begin{figure}[!t]
\centerline{\includegraphics[width=7cm]{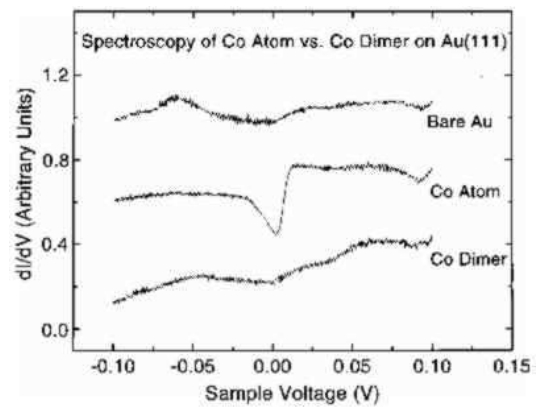}}
\caption{$dI/dV$ spectra obtained with an STM tip held over a
single cobalt atom, an atomically fabricated dimer, and the clean
gold surface. The curves are offset vertically for clarity. The
Kondo resonance can be seen for the individual cobalt atom, but is
absent for the dimer. Reprinted with permission from
\protect\cite{chen99}. \copyright 1999 American Physical Society.}
\label{f.cobalt-dimer}
\end{figure}

The use of STM to study the Kondo effect on magnetic adatoms is
even allowing us to analyze magnetic interactions in man-made
nanostructures obtained by atomic manipulation on a metal surface.
Chen et al. \cite{chen99} studied the case of artificially
fabricated Co dimers on a Au (111) surface. They found an abrupt
disappearance of the Kondo resonance for cobalt-cobalt separations
less than 6\,\AA, a behavior attributed to the reduction of the
exchange coupling between Au conduction electrons and the magnetic
Co dimers (see Fig.\,\ref{f.cobalt-dimer}). Kondo features in
scanning tunneling spectroscopy have recently provided the basis
for a striking demonstration of quantum coherence on a metal
surface by Manoharan \ea  \cite{manoharan00}. They used atomic
manipulation to create an elliptical `quantum corral' of Co atoms
on a Cu surface. When a Co atom was placed at one focus of the
ellipse, the Kondo feature was detected not only at the atom but
also at the empty focus. This focusing effect has been described
as a `quantum mirage'.

\subsection{Non-magnetic Kondo scattering: the 2-channel Kondo problem}
\label{ss.2-channel-Kondo}

In two seminal papers Ralph \ea  \cite{ralph92,ralph94a} reported
the observation of a Kondo-like zero-bias anomaly in the
differential conductance  that could not be attributed to the
presence of magnetic impurities. They studied Cu contacts, of the
type described in Sect.\,\ref{ss.nanofab contacts}. The
non-magnetic origin of the signals was illustrated by the absence
of a splitting of the feature in an applied magnetic field. The
shape of the dip in the conductance was not logarithmic in bias
voltage $V$, or temperature $T$, as would be expected for a Kondo
minimum. In stead, a $T^{1/2}$ and $V^{1/2}$ dependence was
observed, with the appropriate scaling behavior that is expected
for a 2-channel Kondo system \cite{nozieres80}. It was argued that
defects in the contact that act as fast (low transition barrier)
two-level tunneling systems form the active scattering systems. It
had been proposed earlier that such two level-systems are
candidates to show 2-channel Kondo behavior
\cite{zawadowski80,vladar83}.

A strong argument in favor of an interpretation in terms of fast
TLF was found in the fact that the signals disappear when the
samples are kept (annealed) at room temperature for a few days.
Further support comes from experiments on point contacts made by
the MCBJ technique on metallic glasses, where a high concentration
of TLF centers is expected \cite{keijsers96b,balkashin98}.
However, Wingreen \ea  \cite{wingreen95} raised a number
objections against this interpretation, and proposed an
alternative mechanism in terms of electron-electron interactions
enhanced by defect scattering. Although strong arguments were
given in the reply \cite{ralph95} that refute the alternative
model, the discussion is still not completely settled. The
amplitude of the signal implies that many (of order 10 or more)
TLF should contribute to the signal. Furthermore, the two-level
systems should have a very narrow distribution of separation of
the energy levels in the two available states, at a very low
value. It remains to be demonstrated that such systems exist in
large concentrations.

The nature of the defects involved in producing the signals is
also not yet established. A serious candidate was proposed by
Vegge \ea  \cite{vegge01}, who showed by molecular dynamics
calculations that dislocation kinks in a copper crystal have
appropriately low energy barriers and low effective masses to
allow fast quantum tunneling. For further information we refer the
reader to two recent extensive review papers by von Delft \ea
\cite{vondelft98,vondelft99}.

\subsection{Environmental Coulomb blockade}
\label{ss.ECB}

A common simplifying assumption in the analysis of electron
transport in quantum coherent structures is that the system is
connected to an ideal voltage source. In practice the voltage
source is never ideal but contains a finite internal impedance
$Z(\omega)$. At the same time, an atomic-sized contact will have a
certain capacitance $C$. Although the capacitance associated
strictly with the atomic-sized conductor is in theory extremely
small (of the order of an aF, see Ref.\,\cite{wang98a}), the
capacitance $C$ will be dominated by the contribution of the much
wider leads in which the atomic conductor is embedded. The value
of $C$ depends thus greatly on the fabrication technique that is
used. In the case of microfabricated break junctions $C$ is
typically of the order of one fF.

In principle, one should take both $C$ and $Z(\omega)$ into
account in determining the transport properties of the contact. At
low temperatures the voltage across the contact develops quantum
fluctuations and the electrical properties of the circuit cannot
be inferred from the conductance of the separate elements. This
problem has been extensively studied in the case of small tunnel
junctions, where the conductance of the series circuit can be
completely suppressed at sufficiently low voltages and
temperatures. This phenomenon is called {\em environmental}
Coulomb blockade (for a review see Ref. \cite{grabert92}).
Qualitatively, it arises when the impedance of the environment is
high enough that the charge of a single electron tunneling across
the junction leaks away only slowly. In that case the charging
energy associated with the contact capacitance $E_{\rm C} =
e^2/2C$ starts to play a role, which results in a suppression of
the tunneling current when both the applied bias, $eV$ and the
temperature $k_{\rm B} T$ are smaller than $E_{\rm C}$.

A question which has been recently addressed in the literature
\cite{golubev01,levy01} considers how this phenomenon is modified
when one replaces the tunnel junction by a generic quantum
coherent structure. The case of an atomic-sized contact is
particularly interesting as it provides a system characterized by
a few conduction channels whose transmissions can be determined
using the techniques discussed in Sect.\,\ref{ss.sgs_exp}.
Moreover, the impedance of the environment embedding such contacts
can be tuned within a desired range using nanolithography
\cite{goffman00}.

A  simple argument can be used to demonstrate that the
environmental Coulomb blockade in such systems should disappear
when the perfect transmission limit is reached. As we have seen, a
quantum point contact is characterized by a current fluctuation
spectrum, which at low frequency and zero temperature is given by
$S=2eVG_{0}\sum \tau _{i}(1-\tau _{i})$. One may then speculate
that when the transmissions approach unity the contact cannot be
`felt' by the series impedance $Z(\omega)$ and, conversely, the
transport properties of the contact should not be affected by the
presence of a series impedance. In fact, rigorous calculation
predicts that the Coulomb blockade features in the current-voltage
characteristics should vanish for perfect transmission in the same
way as shot noise does \cite{levy01}.

In Ref.\,\cite{levy01} environmental Coulomb blockade for a single
channel contact of transmission $\tau$ in series with an arbitrary
frequency-dependent impedance $Z_t(\omega)$ (including the contact
capacitance, i.e. $Z^{-1}_t(\omega) = Z^{-1} + i\omega C$) was
studied starting from a model Hamiltonian and using the Keldysh
Green function technique. It was shown that the correction to the
contact conductance, $\delta G$, to the lowest order in
$Z_t(\omega)$ is given by
\begin{equation}
\frac{\delta G}{G} = - G_0 (1 - \tau) \int_{eV}^{\infty} {\rm
d}\omega \frac{\mbox{Re}[Z_t(\omega)]}{\omega} .  \label{eq.ohmic}
\end{equation}
It should be noted that the correction to the conductance is
affected by the same reduction factor, $(1-\tau)$, that applies
for shot noise (see Sect.\,\ref{sss.theory_shot_noise}). In the
simple but realistic case for which the impedance $Z_t(\omega)$ is
composed by the resistance $R$ of the leads embedding the contact
in parallel with the capacitance $C$ of the contact itself, the
integral in (\ref{eq.ohmic}) yields
\begin{equation}
\frac{\delta G}{G} = - G_0 R (1-\tau) \ln\sqrt{1 +
\left(\frac{\hbar\omega_R}{eV} \right)^2 } , \label{eq.ohmic2}
\end{equation}
where $\omega_R = 1/RC$. At finite temperature the singularity in
Eq.\,(\ref{eq.ohmic2}) at $V=0$ becomes progressively rounded. The
finite temperature version of this equation can be found in
\cite{cron01a}.

\begin{figure}[!t]
\centerline{\includegraphics[width=8cm]{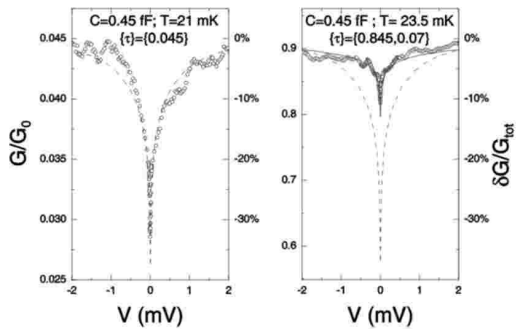}}
\caption{Measured differential conductance curves for two atomic
contacts. The scale of the left axis applies for the measured data
points (circles), in units of \g0 . In the left panel the contact
consists of a single weakly transmitted channel, and the
conductance around zero bias is well-described by the standard
theory of environmental Coulomb blockade, valid for tunnel
contacts, as expected (dashed curve, right axes, relative to the
total conductance) \protect\cite{ingold92}. In the right panel,
the contact has a well-transmitted channel with $\tau_1 =0.835$
plus a second smaller contribution $\tau_2 =0.07$. In this case,
the relative reduction of conductance is much less than expected
from the expression for tunnel junctions and is in agreement with
the predictions of Eq.\,(\ref{eq.ohmic}) (full curve). The wiggles
and asymmetry seen in the experimental curves are reproducible
conductance fluctuations due to the interference effects discussed
in Sect.\,\ref{ss.cond-fluct}. Reprinted with permission from
\protect\cite{cron01a}. \copyright 2001 EDP Sciences.}
\label{f.expECB}
\end{figure}

In order to verify these predictions Cron \ea\  \cite{cron01a}
fabricated an atomic contact embedded in an electromagnetic
environment essentially equivalent to a pure ohmic resistor of the
order of 1\,k$\Omega$, defined by e-beam lithography. The material
chosen for both the atomic contact and the series resistor was
aluminum, which allowed them to extract the channel composition,
or the `mesoscopic PIN code' $\{\tau_1\dots\tau_N\}$ for the
contacts, using the techniques discussed in
Sect.\,\ref{ss.sgs_exp}. The environmental Coulomb blockade was
then measured in the presence of a $0.2$\,T magnetic field which
brings the sample in the normal state. The results for two
contacts with very different transmissions are shown in
Fig.\,\ref{f.expECB}. The standard theory of environmental Coulomb
blockade \cite{devoret90} is able to account for the results
obtained for the contact in the left panel, which has a single
weakly transmitted channel. On the other hand, for the contact in
the right panel, with one well-transmitted channel ($\tau \simeq
0.83$) the relative amplitude of the dip at zero bias is markedly
reduced with respect to the tunnel limit predictions (dashed
line). The experimental results are in good agreement with the
predictions of Ref.\,\cite{levy01} summing the contributions of
all channels (solid line).

In spite of this qualitative understanding, the issue of Coulomb
blockade in atomic size contacts remains to be explored in further
detail.

\section{Superconducting quantum point contacts}
\label{s.superconductors}

In Sect.\,\ref{s.exp_modes} we discussed the experiments on the
$IV$ characteristic of superconducting atomic contacts and their
use to extract information on the conductance modes. In this
section we shall analyze another series of experiments in
superconducting contacts. Most of these experiments have been
conducted in order to test some of the theoretical predictions
presented in Sect.\,\ref{s.sctransporttheory}, like the
quantization of the supercurrent through a narrow constriction,
the supercurrent-phase relation for arbitrary transmission and the
increase of shot noise associated with multiple Andreev
reflections.

\subsection{Supercurrent quantization}
\label{ss.supercurrent}

A rather straightforward consequence of the quantization of the
conductance in a smooth constriction with a cross section
comparable to the Fermi wavelength is the quantization of the
supercurrent when the constriction connects two superconducting
leads \cite{beenakker91a}. For such ideal system the supercurrent
should be quantized in units of $e\Delta/ \hbar$.

\begin{figure}[!b]
\centerline{\includegraphics[height=8cm]{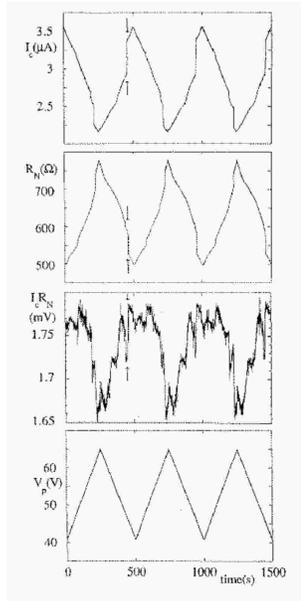}}
\caption{Critical current and normal-state resistance measurement
for a periodic variation of the piezo voltage. The fine structure
in $I_c$ lines up with that in $R_N$. The relatively large changes
in $I_c$ and $R_N$ at the steps almost compensate the other in the
product $I_c R_N$. Reprinted with permission from
\protect\cite{muller92}. \copyright 1992 American Physical
Society.} \label{f.muller}
\end{figure}

Although this prediction was originally proposed for a 2DEG device
it was soon realized that it should be also valid for a
superconducting atomic contact provided the condition of
conductance quantization was reached. The first experiments in
this direction were performed by Muller \ea \cite{muller92} who
studied Nb contacts made by the MCBJ technique. They measured the
critical current (defined as the current value at a set point
voltage near $V=0$) at 1.2 K while the piezo voltage was varied
periodically. In the scans for rising piezo voltage they observed
steps in the critical current of a size close to $e\Delta/ \hbar$.
In another experiment, reducing the range of the piezovoltage
scans they measured simultaneously $I_c$ and $R_N$. Their results
are shown in Fig.\,\ref{f.muller}. As can be observed, the steps
in $I_c$ are correlated with steps in $R_N$ at the same positions.
The variations of 50\% in $I_c$ and $R_N$ separately are reduced
to variations of only 7\% in the product. The average value for
$I_c R_N$ was $1.75 \pm 0.05$\,mV,  which is of the same order but
considerably smaller than the predicted value $\pi \Delta/e$. A
reduction of $I_c R_N$ below the theoretical value was also
observed in a systematic study \cite{muller92a}.
The authors suggested thermal or quantum fluctuations as a
possible source for this discrepancy. In a subsequent study Muller
and de Bruyn Ouboter \cite{muller94} analyzed the slope in the
supercurrent branch and the maximum current near zero voltage that
they called threshold current. As discussed in
Sect.\,\ref{ss.RSJ}, the slope and the actual value of this
maximum current is very sensitive to the electromagnetic
environment in which the contact is embedded.

Finally, in Ref.\,\cite{vleeming94} it was shown that the
threshold current for a one-atom contact exhibits large variations
even when the normal conductance is nearly constant. As already
pointed out in Sect.\,\ref{ss.sgs_exp}, even when the total
conductance is fixed the channel decomposition can fluctuate. This
points towards the need of more controlled experiments in which
the channel content of the contact be determined while measuring
the supercurrent.

\subsection{Current-phase relation}
\label{ss.cpr}

An even more ambitious goal than the measurement of the critical
current is the measurement of the whole current-phase relation
(CPR) in a superconducting point contact. In
Sect.\,\ref{s.sctransporttheory} we have already discussed the
different theoretical predictions for the CPR ranging from tunnel
junctions to ballistic contacts. In this last case one expects a
non-sinusoidal behavior with a maximum at $\phi = \pi$
\cite{kulik75}. Again, atomic contacts have been viewed as ideal
tools to test these predictions in the quantum regime. The
measurement of the CPR requires, however, a rather sophisticated
setup in which the phase can be fixed by an external magnetic
flux.

\begin{figure}[!t]
\begin{center}
\includegraphics[height=7cm]{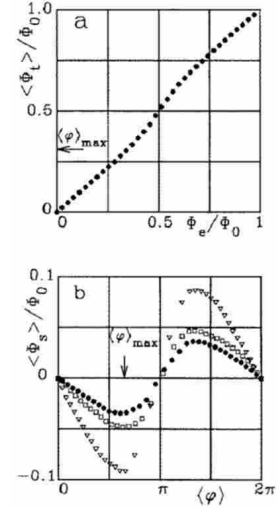}
\end{center}
\caption{Determination of the current-phase relation in the
experiment by Koops \ea The relation between the total flux
$\langle\Phi_t\rangle$ and the applied flux $\Phi_e$ (shown in
(a)) allows to determine the self-induced flux
$\langle\Phi_s\rangle$ which is proportional to the current.
Reprinted with permission from \protect\cite{koops96}. \copyright
1996 American Physical Society.} \label{koops}
\end{figure}

To this end, Koops \ea \cite{koops96} fabricated superconducting
loops of micrometric size in which a MCBJ was placed. They used Nb
and Ta foils and laser cutting techniques. On top of the loop they
placed a flux-detection coil which allowed for measuring the CPR
inductively. The enclosed area of the loop was chosen such as to
have a small self-inductance (smaller than 1 nH). This condition
is necessary to prevent a multivalued relation between the
external flux, $\Phi_e$, and the total flux through the loop
$\langle\Phi_t\rangle$ (observed mean value). $\Phi_e$ and
$\langle\Phi_t\rangle$ differ due to the presence of the
self-induced flux $\langle\Phi_s\rangle = L \langle I_s\rangle$,
where $L$ is the self-inductance of the loop, i.e.
$\langle\Phi_t\rangle = \Phi_e + \langle\Phi_s\rangle$.   The mean
phase difference $\langle\varphi\rangle$ over the contact is equal
to $-2\pi\langle\Phi_t\rangle/\Phi_0$ where $\Phi_0 = h/2e$ is the
flux quantum. By measuring $\langle\Phi_t\rangle$ for a given
value of $\Phi_e$ the self-induced flux, which is proportional to
the current, can be detected. An example of the measured
$\langle\Phi_t\rangle$ vs. $\Phi_e$ relation is shown in
Fig.\,\ref{koops}a for Nb at $1.3K$. Fig.\,\ref{koops}b shows the
corresponding $\langle\Phi_s\rangle$ vs. $\langle\varphi\rangle$
relation.

Although the measurements were performed on the last contact
before jump to the tunnel regime, none of the measured CPRs was
found to correspond to the theoretical predictions for perfect
transmission. In particular, the position of the maximum current
in the CPR ($\phi_{max}$) was found to be displaced towards lower
values with respect to the theoretical predictions. The authors
attributed this fact to thermal fluctuations and to the
impossibility to reach perfect transmission for such contacts. A
theory of the displacement of the maximum in the CPR of a
ballistic contact due to thermal fluctuations was presented in
\cite{ouboter96}. In this theory the amplitude of the phase
fluctuations is controlled by the self-inductance of the loop and
decreases for decreasing inductance. Koops \ea observed that
$\phi_{max}$ tended to the expected theoretical value at perfect
transmission for decreasing inductance. Their analysis suggested
that the contact transmissions should be somewhere between 0.9 and
1.0.

The impossibility to determine the contact transmissions
independently was clearly one of the main limitations of the
experiments by Koops \ea \cite{koops96}. The more recently
developed techniques to extract the information on the conductance
modes, discussed in Sect. 8, could ideally be combined with loop
measurements to test the predictions of the theory for the CPR in
atomic contacts. Although this is still an open challenge for the
experimentalists, recent work by Goffman \ea \cite{goffman00}
constitutes an important step in this direction. This work will be
discussed below.

\begin{figure}[!t]
\centerline{\includegraphics[height=4.5cm]{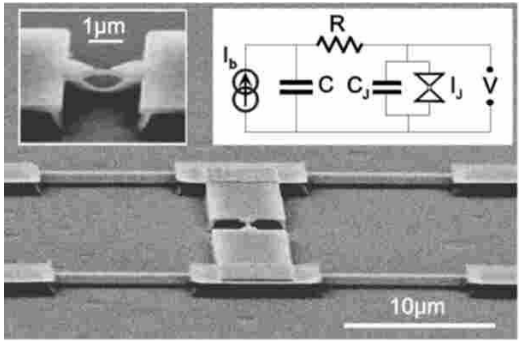}}
\caption{Micrograph of the experimental setup used in Ref.
\,\cite{goffman00} to study the supercurrent in an atomic contact.
Each probe contains a AuCu resistor (thin lines of 10\,$\mu$m
length) and a large capacitor with the metallic substrate. The
left inset shows a close-up of the microfabricated MCBJ and the
right inset illustrates the equivalent circuit. Courtesy M.
Goffman. } \label{goffmansetup}
\end{figure}

Goffman \ea studied the supercurrent in aluminum microfabricated
MCBJs. In order to have good control of the thermal and quantum
fluctuations they designed an on-chip dissipative environment with
small resistors of known value placed close to the atomic contact
in a four probe geometry. A micrograph of the device is shown in
Fig.\,\ref{goffmansetup}.
With the correct choice of the environment parameters the
current-voltage curve becomes hysteretic, which allows detecting
the supercurrent and the dissipative branch simultaneously. This
in turn permits the determination of the the set of tranmission
values $\{\tau_i\}$ for the modes in each atomic contact by the
analysis of the subgap structure as discussed in
Sect.\,\ref{ss.sgs_exp}. This `mesoscopic PIN code' \cite{cron01}
fully characterizes the junction. A typical IV recorded in this
work is shown in Fig.\,\ref{f.expsgs}.

\begin{figure}[!t]
\begin{center}
\includegraphics[height=55mm,angle=270]{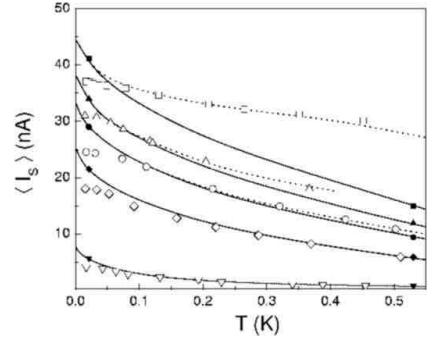}
\end{center}
\caption{Experimental (open symbols) and theoretical (lines)
results for the switching current obtained in
Ref.\,\cite{goffman00}. The results correspond to one atom
contacts
with different channel content. ($\bigtriangledown $)%
$~\{\tau _{i}\}$ = \{0.21,0.07, 0.07\}. From the fit a
zero-temperature supercurrent of $I_{0}$ = 8.0$\pm $0.1$~{\rm nA}$
is obtained. ($\diamondsuit $)~$\left\{ \tau _{i}\right\} $ =
\{0.52,0.26,0.26\}, $I_{0}$=25.3$\pm $0.4~${\rm nA}$. ($\circ $)%
$\{\tau_{i}\}$ = \{0.925,0.02,0.02\}, $I_{0}$=33.4$\pm $0.4~{\rm nA}.%
($\bigtriangleup $) $\left\{ \tau _{i}\right\} $ =
\{0.95,0.09,0.09,0.09\},%
$I_{0}$ = 38.8$\pm $0.2~${\rm nA}$. ($\Box $)%
$~\left\{ \tau_{i}\right\} $ = \{0.998,0.09,0.09,0.09\}, $I_{0}$ =
44.2$\pm $0.9$~{\rm nA}$. The full lines are the predictions of
the adiabatic theory and the dotted lines correspond to the
non-adiabatic theory, which allows for Landau-Zener transitions
between Andreev states. Reprinted with permission from
\protect\cite{goffman00}. \copyright 2000 American Physical
Society. \label{f.supercurrent}}
\end{figure}

Goffman \ea concentrated in the analysis on the threshold or
switching current $I_S$ at which the jump from the supercurrent
branch to the dissipative branch takes place. The switching takes
place close to the maximum in the supercurrent branch just before
the region of negative differential resistance. This value is very
sensitive to thermal fluctuations and decreases with increasing
temperature. The experimental results for the switching current
can be analyzed in terms of a generalized RSJ model discussed in
Sect.\,\ref{ss.RSJ}. The corresponding Langevin equations were
numerically integrated and using the set $\{\tau_i\}$ that
characterizes the contact this was shown to fit the experimental
results without any adjustable parameter, as is illustrated in
Fig.\,\ref{f.supercurrent}. An excellent agreement is found except
for contacts having a well transmitted channel (with transmissions
between 0.95 and 1.0). In this last case it is found that the
switching current is less sensitive to thermal fluctuations than
predicted by the theoretical model. Landau-Zener transitions
between the lower and the upper Andreev states were pointed out as
a possible source for this effect.

Although quite indirectly, the experiments by Goffman \ea provide
a test of the theoretically predicted CPR in atomic contacts. An
interesting aspect of these experiments is that all the relevant
parameters in the problem could be determined independently, which
opens very promising perspectives for future studies.

\subsection{Shot noise in the subgap regime}
\label{ss.shotnoise in subgap}

\begin{figure}[!t]
\centerline{\includegraphics[height=6cm]{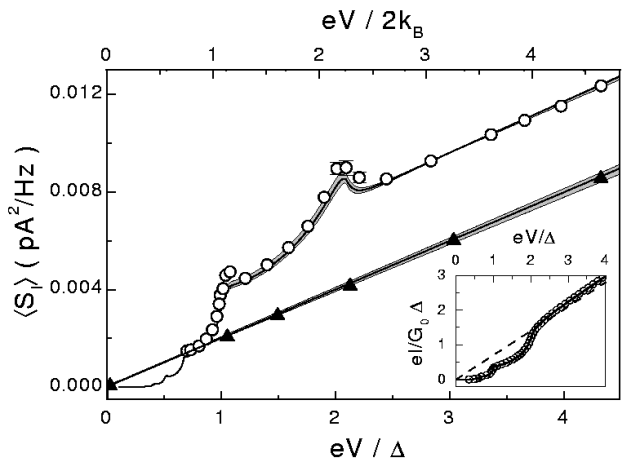}}
\caption{Measured current noise power density versus voltage for a
typical contact both in the normal (triangles) and in the
superconducting state (circles) for a single-atom Al contact at
$T=20$\,mK. The solid lines are predictions using
Eq.\,(\ref{eq.shotnoise}) for the normal state and using the
theory of Ref.\,\cite{cuevas99} for the superconducting state.
Inset: $IV$ in the superconducting state. The solid line is a fit
which provides the values $\left\{\tau_i\right\}=\left\{0.40,
0.27, 0.003\right\}$ for the transmissions. Reprinted with
permission from \protect\cite{cron01}. \copyright 2001 American
Physical Society.} \label{cronfig4}
\end{figure}

As discussed in Sect.\,\ref{ss.SNS2}, the subgap structure in the
$IV$ curve of a SNS junction can be understood in terms of
multiple Andreev reflection processes. A fundamental question,
which is attracting growing attention in recent years, is that
regarding the statistics of the transferred charge associated with
these quantum mechanical processes. Shot noise measurements can
provide a first insight into this problem by the determination of
the effective transferred charge $q$ as the ratio $q=S/2I$ between
the shot noise value $S$ and twice the average current $I$. Taking
into account that at a given subgap voltage $V$ the current is
mainly due to MAR processes of order $n \sim 2\Delta/eV$ in which
a net charge $ne$ is transferred, one can expect that $q$
increases at low bias roughly as $1/V$.

These ideas where first explored in an experiment by Dieleman \ea
in 1997 \cite{dieleman97}. They measured the shot noise in
NbN/MgO/NbN tunnel junctions with small defects in the oxide
barrier acting as `pinholes'. Due to these defects the system
consisted basically of a set of SNS point contacts in parallel.
This interpretation was confirmed by the observation of a finite
subgap current exhibiting the typical structure at $eV =
2\Delta/n$. In spite of the rather large error bars in the noise
determination, it was possible to observe a clear increase of the
effective charge $q = S/2I$ at low bias voltage. Dieleman \ea
developed a qualitative explanation of their experimental data
within the framework of the semiclassical theory of MAR given in
Ref.\,\cite{klapwijk82}.

The increase of the effective charge at low voltages was also
observed by Hoss \ea \cite{hoss00} in diffusive SNS junctions.
They used high transparency Nb/Au/Nb, Al/Au/Al and Al/Cu/Al
junctions prepared by lithographic techniques. Although being
diffusive, the normal region in these junctions was smaller than
the coherence length $L_{\phi}$, which allows to observe the {\it
coherent} MAR regime. On the other hand, the junctions presented a
very small critical current, which permitted to reach the low
voltage regime. The excess noise in these experiments exhibited a
pronounced peak at very low voltages (of the order of a few $\mu
V$) which leads to an effective charge increasing much faster than
$1/V$. It should be pointed out that there is at present no clear
theory for the shot noise in diffusive SNS junctions in the
coherent MAR regime.

As we have emphasized throughout this review, atomic-sized
contacts provide an almost ideal situation where theory and
experiments can meet. Fully quantum mechanical calculations are
available for the low frequency noise in a single channel
superconducting point contact with arbitrary transmission
\cite{cuevas99,naveh99}. On the other hand the channel content of
an actual contact can be extracted using the technique discussed
in Sect. \ref{ss.sgs_exp} allowing for a direct comparison between
theory and experiments without any fitting parameter. This was the
strategy followed in the work by Cron \ea \cite{cron01}, already
discussed in connection with shot noise in the normal state in
Sect. \ref{ss.exp_shotnoise}.

\begin{figure}[!t]
\centerline{\includegraphics[height=6cm]{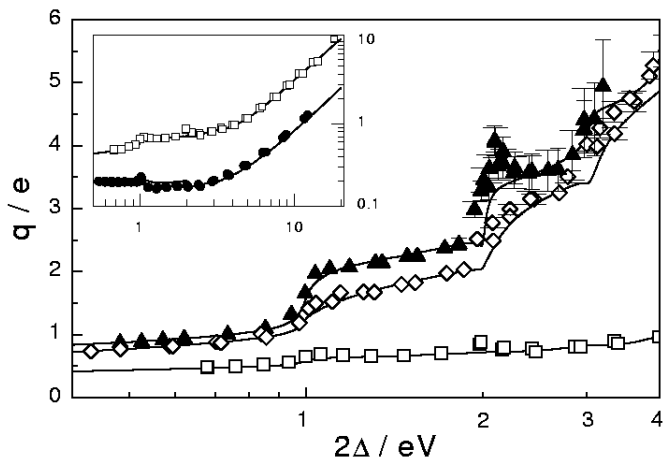}}
\caption{Effective charge $q = S/2I$ versus reduced inverse
voltage for three different single-atom Al contacts in the
superconducting state. The symbols are experimental results and
the solid lines are the predictions of the theory of MAR for
noise. From top to bottom the corresponding set of transmissions
are $\left\{ 0.40, 0.27, 0.03 \right\}$, $\left\{ 0.68, 0.25,
0.22\right\}$ and $\left\{ 0.98, 0.55, 0.24, 0.22\right\}$. Inset:
data for two contacts containing a channel close to perfect
transmission (top $\left\{ 0.98, 0.55, 0.24, 0.22\right\}$, bottom
$\left\{ 0.996, 0.26 \right\}$) shown over a wider range.
Reprinted with permission from \protect\cite{cron01}. \copyright
2001 American Physical Society.} \label{cronfig5}
\end{figure}

In a second step, Cron \ea measured the noise in the
superconducting state. Fig.\,\ref{cronfig4} shows the comparison
between the experimental results and the theoretical predictions
of Refs. \cite{cuevas99,naveh99}, using the measured PIN code
$\{\tau_i\}$ as the input parameters. As can be observed, the
agreement between theory and experiment is quantitative. The
structure in the noise as a function of voltage has the same
physical origin as the subgap structure in the $IV$, i.e. it is
due to multiple Andreev reflection processes having a threshold at
$eV = 2 \Delta/n$. This is better visualized by analyzing the
effective transmitted charge $q = S/2I$ as a function of the
inverse voltage, as shown in Fig.\,\ref{cronfig5}.  As can be
seen, $q/e$ does not necessarily correspond to integer values and
for a given voltage it strongly depends on the set of
transmissions. Only in the tunnel limit, $\tau_i \rightarrow 0$,
the theory \cite{cuevas99} predicts $q/e \rightarrow \rm{Int}
\left[1 + 2e\Delta/V \right]$. Although this limit cannot be
reached experimentally, the emergence of a staircase pattern in
$q$ for decreasing values of the transmissions can be clearly
recognized in Fig.\,\ref{cronfig5}. This work thus provides strong
support for the quantum theory of electronic transport in
superconducting point contacts developed in recent years.

\section{Formation of a conducting wire of single atoms} \label{s.chains}

In the October 22 issue of Nature 1998 two independent groups
demonstrated that chains of atoms self-assemble when thinning the
contact diameter for gold nanocontacts \cite{yanson98,ohnishi98}.
The atomic wires have a conductance very close to the quantum unit
\g0, sustain very large currents, and can be held stable for very
long times at low temperatures. The formation of these atomic
structures was unexpected, and many new properties are predicted.

One-dimensional conductors of different kinds have been
investigated intensively in recent years. Foremost are the carbon
nanotubes \cite{dekker99}, which have a structure that can be
viewed as a rolled-up graphite sheet. Other types of molecular
conducting wires have been investigated, including
Mo$_{6}$Se$_{6}$ \cite{venkataraman99}. Even electrical conduction
across more complicated molecular wires, notably strands of DNA
molecules, has been reported \cite{fink99,porath00}. The molecular
structure of all these systems provides stability at room
temperature and above, and a rich spectrum of physical properties
has been investigated, in particular for the carbon nanotubes.

Ultimately-thin wires of individual carbon atoms have been
prepared by chemical methods. The fabricated substances contain
carbon chains up to 20 atoms in length (see \cite{roth96} and
references therein). Similarly, inorganic chemistry has allowed
the preparation of a compound containing regular arrays of silver
metallic wires \cite{hong01}. As yet another example, chains of
metal atoms have been found to self-assemble when adsorbed at the
surface of other metals or semiconductors, in many cases at step
edges \cite{segovia99,pampuch00,koh01}. For these metallic wire
systems it has not yet been possible to contact individual wires,
to our knowledge.

Although each of these one-dimensional conductors is of great
interest, the metallic wires discussed here have a number of
aspects that make them particularly attractive. First, they are
freely suspended so that there is no complicating interaction with
a substrate, which facilitates comparison with theory and enhances
the one-dimensional character. Second, by their nature they are
already connected to metallic leads, allowing straightforward
measurement of the electrical transport properties of an
individual atomic chain.

Among the experiments preceding the two 1998 papers the experiment
by Yazdani \ea \cite{yazdani96} comes closest to this ideal. These
authors used STM-manipulation techniques to fabricate a two-atom
chain of Xe atoms between the tip of an STM and a metallic
substrate. Although the results showed a favorable agreement with
calculated conductance characteristics, the electronic structure
of Xe leads to a rather poor transmission, i.e. a conductance
several orders of magnitude below the conductance quantum, and the
method is not easily extended to longer wires or other materials.

\subsection{Atomic chains in Transmission Electron Microscopy}

\begin{figure}[!t]
\centerline{\includegraphics[width= 8cm]{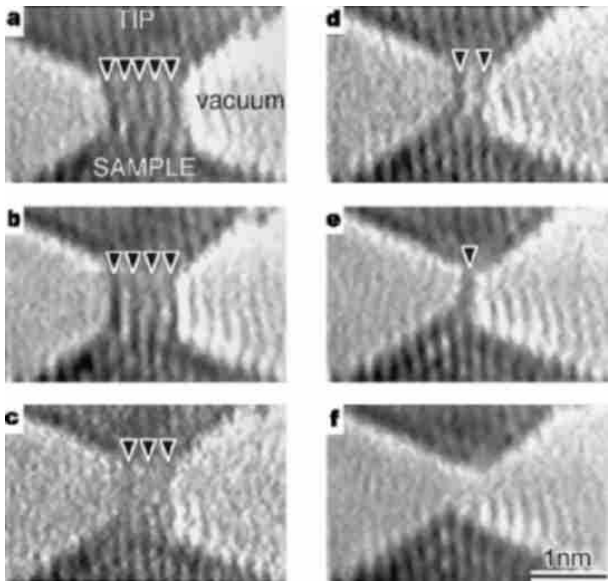}}
\caption{TEM images of a gold contact recorded while withdrawing
the tip from the sample. A gold bridge formed between the gold tip
(top) and the sample (bottom) thins down when going from (a) to
(e), where the conductance drops to ~2\g0. The contact finally
ruptures at (f), for which the conductance drops to zero. The
recording takes 33\,ms per frame and the images are taken at times
0, 0.47, 1.23, 1.33, 1.80 and 2.17\,s, respectively. Reprinted
with permission from Nature \protect\cite{ohnishi98}. \copyright
1998 Macmillan Publishers Ltd. \label{f.TEM-break sequence} }
\end{figure}

By High Resolution Transmission Electron Microscopy (HR-TEM)
imaging it is possible to resolve individual atoms for the heavier
elements. Ohnishi, Kondo and Takayanagi \cite{ohnishi98} exploited
this capability by combining their ultra-high vacuum HR-TEM setup
with two different manipulation techniques to produce atomic
wires. First, they constructed a miniature STM that fits into the
specimen space of the TEM. It is fascinating to see the
atomically-resolved video images they show of a tip scanning a
sample surface, and subsequently indenting it. When retracting a
gold tip from a gold sample the team observed that the connecting
bridge gradually thins down, see Fig.~\ref{f.TEM-break sequence}.
All experiments are performed at room temperature, giving the
atoms enough mobility to optimize the configuration, and as a
result it is seen that the bridge connecting the two electrodes,
oriented along the [110] direction, often consists of a straight
wire section. As the number of atomic rows in the connecting
nanowire decreases the conductance is also seen to decrease in a
step-wise fashion, as expected. The conductance of a one atom
strand in the images is close to $2e^2/h$. However, twice this
value is also found, and it is argued that this is due to a double
strand overlapping in the viewing direction. Evidence for this
interpretation is obtained by analyzing the contrast profile in
the images.

\begin{figure}[!t]
\centerline{\includegraphics[width=6cm]{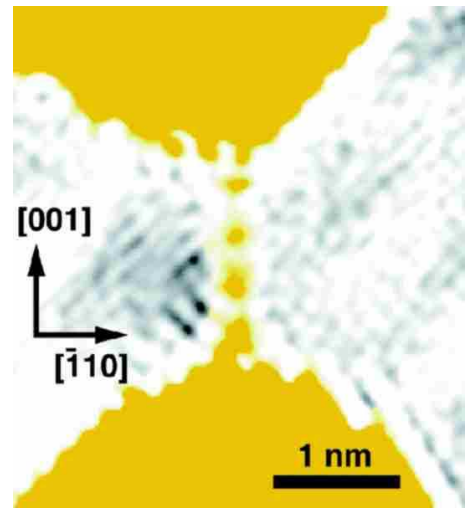}}
\caption{TEM image of a gold atomic chain (four gray dots) forming
a bridge between two gold banks (gray areas above and below). The
structure to the left and right of the chain results from electron
diffraction and interference in the TEM. Reprinted with permission
from Nature \protect\cite{ohnishi98}. \copyright 1998 Macmillan
Publishers Ltd. \label{f.TEM-atomic chain}}
\end{figure}

In order to resolve the individual atoms in the chain a second
technique was employed. Here, the STM was replaced by a very thin
gold film specimen, where an intense electron beam current was
used to melt two adjacent holes in this film. For (110) oriented
films a gold bridge along the [001] direction between these two
holes was seen to evolve into an atomic chain that survived for
about two minutes, see Fig.~\ref{f.TEM-atomic chain}. Note that in
this configuration the conductance of the chain cannot be
measured. Remarkably, the spacing between the atoms in the chain
was found to be 0.35--0.40\,nm, much larger than the nearest
neighbor distance in bulk gold (0.288\,nm). This is much larger
than any model calculation predicts, since the overlap between the
electron clouds of the gold atoms is too small to provide
sufficient stability for the atomic chain. Several explanations
have been put forward. One proposal is based on the observation
that the calculated equilibrium structure for a Au monatomic chain
appears to have a zigzag geometry, as will be discussed in
Sect.~\ref{ss.theory-chains} below.  S{\'a}nchez-Portal \ea
\cite{sanchez99} proposed that every second atom in the zigzag
chain could be thermally excited into a spinning motion around the
chain axis, which would blur their image. Koizumi \ea
\cite{koizumi01} show by comparison of the experimental images to
simulations that a spinning zigzag geometry can be excluded based
on the expected smeared image of the spinning atoms. Other
explanations involve the inclusion of `glue atoms', such as C, O,
or S \cite{hakkinen99,bahn02,bahn01b,legoas02}. The simulations
\cite{koizumi01} suggest that adatoms of Si and S would be
resolved, but the contrast for C (and presumably O) would not
exceed the noise level. Only minute amounts of contaminants are
required, since the regular gold surface is not very reactive,
while the low-coordination gold atoms in the chain bind strongly
to different species, as shown by first-principles calculations
\cite{hakkinen99,bahn02,bahn01b}. Despite the high vacuum
conditions of the experiment, there will be small amounts of
adsorbed molecules running over the surface, and these will stick
preferentially at the strong binding sites in the gold chain.
Oxygen would be a good candidate, since it would not be resolved
in the images and the calculations suggest that a Au-O-Au-O chain
would have Au-Au distance close to the observed values and the
chain would be conducting, with a single open channel
\cite{bahn02,bahn01b}. It would be interesting to test this
suggestion experimentally.

The experimental observation of the atomic chain formation and the
long inter-atomic distances have been confirmed in an independent
experiment by Rodrigues and Ugarte \cite{rodrigues01}, using the
thin-film double-hole technique at $10^3$ times higher residual
gas pressure. Similarly, these authors claim that the spinning
zigzag structure can be excluded based on the absence of ghost
features. Short chains have also been seen for silver
\cite{rodrigues02}, although much less frequent than for gold, and
the large and irregular bond lengths observed suggest the presence
of light interstitial atoms that may stabilize the chains
\cite{legoas02}. In high resolution images taken with a new
generation defocus-imaging modulation processing electron
microscope by Takai \ea \cite{takai01} much smaller Au-Au
distances of 0.25-0.29\,nm were found. The vacuum pressure was
comparable to that of \cite{rodrigues01}.

A further study by Kizuka \ea \cite{kizuka01} appears to be at
variance with most of the earlier results. Here, a miniature STM
is operated inside a HR-TEM at regular-vacuum conditions. Again,
for gold it is frequently observed that upon separation of the
contact between tip and sample it ends with the formation of a
chain of atoms. Similar to the results of Takai \ea\ , a distance
of only $0.27\pm 0.02$\,nm was obtained between the atoms in the
chain, which can be stretched at the break point to $0.30\pm
0.02$\,nm. Surprisingly, the atomic wires were found to be
insulating. At the moment when the structure is seen to jump from
a multi-atom cross section to a single-atom chain the conductance
drops to zero. Moreover, the chains were found to be bent even
under stretched conditions. Very long atomic chains, up to 10
atoms in a row, were observed, that were stable for longer times
than reported before. Although the authors make a few suggestions
to explain these observations, the discrepancy with the other
experiments was not addressed.

We propose that these observations can be understood if we assume
the presence of specific adsorbates. As shown by Bahn \ea
\cite{bahn02,bahn01b}, CO binds strongly to the gold chain, turns
it into an insulator, introduces kinks in the wire, and the CO
bonded gold chain has the lowest energy among all the structures
investigated. This would suggest that CO, or another contaminant
of similar nature, is present in the vacuum space. At a typical
pressure of $10^{-5}$\,Pa the probability of this mechanism is
high.

\subsection{Atomic chains in low-temperature experiments}

The second paper in the 1998 issue of Nature used different
techniques in several important aspects \cite{yanson98}. The
atomic structure was not imaged directly, but the formation of
chains was deduced from the experimental observations of the
conductance as a function of stretching. The advantages, on the
other hand, are the low temperature (4.2\,K) at which the
experiment is performed. This allows for a long-term stability of
the gold atomic chains so that detailed spectroscopy can be done.
In addition, the cryogenic vacuum conditions avoid any
contamination on the nanowires.

\begin{figure}[!t]
\centerline{\includegraphics[width=7cm]{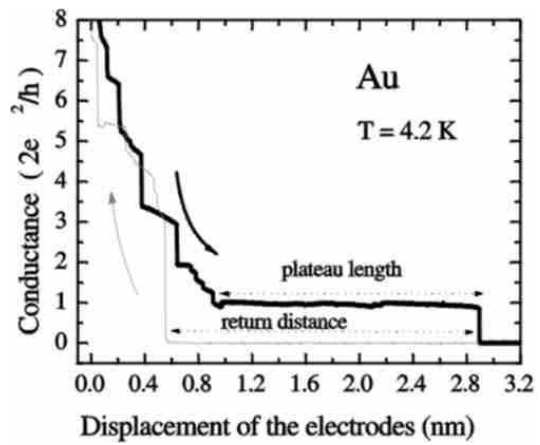} }
\caption{The conductance as a function of the displacement of the
two gold electrodes with respect to each other in an MCBJ
experiment at 4.2\,K. The trace starts at the upper left, coming
from higher conductance values (thick curve). A long plateau with
a conductance near 1\,G$_{0}$ is observed and after a jump to
tunneling one needs to return by a little more than the length of
the long plateau to come back into contact (thin curve). Data
taken from \protect\cite{yanson98}.\label{f.long-plateau-Au}}
\end{figure}

By standard low-temperature STM  and MCBJ techniques atomic-sized
contacts of gold were produced. In contrast to many other metals,
for Au it was found that the last conductance plateau, at a value
of $\sim 1$\,\g0, can often be stretched far beyond a length
corresponding to an atomic diameter. An example is presented in
Fig.~\ref{f.long-plateau-Au}, where a plateau of about 2\,nm
length is found. Since it has been established that the
conductance is predominantly determined by the narrowest cross
section, and that a single-atom contact for Au has a conductance
near 1\,\g0 (Sect.~\ref{s.exp_modes}), this observation led Yanson
\ea to speculate that a chain of atoms was being formed. This is
indeed very surprising, even more so than in the case of the room
temperature TEM experiments. For the latter, the atoms have enough
mobility to produce at an earlier stage a stable, straight
nanowire several atoms in cross section, and the atomic rows in
the wire are removed one after the other by thermal diffusion of
the atoms on the surface. This leaves a single atomic row standing
before contact is finally lost. However, at low temperatures the
atomic structure is frozen into the configuration in which it
lands after an atomic rearrangement, forced by the stretching of
the contact. When arriving at a single-atom contact one would
expect the contact to break at this weakest spot. Instead, atoms
are apparently being pulled out of the banks to join in the
formation of a linear atomic arrangement. Clearly, it is important
to critically evaluate the interpretation of atomic chain
formation. By now, a large set of experiments has been performed
which confirm the picture, and we will now summarize this
evidence.

\subsubsection{Return distance}
A simple test involves recording the distance required to bring
the electrodes back into contact after the conductance has
suddenly dropped to zero, as at the end of the plateau in
Fig.~\ref{f.long-plateau-Au}. We imagine that a chain has formed,
which finally ruptures at this moment. The atoms in the chain are
then expected to fall back onto the banks, which implies that the
separation between the electrodes should be approximately equal to
the length of the chain, being approximately the length of the
plateau. Fig.~\ref{f.long-plateau-Au} illustrates that this is
indeed the case for this particular example, although one may
anticipate variations in the return length according to the actual
arrangement of the atoms after the collapse. By recording many
curves similar to the one in Fig.~\ref{f.long-plateau-Au}, Yanson
\ea obtained an average return distance as a function of the
length of the last plateau. They observed a linear dependence of
the return distance on the plateau length, with a slope between
1.0 and 1.3 and an offset of about 0.5\,nm. The latter can be
understood in terms of the elastic deformation of the banks: Even
when no chain is formed and the contact breaks at a single-atom,
the atomic structure relaxes after rupture of the contact, giving
rise to a finite return length.

\subsubsection{Length histograms}

\begin{figure}[!t]
\centerline{\includegraphics[width=7cm]{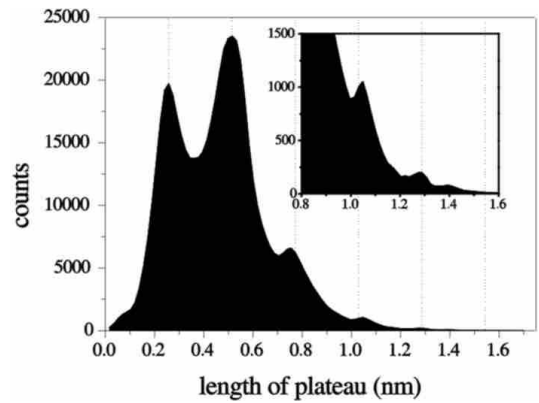} }
\caption{The distribution of lengths for the last conductance
plateau for Au, obtained from 10\,000 experiments similar to those
described in Fig.~\ref{f.long-plateau-Au}. It shows a number of
equidistant maxima, at multiples of 0.257\,nm. The data were
recorded with an MCBJ at 4.2\,K, in crygenic vacuum. The length of
the last plateau was defined as the distance between the points at
which the conductance drops below 1.2\,$G_{0}$ and 0.8\,$G_{0}$,
respectively. The inset shows the tail of the distribution on a
$\sim10\times $ expanded scale. A smoothing function that averages
over three bins has been applied to the data. The accuracy for the
calibration of the length (horizontal scale) is 10\%. Data taken
from \protect\cite{untiedt02}.\label{f.length-histo-Au}}
\end{figure}

Further evidence for the chain structure comes from an analysis of
the distribution of lengths of the last conductance plateaus for
many cycles of contact breaking. Fig.~\ref{f.length-histo-Au}
shows a histogram of plateau lengths. We see that the probability
for early breaking is very low, it then rises to a first peak at
0.25\,nm length, after which it drops before rising to a second
peak, which is usually higher than the first. After the second
peak the distribution of lengths drops steeply, but shows three
additional peaks in the tail. The peak distance of 0.25--0.26\,nm
agrees with the expected bond distance in a chain of gold atoms
(see Sect.~\ref{ss.theory-chains}) and the natural interpretation
of the peaks in the length histogram is in terms of a preferential
breaking of the chain at lengths corresponding to an integer
number of atoms in the chain. The peaks in the distribution are
broadened by the variation in starting and end configurations of
the banks connecting the chain. In fact, a strict periodicity of
the peaks would not be expected to continue much further than the
first few, because the atoms making up the chain are removed from
the banks, which then become shorter. Occasionally plateaus of up
to 2\,nm in length have been found, which suggests that the system
can self-assemble chains of up to 7--8 atoms long. It is often
possible to obtain similar peak structure in a histogram of return
distances \cite{yanson01}.

In the original paper \cite{yanson98} the distance between the
peaks was reported to be larger, 0.36\,nm($\pm$30\%). The larger
value, and the rather large uncertainty, later turned out to
result from the presence of He thermal exchange gas in the vacuum
space. As was recently shown by Kolesnychenko \ea
\cite{kolesnychenko99a}, adsorbed He gas has an unexpectedly large
influence on the work function of metal surfaces. This introduces
an error in the calibration of the displacement of the MCBJ and
STM, when using the exponential tunneling dependence, as pointed
out in Sect.~\ref{sss.calibration}. More recently, Untiedt \ea
\cite{untiedt02} have combined several calibration techniques to
obtain a more reliable value for the inter-peak distance in the
length histograms, and the value obtained for Au,
$0.26\pm0.02$\,nm, is in excellent agreement with the calculated
Au-Au distance in the chains
\cite{sanchez99,sorensen98,dasilva01,okamoto99,torres99,hakkinen00,demaria00,bahn01,nakamura01}.

\subsubsection{Evolution of the force in atomic chains}

\begin{figure}[!t]
\centerline{\includegraphics[width=8cm]{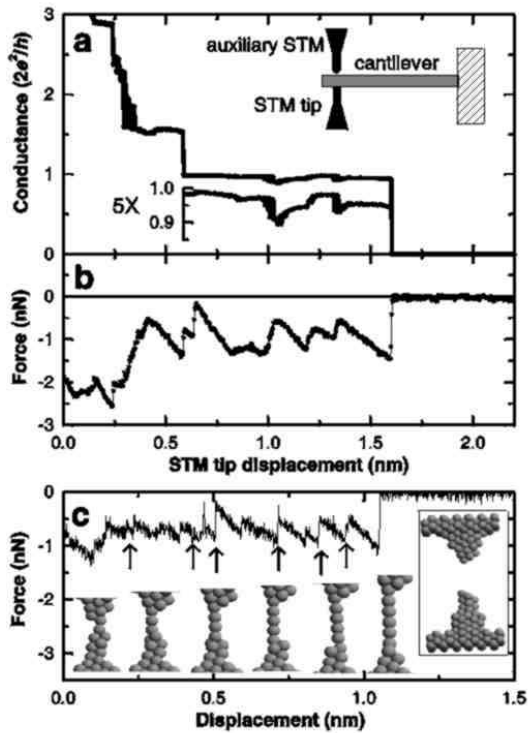} }
\caption{Simultaneous conductance (a) and force (b) measurements
during chain fabrication and breaking. The conductance on the last
plateau is shown on an expanded scale to illustrate small
variations in the conductance. The inset shows a schematic drawing
of the experimental setup. (c) Calculated force evolution obtained
from molecular dynamics simulations. The arrows indicate the
points at which a new atom pops into the chain and snapshots of
the structure at these positions are shown. Reprinted with
permission from \protect\cite{rubio01}. \copyright 2001 American
Physical Society. \label{f.force-Au-chain}}
\end{figure}

Rubio-Bollinger \ea measured the force evolution simultaneously
with the conductance while drawing out a chain of atoms at 4.2\,K
\cite{rubio01}. They employed an auxiliary STM at the back of a
cantilever beam, on which the sample was mounted, in order to
detect the deflection, and therewith the force on the sample
(Sect.~\ref{ss.force techniques}). An example of such a
measurement is shown in Fig.~\ref{f.force-Au-chain}, where the
contact is stretched at a constant speed of 0.5\,nm/s. The force
shows a saw-tooth-like pattern corresponding to elastic
deformation stages interrupted by sudden force relaxations. The
conductance on the last plateau remains nearly constant and just
below 1\,\g0, but note that the force jumps are accompanied with
simultaneous jumps in the conductance with a magnitude of only a
small fraction of \g0.

In each measurement, the largest force on the last conductance
plateau is reached at the end, as expected. For a series of 200
experiments this final breaking force shows a narrowly peaked
distribution, centered at 1.5\,nN, with a standard deviation of
only 0.2\,nN. The force calibration has an accuracy of 20\%. The
break force was found to be independent of the chain length. The
force is considerably larger than the force required to break
individual bonds in bulk gold, which is estimated at only
0.8--0.9\,nN, and this large force agrees very well with theory,
as will be discussed below.

It was at first sight surprising to find that the slopes of the
force as a function of displacement are nearly constant in
experiment. One would expect a smaller force constant for longer
chains. This observation is explained by the fact that the chain
is unusually stiff. The bonds are much stronger than bulk bonds,
and the largest elastic deformation takes place in the banks next
to the chain. The calculated deformation of the banks amounts to
0.5--1.0\,nm, which agrees well with the offset observed in the
return distance.

\subsubsection{Phonon modes in atomic chains}

\begin{figure*}
\centerline{\includegraphics[width=12cm]{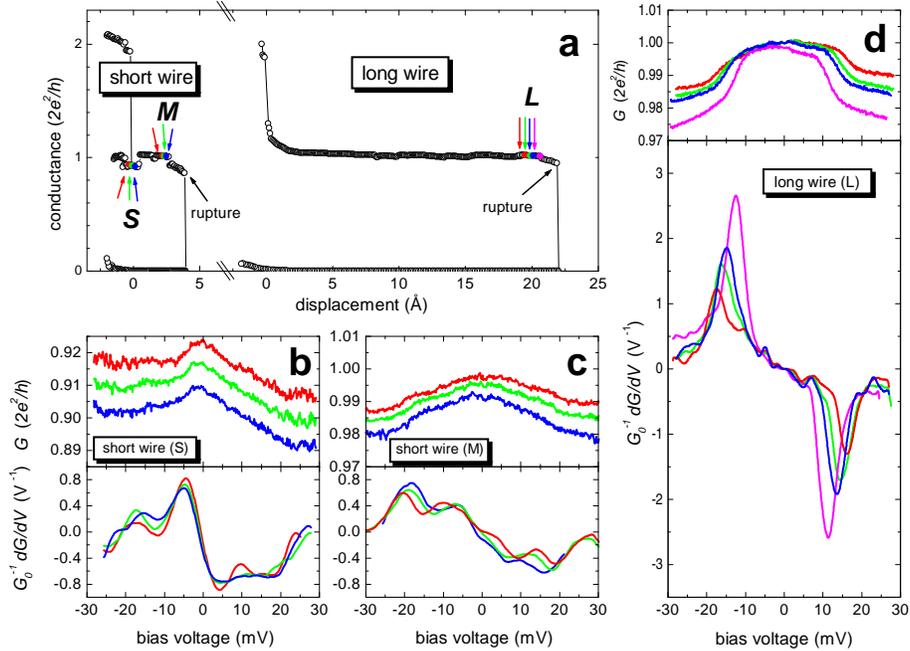} }
\caption{(a) Short and long atomic wire, $\sim0.4$ and
$\sim2.2$\,nm, respectively, as given by the length of the
conductance plateau. Panels (b--d) show the differential
conductance and its derivative at points S, M, and L,
respectively. The various curves in (b--d) were acquired at
intervals of 0.03, 0.03 and 0.05\,nm, respectively. Note that the
vertical scales for the last thee panels are chosen to be
identical, which brings out the relative strength of the
electron-phonon interaction for the longer chains. The  wire in
(d) has a length of about 7 atoms. Reprinted with permission from
\protect\cite{agrait02a}. \copyright 2002 American Physical
Society. \label{f.phonons-Au-chain}}
\end{figure*}

The phonon modes in Au atomic chains have been investigated
experimentally \cite{agrait02,agrait02a} by means of point contact
spectroscopy (see Sects.~\ref{ss.spear-anvil} and
\ref{ss.inelastic}). The differential conductance $dI/dV$ was
measured using lock-in detection with a 1\,mV modulation voltage,
from which $dG/dV$ was calculated numerically. The energy
resolution was limited by the temperature of 4.2\,K to 2\,meV. The
results are shown in Fig.~\ref{f.phonons-Au-chain}. At $\pm$15\,mV
bias the differential conductance of a chain of atoms shows a
rather sharp drop by about 1\% (top panel in
Fig.\,\ref{f.phonons-Au-chain}). In the second derivative
$d^2I/dV^2$ this produces a pronounced single peak,
point-symmetric about zero bias. The chains of Au atoms have the
fortuitous property of having a single nearly perfectly
transmitted conductance mode, which suppresses conductance
fluctuations that would otherwise drown the phonon signal, see
Sect.~\ref{ss.inelastic}. Some asymmetry that can still be seen in
the conductance curves is attributed to the residual elastic
scattering and interference contributions.

Bulk gold point contact spectra have peaks at 10 and 18\,mV
(Fig.~\ref{f.PCS}), due to transversal and longitudinal acoustic
phonon branches, respectively. The first of these has a higher
intensity. At point S in Fig.~\ref{f.phonons-Au-chain} the contact
consists of a single atom and the spectrum still resembles the
spectrum for bulk contacts, although the energy of the first peak
is significantly shifted downward due to the reduced coordination
of the atom.

For the linear chain configuration the electron-phonon interaction
simplifies considerably. By energy and momentum conservation the
signal only arises from electrons that are back scattered,
changing their momentum by $2k_{\rm F}$. With $\hbar
\omega_{2k_{\rm F}}$ the energy for the corresponding phonon, the
derivative of the conductance is expected to show a single peak at
$eV=\pm \hbar \omega_{2k_{\rm F}}$. The transverse phonon mode
cannot be excited in this one-dimensional configuration and only
the longitudinal mode is visible.

The position of the peak shifts as a function of the strain in the
wire. It is somewhat like the pitch of a guitar string that
changes as a function of the tension. Except that for atomic wires
the frequency decreases as a function of tension because of the
decreasing bond strength between the atoms. The frequency
decreases, and the amplitude increases, until an atomic
rearrangement takes place, signaled by a small jump in the
conductance. At such points the amplitude and energy of the peak
in $dG/dV$ jump back to smaller and larger values, respectively.
This is consistent with the phonon behavior of Au atomic chains
found in {\it ab initio} calculations \cite{sanchez99}. The
growing amplitude is due, in part, to the softening of the phonon
modes with tension. The 1\% drop in conductance for a wire of
2\,nm length implies a mean free path for the electrons of about
200\,nm, much longer than the nanowire itself.

\subsection{Other properties of atomic chains at low temperatures}

The most striking properties are the long-time stability of the
chains and the large current densities that the chains can
survive. Even the longest ones can be held stable for times of at
least an hour (no real record-breaking attempts have been made so
far). Bias voltages as high as 2\,V have been applied without
damage, although the maximum decreases with the length of the
chain \cite{untiedt00,nielsen02}. With a conductance equal to
$2e^2/h$ this corresponds to 150\,$\mu$A, and with the cross
section equal to one atom we obtain a current density of $2.1\cdot
10^{15}$\,A/m$^2$. This current density is more than seven orders
of magnitude larger than the current density that turns the
tungsten wire inside a light bulb white-hot. This is an amazing
fact. Although we understand that it is related to the ballistic
nature of the transport, a significant amount of inelastic
scattering inside and near the wires should be expected at such
high biases. More quantitative studies of this phenomenon are
needed. Preliminary studies suggest that the limiting current
values decrease roughly proportional to the inverse of the length
of the chains, $\sim 1/L$ \cite{untiedtTBP}. Todorov \ea
\cite{todorov01} have calculated the current-induced forces in
atomic chains, and find that the forces on the atoms weaken the
atomic structure, in particular at the junction with the banks.
This was confirmed by {\it ab initio} calculations by Brandbyge
\ea \cite{brandbyge02}. Nevertheless, it is most likely it is the
heat generated by the current that causes the wire to break
\cite{todorov01}. A phenomenon related to the high current
carrying capacity is the fact that the current-voltage
characteristics for the chains is highly linear. This is confirmed
by tight-binding calculations under application of a large voltage
bias \cite{brandbyge99}. More recently, non-equilibrium DFT
calculations \cite{brandbyge02} predict a significantly reduced
conductance above 0.5\,V for a 3-atom Au chain, but this is
probably a specific feature for the (111)-oriented substrate.

When pulling a contact in forming a chain the conductance on a
plateau, once it drops below 1\,\g0 never rises above it. The
conductance stays almost invariably very close to 1\,\g0. This
implies that the conductance of any additional channels is
negligible and that a nearly perfect adiabatic coupling with the
banks is attained. This is surprising, since the atomic structure
at the coupling with the banks is not controlled and should vary
quite a bit, but it agrees with results of calculations
\cite{kobayashi00,sim01,okamoto99,hakkinen00,emberly99}. Sim \ea
\cite{sim01} argue that a conductance of exactly 1\,\g0 is a
robust property for a contact having a single conductance channel,
when imposing charge neutrality and inversion symmetry.

Still, deviations of the conductance below 1\,\g0 are observed. In
particular jumps are seen due to atomic rearrangements when the
chain is being built up, see e.g. Fig.~\ref{f.force-Au-chain}.
This implies that there is a finite amount of scattering at the
interface between chain and banks, or inside the banks close to
the interface. Such small jumps are also seen when the chain is
swung sideways in an STM experiment \cite{yanson98}, in which case
these can be due to jumps in the anchoring point of the chain on
the banks. The latter experiment also illustrates the robustness
of the chains, and the fact that the freedom for lateral
displacement is over a distance comparable to the length of the
chains forms an additional element in verifying the
chain-interpretation of the experiments.

\subsection{Numerical calculations of the stability and conductance of Au chains}\label{ss.theory-chains}

Several molecular dynamics simulations have preceded these
experiments in suggesting the formation of chains
\cite{brandbyge97,sorensen98,finbow97,SuttonUNPUB}. However, the
effective potentials employed in these simulations were not
regarded to be sufficiently reliable for such exceptional atomic
configurations to be predictive. Nevertheless, in many cases the
simulations agree with the observations, and they are very helpful
in visualizing the mechanism by which the chains unfold from the
banks. Full DFT molecular dynamics modeling of this process is
still too demanding. More recently da Silva \ea \cite{dasilva01}
have used a method that forms a compromise between accuracy and
computational efficiency, involving tight-binding molecular
dynamics. The results for gold are generally consistent with the
previously employed methods. First-principles DFT molecular
dynamics was used by H\"akkinen \ea \cite{hakkinen00} by limiting
the number of possible configurations. Taking a starting
configuration of two atomic gold tips connected by two parallel
two-atom long strands, the stretching of this double-chain was
seen to evolve into a four-atom long single-atom chain via a bent
chain structure.

In order to investigate the equilibrium structure, bond length and
breaking force many model systems of moderate size have been
considered, using first principles calculations based on the local
density approximation. The structures considered are infinite
chains, using periodic boundary conditions
\cite{demaria00,sanchez99,torres99,bahn01},  finite isolated chain
sections \cite{demaria00,sanchez99}, or finite wires connected to
an atomic base on either side \cite{okamoto99,sanchez99,rubio01}.
Sanchez-Portal \ea \cite{sanchez99} have investigated all
structures by various computational approximations and find only
minor quantitative and no qualitative differences. All
calculations agree on the equilibrium bond length, ranging only
between 0.250 and 0.262\,nm, and agree on the maximum bond
distance at which the chain breaks, 0.28 -- 0.30\,nm. The break
force is more sensitive to the type of approximations involved,
ranging from 0.91\,nN in Ref.~\cite{okamoto99} to 2.2\,nN in
Ref.~\cite{sanchez99}. Rubio-Bollinger \ea \cite{rubio01} made the
most extensive analysis of the breaking force and obtain a force
between 1.55 and 1.68\,nN, in good agreement with the experimental
value of $1.5\pm0.3$\,nN.

Sanchez-Portal \ea \cite{sanchez99} found for the optimized
geometry a planar zigzag structure with two atoms per unit cell.
The zigzag deformation was even found for free standing wire
sections and the origin was argued to be related to a reduction in
the transverse kinetic energy for the electrons due to the
increased effective wire width. This mechanism is of the same
nature as the shell structure observed for alkali metal wires, as
discussed in Sect.~\ref{s.shells}. The chain is stretched to a
linear configuration only for bond lengths above about 0.275\,nm,
shortly before breaking. The zigzag structure is confirmed in the
work of Refs.~\cite{demaria00,bahn02}. On the other hand,
H\"akkinen \ea \cite{hakkinen00} find for a four atom chain
between two tips that, before the chain is fully stretched, it
assumes a bent configuration, that appears to be lower in energy
than the zigzag conformation.

For a linear chain with a single half-filled conduction band a
Peierls distortion towards a string of dimers is generally
expected to occur. The majority of the calculations suggest that
this dimerization only sets in just before stretching to the point
of breaking. The variation in bond lengths observed for a
four-atom chain by H\"akkinen \ea \cite{hakkinen99} that was
argued to be related to a Peierls distortion may also be due to
end-effects \cite{okamoto99}. De Maria and Springborg
\cite{demaria00} provide fairly general arguments that the
half-filled band for this chain system should lead to dimerization
when the bond length becomes larger than 0.29\,nm. Below 0.27\,nm
a second band was found to cross the Fermi level and the $\sigma$
orbital becomes partially depleted. Since the $\sigma$ orbital is
no longer half filled the driving mechanism to dimerization is
suppressed. A second band crossing the Fermi level for short
distances was also found for linear chains by Sanchez-Portal.
However, in their calculations it is removed by the zigzag
deformation. The presence of a second conduction band should be
visible in the conductance. Calculations of the conductance by
other groups \cite{brandbyge99,okamoto99,hakkinen00} consistently
find a conductance equal to 1\g0 or slightly below, in agreement
with the experiments.

The atomic chain configuration is clearly a meta-stable structure.
Bahn \cite{bahn01b} calculated the time to break a chain for
various temperatures by the EMT molecular dynamics method. He
found that the chains would be unstable on a time scale of
nanoseconds at room temperature. The barrier to breaking is only
about 0.03\,eV, with an attempt frequency of $5\cdot
10^{11}$\,s$^{-1}$. Only higher temperature break times could be
obtained in the time span accessible by these calculations. The
mean time to breaking at 200\,K is found to be $\sim 0.1$\,ns,
while extrapolation of the numbers obtained gives a lifetime of
hours or even days at 4.2\,K. The lack of predicted long-time
stability poses a second challenge to understanding of the
room-temperature TEM results.

Again, the presence of other chemical species, such as CO and O,
may resolve the problem, since they would provide stronger bonding
\cite{bahn02,bahn01b}. A chain with oxygen atoms inserted between
the Au atoms would, surprisingly, still be conducting, with a
single conductance channel. A similar result was obtained by
H\"akkinen \ea \cite{hakkinen99} for the insertion of methylthiol,
SCH$_3$, into a gold chain.

\subsection{The mechanism behind atomic chain formation: Ir, Pt and Au}

All the discussion above has been limited to gold chains.
Surprisingly, Au appears to be favorable for chain formation while
Ag an Cu do not (or to a very limited extent) have this property.
This led Smit \ea \cite{smit01} and Bahn and Jacobsen
\cite{bahn01} to investigate the mechanism behind this phenomenon.
It turns out that gold is distinct from the other noble metals in
another surface property: Clean gold surfaces reconstruct in
remarkable ways (see \cite{bartolini88} and references therein).
The (110) surface shows a `missing row' reconstruction, where
every second row along the [001] direction is removed. The (001)
surface has a hexagonally packed top layer, that is more densely
packed than the bulk. Even the (111) surface has a herringbone
reconstruction that is slightly more densely packed than the bulk.
It turns out that the three end-of-series 5$d$ elements Ir, Pt,
and Au have similar surface reconstructions, which are absent in
the related 4$d$ elements Rh, Pd and Ag, suggesting that the
explanation for the reconstructions cannot lie in any particular
detail of $d$ band electronic structure. There appears to be a
growing consensus that an explanation can be found in a stronger
bonding of low-coordination atoms of the 5{\it d} metals with
respect to the 4{\it d} metals as a result of relativistic effects
in the electronic structure
\cite{ho87,takeuchi89,takeuchi91,filippetti97}. From numerical
work that uses relativistic local-density-functional calculations
to evaluate the various contributions to the atomic binding
energies qualitatively the following picture emerges. The
effective Bohr radius for 1$s$ electrons of the heavier (5$^{th}$
row) elements contracts due to the relativistic mass increase.
Higher $s$ shells also undergo a contraction, both because they
have to be orthogonal against the lower ones and because they feel
the same mass-velocity terms directly. As Takeuchi {\it et al.}
explain \cite{takeuchi89} the contraction of the  $s$ shell allows
those electrons to profit more from the Coulomb interaction with
the positively charged core and reduces the energy of the $s$
shell, increasing the $s$ occupation at the expense of the $d$
electrons. Since the top of the $d$ band consists of states with
anti-bonding character that are now partially depleted, the $d$
bond becomes stronger. While the $d$ electrons thus tend to
compress the lattice the $s$ electrons exert an opposing Fermi
pressure. At the surface, the spill-out of the $s$ electron cloud
into the vacuum relieves some of the $s$ electron pressure, and
allows a contraction of the inter-atomic distance and a
strengthening of the bonds at the surfaces, giving rise to the
observed reconstructions. Since the relativistic corrections grow
roughly as the square of the nuclear charge \cite{pyykko88},
$Z^2$, they are more important for the 5$d$ elements than for the
equivalent 4$d$ and 3$d$ transition metals.

As for the reconstructions, the bonding in atomic chains will be
influenced by a tilting of the balance between $s$ and $d$
electrons by relativistic effects. This difference in bonding
becomes revealed when the Fermi pressure of the {\it s} electrons
can be released by spill-out of the wave functions into the
vacuum. As a result, there is a gain in energy from the stronger
{\it d} bonds and a reduction of the inter-atomic distance.
Clearly, the 1D chain geometry allows for an even larger {\it s}
pressure release than at a flat surface. This provides an
explanation for the difference observed between the tendency to
chain formation for Au, Ag and Cu.

De Maria and Springborg \cite{demaria00} argue that including
relativistic effects in numerical calculations for the chains is
important. DFT calculations commonly include the scalar
relativistic effects in the choice of the pseudo-potentials and in
the molecular dynamics simulations they are effectively included
by adjustment of the parameters to known properties of the
materials. Bahn and Jacobsen \cite{bahn01} investigated by these
numerical methods the mechanism and energy balances for chain
formation in a series of transition metals: Ni, Pd, Pt, Cu, Ag and
Au. By performing a large number of molecular dynamics simulations
of contact breaking, using the EMT method (see
Sect.~\ref{s.models}), they found that among this selection of
elements only Au and Pt spontaneously form atomic chains. The
break of other metals occurred at best with two atomic tips
touching, which can then be viewed as a chain of two atoms.

\begin{figure}[!t]
\centerline{\includegraphics[width=6cm]{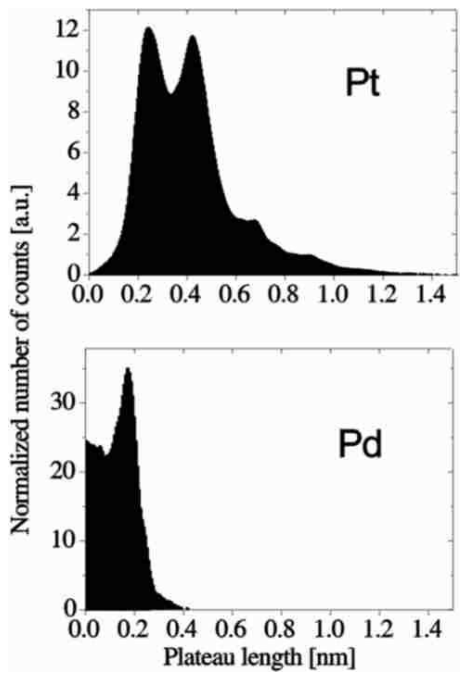} }
\caption{Histogram of the distribution of lengths for the last
conductance plateau for platinum contacts (top panel) and for its
iso-electronic neighbor palladium (bottom panel). Pd breaks
essentially at the level of a single atom, while Pt spontaneously
forms chains of atoms, which is apparent from the tail of the
distribution towards long plateau lengths, and from the peaks at
multiples of the inter-atomic distance. Data taken from
\protect\cite{smit01}.} \label{f.PtandPd-lhisto}
\end{figure}

Smit \ea\ \cite{smit01} used the MCBJ technique for a similar set
of transition metals, namely the 4$d$ metals Rh, Pd, and Ag, and
the 5$d$ series Ir, Pt, and Au, to demonstrate experimentally that
all of the latter form chains of atoms, in contrast to the
corresponding 4$d$ elements. In order to show this the length of
the last conductance plateau was measured in a length histogram
similar to the one shown for Au in Fig.~\ref{f.length-histo-Au}.
Since only Au and Ag have a sharply defined conductance for a
single-atom contact or wire of nearly 1\,\g0, there is an
additional difficulty for the other metals. They are expected to
have a partially filled $d$ shell giving rise to five conduction
channels with transmission values that are smaller than unity and
depend sensitively on the coupling to the banks, see
Sect.~\ref{s.exp_modes}. The procedure followed was as follows:
first a conductance histogram was recorded for each of the metals.
As for the niobium histogram shown in Fig.~\ref{f.histoNb} they
all show a single rather broad peak, with little other structure
in most cases. The peak is taken to represent the typical
conductance of a single-atom contact. Taking the high-conductance
tail of the peak as the starting-point for a chain, and the
low-conductance tail as the end value, a large series of
conductance traces was analyzed. The length of the `plateaus'
between the start and stop values for the given metal were
accumulated into a length-histogram. In addition to Au, also Pt
and Ir showed a plateau-length distribution with a tail toward
long lengths, up to 1.5 to 2\,nm, with well-defined peaks at
regular spacing. Neither Ag, nor Pd or Rh showed these properties.
Fig.~\ref{f.PtandPd-lhisto} shows a comparison of the length
histograms measured for Pt and Pd. Although the criteria taken as
the start and end of a chain seem to be a bit arbitrary, the
results were not very sensitive to variations in the definition of
these values. The experiment demonstrates a perfect correspondence
between the metals that have surface reconstructions and those
that form chains, in agreement with a common mechanism of
relativistic shifts in the $sp$ and $d$ bonding, described above.

It still needs to be explained why a single bond in a linear chain
wins out against many bulk bonds when a chain is being formed by
low-temperature pulling. It is typical for metallic bonding that
the bond strength increases as the coordination number is
decreased. Bahn and Jacobsen \cite{bahn01} calculated for the
series of six transition metals the force required to break a bond
in a linear chain relative to the force to break a bulk bond,
using density functional theory. They discovered that this ratio
is larger for Au and Pt than for any of the other metals in the
study, and attains a value of about 3 for Au and 3.2 for Pt. Thus
a single linear bond is able to compete with three bulk bonds in
these metals. When the structure of the banks is such that the
atoms can roll-over to break only one or two bonds at a time, then
the chain atoms are bound to prevail and pull new atoms into the
linear arrangement.

\subsubsection{Odd-even behavior in the conductance of atomic chains}

Lang reported calculations that suggest that the conductance of an
atomic chain of simple monovalent metals, where he used sodium as
a model system, depends in a non-monotonous and even oscillatory
way on the length of the chain \cite{lang97a}. His model consists
of two semi-infinite jellium electrodes connected by a chain of
$N$ sodium atoms. For $N=1$, a single atom contact he reported a
surprisingly small conductance of only 0.39\,\g0. The conductance
was then seen to increase to 0.79\,\g0 for $N=2$ and then
oscillating from 0.66 at $N=3$ to 0.75\,\g0 at $N=4$. Such
odd-even oscillations in the conductance of an atomic chain had
already been reported from tight binding calculations by Pernas
\ea \cite{pernas90} and more recently several other workers
reported odd-even oscillations in model calculations
\cite{lang98,lang00,sim01,emberly99,gutierrez01}. Based on earlier
work by Datta \ea \cite{datta97}, Sim \ea \cite{sim01} show, using
the Friedel sum rule, that a perfect conductance of 1\,\g0 is
expected for an odd number of monovalent atoms in the chain, when
the system obeys inversion symmetry and charge neutrality. For an
even number of atoms the conductance is lower, depending on the
structure of the banks on either side of the chain. The effect is
corroborated by a DFT calculation of a chain of Na atoms between
two Na-atom tips.

There appears to be a contradiction with the results obtained
earlier by Lang \cite{lang97a}, since the more recent works find
the minima in the conductance for the even-atom chains. However,
Guti{\'e}rrez \ea \cite{gutierrez01} repeated calculations by the
same method as Lang, for Na chain in between jellium banks, and
showed that the proper odd-even behavior is obtained when changing
the coupling-distance between the chain and the banks. For a
slightly larger distance than used by Lang the odd chains,
including the single-atom junction, obtain a conductance of 1\,\g0
and the even-atom chains have a lower conductance.

The reason for the anomalous dependence observed by Lang has been
attributed to the fact that he effectively considers a
heterogeneous system. Kobayashi \ea \cite{kobayashi00} showed that
the conductance for a chain of 3 Na atoms coincides with \g0,
provided that the jellium electrodes are chosen to have an
electron density that matches that for sodium, and a small atomic
pyramid structure is inserted to connect the chain with the
jellium electrodes. Both features bring the model closer to the
experimental conditions, and are important to avoid reflection of
the electron wave function at the interfaces. Moreover, a mismatch
between the charge density of the Na atoms and the jellium
electrodes results in charge transfer, which moves the Fermi
energy away from the central position in the DOS resonance of the
wire.

In addition, Emberly and Kirczenow \cite{emberly99} showed that
the effective wire length may be greater than suggested by the
atomic structure, because a heterogeneous system, having a chain
composed of a different atomic element than the banks, effectively
gives a longer wire. They show that the oscillations can be
interpreted in terms of a standing wave pattern that develops
between the two ends of the chain, due to the finite reflection
amplitude of the waves at the connections. When the chain has
strictly one conduction channel it reduces the problem effectively
to a one-dimensional system. In this way, the transmission of
conduction electrons through an atomic chain is a problem
equivalent to that of the Fabry-Perot interferometer in optics.

The odd-even oscillations are not restricted to monovalent metals.
As can be judged from the calculations on carbon atom chains
\cite{lang98,emberly99,lang00,palacios02,larade01} the amplitude
can be a lot stronger, going from nearly 1\,\g0 to about twice
this value between even and odd chains, respectively. The actual
periodicity and amplitude, however, is expected to depend
sensitively on the type of atoms of the chain.

Atomic chains can be regarded as unusual atomic configurations
belonging to a larger class of recently discussed ``weird wire''
structures \cite{gulseren98,tosatti01} that will be discussed in
Sect.~\ref{ss.geometric} below.

\section{Shell-filling effects in metallic nanowires} \label{s.shells}

Apart from their role in determining the electronic transport
properties of the contacts, the quantization of the conductance
modes also affects the cohesion energy of the contact. In
Sect.\,\ref{ss.force_models} we discussed the fluctuations in the
tensile force on a nanowire as a result of the successive
occupation of individual quantum modes. While the relevance of
this free-electron model approach to the actual cohesion of
metallic atomic-scale wires is still being debated, there is a
related quantum effect on the cohesion for which already clear
experimental evidence has been obtained. This can be seen for
alkali metal contacts at contact diameters well above those for
which the quantization peaks in the histograms are observed (see
e.g. Fig.\,\ref{f.histK}), and for temperatures well above the
temperature of liquid helium. During the stretching of the contact
under these conditions a nanowire is formed, which shows a
periodic structure of stable diameters. The periodic structure is
closely related to the well-known magic number series for metal
clusters, and in the next section we will briefly summarize these
earlier results from the field of cluster physics. The initial
series of magic numbers arise as a result of fermionic
shell-closing effects similar to those leading to the periodic
table of the elements and the stable atomic nuclei, and the theory
for electronic shell effects in clusters will be briefly
summarized. In Sect.\,\ref{ss.theory-shell} the theory will be
extended for application to nanowires. The subsequent sections
then present the experimental evidence for electronic shell
effects and supershell effects in metallic nanowires. At still
larger diameters of the clusters a new set of magic numbers is
observed, which derives from geometric shell closing of atomic
layers at the surface. This will also be included in the next
section, and the analogous atomic shell structure in nanowires is
presented in Sect.\,\ref{ss.geometric}. For a more complete
discussion of shell closing effects in metallic clusters we refer
to excellent recent review papers on this subject
\cite{deheer93,brack93,martin96,deheer00}.

\subsection{Introduction: shell effects in metallic clusters} \label{ss.clusters}
\begin{figure}[!t]
\centerline{\includegraphics[width=7cm]{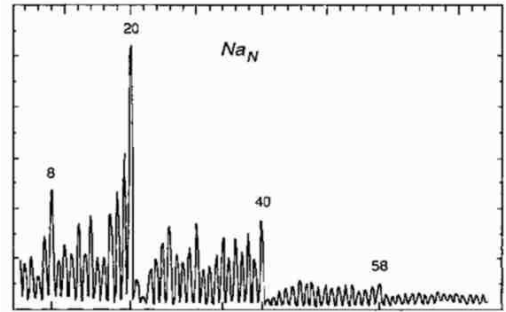}}
\caption{Cluster abundance spectrum for Na. Reprinted with
permission from \protect\cite{knight84}. \copyright 1984 American
Physical Society.} \label{f.cluster-exp}
\end{figure}

When a hot metal vapor is allowed to expand through a nozzle into
vacuum in the presence of an inert carrier gas the metal atoms
fuse together to form clusters of various sizes. The size
distribution for these clusters can be measured using a mass
spectrometer. In a seminal experiment Knight \ea \cite{knight84}
discovered that this distribution for sodium is not a smooth
bell-type shape, as one might naively expect, but rather shows a
series of so-called magic numbers. Those clusters that are
composed of a number of atoms corresponding to one of these magic
numbers are found to be exceptionally abundant.
Fig.\,\ref{f.cluster-exp} shows the experimental abundance
spectrum for sodium. One recognizes peaks or discontinuities in
the abundance spectrum at 8, 20, 40, 58, 70 and 92 atoms. The
explanation for these numbers is based on a shockingly simple
model. The clusters are regarded as spherically symmetric
potential wells containing a number of free and independent
electrons equal to the number of atoms forming the cluster (since
we consider monovalent metals). For hard-wall boundary conditions
the wave functions for the electrons are given by the product of
the spherical harmonics $Y_{lm}(\theta ,\phi)$, and the spherical
Bessel functions, $j_{l}(\gamma_{nl}r/r_0)$, with $r_{0}$ the
radius of the cluster. The ground state of the system is
determined by the filling of the energy levels starting at the
bottom. The levels are given by $E_{nlm}=\hbar^2 \gamma_{nl}^2/2
m_e r_0^2$, where $m_e$ is the electron mass and $\gamma_{nl}$ is
the $n^{\rm th}$ zero of the $l^{\rm th}$ order Bessel function:
$j_{l}(\gamma_{nl})=0$. The lowest energy level corresponds to the
first zero of the $l=0$ Bessel function, the only possible value
for the quantum number $m$ is 0, and including spin degeneracy two
electrons can be accommodated in the lowest level. The next level
is three-fold degenerate ($m=0, \pm 1$) so that it is completed
with a total number of 8 electrons in the cluster. For the
monovalent metal considered in the experiment, this implies that
also 8 atoms are required to allow filling the electronic shell.
The next shell of electrons is characterized by the quantum number
$l=2$, which is five-fold degenerate and would be complete with a
total of 18 electrons in the cluster. However, there is only a
small distance to the next electron level that corresponds to the
second zero of the zeroth order Bessel function, $(n,l)=(2,0)$.
Since the latter is not $m$-degenerate it takes only two electrons
to fill it, and the completion of the shell is observed at 20
electrons (Fig.\,\ref{f.cluster-exp}), and so forth. The
difference in the total energy of the cluster makes a large jump
each time a degenerate level is completed, and the energy gap to
the next level needs to be bridged. This explains the stability of
the filled-shell clusters, just as the stability of the noble
gasses is explained by the filling of the electronic shells for
the elements in the periodic table.

Thus the dominant structure in the experimental abundance spectrum
can be understood by considering the clusters as smooth spheres
filled with free and independent electrons. This interpretation in
its simplest form works best for the alkali metals, for which the
quasiparticle spectrum in the bulk metals is known to be
well-approximated by a free and independent electron gas.
Moreover, the outer $s$-electron of the alkali metal atoms is
weakly bound to the atom core and orbits at a rather large
distance around the nucleus. This leads to a smoothening of atomic
surface corrugation, to such an extent that the electron
distribution around a di-atomic molecule is not a dumbbell, as one
would naively expect, but is rather more closely approximated by a
spherical electron cloud.

Evidently, an explanation of cluster properties such as ionization
energies, static polarizability, optical properties, etc. requires
refinements of the model. To begin with, the walls confining the
electrons are not infinitely hard potentials, but smooth
potentials that result from self-consistent evaluation of the
interaction with the ion cores and with all other electrons in the
cluster. Various groups have performed such calculations, and
although the electronic levels are shifted, the principle
described above largely survives \cite{deheer93,brack93}.
Furthermore, ellipsoidal deformations of the clusters have to be
taken into account in order to explain structure observed in
surface plasma resonances for non-magic cluster sizes. We will not
discuss these complications here, but refer the interested reader
to the review papers on this subject.

\begin{figure}[!t]
\centerline{\includegraphics[width=7cm]{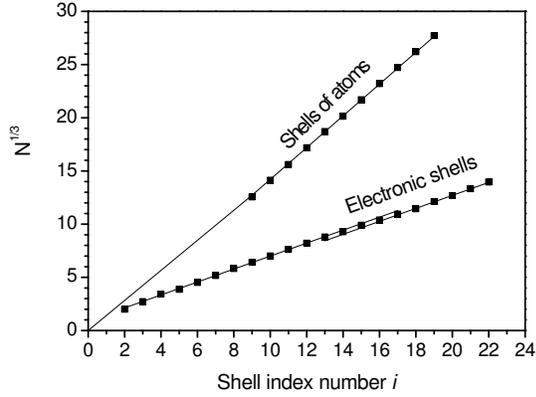}}
\caption{Cube root of the cluster magic numbers obtained from
abundance spectra for Na, plotted against the index of the magic
number. The data set labeled as 'Electronic shells' is taken from
\protect\cite{pedersen91}. The break in the linear dependence
signifies a phase shift in the points at a beat minimum due to the
supershell structure. The data set labeled 'Shells of atoms' is
taken from  \protect\cite{martin91} and is obtained for the
largest clusters in a relatively cold cluster beam.}
\label{f.linear_magic_number_plot}
\end{figure}

When the magic numbers are indicated as $N_i$, with $i$ an index
giving the sequential number of the magic number, e.g. $N_1=2$,
$N_2=8$, $N_3=20$ etc., then a plot of the cube root of the magic
numbers against the index number shows a linear dependence, see
Fig.\,\ref{f.linear_magic_number_plot}. In other words, the magic
numbers form a periodic series as a function of the cube root of
the number of atoms in the cluster, which is proportional to the
cluster radius. The explanation for this periodic appearance of
the magic numbers is given in terms of the bunching of energy
levels, which is conveniently described in a semi-classical
approximation. We will briefly summarize the ideas behind this.

\begin{figure}[!t]
\centerline{\includegraphics[width=7cm]{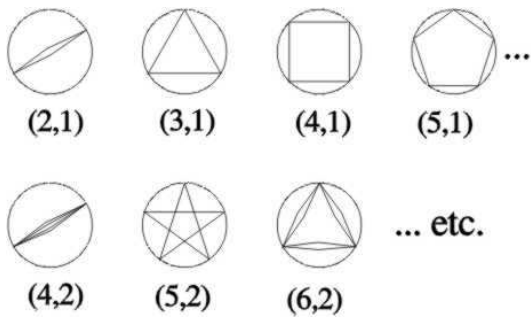}} \caption{The
lowest-order semi-classical trajectories for a particle inside a
steep-walled spherical or cylindrical cavity. The orbits are
labeled by two integers $(p,t)$, where $p$ is the number of
vertices of the orbit and $t$ is the winding number, i.e. the
number of times the particle encircles the center before the orbit
is closed.} \label{f.semi-classical_orbits}
\end{figure}

The individual-level picture described above only works for the
lowest quantum numbers, and even there we observe already that due
to the closeness of the levels $(n,l)=(1,2)$ and $(2,0)$, they are
not separately visible in the experiment. The experiment shows
shell structure up to very large diameters, much further than is
visible in the experiment of Fig.\,\ref{f.cluster-exp}, but for
larger quantum numbers we cannot even approximately identify the
shells with individual quantum numbers. For these large quantum
numbers, i.e. in the limit of large cluster sizes, it is
instructive to approximate the true quantum properties of the
clusters by a semi-classical description. This makes use of the
work on the distribution of the resonant modes in a cavity that
has been discussed in terms of classical trajectories by Guzwiller
\cite{guzwiller71} and by Balian and Bloch \cite{balian72}. In
this approximation the allowed trajectories for the electrons
inside a steep-walled potential well can be identified with the
closed classical trajectories of the electron as it bounces back
and forth between the walls. These semi-classical trajectories can
be constructed from the full quantum orbits for the electrons by
combining an infinite series of them. For a circular system the
lowest order trajectories are given in
Fig.\,\ref{f.semi-classical_orbits}. The allowed wave vectors are
then obtained by requiring the phase around the trajectories to be
single-valued, which gives a quantization condition for each
closed trajectory. It is then immediately obvious that for each
solution at a given radius $R$ of the system a new solution will
be found when the length of the trajectory is increased by a full
electron wavelength. Therefore, the distribution of modes (as a
function of $R$, or as a function of wave vector $k_{\rm F}$, or
more generally as a function of $k_{\rm F}R$) has a periodic
structure. However, we do not have a single period, since each of
the classical trajectories (Fig.\,\ref{f.semi-classical_orbits})
gives rise to its own periodicity. From the Gutzwiller trace
formula the oscillating part of the semi-classical density of
electronic states within a spherical cavity can be expressed as
\cite{brack97}
\begin{eqnarray}
\delta g_{\rm scl}(E) =  {2 m R^2\over \hbar^2} \sqrt{kR \over
\pi}\sum_{p,t}\left[ (-1)^{t}\sin({2\pi t \over p})
\sqrt{\sin({\pi t
\over p})\over p} \right. \nonumber \\
\left. \times\sin(kL_{pt}-{3\pi p \over 2}+{3\pi \over 4})\right]
 -  {2 m R^2\over \hbar^2}\sum_{t=1}^\infty {1\over 2\pi
t}\sin(4tkR). \nonumber
\\
\label{eq.gutzwiller-sphere}
\end{eqnarray}
Here $p$ and $t$ are the number of vertices and the winding number
for the orbits, respectively, as indicated in
Fig.\,\ref{f.semi-classical_orbits}; $R$ is the radius of the
sphere, the wavevector is given by $k=\sqrt{2mE}/\hbar$ and the
lengths of the orbits are given as
\begin{equation}
L_{pt}=2 p R\sin({\pi t\over p}). \label{eq.Lpt}
\end{equation}
The amplitude factors in this expression account for the fact that
the dominant contributions to the level distribution for a
spherical system come from the triangular and square orbits. The
diameter orbit is singled out by the second term in
Eq.\,(\ref{eq.gutzwiller-sphere}) since it has a lower degeneracy:
rotation of the orbit around the vertices does not give rise to a
new orbit. The difference in the periods for the level
distribution belonging to the triangular and square orbits gives
rise to a beating pattern: in some range of $k_{\rm F}R$ the two
orbits will give rise to levels that nearly coincide, thus
amplifying the shell structure in this range. On the other hand at
other ranges of $k_{\rm F}R$ the levels of the square orbits will
fall in between the levels of the triangular orbit, thus
smoothening out the shell effects. This beating pattern in the
shell structure is known as supershell structure, and it has been
observed for Na, Li and Ga metallic clusters
\cite{pedersen91,brechignac92,pellarin95}. The length of the
triangular orbit is $3\sqrt{3}R$ while that of the square orbit is
$4\sqrt{2}R$. The ratio of these two lengths is 1.09 from which we
see that it requires about 11 periods of the individual
oscillations to complete a beat of the combined pattern. For
metallic clusters one or two beat-minima have been observed.
Nature does not provide enough chemical elements to observe
supershell structure in the periodic table or in the stability
pattern of the nuclei.

For the simple, spherical, independent-electron quantum well
system under consideration one can, of course, obtain all the
quantum levels directly from the exact solutions of the
Schr\"odinger equation. However, one cannot identify the periodic
structure with any particular quantum numbers. The periodic
structure of the levels, in other words the modulation in the
density of levels, is a result of a bunching of individual levels.
A bunch of levels together forms a wave packet that approximates
the motion along the classical trajectories.

At still larger cluster sizes one observes a crossover to a new
series of magic numbers, having a different periodicity, which is
seen in the plot of Fig.\,\ref{f.linear_magic_number_plot} as a
crossover to a new slope in the data points. The new series is a
result of the filling of {\it geometric} shells of atoms, rather
than electronic shells \cite{martin96}. The clusters form regular
facetted crystalline structures, which attain an energy minimum
each time a new layer of atoms filling all facets is completed.
For some metals (e.g. Al and Ca) one observes additional structure
due to the filling of partial atomic shells, corresponding to the
completion of individual facets. The facet-completing sub-shell
structure can in some cases be as strong as the full atomic shell
structure.

\subsection{Theory for electronic shell effects in nanowires}
\label{ss.theory-shell}

\begin{figure}[!t]
\centerline{\includegraphics[width=7cm]{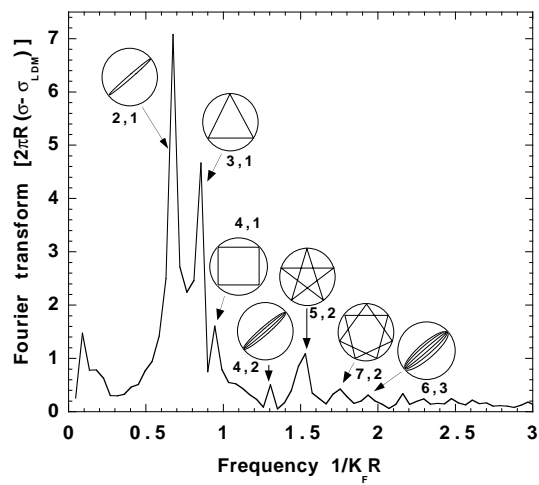}}
\caption{Fourier transform of the oscillating part of the total
energy per unit length for jellium nanowires with a electron
density adjusted to that for Na. The frequencies are given in
units of $(k_{\rm F}R)^{-1}$ and the semi-classical orbits
responsible for the peaks are indicated. } \label{f.PuskaFFT}
\end{figure}

The shell effects discussed previously concerning atomic nuclei,
the elements or metallic clusters, are all based on models of
fermions that are completely confined inside symmetric potential
wells. We now want to apply these concepts to nanowires, for which
one of the three dimensions is open and connected to infinitely
long metallic leads. In this case there are no true gaps in the
density of states, since the conductance modes have a continuous
degree of freedom along the current direction. However, a
modulation of the density of levels is still present. At the
points where a new mode (the bottom of a 1-dimensional sub-band)
crosses the Fermi energy, the density of states shows a
$1/\sqrt{E-E_n}$ singularity. These singularities are smeared out
by the finite length of the wire and would not have a very
pronounced effect if they were homogeneously distributed. However,
by the same mechanism that leads to the shell structure for closed
fermion systems the symmetry of the wire gives rise to a bunching
of the singularities, and these bunches of singularities can be
associated with the electronic shells in metal clusters. The
resulting density of states oscillations have been analyzed using
independent-electron models
\cite{stafford97,ruitenbeek97,kassubek01,kassubek99,hoppler99} and
by local-density functional methods \cite{yannouleas98,puska01}.
In Sect.\,\ref{ss.force_models} these models have already been
discussed in view of quantum contributions to the cohesion of
nanowires. All models reproduce the basic features, namely the
fluctuations in the density of states resulting in local minima in
the free energy for the nanowire. The minima give rise to the
preferential appearance self-selecting magic wire configurations.
As for metallic clusters, one can make a semi-classical expansion
of the density of states and this was done by Yannouleas \ea
\cite{yannouleas98} and in spirit of Balian and Bloch by H\"oppler
and Zwerger and by Kassubek \ea \cite{kassubek01,hoppler99}. The
Gutzwiller trace formula for the density of states of the
perpendicular component of the electron wavefunctions in a
cylindrical cavity can be expressed as \cite{brack97}
\begin{eqnarray}
\delta g_{\rm scl}(E)& = & {2 m R^2\over \hbar^2} \sqrt{1 \over
\pi k R}\sum_{t=1}^{\infty} \sum_{p=2t}^{\infty} f_{pt}
{\sin^{3/2}({\pi t \over p}) \over \sqrt{p}} \nonumber \\
& & \times \sin(kL_{pt}-{3\pi p \over 2}+{3\pi \over 4}) .
\label{eq.gutzwiller-cylinder}
\end{eqnarray}
Here, the factor $f_{pt}=1$ for $p=2t$ and $f_{pt}=2$ for $p>2t$.
The stable wire diameters predicted from this model are in fairly
good agreement with the observed periodic peak structure in the
histograms \cite{yanson99}, as we will discuss below. The periodic
pattern has again a supershell modulation due to the contribution
of several types of orbits. The types of orbits are the same as
for spherical clusters, Fig.\,\ref{f.semi-classical_orbits}, but
the relative weight for each is different. The diametric orbit
$(2,1)$ has only a minor contribution to the structure in the
density of states for clusters, as pointed out below
Eq.\,(\ref{eq.gutzwiller-sphere}), while it is of comparable
weight as the triangular and square orbits for a cylindrical
system. The result is a dominant supershell structure due to the
beating of the period corresponding to the diametric orbit, with
the combined period of the triangular and square orbits. The
length of the diametric orbit is 4R and the ratio to the mean of
the other two is 1.36. This implies that the beating pattern is
only about 3 periods of the principal oscillation wide, and as a
result the supershell structure is more readily observable. This
can be verified by a straightforward calculation of the density of
states for a free electron gas inside a hard-wall cylindrical wire
\cite{yanson01} which has the same dominant features as a more
complete DFT calculation for a jellium-model cylindrical nanowire
\cite{puska01}, for which a Fourier transform of the oscillating
part of the total energy is shown in Fig.\,\ref{f.PuskaFFT}. The
first peak at 0.65 can be identified with the diametric orbit in
the semi-classical analysis, while the triangular and square
orbits give rise to the peaks at 0.83 and 0.90. Indeed, $L_{pt}$
given in (\ref{eq.Lpt}) is the length of a particular orbit. For
the diametric orbit we then have $L_{2,1}/\lambda_{\rm
F}=4R/\lambda_{\rm F}=(2/\pi)(k_{\rm F}R)\simeq 0.637(k_{\rm
F}R)$. This is just the frequency of the first peak in the Fourier
spectrum. Similarly, we obtain for the triangular orbit
$L_{3,1}/\lambda_{\rm F}=3\sqrt{3}R/\lambda_{\rm F}\simeq
0.827(k_{\rm F}R)$ and for the square orbit $L_{4,1}/\lambda_{\rm
F}=4\sqrt{2}R/\lambda_{\rm F}=\simeq 0.900(k_{\rm F}R)$. The peaks
above 1 are attributed to orbits with higher winding numbers.

\subsection{Observation of electronic shell effects in nanowires}
\label{ss.electronic-shell} 
\begin{figure}[!t]
\centerline{\includegraphics[width=8cm]{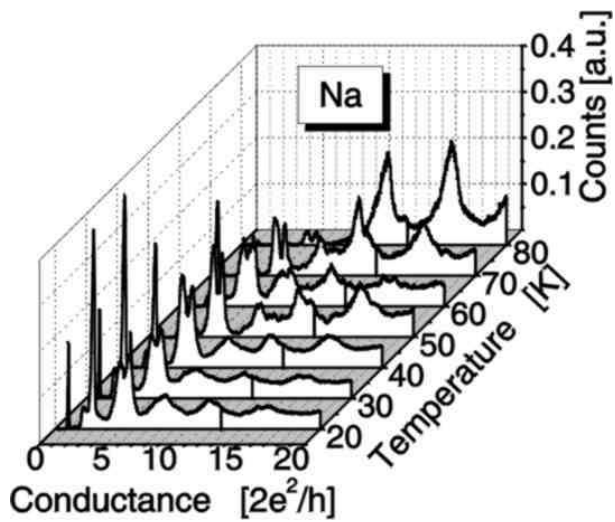}}
\caption{Temperature evolution of sodium histograms in the range
from 0 to 20\,$G_0$. The voltage bias was 10\,mV and each
histogram is constructed from 1000-2000 individual scans. The
amplitude has been normalized by the total area under each
histogram. Reprinted with permission from Nature
\protect\cite{yanson99}. \copyright 1999 Macmillan Publishers Ltd.
} \label{f.T-dependence}
\end{figure}

\begin{figure}[!b]
\centerline{\includegraphics[width=7cm]{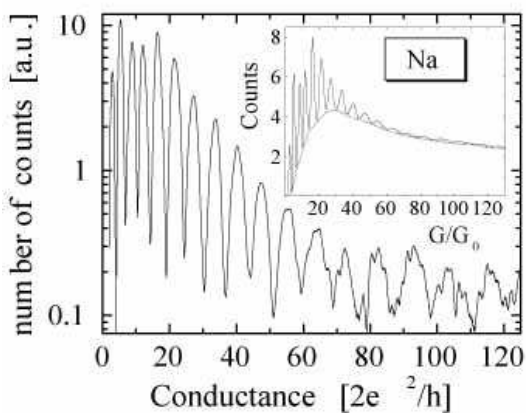}}
\caption{Histogram of the number of times each conductance is
observed versus the conductance in units of \g0 for sodium at
T=80\,K and at a bias voltage of $V$=100\,mV, constructed from
over 10\,000 individual scans. The logarithmic scale for ordinate
axis helps to display the smaller amplitude features at high
conductance values. The inset shows the raw data and the smooth
background (dashed curve), which is subtracted in the main graph.
Reprinted with permission from Nature \protect\cite{yanson99}.
\copyright 1999 Macmillan Publishers Ltd.  }
\label{f.Histogram80K}
\end{figure}

\begin{figure}[!b]
\centerline{\includegraphics[width=7cm]{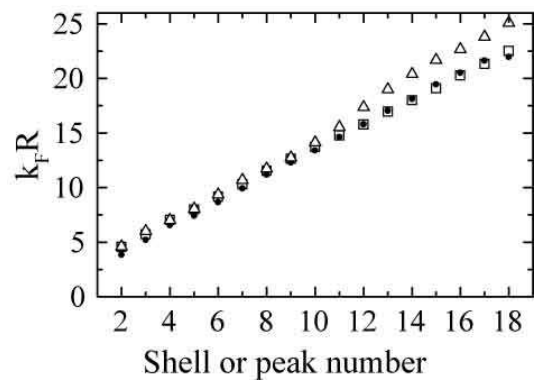}}
\caption{Radius of the nanowire at the positions of the maxima in
Fig.~\protect\ref{f.Histogram80K} versus peak number ($\Box$),
where $R$ is given in units $k_F^{-1}$. The radii at the peak
positions are compared to the radii corresponding to the magic
numbers for sodium metal clusters ($\bullet$)
\protect\cite{martin91a} and to those expected from a
semi-classical description for the fluctuations in the free energy
for the nanowire ($\triangle$)
\protect\cite{yannouleas98,hoppler99}.} \label{f.MWDplot}
\end{figure}

The shell effects nanowires have been
observed most clearly for the alkali metals Li, Na, K, and Cs
\cite{yanson99,yanson01}. The experiments were performed using the
MCBJ technique described in Sect.\,\ref{sss.special MCBJ
techniques}. For atomic-sized contacts at low temperatures the
conductance histograms for the alkali metals show the
characteristic series of peaks at 1, 3, 5, and 6 times \g0. As was
discussed in Sect.\,\ref{sss.alkali}, this observation strongly
suggests that one forms nanowires of nearly-perfect
circularly-symmetric shape while stretching the contact, and the
conduction electrons behave nearly free-electron like. When the
experiment is extended to higher conductance one observes a number
of rather broad but reproducible peaks. The peaks are not as sharp
as the ones associated with conductance quantization at low
conductance (Fig.\,\ref{f.histK}) and cannot be identified with
multiples of the conductance quantum. However, these peaks become
more pronounced as the temperature is raised to about 80~K
(Fig.\,\ref{f.T-dependence}) and more of them become visible at
still higher conductance. At 80\,K up 17 peaks can be seen for
sodium in Fig.\,\ref{f.Histogram80K}. The separation between the
peaks grows for increasing conductance, but a regular periodic
peak structure is obtained when plotting the spectrum as a
function of the square root of the conductance. The square root of
the conductance is to a good approximation proportional to the
radius of the nanowire, see Eq.~(\ref{eq.sharvin}). A
semi-classical relation between the conductance and the radius
taking higher-order terms into account is given by
\cite{torres94,hoppler98},
\begin{equation}
G=gG_0\simeq G_0 \bigl[ \bigl( {k_{\rm F}R\over 2} \bigr)^2 -
{k_{\rm F}R\over 2} + {1 \over 6} + \cdots \bigr]
\label{eq:g-expansion} ,
\end{equation}
as we have seen in Sect\,\ref{sss.sharvin}.
Using the radius of the nanowire obtained from the conductance
through Eq.\,(\ref{eq:g-expansion}) we plot the values of $ k_{\rm
F}R $ at the positions of the peaks in the histogram versus the
index of the peaks, as shown in Fig.\,\ref{f.MWDplot}. The linear
dependence illustrates the fact that the peaks in the histogram
are found at regular intervals in the radius of the nanowire and
this periodicity suggests an analogy with the shell structure
observed for clusters. But it is more than an analogy: when we
take the cluster magic numbers for the electronic shell structure
from Fig.\,\ref{f.linear_magic_number_plot} and convert the number
of atoms $N_i$ into the cluster radius through
$k_FR_i=1.92N_i^{1/3}$, appropriate for sodium, we obtain the
black dots in Fig.\,\ref{f.MWDplot}. The close agreement between
the two sets of data suggests an intimate relation between the
cluster magic numbers and the peaks in the conductance histogram.
Applying the semi-classical theory outlined in
Sect.\,\ref{ss.theory-shell} for shell structure in nanowires
produces the open triangles in the figure. The agreement with the
experimental points is quite good, especially in the beginning.
The deviation for larger cluster sizes requires further study, but
it may be due to approximations in the calculation, where hard
wall boundaries have been assumed, and to ignoring contributions
from electron trajectories beyond the lowest three.

The shell structure only clearly appears when the temperature is
raised well above liquid helium temperature, see
Fig.\,\ref{f.T-dependence}. This finds a natural interpretation
within the framework of the shell effect. The shell structure
makes the free energy for specific wire configurations favored
above the average. When a contact is stretched at very low
temperatures, the shape and size of the contact evolves along a
local energy minimum trajectory. Only when the system is exposed
to large enough thermal fluctuations it can explore the various
contact configurations in order to find the deeper minimum at the
magic wire configurations. Also for metallic clusters temperature
has a significant influence on the degree to which shell structure
is expressed in the experiments \cite{deheer93}.

The interpretation of the periodic peak structure in the
conductance histograms is based on the relation $G \propto R^2$,
Eq.~(\ref{eq:g-expansion}), in order to extract the radius of the
wire. Modifications of this relation are expected to arise from
back scattering on defects (Sect.\,\ref{s.defect_scattering}).
However, the data suggest that this shift of the conductance
values is small otherwise the peaks would be smeared over a much
wider range. An important series resistance correction would also
lead to a deviation from the linear dependence in
Fig.\,\ref{f.MWDplot}. The scattering on defects is probably
suppressed also by the higher temperatures. As pointed out in
Sect.\,\ref{s.defect_scattering}, surface roughness is expected to
be one of the major sources for electron scattering in the
contacts. A higher temperature in the experiment promotes a
smoother surface and a gradual variation of the contact
cross-section along the wire length.

\begin{figure}[!t]
\centerline{\includegraphics[width=6cm]{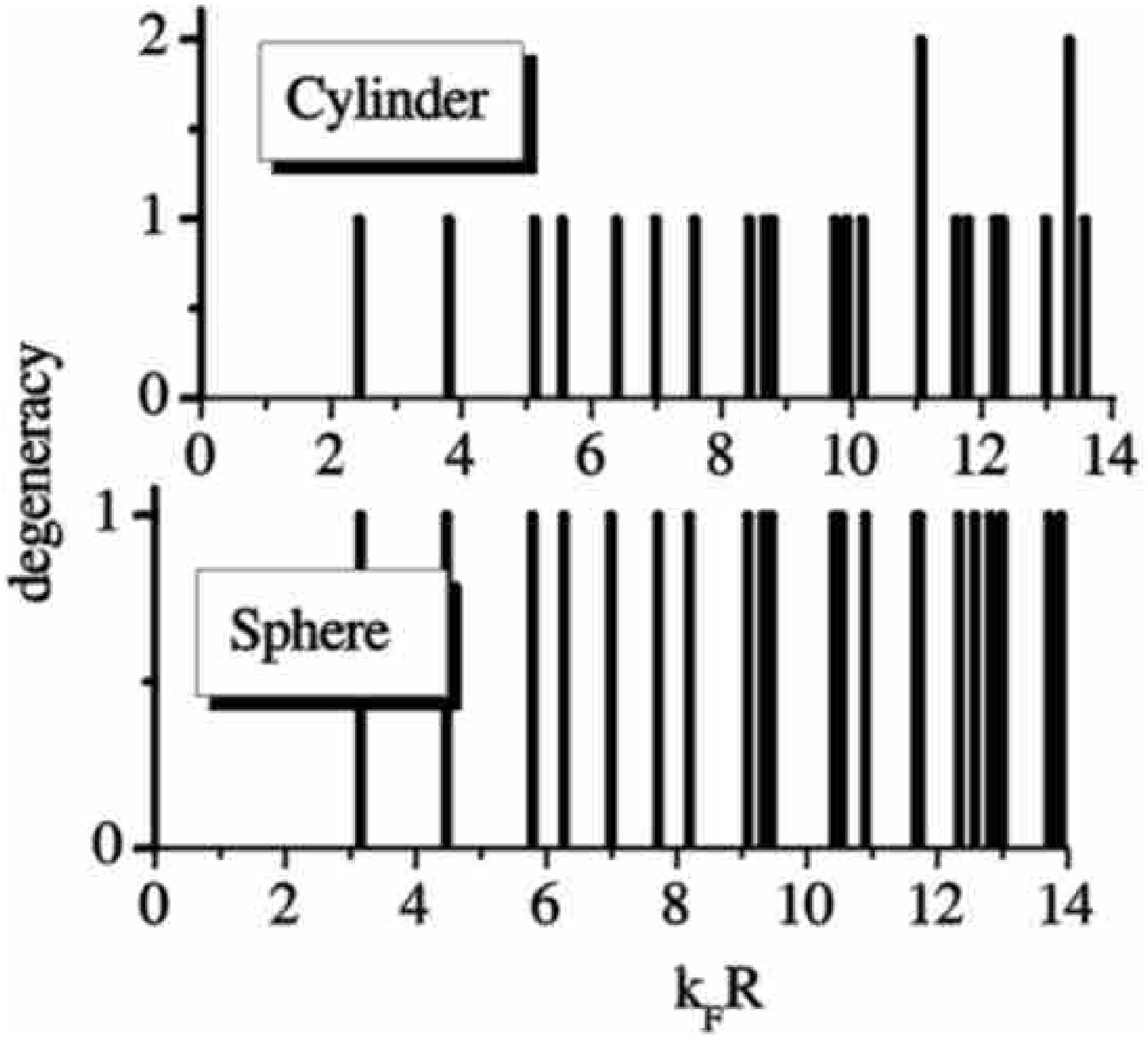}}
\caption{Distribution of zeros for the Bessel functions (giving
the solutions for a cylindrical nanowire) and the spherical Bessel
functions (giving the solutions for spherical clusters). After
application of a small shift it can be seen the the two spectra
have a very similar structure.} \label{f.Bessel_spectra}
\end{figure}

The use of Eq.\,(\ref{eq:g-expansion}) further ignores all quantum
modulation, which is clearly not entirely justified. Indeed, for
low conductance this is just what is believed to be responsible
for the peaks at 1, 3, 5, and 6 multiples of the conductance
quantum. However, this modulation rapidly smears out due to
tunneling contributions of additional conductance channels as the
cross section of the contact grows. Any residual modulation would
lead to a peak structure of the same nature as the shell structure
observed here, and it would have the same origin. Indeed, this
effect has been seen in simulations of conductance histograms that
take quantum effects in the electronic states into account for a
free-electron gas model \cite{burki01}. However, the temperature
dependence observed in the experiment clearly demonstrates that
the shell structure is mainly a result of the enhanced stability
at the magic wire configurations. The stability was included in
the calculations in an approximate way by Ksuubek \ea
\cite{kassubek01} and the simulated conductance histogram
reproduces the experimental data surprisingly well.

The similarity between the magic wire radii for clusters and
nanowires observed in Fig.\,\ref{f.MWDplot} is striking, but it is
also surprising in the sense that one would at first sight expect
a very different series of magic numbers for a spherical system
compared to a cylindrical one. The reason for the correspondence
is in the similarity of the distribution of zeros for Bessel
function and spherical Bessel function. As illustrated in
Fig.\,\ref{f.Bessel_spectra} the bunching of the zeros and the
gaps in the distributions of zeros are very similar, apart from a
small shift.

The electronic shell structure has so far only been observed for
the alkali metals Li, Na, K and Cs. From shell effects observed
for metal clusters one may infer that shell effects in nanowires
may be observable in other metals as well. In view of the much
higher melting point for some of the favorable candidates (Au, Al,
...) observation of such effects will probably require experiments
under UHV conditions at elevated temperatures.

\subsubsection{Supershell effects} \label{sss.supershell}

\begin{figure}[!t]
\centerline{\includegraphics[width=8cm]{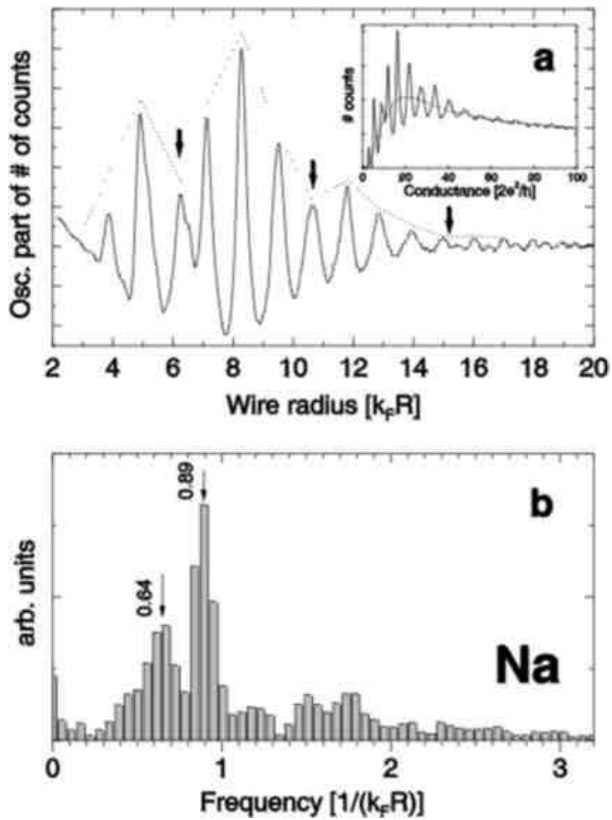}}
\caption{(a) Conductance histogram for sodium after subtraction of
a smooth background. $T=90$\,K, bias voltage $V=100$\,mV, and the
number of curves included in the histogram is 13\,800. The
original curve and the background are shown in the inset. The
arrows indicate the positions of minimal amplitude, which are the
nodes of the supershell structure. When taking the Fourier
transform of the curve in (a) the spectrum in panel (b) is
obtained, which shows two dominant frequency components. Data
taken from \protect\cite{yanson01,yanson00a}.}
\label{f.NaSupershells}
\end{figure}

On top of the shell structure a pronounced modulation due to
supershell structure can be observed for metallic nanowires.
Fig.\,\ref{f.NaSupershells} shows a conductance histogram for Na,
similar to the one in Fig.\,\ref{f.Histogram80K}, for which the
periods in the structure are made visible by taking a Fourier
transform. Similar results have been obtained for Li and K
\cite{yanson01}. The modulation in the histogram peak amplitudes
already suggests a beating pattern of two dominant frequencies,
which is confirmed by the Fourier spectrum.  The two peaks in the
spectrum correspond very well to the ones in
Fig.\,\ref{f.PuskaFFT} for the free energy in a cylindrical wire.
We conclude that the first peak is due to the diametric orbit,
which is prominent in the spectrum in contrast to that for
metallic clusters. The second peak is due to a combination of the
triangular and the square orbits, which are not resolved here. The
larger separation of the period of the diametric orbit from that
of the triangular/square orbits results in a shorter beat-period
compared to the situation for clusters, for which the beat is due
to interference of the square with the triangular orbits. As a
result, for the nanowires the two main periods can be directly
resolved in the Fourier transform. The small peak at 1.5 possibly
results from the $(5,2)$ orbit, as was suggested by Puska \ea
\cite{puska01}. The simulations of the latter work reproduce the
supershell features, not only as regards the periods involved, but
also the phase, i.e. the sizes of the nanowires that produce a
minimum in the beating pattern.

\subsection{Geometric shell effects} \label{ss.geometric}

There is evidence for geometric atomic shell structure for the
alkali metals and, of a very different nature, for gold. The
results for the alkali metals extend the results discussed above
and it is convenient to present this work first.

Beyond a certain diameter a new periodic structure was encountered
in the conductance histograms for Na, K and Cs, which points at a
transition from electronic shell effect to atomic shell structure
\cite{yanson01,yanson01a}, in close analogy to what has been
observed for the clusters (Fig\,\ref{f.linear_magic_number_plot}).
The position of the transition region between the two periodic
features depends strongly on experimental parameters such as the
sample temperature, the voltage bias and the depth of indentation.
The transition shifts to smaller diameters when the metal is found
further down the first column of the periodic table of the
elements. The new oscillating phenomenon has approximately a three
times smaller period than that for the electronic shell structure.
The points for the electronic shell effect in sodium nanowires
plotted in Fig.\,\ref{f.MWDplot} have a slope of $0.56\pm0.01$.
Beyond the cross-over (not shown in the figure) the points obey a
linear relation with a much smaller slope of $0.223\pm0.001$. By
varying the experimental conditions one can greatly extend the
conductance range over which the electronic shell structure is
observed, while the slope always agrees with the periodicity
expected for this effect. As an interpretation for the new
periodic structure it was proposed that geometric shell closing
takes over, where the dominant effect comes from a wire atomic
arrangement with only close-packed surfaces. For a bcc stacking a
hexagonal wire with exclusively low-energy [110] facets can be
formed having its axis along [111]. Similar geometrical shell
configurations were obtained in a simplified model calculation by
Jagla and Tosatti \cite{jagla01}, where even the facet-related
structure can be recognized.

\begin{figure}[!t]
\centerline{\includegraphics[width=6cm]{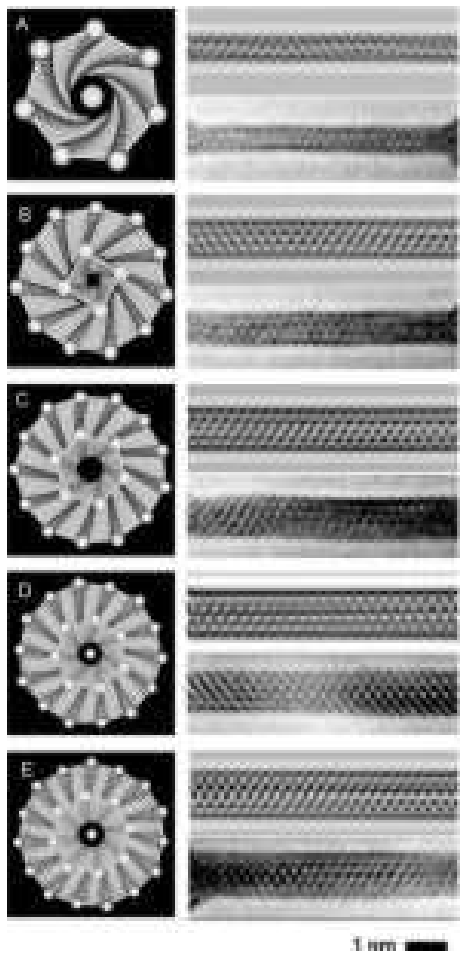}}
\caption{Panels A--E give an axial view of the most prominent
helical shell arrangements proposed as an interpretation of the
gold nanowire images obtained in HRTEM. The observed images are
shown for each structure in the lower right panel, while the upper
right panel shows the corresponding simulated images for
comparison. The atomic row numbers for the structures are 7-1 (A),
11-4 (B), 13-6 (C), 14-7-1 (D), and 15-8-1 (E). Reprinted with
permission from Science \protect\cite{kondo00}. \copyright 2000
American Association for the Advancement of Science. }
\label{f.kondo_gold_shells}
\end{figure}

By the MCBJ technique shell effects have only been uncovered for
alkali metals. However, they are expected to be more general and
by a completely different approach geometric shell effects have
been recently discovered for gold. Using the same technique as
used in the discovery of chains of atoms (Sect.\,\ref{s.chains})
Kondo and Takayanagi \cite{kondo00} found helical atomic shells in
their images of gold nanowires by HRTEM at room temperature. By
electron-beam thinning two adjacent holes are formed {\it in situ}
in a thin gold film. As the gold bridge separating the two holes
is narrowed further 3--15\,nm long nanowires are formed having
unusual structure for diameters in the range of 0.6 to 1.5\,nm.
The unusual structures are described as coaxial tubes of atoms,
each tube consisting of linear arrangements of atoms, that are
coiled around the wire axis. It bears some resemblance to the
structure of multiwalled carbon nanotubes. The various diameters
observed in many images were collected in a histogram, which
showed a number of distinct preferred diameters. These peaks in
the histogram are interpreted in terms of a special geometric
shell structure, which is illustrated in
Fig.\,\ref{f.kondo_gold_shells}. The structures are characterized
by the number of atomic rows in each coaxial tube, e.g. as
$n-n'-n''$ for a nanowire having an outer tube with $n$ rows,
inside of which we find a second tube with $n'$ rows and in the
center a tube of $n''$ rows of atoms. The difference in the number
of atomic rows per layer is 7, with the provision that when this
ends in a central tube with 0 rows (next tube has 7) then the
central tube will be a linear chain of atoms, $n''=1$. The magic
wire structures observed then correspond to $7-1, 11-4, 13-6,
14-7-1$ and $15-8-1$ (Fig.\,\ref{f.kondo_gold_shells}) and there
are two less-well resolved structures that may fit to $12-5$ and
$16-9-2$. For larger diameters there is a cross-over to bulk fcc
stacking. Similar nanowire structures have been found for Pt
\cite{oshima02}.

Novel types of atomic packing, including helical arrangements had
already been observed in molecular dynamics simulations for
unsupported wires of Pb and Al by G{\"u}lseren \ea in 1998
\cite{gulseren98}. Structures they refer to as `weird wires' are
found below a characteristic radius $R_{\rm c}$, which is of the
order of thee times the inter-atomic spacing. These structures are
formed in order to minimize the surface energy, and the shape of
the wires depends sensitively on the anisotropy of the surface
energy for the metallic elements considered. Hasmy \ea
\cite{hasmy01} have made simulations of histograms of nanowire
diameters by repeated molecular dynamics simulations for aluminum
\footnote{The use of the word supershell structure in
Ref.~\protect\cite{hasmy01} is not appropriate in this context. In
the molecular dynamics type calculation employed the electronic
density of states is not evaluated, so that effects of electronic
(super)shell structure cannot be observed.}. For higher
temperatures, 450\,K, broader peaks are observed in the simulation
at 5, 8--9, 12--13 and 15 atoms. These are attributed to shell
structure. The nature of the atomic packing at these magic numbers
was not reported.

The seminal work of G{\"u}lseren \ea was extended to gold by
several groups \cite{tosatti01,bilbalbegovic98,wang01}, partly
stimulated by the experimental observations. The simulations
confirm that the structure of the thinnest gold nanowires can be
viewed as consisting of concentric cylindrical sheets of atoms.
Tosatti \ea \cite{tosatti01} have made a series of successive
refinements of the calculations finishing with a full density
functional calculation of the electronic structure and the wire
tension. They point out that the appropriate thermodynamic
function to consider is the wire {\it tension} $f=(F-\mu N)/L$,
rather than the free energy $F$. Here, $\mu$ is the chemical
potential, $N$ the total number of atoms in the wire and $L$ its
length. The chemical potential takes into account the fact that
the wire diameter changes by diffusion of atoms to and fro between
the nanowire and the banks between which it is suspended. They
find for the thinnest structure an optimal arrangement in perfect
agreement with the experimental observation by Kondo and
Takayanagi \cite{kondo00}. They explicitly verify that there is no
contribution of {\it electronic} shell closing for these
structures. The particular type of magic wire structure involves a
mechanism of $s$-$d$ competition that is also believed to be
responsible for the formation of atomic chains
(Sect.\,\ref{s.chains}), and may therefore apply to Au, Pt and Ir.
For diameters larger than about 2.5\,nm the core of the nanowires
is fcc crystalline, with a few layers of non-crystalline
arrangement covering the surface, referred to as curved surface
epitaxy by Wang \ea \cite{wang01}.

It is to be expected that shell effects will be observable for
many other metals, as suggested by the observations for metal
clusters, and new surprising structures may yet be found. The
implications of the chiral atomic ordering in gold nanowires on
observable effects in electron transport properties deserve
further investigation.

\section{Conclusion and Outlook}

Let us summarize what we view as the central achievement of the
research described above, and what we feel deserves to be known:
Metallic atomic-sized contacts can be characterized by a finite
set of conductance eigenchannels with transmission probabilities
$\{\tau_i\}$; this set $\{\tau_i\}$ has been nicknamed `the
mesoscopic PIN code' by Urbina's group in Saclay \cite{cron01b}.
For a point contact of just a single atom in cross section the
number of valence orbitals of the atom fixes the number of
eigenchannels. A single atom can thus carry many channels, up to 5
or 6 for $sd$ metals, and they are generally only partially open.
By measuring the current-voltage characteristics in the
superconducting state we can obtain the PIN code experimentally.
When the metal is not a superconductor this is somewhat less
straight forward, but making use of the proximity effect it can
still be done. The knowledge of the PIN code allows predicting
other transport properties such as shot noise, the Josephson
supercurrent or the dynamical Coulomb blockade effect, that can
all be expressed in terms of the set $\{\tau_i\}$. Atomic-sized
contacts have thus become an ideal test bench for mesoscopic and
nanophysics, where a high degree of accuracy can be reached. This
accuracy is remarkable for a system whose geometrical
characteristics cannot be completely controlled.

There are still many open questions to be answered. Other
properties, such as the mean amplitude of conductance
fluctuations, are also expected to be described by an average over
sets $\{\tau_i\}$, but this has not yet been verified. Similarly,
the experiment by Koops \ea (Sect.\,\ref{ss.cpr}) on the current
phase relation of an atomic contact in a superconducting loop
deserves to be taken one step further by first establishing the
PIN code of the atomic contact. This problem has attracted
interest from a different perspective since the two Andreev bound
states for a single-channel contact inside a superconducting loop
have been proposed as the basic qubit element for applications in
quantum computing \cite{desposito01,lantz02}. The question as to
the limits in accuracy of the PIN code determination has not been
addressed very generally. How many channels can be determined
unambiguously? An experiment is under way (G. Rubio-Bollinger,
private communication) to attempt to determine the distribution of
channel transmissions for large contacts, having thirty or more
channels. Is this procedure reliable and does the distribution
obtained correctly predict other properties, such as shot noise,
for the same contact?

At the start of this field the question of whether or not the
concept of conductance quantization applies for atomic-sized
contacts has played a central role. Although there are many ways
of interpreting this concept, it has become clear that it makes no
sense to speak about conductance quantization for metals other
than the monovalent metals. And even for those it requires some
specification of the concept when discussing the data in such
terms. What we have learned is that for a one-atom contact the
monovalent metals have a single conductance channel that is nearly
perfectly transmitted, and for larger contacts (up to three to six
atoms) the next channels show a tendency to open one after the
other. This phenomenon has been referred to as the saturation of
channel transmission. For still larger contacts several partially
open channels are active simultaneously.

We can see many directions in which this field of research may be
further developed in the near future. A prominent problem to be
attacked is the investigation, both from theory and experiment, of
many-body interactions in atomic-sized contacts. These are not
included in the conventional scattering approach and new
descriptions are required. One type of interaction is the
electron-phonon scattering. Some level of perturbative description
of the interaction has been given (Sect.\,\ref{sss.phonons}), but
a more detailed understanding of electron-phonon interactions in
atomic-sized conductors is still lacking. The phonons in this
description should be replaced by the local vibration modes of the
atomic structure, coupled to the continuum modes of the banks.
Experimentally, only very few results are available, mostly for
Au. Many-body effects may also arise as a result of
electron-electron interactions through the charge or spin degrees
of freedom. Charge effects are believed to be small as a result of
the effective screening provided by the metallic electrodes on
either side of the atomic contact. Still, the screening is less
effective than in bulk metals. The spin degrees of freedom may
give rise to Kondo-like physics. Experimentalists should keep an
open eye to characteristics that may be indications of these
effect, which are most likely seen as anomalies around zero bias
in the differential conductance. We may need to explore different
metals and we are likely to find new surprises in the uncharted
parts of the periodic table. In particular the lanthanides and
actinides have not received much attention, to date.

New directions in experiment will be found by adding parameters to
be measured or controlled. Force measurements in atomic contacts
have been very successful, but only very few results are
available, exclusively for Au. New experiments are under way,
which combine a tuning-fork piezo resonator, as a sensor of the
force constant, with the advantages of the MCBJ technique. This
will open the way to explore quantum properties in the force, and
its relation to the conduction, in a wide range of materials. A
challenge that has been recognized by many is introducing a gate
electrode to a one-atom contact. It would allow controlling the
transmission of the eigenchannels, and possibly even the number of
eigenchannels, by adjusting an external electrical potential.
Implementing this will require new experimental approaches,
possibly similar to the techniques used in two recent papers to
demonstrate Kondo behavior in a single atom embedded between
electrodes inside a molecular structure \cite{park02,liang02}.

Chains of atoms constitute the ultimate one-dimensional metallic
nanowires. For gold, the current is carried by a single mode, with
nearly unit transmission probability. They may form a source for
many discoveries in the near future. New properties may be
discovered when considering, for example, the thermal conductance
\cite{ozpineci01} or magnetic order, that is predicted for Pt
chains under strain \cite{bahn01}. One may expect that mixed
chains, e.g. Au chains containing Pt impurity atoms, will give
access to interesting Kondo-type physics in a purely
one-dimensional system. More work is needed to elucidate the
mechanism of chain formation, and in particular to understanding
what limits the length of the chains. It is most likely that the
length is currently limited by having sufficient weakly bonded
atoms available. Once the structure in the tips, between which the
chain is suspended, acquires a regular and stable configuration
new atoms cannot be coaxed to join in the formation of a longer
chain, as was illustrated by the simulations by da Silva \ea
\cite{dasilva01}.

If we can develop methods to produce much longer chains
electron-electron interactions may start to dominate, converting
the electron spectrum in the chain into that of a Luttinger liquid
\cite{fisher97}. Evidence for Luttinger-liquid behavior has
already been obtained for carbon nanotubes \cite{bockrath99} and
for chains of gold atoms self-assembled on Si surfaces
\cite{segovia99}. A Peierls transition is predicted for gold
atomic chains when stressed close to the breaking point. No
evidence for this phenomenon has been reported yet, but it is
worthwhile investigating this further.

One of the least understood aspects of atomic-sized contacts is
the role of heating by the electron current and possible forces on
the atoms resulting from the large current density. Joule heating
and electromigration are important problems in integrated circuit
technology and atomic contacts are probably ideal model systems to
study these effects, since the maximum current density is larger
than in any other system. A large current may drive the contacts
over time towards the more stable configurations. This brings us
immediately to the related question whether shell effects exist
for other metals than the alkali. In a limited sense, atomic shell
configurations have already been observed for Au in HRTEM studies,
but the electronic and atomic shell effects are expected to be
found for a wide range of elements. The challenge is to find the
proper experimental conditions that will bring them out.

We want to end this review by pointing at one of the most
promising directions of future research: The study of conductance
through individual organic molecules is part of a world-wide
effort towards building electronic circuits that exploit the
intrinsic functionally of specially designed molecules. The
possibility of having molecules that function as diodes
\cite{metzger98}, electronic mixers \cite{chen-j99} or switching
elements \cite{collier99,gao00,reed01}  has inspired hope of
developing entirely new, molecular based electronics. The first
steps have been made
\cite{smit02,park02,liang02,reed97,kergueris99,reichert02} and the
developments in this field are so rapid that they deserve a
separate review paper, while some reviews have already appeared
\cite{joachim00,nitzan01}. The subject is intimately connected
with that of the present review: the STM and MCBJ are the most
widely employed tools for contacting the molecules. Some of the
fundamental concepts described above for atomic contacts will
apply directly to molecular systems, while on the other hand new
and unexpected physical phenomena will surface.

\subsection*{Acknowledgements}
We have profited from many discussions with Gabino
Rubio-Bollinger, Helko van den Brom, Juan Carlos Cuevas, Michel
Devoret, Daniel Esteve, Fernando Flores, Bas Ludoph, Alvaro
Mart{\'\i}n Rodero, Yves Noat, Elke Scheer, Roel Smit, Carlos
Untiedt, Cristian Urbina, Sebastian Vieira, Alex Yanson and Igor
Yanson. This work has been supported by Spanish CICyT under
contract PB97-0044, the DGI under contract MAT2001-1281 and by the
Dutch `Stichting FOM', that is financially supported by NWO.

\bibliographystyle{review}
\bibliography{QPC_v24}
\end{document}